\documentclass[11pt,a4paper,twoside]{report}
\usepackage[utf8]{inputenc}
\usepackage{amsmath}
\usepackage{amsthm}
\usepackage{amsfonts}
\usepackage{amssymb}
\usepackage{setspace}
\usepackage{graphicx}
\usepackage{fancyhdr}
\usepackage{comment}
\usepackage{bibentry}
\usepackage{wasysym}
\usepackage[export]{adjustbox}
\usepackage{tikz}
\usepackage{graphbox}
\usepackage{bibentry}
\usepackage[toc,page]{appendix}
\usepackage[inner=4cm,outer=2.5cm,top=2.5cm,bottom=2.5cm]{geometry}
\usepackage{caption}
\usepackage{subcaption}

\theoremstyle{plain}

\theoremstyle{definition}
\newtheorem{definition}{Definition}

\makeatletter
\renewcommand{\@chapapp}{}
\newenvironment{chapquote}[2][2em]
  {\setlength{\@tempdima}{#1}%
   \def\chapquote@author{#2}%
   \parshape 1 \@tempdima \dimexpr\textwidth-2\@tempdima\relax%
   \itshape}
  {\par\normalfont\hfill--\ \chapquote@author\hspace*{\@tempdima}\par\bigskip}
\makeatother

\usepackage[colorlinks,bookmarks=true,citecolor=blue,urlcolor=blue]{hyperref}

\usepackage[backend=biber,style=numeric-comp,sorting=none,giveninits=true,url=false]{biblatex}
\begin{document}

\begin{titlepage}
    \begin{center}
            
        \LARGE
        \textbf{Emergent Spacetime in \\ Quantum Lattice Models}
            
		\vspace{2cm}
		
		        \includegraphics[width=0.25\textwidth]{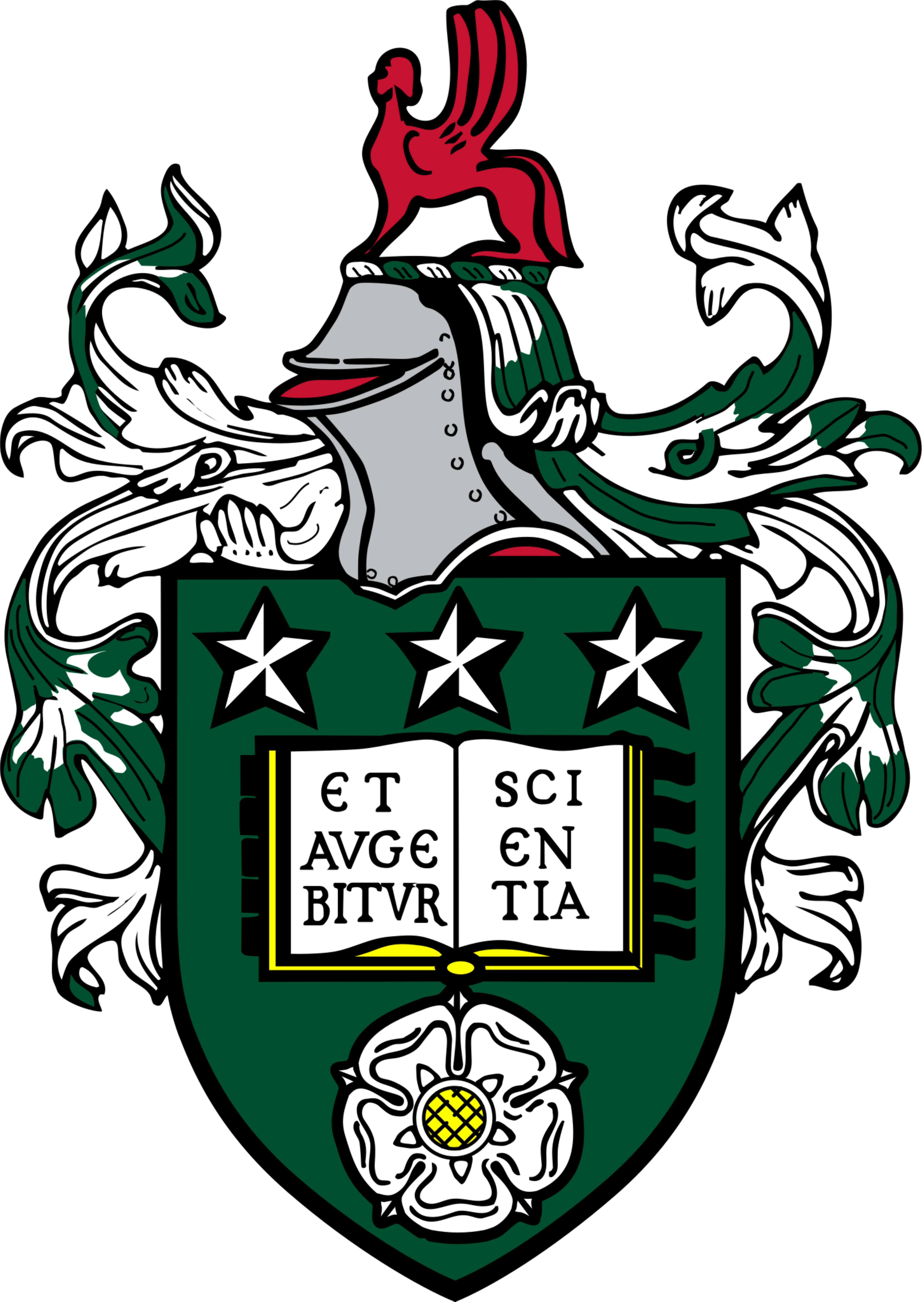}
            
        \vspace{2cm}
        \LARGE
        Matthew Donald Horner \\
        \Large 
        \vspace{0.25cm}
        School of Physics and Astronomy\\
        \vspace{0.25cm}
        The University of Leeds\\
            
        \vfill
            
        Submitted in accordance with the requirements for the degree of\\

\vspace{0.25cm}        
        
        \textit{Doctor of Philosophy}
            
        \vspace{0.8cm}

        September 2022
            
    \end{center}
\end{titlepage}

\newpage
\renewcommand{\thepage}{\Roman{page}}
\
\newpage 

\begin{center}

	\vspace*{5cm}
	\Large
	\textit{To my parents.}

\end{center}

\newpage
\
\newpage

\addcontentsline{toc}{chapter}{Declarations}

\begin{center}
\LARGE
\textbf{Declarations}
\end{center}

The candidate confirms that the work submitted is his own, except where work which has formed part of jointly authored publications has been included. The contribution of the candidate and the other authors to this work has been explicitly indicated below. The candidate confirms that appropriate credit has been given within the thesis where reference has been made to the work of others.

\section*{Chapter summary}
\begin{itemize}
\item Chapter~\ref{chapter:nanotube} includes work from the publication: \begin{refsection} \fullcite{nanotube_paper}\end{refsection}, where the entire paper is directly attributable to me, under general guidance and supervision by Jiannis K. Pachos.

\item Chapter~\ref{chapter:RC_background} contains only background material and no published work.

\item Chapter~\ref{chapter:kitaev} includes work from the publication: \begin{refsection} \fullcite{C2_Farjami}\end{refsection}, where analytic calculations are attributed to me whilst numerical simulations and figures are attributed to Ashk Farjami and Chris Self. This work was conducted under general guidance and supervision by Zlatko Papić and Jiannis K. Pachos. Figure~\ref{fig:C2_kekule} was produced by me and is an updated and expanded version compared to the paper. Section~\ref{sec:spin_densities} contains unpublished work.

\item Chapter~\ref{chapter:chiral} includes work from the publication: \begin{refsection} \fullcite{chiral_paper}\end{refsection}, where analytical calculations are attributed to me whilst numerical simulation is attributed to Ashk Farjami. This work was conducted under general guidance and supervision by Jiannis K. Pachos.

\item Chapter~\ref{chapter:black_hole} includes work from the publication: \begin{refsection} \fullcite{black_hole_paper}\end{refsection}, which has been submitted to Physical Review Letters, where all analytical calculations and all numerical simulation of the mean field (MF) theory are attributed to me, under guidance from Andrew Hallam. Matrix product state (MPS) numerics is attributed to Andrew Hallam. This work was conducted under general supervision and guidance by Jiannis K. Pachos.
\end{itemize}

\vfill
This copy has been supplied on the understanding that it is copyright material and that no quotation from the thesis may be published without proper acknowledgement.

\vspace{0.5cm}

The right of Matthew Donald Horner to be identified as Author of this work has been asserted by Matthew Donald Horner in accordance with the Copyright, Designs and Patents Act 1988.

\newpage
\
\newpage

\addcontentsline{toc}{chapter}{Acknowledgements}
\onehalfspacing 
\begin{center}
\LARGE
\textbf{Acknowledgements}
\end{center}

I would like to begin by thanking my supervisor Professor Jiannis K. Pachos for his wisdom, guidance, support and friendship which has been a constant throughout the four years of my Ph.D., who shaped me into the researcher I am today. He instilled a passion for physics and research in me that I could not have achieved without him. In particular, I am grateful for Jiannis' achievement of converting me from a high-energy theorist with an inherent dislike for numerics and solid state physics, to a condensed matter computational physicist. 

In addition, I am truly in debt with Ashk Farjami who I collaborated with for two of my publications. I thank him for his friendship during the early days of my Ph.D. which helped me endlessly, with highlights including Mario Kart, discussion of Westerosi politics and our trip to Benasque, where he revealed to me that a physics conference is more about running up mountains in the Pyrenees than the physics. I would also like to thank Andrew Hallam for his collaboration on my final publication. This is the piece of work I am most proud of and I would not have achieved this without him. Thanks is also extended to Giandomenico Palumbo for sharing his knowledge and hosting me in Belgium; Omri Golan for our frequent Zoom meetings and teaching me practically everything I needed to know for my Ph.D.; and Patricio Salgado-Rebolledo who has been the go-to person for anything gravitational.

It is not only scientists who I directly worked with that warrant an acknowledgement. I thank Dr.~Robert Purdy, Daniel Hodgson, Jake Southall and Kieran Bull for their friendship both in and out of the office. In addition to scientists, it goes without saying that without my parents I would not have achieved this and I am eternally grateful for them. Their unwavering love and support (and supply of food) was key to getting out the other side in one piece. I also thank Emma for her love, enthusiasm for what I do and reminding me that a Ph.D. need not take up every waking hour. Finally, I thank Chris, Alex, Amy, Tim and the members of 32 Haddon Road for being great friends and giving me wonderful memories of my time in Leeds.

\newpage
\
\newpage

\addcontentsline{toc}{chapter}{Abstract}
\begin{center}
\LARGE
\textbf{Abstract}
\end{center}

Many quantum lattice models have an emergent relativistic description in their continuum limit. The celebrated example is graphene, whose continuum limit is described by the Dirac equation on a Minkowski spacetime. Not only does the continuum limit provide us with a dictionary of geometric observables to describe the models with, but it also allows one to solve models that were otherwise analytically intractable. In this thesis, we investigate novel features of this relativistic description for a range of quantum lattice models. In particular, we demonstrate how to generate emergent curved spacetimes and identify observables at the lattice level which reveal this emergent behaviour, allowing one to simulate relativistic effects in the laboratory. We first study carbon nanotubes, a system with an edge, which allows us to test the interesting feature of the Dirac equation that it allows for bulk states with support on the edges of the system. We then study Kitaev's honeycomb model which has a continuum limit describing Majorana spinors on a Minkowski spacetime. We show how to generate a non-trivial metric in the continuum limit of this model and how to observe the effects of this metric and its corresponding curvature in the lattice observables, such as Majorana correlators, Majorana zero modes and the spin densities. We also discuss how lattice defects and $\mathbb{Z}_2$ gauge fields at the lattice level can generate chiral gauge fields in the continuum limit and we reveal their adiabatic equivalence. Finally, we discuss a chiral modification of the 1D XY model which makes the model interacting and introduces a non-trivial phase diagram. We see that this generates a black hole metric in the continuum limit, where the inside and outside of the black hole are in different phases. We then demonstrate that by quenching this model we can simulate Hawking radiation.

\newpage
\
\newpage

\singlespacing

\begin{center}
\LARGE
\textbf{Abbreviations and conventions}
\end{center}

\section*{Abbreviations}

\begin{tabular}{ r  l  }

B.Z. & Brillouin Zone \\ 
c.c. & Complex Conjugate \\
DMRG & Density Matrix Renormalisation Group \\
GE & Gibbs ensemble \\
GGE & Generalised Gibbs Ensemble \\
H.c. & Hermitian conjugate \\
LDOS & Local Density of States \\
MF & Mean Field \\ 
MPS & Matrix Product State \\  
\end{tabular}

\section*{Conventions}
The Pauli matrices are given by
\begin{equation*}
\sigma^x = 
\begin{pmatrix} 
0 & 1 \\ 
1 & 0 
\end{pmatrix}, \quad \sigma^y = \begin{pmatrix} 0 & -i \\ i & 0 \end{pmatrix}, \quad \sigma^z = \begin{pmatrix} 1 & 0 \\ 0 & -1 \end{pmatrix}.
\end{equation*}
The Einstein summation convention for repeated indices is assumed
\begin{equation*}
a_i b^i \equiv \sum_i a_i b^i.
\end{equation*}
We use the mostly-minus metric signature
\begin{equation*}
\eta_{ab} = \mathrm{diag}(+1,-1,-1,\ldots,-1).
\end{equation*}
We use natural units: $\hbar = c = k_\mathrm{B} = 1$.

\newpage
\
\newpage

{
\hypersetup{linkcolor=black}
  \tableofcontents
  
  \newpage
  \
  \newpage
}

{
  \hypersetup{linkcolor=black}
  \addcontentsline{toc}{chapter}{List of Figures}
  \listoffigures
}

\onehalfspacing

\newpage
\
\newpage

\renewcommand{\thepage}{\arabic{page}}
\setcounter{page}{1}

\chapter{Introduction}
\section{Motivation}
Condensed matter is the study of \textit{emergence}: large collections of particles, which individually follow simple laws of physics, exhibit a wealth of emergent complex phenomena that we exploit in nearly every aspect of our day-to-day lives. An example of which is the humble semi-conductor which has provided us with practically every electronic device we take for granted. On the other hand, more exotic \textit{topological} phases can emerge, such as topological insulators~\cite{Xiao-Liang,Asboth_2016}, topological superconductors~\cite{Xiao-Liang,C1_Bernevig} and the quantum hall effect~\cite{fradkin_2013,Prange} which will likely contribute the technologies of the future and improve quantum computing as we know it.

As a theorist, I cannot deny that condensed matter is beautiful from a theoretical perspective. Condensed matter overlaps heavily with many cornerstones of physics such as high-energy physics, general relativity and quantum information; even going beyond them in some instances as we shall see in this thesis. We will discover particles that are not described by the Standard Model and geometries that must be described by a generalisation of general relativity, as general relativity was \textit{not quite general enough} for us.

Many systems in condensed matter are described by \textit{lattices}, such as the honeycomb lattice of ions in graphene~\cite{Wallace,C1_Neto,C1_Altland}, or the lattice of spins in a magnetic material~\cite{doi:10.1142/2945,C4_DePasquale}. However, these are analytically intractable due to the immense degrees of freedom required to simulate them. For example, a set of $N$ spin-$1/2$ particles has a Hilbert space that grows exponentially as $2^N$, which requires sophisticated numerical algorithms such as matrix product states~\cite{C4_schollwock2011density} to tackle numerically. This is hard numerically, but analytically this would be even harder. For this reason, we resort to continuum limit techniques to solve these. This is where we ``zoom out" and assume the lattice is a \textit{continuum}, mapping the problem to  something that should be easier to solve. This technique is equivalent to focussing just on the low-energy properties of the system. Many systems have been tackled this way, such as graphene~\cite{Wallace,C1_Neto,C1_Altland}, the Ising model~\cite{CFT}, topological superconductors~\cite{C2_Golan,C1_Bernevig}, the SSH model~\cite{Asboth_2016}, and so on. 

A general theme is that the continuum limit is \textit{relativistic}, that is, the equation of motion for the continuum limit is the famous Dirac equation on a particular spacetime, where the emergent spacetime can be either flat or curved, depending on the underlying model.

The motivation for the continuum limit is the following:  
\begin{enumerate}
\item It allows one to arrive at an analytic solution, albeit an approximation, allowing one to calculate observables, critical points, and discover novel features, e.g., edge modes in the SSH model; 
\item It allows one to use our knowledge of high-energy physics and general relativity to provide us with an explanation of many observed phenomena that we would not have at the lattice level and a dictionary of geometric quantities to hunt for;
\item It allows us to devise condensed matter systems in the laboratory that can simulate relativistic phenomena, including black holes.
\end{enumerate}
This thesis will be investigating these three points.
\section{Structure of thesis}
The thesis is structured as follows. 
\subsubsection*{Chapter \ref{chapter:nanotube}}
Graphene, the famous honeycomb lattice of carbon atoms, has a relativistic continuum limit~\cite{Wallace,C1_Neto,C1_Altland}. In this chapter we shall investigate how relativistic effects manifest in graphene. In particular, we shall be interested in the edge effects of a carbon nanotube. Many systems in condensed matter, such as topological insulators~\cite{Xiao-Liang,Asboth_2016,C1_Bernevig}, exhibit localised zero-energy states at their edge which are distinct from states that live in the bulk, whereas bulk states do not have any support on the edge. Here, we investigate the interesting feature of the Dirac equation that its bulk wavefunctions $\psi$ are allowed to be non-zero on the boundaries. As the Dirac equation describes the low-energy properties of graphene, we investigate under what conditions this edge effect applies for zig-zag carbon nanotubes and we also provide an alternative view of how relativistic effects emerge in a non-relativistic model. 

\subsection*{Chapter \ref{chapter:RC_background}}
In this chapter we present relevant background material for chapter~\ref{chapter:kitaev}. We provide an introduction to Kitaev's honeycomb model~\cite{C3_Kitaev}, specifically reviewing how one solves the model and how to take its continuum limit. We then introduce the theory of Riemann-Cartan geometry for spacetimes with both curvature and torsion.

\subsubsection*{Chapter \ref{chapter:kitaev}}
Kitaev's honeycomb lattice model, like graphene, has a relativistic continuum limit described by the Dirac equation~\cite{C3_Kitaev,C3_PachosBook}. In this chapter, we demonstrate that one can generate a continuum limit describing a Majorana spinor propagating on a \textit{curved} Riemann-Cartan spacetime. We identify the couplings of the model as generating a metric and torsion in the continuum limit, allowing one to use the model to simulate Riemann-Cartan spacetimes. We discuss how the metric encodes the phase diagram of the model, in particular, we show that the metric is encoded in Majorana correlation functions and zero-mode wavefunctions, allowing us to detect the presence of the metric at the lattice level; we then show that we can generate curvature in the model by introducing inhomogeneity in the couplings; and finally we identify the spin density as the observable that we can measure to detect the curvature.

\subsubsection*{Chapter \ref{chapter:chiral}}
In this chapter, we propose to build upon chapter~\ref{chapter:kitaev} by considering chirality and chiral gauge fields. Massless fermions in $(3 + 1)$D can be described by spinors which are reducible into a pair of Weyl fermions of opposite chirality. This chirality, either left-handed or right-handed,
signals how these objects transform under Lorentz transformations~\cite{C2_Maggiore}. In Kitaev's honeycomb, the model contains two Fermi points in the dispersion, therefore two branches of excitations in the spectrum. This manifests itself as a pseudo-chirality in the continuum limit, allowing us to apply ideas from high-energy physics. We demonstrate that the $\mathbb{Z}_2$ gauge field of Kitaev's honeycomb model translates to a chiral gauge field in the continuum limit which encodes the $\pi$-fluxes of the $\mathbb{Z}_2$ gauge field. We also demonstrate that these continuum limit gauge fields can also be induced by inserting deformations to the lattice, and we demonstrate an adiabatic equivalence between deformations.

\subsubsection*{Chapter \ref{chapter:black_hole}}
In this chapter, we discuss how one can simulate a black hole in the continuum limit by studying a \textit{chiral} modification of the spin-$1/2$ XY model, by inserting three-spin interactions into the Hamiltonian, corresponding to an \textit{interacting} Hamiltonian at the fermionic level~\cite{C4_Pachos1,C4_Pachos2}. We tackle this model from two perspective: the condensed matter perspective concerned with the phase diagram of the model, and the high-energy physics perspective concerned with the continuum limit. We first study the phase properties of the model by applying mean field theory and comparing it to matrix product techniques from Ref.~\cite{black_hole_paper} and demonstrate the model has two distinct phases and tilting Dirac cones. We then bosonise the system which provides a more accurate analytic description of the model, and explains some of the features of the phase diagram. We then discuss the high-energy physics, specifically how one obtains a black hole in the continuum limit. We demonstrate that the boundary between the two phases of the model corresponds to the event horizon of the model and we show numerically that the interface thermalises wavefunctions that propagate through it, giving a temperature of the Hawking temperature. This study allows one to simulate the Hawking effect in a simple spin chain.

\subsubsection*{Chapter \ref{chapter:conclusion}}
We then close the thesis with a conclusion and discussion.

\chapter{Edge density of bulk states in carbon nanotubes \label{chapter:nanotube}}

\section{Introduction}
Several materials have low-energy quantum properties that are faithfully described by the relativistic Dirac equation. The celebrated example of graphene owes some of its unique properties, such as the half-integer quantum Hall effect \cite{C1_Novoselov_2005,C1_Novoselov_2007,C1_Fujita} and the Klein paradox effect \cite{C1_Katsnelson_2006,C1_Neto}, to the relativistic linear dispersion relation describing its low-energy sector~\cite{Wallace,C1_Neto}. This is by no means a singular case. A wide range of materials have been recently identified that admit 1D, 2D or 3D relativistic Dirac description, including many topological insulators and $d$-wave superconductors \cite{C1_Wehling_2014,C1_Moore, C1_Jia, C1_Hasan, C1_Hasan2,C1_Bernevig,Xiao-Liang}. The unusual dispersion relation of Dirac materials gives rise to effective spinors, where the sublattice degree of freedom is encoded in the pseudo-spin components. Nevertheless, the emerging excitations are spinor quasiparticles that can exhibit novel transport properties or responses to external fields akin only to relativistic physics \cite{C1_Neto}.

Here we present another counter-intuitive aspect of the emergent relativistic description by studying the possibility of \textit{bulk} states that have support on the edges of the system. In general, the choice of boundary conditions one imposes on single-particle wavefunctions of a system must ensure its Hamiltonian remains Hermitian. For the example of a non-relativistic particle in a box obeying the Schr\"odinger equation, the boundary conditions are simply that the wavefunction vanishes on the walls of the box. However, for spin-$1/2$ particles of mass $m$ obeying the $(2+1)$D Dirac equation
\begin{equation}
\left( -i \alpha^i \partial_i  + \beta m \right) \psi = E\psi,  \quad \psi = \begin{pmatrix} \psi_{\uparrow} \\ \psi_\downarrow \end{pmatrix},
\label{eq:Dirac}
\end{equation}
on some domain $D$, where $\alpha^i$ and $\beta$ are the two-dimensional Dirac alpha and beta matrices, vanishing of the spinor $\psi$ is not possible on all boundaries without the solution being trivially zero everywhere. The requirement that the Dirac Hamiltonian $h = -i \alpha^i \partial_i + \beta m$ is Hermitian with respect to the inner product $\langle \psi | \phi \rangle = \int_D \mathrm{d}^2 x \psi^\dagger(\mathbf{x}) \phi(\mathbf{x})$ reads $\langle \psi | h \phi \rangle = \langle h \psi | \phi \rangle$. This implies 
\begin{equation}
\begin{aligned}
\langle \psi | h \phi \rangle & = \int_D \mathrm{d}^2 x \psi^\dagger\left( -i \alpha^i \partial_i \phi+ \beta m \phi \right) \\
& = \int_D \mathrm{d}^2 x \left(i \partial_i \psi^\dagger \alpha^i  + m \psi^\dagger \beta \right) \phi + i\int_D \mathrm{d}^2x \partial_i \left( \phi^\dagger \alpha^i \psi \right) \\
& = \langle h \psi| \phi \rangle+ i \int_{\partial D} \mathrm{d}y \phi^\dagger \alpha^i \psi n_i, 
\end{aligned}
\end{equation}
where $n_i$ is the outward-pointing normal vector to the boundary $\partial D$. Therefore, we require the boundary term to vanish for all fields $\psi$ and $\phi$ to ensure hermiticity of $h$. In particular, we require this to hold for the case where $\psi = \phi$ which implies that on a finite domain $D$, the U(1) charge current $J^i = \psi^\dagger \alpha^i \psi$ normal to the boundary $\partial D$ is zero for all spinors $\psi$. In other words, if $n_i $ is the outward pointing normal to the boundary $\partial D$, then 
\begin{equation}
n_i J^i = 0 \label{eq:boundary_flux}
\end{equation}
for all points on $\partial D$~\cite{C1_Berry}. In contrast to the non-relativistic case, this zero-flux condition allows for bulk solutions $\psi$ whose charge density $\rho = \psi^\dagger \psi$ is non-zero on the boundaries~\cite{C1_Alberto,C1_Alonso}. This is consistent with our treatment of boundaries in the non-relativistic case of the Schr\"odinger equation, as the zero-flux condition of Eq.~(\ref{eq:boundary_flux}) has the physical interpretation that it ensures all particles are trapped inside $D$ and that there is no current across the boundary $\partial D$. This is of course exactly what the vanishing of the Schr\"odinger wavefunctions on the boundary achieves. The only difference in the relativistic case of the Dirac equation is we have more freedom to satisfy this constraint.

On the other hand, edge states are nothing new as many systems have edge states, such as topological insulators~\cite{C1_Bernevig,Asboth_2016,Xiao-Liang}, however these states are localised on the edge and do not extend into the bulk. Here, we investigate \textit{bulk} states that extend fully to the edges.

To exemplify our investigation, we consider how bulk spinor states behave at the edges of a zig-zag carbon nanotube---a system which is described by the Dirac equation of Eq.~(\ref{eq:Dirac}). We find that bulk states have support on the edges of the nanotube depending on the size of the system. Importantly, these relativistic effects become more dominant for gapless nanotubes, corresponding to systems with a multiple of three unit cells in circumference, or when the length of the nanotube is small. Such relativistic properties of spinor eigenstates are expected to be present in all Dirac-like materials and are complementary to the typically linear dispersion relation they exhibit. Bulk states with non-zero density at the boundaries are expected to impact the coupling of Dirac materials to external leads, their transport properties or their response to external magnetic fields.

The structure of this chapter is the following. First, in Sec.~\ref{section:nanotube_background}, we will introduce the basics of tight-binding models and Bravais lattices. This will provide us with the tools required to describe a material and will be used extensively throughout this thesis. Then we will introduce the tight-binding model of graphene and demonstrate its effective relativistic description at low energy in terms of a Dirac equation. Then, in Sec.~\ref{section:carbon_nanotubes}, we will focus on a model of graphene with a boundary, namely a zig-zag carbon nanotube, which provides us with a simple system containing a boundary and allows us to explore the requirements and consequences of a non-zero edge density, which will give us an alternative explanation for how relativistic effects appear in lattice models. We close the chapter with a conclusion in Sec.~\ref{nanotube_conclusion}.

\section{Background \label{section:nanotube_background}}
\subsection{Bravais lattices and tight-binding models \label{sec:bravais_lattice}}
The tight-binding model is a simple model that allows one to approximate the electronic properties of a material~\cite{C1_Altland,Steve_simon,Ashcroft,coleman_2015}. This has been successfully applied to a range of systems in condensed matter physics, from traditional systems such as metals,  insulators and superconductors~\cite{Ashcroft,Steve_simon,coleman_2015}, to more exotic phases of matter such as topological insulators and superconductors~\cite{Xiao-Liang,Asboth_2016,C1_Bernevig}. The tight-binding model also arises in the study of spin lattices after a mapping to fermions or bosons, as will be done in Chap.~\ref{chapter:black_hole} of this thesis. In this section, we review the background material.

We begin by considering the simplest picture of a material situated in $n$-dimensional Euclidean space $\mathbb{R}^n$ as a periodic lattice of positive ions arranged into a Bravais lattice $\Lambda$. A Bravais lattice is defined as a subset of points of $\mathbb{R}^n$ which is generated by integer multiples of a set of linearly independent vectors $\{ \mathbf{n}_i\}_{i=1}^n$ known as the generators of the lattice~\cite{C1_Altland,Steve_simon,Ashcroft}. In other words, the position vector of a given point $\mathbf{r} \in \Lambda$ is given by
\begin{equation}
\mathbf{r} = \sum_{i=1}^n x_i \mathbf{n}_i,
\end{equation}
where $x_i \in \mathbb{Z}$ are the integer coordinates of point $ \mathbf{r}\in \Lambda$ and $|\mathbf{a}_i| = a$ for all $i$, where $a$ is the lattice spacing. In addition, the basis of generators is not unique. From the Bravais lattice $\Lambda$ we can define a second Bravais lattice $\Lambda^*$ called the reciprocal lattice. The reciprocal lattice is defined as the lattice dual to $\Lambda$, that is, if the set of vectors $\{ \mathbf{n}_i \}_{i=1}^n$ generates $\Lambda$, then we define another set of generators $\{ \mathbf{G}_i \}_{i=1}^n$ obeying $
\mathbf{n}_i \cdot \mathbf{G}_j = 2\pi \delta_{ij}$ which generates $\Lambda^*$.

The lattice of ions generates a potential which is periodic with respect to discrete translations on $\Lambda$. Therefore, the Hamiltonian $H$ describing a single charged particle moving through this lattice has discrete translational symmetry with respect to $\Lambda$, so $[H , T(\mathbf{r})] = 0 $ for all $\mathbf{r} \in \Lambda$, where $T(\mathbf{r}) = \exp(-i \hat{\mathbf{p}} \cdot \mathbf{r})$ is the translation operator and $\hat{\mathbf{p}}$ is the canonical momentum operator. According to Bloch's theorem, the eigenstates of the Hamiltonian are therefore given by $|\mathbf{k}\rangle$, where $T(\mathbf{r}) |\mathbf{k}\rangle = e^{i \mathbf{k} \cdot \mathbf{r}} |\mathbf{k} \rangle$ for all $\mathbf{r} \in \Lambda$. This defines the crystal momentum $\mathbf{k}$ which, from its definition, is only defined uniquely up to a reciprocal basis vector. For this reason, the crystal momentum $\mathbf{k}$ is found within the Brillouin zone (B.Z.) defined as the smallest unit cell of $\Lambda^*$. From the Bloch basis $|\mathbf{k}\rangle$ we can define the Wannier basis of states $|\mathbf{r}\rangle$ by Fourier transforming the Bloch states, where these states are localised on a lattice site $\mathbf{r} \in \Lambda$ and provide the basis for the tight-binding model~\cite{C1_Altland}.

The reciprocal basis is a useful basis to expand our momenta $\mathbf{k} \in \text{B.Z.}$ with respect to, so we write
\begin{equation}
\mathbf{k} = \frac{a}{2\pi} \sum_{i = 1}^n k_i \mathbf{G}^i,
\end{equation}
where $ k_i = \mathbf{k} \cdot \frac{\mathbf{n}_i}{a} \in \mathbb{R}$ are the components of the momentum. Note that the momentum does not necessarily lie on $\Lambda^*$. The requirement that the momentum is only defined up to a reciprocal basis vector is equivalent to saying $k_i$ is defined modulo $2\pi/a$, therefore we identify the Brillouin zone (B.Z.) as the hypercube $\mathrm{B.Z.} = [-\pi/a,\pi/a)^n$.

 With this, the tight-binding model approximates a material as a collection of particles hopping on lattice of ions $\Lambda$ obeying either fermionic or bosonic statistics. In this thesis we study many-body fermionic models using the language of second quantisation whereby the Hamiltonian is expressed in terms of fermionic operators $\{ f_i \}$ on a Fock space $\mathcal{F}$ obeying the anti-commutation relations
\begin{equation}
\{ f_i,f_j^\dagger \} = \delta_{ij}, \quad \{ f_i,f_j\} = \{f_i^\dagger, f_j^\dagger \} = 0,
\end{equation}
where the indices label the possible single-particle quantum states, which for the tight-binding model consist of the Wannier states~\cite{C1_Altland,coleman_2015}. We call the operators $\{ f_i \}$ annihilation operators and their Hermitian conjugates $\{ f_i^\dagger \}$ as creation operators. We assume the existence of a vacuum state $|0\rangle \in \mathcal{F}$, where $|i\rangle = f^\dagger_i |0\rangle \in \mathcal{F}$ is interpreted as the state of a single particle in the $i$th quantum state, whilst $f_i |0\rangle  = 0$ for all $i$. With these properties, we can construct an orthonormal basis of $\mathcal{F}$ via repeated application of the creation operators on the vacuum, giving $\mathrm{dim}(\mathcal{F})= 2^N$, where $N$ is the number of possible single-particle quantum states. 

\subsection{Graphene}

\begin{figure}[t]
\begin{minipage}{\textwidth}
\begin{center}
\includegraphics[scale=0.9,valign=t]{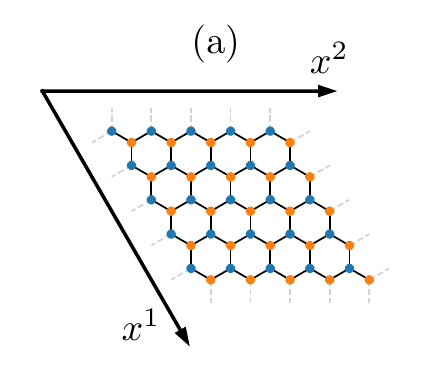}
\includegraphics[scale=0.9,valign=t]{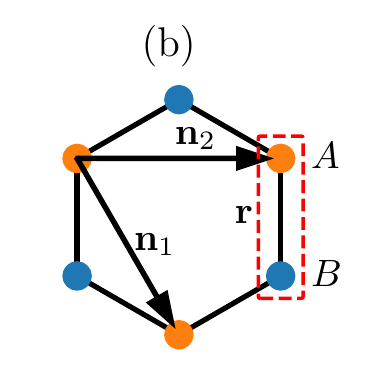}
\includegraphics[scale=0.9,valign=t]{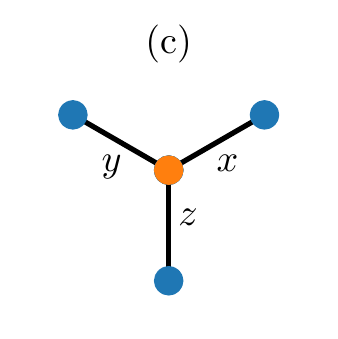}
\end{center}
\end{minipage}
\caption{(a) The honeycomb lattice of graphene. The two coordinates axes, $x^1$ and $x^2$, form a non-Cartesian coordinate system which point along the basis vectors $\mathbf{n}_1$ and $\mathbf{n}_2$ respectively. (b) The basis vectors $\mathbf{n}_1$ and $\mathbf{n}_2$, and the unit cell located at the position $\mathbf{r}$ containing the two sites that form a vertical $z$-link. (c) The three types of nearest-neighbour links, labelled as $x$-, $y$- and $z$-links.}
\label{fig:C1_graphene_honeycomb}
\end{figure}

Graphene is a two-dimensional lattice of carbon atoms arranged into a honeycomb lattice \cite{Wallace,C1_Neto,C1_Altland}. The honeycomb lattice itself is not a Bravais lattice, however one can construct the honeycomb from a Bravais lattice $\Lambda$ with a unit cell containing two sites, $A$ and $B$, as shown in Fig.~\ref{fig:C1_graphene_honeycomb}. Therefore, the lattice is given by $\Lambda_\text{tot} = \Lambda \times \{A,B\}$ and every point $p \in \Lambda_\text{tot}$ can be labelled with the pair $(\mathbf{r},\mu)$, where $\mathbf{r} \in \Lambda$ labels the unit cell and $\mu \in \{A,B\}$ labels the site within the cell. For later convenience, we choose the lattice generators
\begin{equation}
\mathbf{n}_1 = \frac{a}{2}\left(1, -\sqrt{3}\right), \quad \mathbf{n}_2 = a(1,0),
\end{equation}
as shown in Fig.~\ref{fig:C1_graphene_honeycomb}(b), with the corresponding reciprocal basis
\begin{equation}
\mathbf{G}^1 = - \frac{4 \pi}{\sqrt{3} a} \left( 0, -1 \right), \quad \mathbf{G}^2 = \frac{2 \pi}{a} \left( 1, \frac{1}{\sqrt{3}} \right).
\end{equation}
We model graphene using the tight-binding model under the assumption that electrons can only hop between nearest-neighbour sites as shown in Fig.~\ref{fig:C1_graphene_honeycomb}(c), i.e., $\langle \mathbf{r}, \mu |H |\mathbf{r}' , \nu\rangle = -t $ for nearest-neighbour Wannier sites, where $( \mathbf{r},\mu),(\mathbf{r}',\nu) \in \Lambda_\mathrm{tot}$, with all other matrix elements zero, where $t \in \mathbb{R}$ is the hopping parameter~\cite{Wallace,C1_Altland}. If we second quantise this Hamiltonian, with our choice of lattice generators we arrive at
\begin{equation}
H = -t \sum_{\mathbf{r} \in \Lambda} a^\dagger_\mathbf{r} \left( b_\mathbf{r} + b_{\mathbf{r} - \mathbf{n}_1 + \mathbf{n}_2}  + b_{\mathbf{r} - \mathbf{n}_1} \right) + \text{H.c.}, \label{eq:C1_graphene_ham}
\end{equation}
where the sum is over nearest-neighbour pairs, and $a^\dagger_\mathbf{r}$ is the creation operator for the Wannier state at $(\mathbf{r}, A) \in \Lambda_\mathrm{tot}$ and similarly for $b^\dagger_\mathbf{r}$. The Fermionic modes obey the anti-commutation relations $\{ a_\mathbf{r} , a_{\mathbf{r}'}^\dagger \} = \delta_{\mathbf{r}\mathbf{r}'}$ and $\{ a_\mathbf{r},a_{\mathbf{r}'} \} = \{ a^\dagger_\mathbf{r},a^\dagger_{\mathbf{r}'}\} = 0$ and similarly for $b_\mathbf{r}$, while pairs of fermions from different sub-lattices anti-commute. 

\begin{figure}
\begin{center}
\includegraphics[scale=0.4]{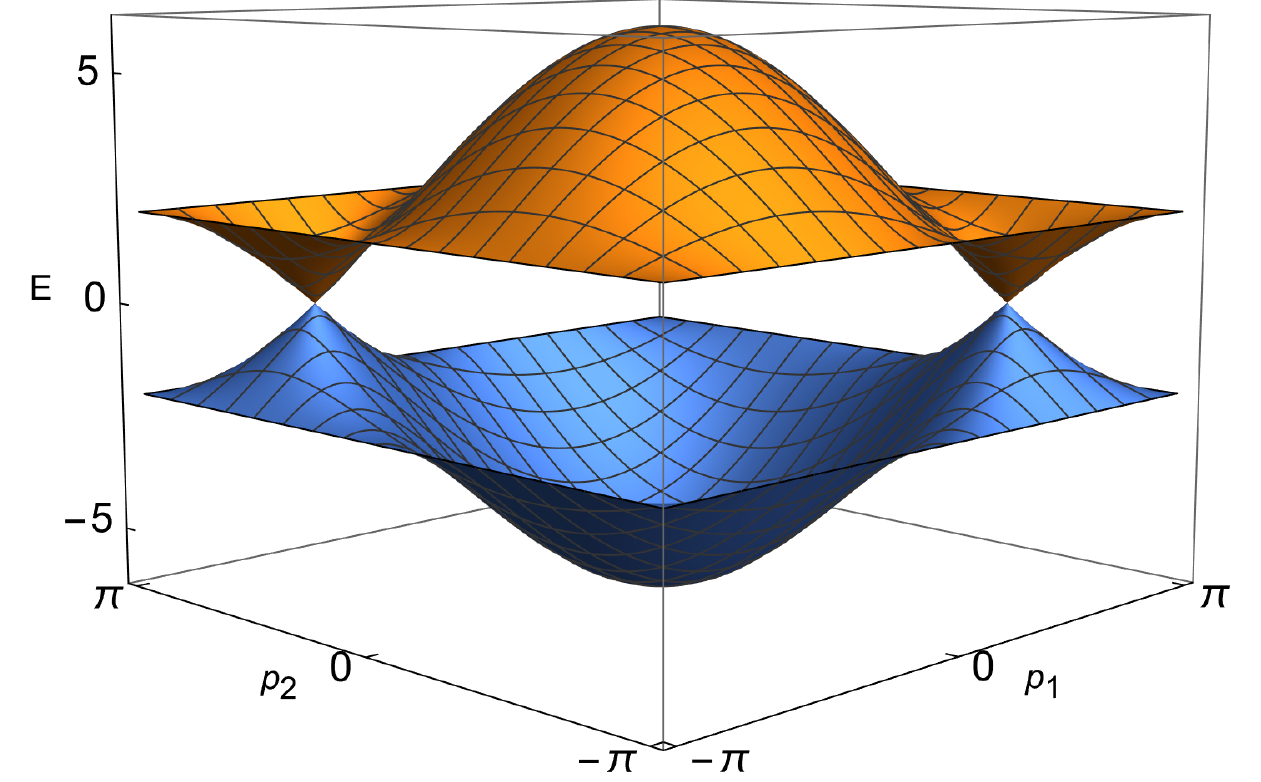}
\end{center}
\caption{The dispersion relation of Eq.~(\ref{eq:C1_graphene_dispersion}) for $t = 2$. There are two Fermi points $\mathbf{K}_1$ and $\mathbf{K}_2$ where the two bands touch, forming a pair of Dirac cones.}
\label{fig:C1_graphene_dispersion}
\end{figure}

The Hamiltonian for graphene will be periodic with respect to $\Lambda$ and not the total lattice $\Lambda_\text{tot}$, therefore we Fourier transform the fermions as
\begin{equation}
a_\mathbf{r} = \frac{1}{\sqrt{N}_\mathrm{c}} \sum_{\mathbf{p} \in \mathrm{B.Z.}} e^{i \mathbf{p} \cdot  \mathbf{r}} a_\mathbf{p},
\end{equation}
and similarly for $b_\mathbf{r}$, where $N_\mathrm{c}$ is the number unit cells of the lattice $\Lambda$. This defines the momentum space fermionic modes which obey $\{ a_\mathbf{p} , a_\mathbf{q}^\dagger \} = \delta_{\mathbf{p}\mathbf{q}}$ and $\{ a_\mathbf{p},a_\mathbf{q} \} = \{ a^\dagger_\mathbf{p},a^\dagger_\mathbf{q}\} = 0$ and similarly for $b_\mathbf{p}$, while pairs of fermions from different sub-lattices anti-commute. Fourier transforming the Hamiltonian of Eq.~(\ref{eq:C1_graphene_ham}) yields
\begin{equation}
\begin{aligned}
H & = -\frac{t}{N_\mathrm{c}} \sum_{\mathbf{r} \in \Lambda} \sum_{\mathbf{p} \in \mathrm{B.Z.}} \sum_{\mathbf{q} \in \mathrm{B.Z.}} e^{-i \mathbf{p} \cdot \mathbf{r}} a^\dagger_\mathbf{p} \left( e^{i \mathbf{q} \cdot \mathbf{r}} + e^{i \mathbf{q} \cdot (\mathbf{r} - \mathbf{n}_1 + \mathbf{n}_2)}  + e^{i \mathbf{p} \cdot (\mathbf{r} - \mathbf{n}_1)}   \right)b_\mathbf{q} + \mathrm{H.c.} \\
& = -t \sum_{\mathbf{p} \in \mathrm{B.Z.}} \sum_{\mathbf{q} \in \mathrm{B.Z.}} \left( 1 + e^{-i \mathbf{q} \cdot (\mathbf{n}_1 - \mathbf{n}_2) } + e^{-i \mathbf{p} \cdot \mathbf{n}_1} \right) a^\dagger_\mathbf{p} b_\mathbf{q} \underbrace{\left( \frac{1}{N_\mathrm{c}} \sum_{\mathbf{r} \in \Lambda} e^{-i (\mathbf{p} - \mathbf{q}) \cdot \mathbf{r}} \right)}_{\delta_{\mathbf{p}\mathbf{q}}} + \mathrm{H.c.} \\
& = -t \sum_{\mathbf{p} \in \mathrm{B.Z.}} f(\mathbf{p}) a^\dagger_\mathbf{p} b_\mathbf{p} + \mathrm{H.c.}, \label{eq:C1_fourier_transform_H}
\end{aligned}
\end{equation}
where 
\begin{equation}
f(\mathbf{p}) =  1 + e^{-i \mathbf{p} \cdot (\mathbf{n}_1 - \mathbf{n}_2) } + e^{-i \mathbf{p} \cdot \mathbf{n}_1} = 1 + e^{-i (p_1 - p_2)a } + e^{-ip_1 a} .
\end{equation}
If we define the two-component spinor $\chi_\mathbf{p} = (a_\mathbf{p},b_\mathbf{p})^\mathrm{T}$, then the Hamiltonian is given by
\begin{equation}
H = \sum_{\mathbf{p} \in \mathrm{B.Z.}} \chi^\dagger_\mathbf{p} h(\mathbf{p}) \chi_\mathbf{p}, \quad h(\mathbf{p}) = 
\begin{pmatrix}
0 & f(\mathbf{p}) \\
f^*(\mathbf{p}) & 0 
\end{pmatrix}, \label{eq:C1_graphene_ham_diag}
\end{equation}
where $h(\mathbf{p})$ is the single-particle Hamiltonian. The Hamiltonian $H$ can be diagonalised by simply diagonalising the single-particle Hamiltonian $h(\mathbf{p})$ with a unitary transformation, whose eigenvalues give us the dispersion relation 
\begin{equation}
E(\mathbf{p}) = \pm |f(\mathbf{p})| = \pm t \sqrt{ 3 + 2 \cos (p_1a ) + 2\cos((p_1 - p_2)a) + 2 \cos (p_2a)}, \label{eq:C1_graphene_dispersion}
\end{equation}
as shown in Fig.~\ref{fig:C1_graphene_dispersion}.

The Fermi points are defined as the points for which $E(\mathbf{p}) = 0$ within the Brillouin zone. For our system, they are located at $\mathbf{p} = \mathbf{K}_1$ and $\mathbf{p} =  \mathbf{K}_2$, where $\mathbf{K}_1 = \left( \frac{2 \pi}{3} , \frac{4 \pi}{3} \right)$ and $\mathbf{K}_2 = \left( \frac{4 \pi}{3} , \frac{2 \pi}{3} \right)$, as shown in Fig.~\ref{fig:C1_graphene_dispersion}, which are expressed with respect to the reciprocal basis $\{ \mathbf{G}_i \}$. At the Fermi points, we see that the valence and conduction bands touch, forming a conical dispersion as $E(\mathbf{K}_i + \mathbf{p}) = v_\mathrm{F} |\mathbf{p}|$ in their neighbourhood, where $v_\mathrm{F} \equiv |\nabla E(\mathbf{K}_i)| = 3ta/2$ is the Fermi velocity. In other words, the dispersion close to the Fermi points is \textit{relativistic} where the Fermi velocity plays the role of an effective speed of light.  We take the continuum limit by Taylor expanding the single-particle Hamiltonian $h(\mathbf{p})$ about the Fermi points and taking the limit that $a \rightarrow 0$ and ensuring the Fermi velocity $v_\mathrm{F}$ remains finite, so we must absorb the lattice spacing into the coupling as $at \rightarrow t$ while keeping $t$ finite. This yields the pair of Dirac Hamiltonians
\begin{equation}
h_i(\mathbf{p}) \equiv h(\mathbf{K}_i  + \mathbf{p}) = v_\mathrm{F} (\sigma^x p_x \pm \sigma^y p_y) + O(\mathbf{p}^2),
\end{equation}
where we have expressed it in Cartesian coordinates and $\pm$ corresponds to $i=1,2$ respectively. This is the celebrated relativistic continuum limit of graphene~\cite{Wallace,C1_Neto}.

\section{Zig-Zag Carbon Nanotubes \label{section:carbon_nanotubes}}
\subsection{Dispersion relation}

\begin{figure}
\begin{center}
\includegraphics[scale=0.9,valign=t]{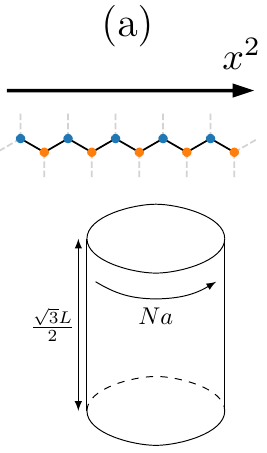}
\includegraphics[scale=1,valign=t]{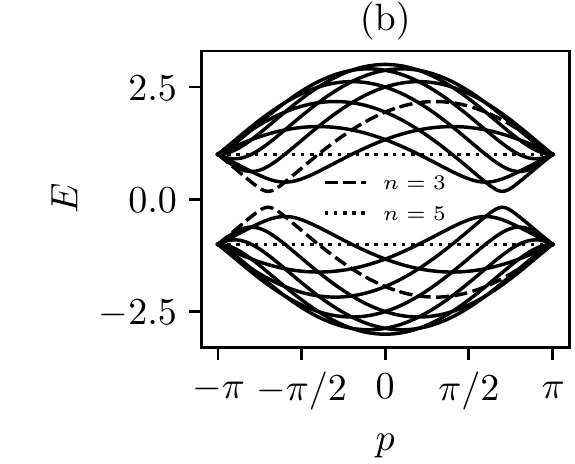}
\includegraphics[scale=1,valign=t]{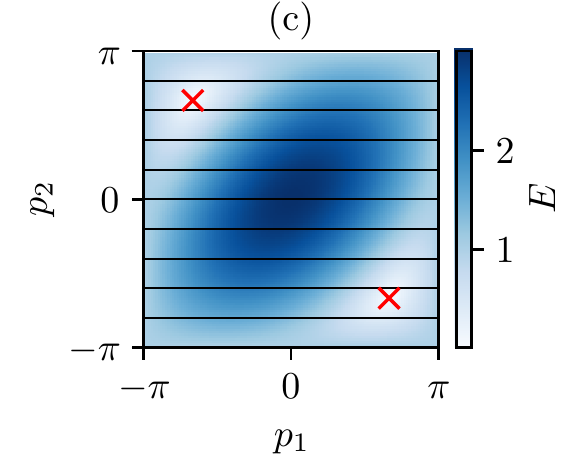}
\caption{(a) We impose open boundary conditions in the $\mathbf{n}_1$ direction with a length $L$ and periodic boundary conditions with $N$ unit cells in the $\mathbf{n}_2$ direction. This generates a cylinder with a zig-zag boundary of circumference $Na$ and height $\frac{\sqrt{3} L}{2}$. (b) The nanotube dispersion relations of Eq.~(\ref{eq:C1_nanotube_dispersion}) for $ N = 10$ and $t = 1$. The system is gapped as $N$ is not a multiple of three. The band $n = N/2 = 3$ is also completely flat. (c) The bands on the left but as seen compared to the full graphene dispersion of Eq.~(\ref{eq:C1_graphene_dispersion}) in the colour plot which form stripes through the Brillouin zone. We see that the nanotube is gapped because, due to the quantisation of $p_2$, none of the bands pass through the Fermi points of graphene, marked with the red crosses.} 
\label{fig:C1_nanotube_dispersions}
\end{center}
\end{figure}

The previous section is concerned with an infinite sheet of graphene. We now consider a system with edges in order to test the zero-flux boundary condition of Eq.~(\ref{eq:boundary_flux}). We consider a zig-zag carbon nanotube which is one of the simplest systems with a boundary that can be created in the laboratory.

The first step in constructing a zig-zag carbon nanotube is to take the limit that the length $L$ of the $\mathbf{n}_1$ direction goes to infinity, whilst imposing periodic boundary conditions in the $\mathbf{n}_2$ direction with a finite length of $N$ unit cells~\cite{C1_Altland}. This generates a cylinder of circumference $Na$ and height $\sqrt{3}L/2$ with zig-zag edges on the top and bottom, as shown in Fig.~\ref{fig:C1_nanotube_dispersions}(a). The periodic boundary conditions for single-particle eigenstates read
\begin{equation}
T(N \mathbf{n}_2 ) |\mathbf{k} \rangle = e^{i \mathbf{k} \cdot (N \mathbf{n}_2)} |\mathbf{k}\rangle = |\mathbf{k}\rangle,
\end{equation}
where the first equality is by Bloch's theorem and the second equality is our periodic boundary condition. This implies our momenta $\mathbf{p} = (p_1,p_2)$ have an unconstrained $p_1$, whilst $p_2$ is quantised as
\begin{equation}
p_2 = \frac{2 n \pi}{Na}. \label{eq:C1_constrained_momenta}
\end{equation}
If we substitute the allowed momenta into the dispersion relation of graphene Eq.~(\ref{eq:C1_graphene_dispersion}), we arrive at a one-dimensional dispersion relation for each value of $n$:
\begin{equation}
E_n(p) = \pm t \sqrt{ 3 + 2 \cos \left( \frac{2 n \pi}{N} \right) + 2\cos \left(\frac{2 n \pi}{N} - p a \right) + 2 \cos (p a)}, \label{eq:C1_nanotube_dispersion}
\end{equation}
where we rename $p_1 \rightarrow p$. This dispersion relation is plotted in Fig.~\ref{fig:C1_nanotube_dispersions}(b)-(c) for each value of $n$. Each band for a fixed $n$ has a single minima.

\subsection{Continuum limit}
We are only interested in the properties of the model near the ground state, which in the many-body picture is the state for which all negative energy single-particle states are occupied. Therefore, the low-lying excitations will be positive-energy states that exist near the minima of the dispersion, so we project the Hamiltonian onto a small neighbourhood of the minima. This process of projecting onto the low-lying states is sometimes called the continuum limit, as it is equivalent to taking the limit that $a \rightarrow 0$ whilst keeping the Fermi velocity $v_\mathrm{F}$ fixed as only the states near the Fermi points remain. In order to search for the minima, we first search for the minima of each band of Eq.~(\ref{eq:C1_nanotube_dispersion}). The minima are located at the same position as the minima of $F_n(p) = E_n^2(p)$, so we use this as it is easier to work with. The turning points $p_0$ obey $F_n'(p_0) = 0$ which gives
\begin{equation}
\sin\left(\frac{2n \pi}{N} - p_0a \right) = \sin\left(p_0a \right).
\end{equation} 
Using the result that $\sin(x) = \sin(y) \Rightarrow x = m \pi + (-1)^m y$ for $m \in \mathbb{Z}$, this implies
\begin{equation}
p_0 = \frac{ \pi}{(1+(-1)^m)a} \left(\frac{2n}{N}-m \right). \label{eq:C1_p_0_first}
\end{equation}
Due to the denominator we have finite solutions only when $m$ is an even number, so we take $m=-2l$ which gives us the turning points 
\begin{equation}
p_0  = \frac{\pi}{a} \left( \frac{n}{N} + l \right) \mod \frac{2 \pi}{a}, \quad l \in \mathbb{Z}.
\end{equation}
Note that as $p_0 \in [- \pi/a, \pi/a)$ and $n/N \in [-1/2,1/2]$, the only possibilities are that $l = -1, 0,1$. We now need to identify which of these turning points are minima, where $F''_n(p_0) > 0$. We have
\begin{equation}
F''_n(p) = - 2a^2 t^2 \left[ \cos \left( \frac{2 n \pi}{N} - pa \right) + \cos(pa) \right].
\end{equation}
Substituting in $p_0$ of Eq.~(\ref{eq:C1_p_0_first}), we require 
\begin{equation}
\begin{aligned}
\cos\left[ \frac{2n \pi}{N} - \pi \left( \frac{n}{N} + l \right) \right] + \cos \bigg[ \pi \left( \frac{n }{N} + l  \right) \bigg] & = \cos\left( \frac{n\pi}{N} - l \pi \right) + \cos \left( \frac{n \pi}{N} + l \pi \right) \\ & = 2(-1)^l \cos\left(\frac{n \pi}{N} \right) < 0.
\end{aligned}
\end{equation}
As $n/N \in [- 1/2,1/2]$, then $ \cos(\frac{n \pi}{N}) \in [0,1]$, therefore in order to satisfy this constraint whilst ensuring that $p_0 \in  [-\pi/a,\pi/a)$, we require
\begin{equation}
l = \begin{cases} -1 & \text{if} \ n > 0 \\ 1 & \text{if} \  n < 0 \end{cases} \quad \Rightarrow \quad  p_0  = \begin{cases} \frac{\pi}{a}\left( \frac{n}{N} - 1 \right) & \text{if} \ n > 0 \\ \frac{\pi}{a}\left( \frac{n}{N} + 1 \right) & \text{if} \ n < 0 \end{cases}.
\end{equation}
Each band has a single minima. If $n = \frac{N}{2}$, there does not exist a minima as the band is completely flat as can be seen in Fig.~\ref{fig:C1_nanotube_dispersions}.

We now take the continuum limit by tackling each band of Eq.~(\ref{eq:C1_nanotube_dispersion}) separately. The Hamiltonian of the $n$th band is obtained by substituting in the constrained momenta of Eq.~(\ref{eq:C1_constrained_momenta}) into the single-particle Hamiltonian of graphene from Eq.~(\ref{eq:C1_graphene_ham_diag}) and Taylor expanding about the minima in $p_1$ only by defining $h_n(p) \equiv h(p_0 + p, 2n \pi/N)$ to first order in $p$. This gives us the effective Hamiltonian
\begin{equation}
h_n(p)    = 
\begin{pmatrix} 0 &   \Delta_n - i e_n p \\
 \Delta_n + i e_n p  & 0  \end{pmatrix} + O(p^2) 
 = e_n \sigma^y p + \Delta_n \sigma^x + O(p^2),
\end{equation}
where $\Delta_n$ is the gap of the $n$th band, given by
\begin{equation}
\Delta_n = at\left[ 2 \cos \left( \frac{n \pi}{N} \right) -1 \right], \label{eq:C1_nanotube_gap}
\end{equation}
and $e_n$ is the spatial component of the zweibein (2D veilbein to be defined later in Sec.~\ref{chapter:RC_background}) given by
\begin{equation}
e_n = 2at \cos \left( \frac{n \pi}{N} \right).
\end{equation}
We see that the continuum limit of the flat band, $n = N/2$, contains no kinetic term as the zweibein vanishes here. The corresponding dispersion relation is given by $E_n(p) = \pm \sqrt{\Delta^2_n + e_n^2 p^2}$ which is a relativistic dispersion relation for a particle of mass $\Delta_n$. Note the gap closes if $n/N = \pm 1/3$, which is only possible if $N$ is a multiple of three. This is an important observation: despite an infinite sheet of graphene being gapped, a carbon nanotube is not in general~\cite{C1_Saito}

\subsection{Solutions of the Dirac equation with zig-zag boundary conditions}

If we inverse Fourier transform back to real space, we arrive at 
\begin{equation}
h_n = -i e_n \alpha^x \partial_x + \Delta_n \beta ,
\end{equation}
which takes the form of a $(1+1)$D Dirac Hamiltonian with mass $\Delta_n$ using the representation $\alpha^x =  \sigma^y$ and $\beta = \sigma^x$, where the continuum coordinate is given by $x \equiv a x^1$ in the limit that $a \rightarrow 0$. Let us take the usual plane-wave ansatz $\psi = \phi e^{i p x} $, where $\phi = (\phi_A , \phi_B)^\mathrm{T}$. The Dirac equation reads
\begin{equation}
 \begin{pmatrix} 0 &   \Delta_n - i e_n p \\   \Delta_n + i e_n p  & 0  \end{pmatrix} \begin{pmatrix} \phi_A \\ \phi_B \end{pmatrix} = E_n \begin{pmatrix} \phi_A \\ \phi_B \end{pmatrix}.
\end{equation}
This yields two equivalent equations, both implying
\begin{equation}
\phi_B = \left( \frac{\Delta_n + i e_n p}{E_n} \right)  \phi_A \equiv s e^{i \theta_{n,p}} \phi_A, \quad \theta_{n,p} = \arg(\Delta_n +ie_n p), \label{eq:C1_phase}
\end{equation}
where $s = \mathrm{sgn}(E_n)$. The corresponding unnormalised eigenvectors in one-dimensional position space are given by
\begin{equation}
\phi_{n,p}(x) = \begin{pmatrix} 1 \\ s  e^{i \theta_{n,p}} \end{pmatrix}  e^{i px},
\end{equation}
where $n$ labels the band and $p$ labels the momentum eigenstate. We interpret the top and bottom components of our spinors as the wavefunction on sublattices $A$ and $B$ respectively.

\begin{figure}[t]
\begin{center}
\includegraphics[scale=1]{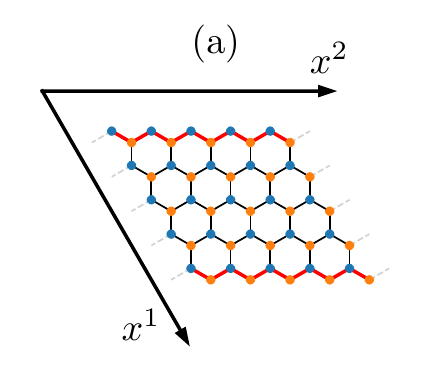}
\includegraphics[scale=1]{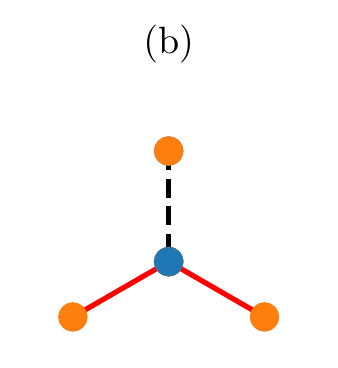}
\includegraphics[scale=1]{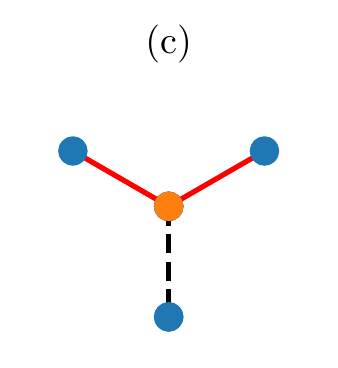}
\end{center}
\caption{(a) The representation of the zig-zag nanotube. We have a finite length in the $x^1$ direction and roll up the lattice in the $x^2$ direction to form a cylinder with a zig-zag boundaries, represented by the red links (b)-(c) Due to the choice of unit cell as the $z$-links (dashed lines), the unit cells overlap with the outside of the system, which sets the zig-zag boundary conditions $\psi_A = 0$ at the top and $\psi_B = 0$.}
\label{fig:C1_nanotube_boundaries}
\end{figure}

With the continuum limit approximation, we now study a nanotube of finite length $L$ in the $\mathbf{n}_1$  direction (the $x$-direction in the continuum) by imposing suitable boundary conditions for the zig-zag edge. First, we build positive energy, $s=1$, standing waves by superimposing forward and backward propagating waves as $\psi_{n,p} = \phi_{n,p} + R \phi_{n,-p}$, for $R \in \mathbb{C}$, as
\begin{equation}
\psi_{n,p} =  \begin{pmatrix} 1 \\   e^{i \theta_{n,p}} \end{pmatrix}  e^{i px} +  R \begin{pmatrix} 1 \\   e^{-i \theta_{n,p}} \end{pmatrix}  e^{-i px},
\end{equation}
where we used the fact that $\theta_{n,-p} = -\theta_{n,p}$. The zig-zag boundary conditions are given by $
\psi_A(0) = \psi_B(L) = 0$. These boundary conditions can be seen clearly in Fig.~\ref{fig:C1_nanotube_boundaries} as the unit cells of the top and bottom row, where $x= 0$ and $x=L$, each contain a ``missing" site that is outside of the system (recall that our coordinates $x$ label the vertical $z$-link unit cells and not the individual sites). Note that, in our representation of the Dirac alpha and beta matrices, the zero-flux condition of Eq.~(\ref{eq:boundary_flux}) reads $\mathrm{Im}(\psi^*_A \psi_B) = 0$ on the boundaries which the zig-zag boundary conditions satisfy. The first boundary condition gives $R = -1$, so our solutions take the form
\begin{equation}
\psi_{n,p}(x) = \mathcal{N}_{n,p} \begin{pmatrix}  \sin(px)  \\ \sin(px + \theta_{n,p}) \end{pmatrix} , \label{eq:C1_nanotube_wavefunctions}
\end{equation}
where $\mathcal{N}_{n,p}$ is a normalisation constant. The second boundary condition gives $\sin(pL + \theta_{n,p}) = 0$, giving the transcendental equation for the allowed momenta
\begin{equation}
pL + \theta_{n,p} = m \pi , \ m \in \mathbb{N}, \label{eq:C1_transcendental_eq}
\end{equation}
which can be solved numerically by minimising the function $f_{nm}(p) = |pL + \theta_{n,p} - m \pi|$ with respect to $p$ for a fixed $n$ and $m$. This gives us a clear example of how a spinor can be non-zero at the edges whilst satisfying the zero-flux condition on the boundaries.
\subsection{Densities at the edge}

The U(1) electric charge density of $(1+1)$D Dirac spinors $\psi = (\psi_A,\psi_B)^\mathrm{T}$  is given by $\rho = \psi^\dagger \psi = |\psi_A|^2 + |\psi_B|^2$. With our interpretation of the pseudo-spin components $\psi_A(x)$ and $\psi_B(x)$ as the sublattice wavefunctions, where $x$ labels the unit cell, $\rho$ is therefore the charge density with respect to the unit cells. For the bulk standing wave solutions of Eq.~(\ref{eq:C1_nanotube_wavefunctions}), we have
\begin{equation}
\rho_{n,p}(x) = |\mathcal{N}_{n,p}|^2 \left( \sin^2(p x) + \sin^2(p x + \theta_{n,p}) \right), 
\label{eq:C1_rho}
\end{equation}
which gives a charge density at the boundaries of
\begin{equation}
\rho_{n,p}(0) = \rho_{n,p}(L) = |\mathcal{N}_{n,p}|^2 \sin^2(\theta_{n,p}). \label{eq:C1_edge_density}
\end{equation}
We see that it is possible to have $\rho_{n,p}(0)\neq 0$ due to the phase difference, $\theta_{n,p}$, which is purely a relativistic effect. The edge charge density of bulk states is maximal when $\theta_{n,p} = \pm \pi/2$. Referring to Eq.~(\ref{eq:C1_phase}), this is achieved when $\Delta_n = 0$, i.e., when the $n$th band is gapless. From Eq.~(\ref{eq:C1_nanotube_gap}) we see that the gap closes if $n/N = \pm 1/3$ which is only possible if $N$ is a multiple of three. Note that, for a gapless band, the charge density of Eq.~(\ref{eq:C1_rho}) is also completely uniform with
\begin{equation}
\rho_{n,p}(x) = \frac{1}{L},  
\label{eq:C1_edge_density2}
\end{equation} 
which is independent of the momentum $p$, where we have chosen a $1$D normalisation. On the other hand, when the system is gapped, then the density oscillates along the length of the nanotube and becomes vanishingly small at the edges. This shift in behaviour of the charge density reflects the expected transition from the relativistic to non-relativistic regime witnessed in confined Dirac particles as their mass increases~\cite{C1_Alberto}.

\begin{figure}[t!]
\begin{center}
\includegraphics[scale=1]{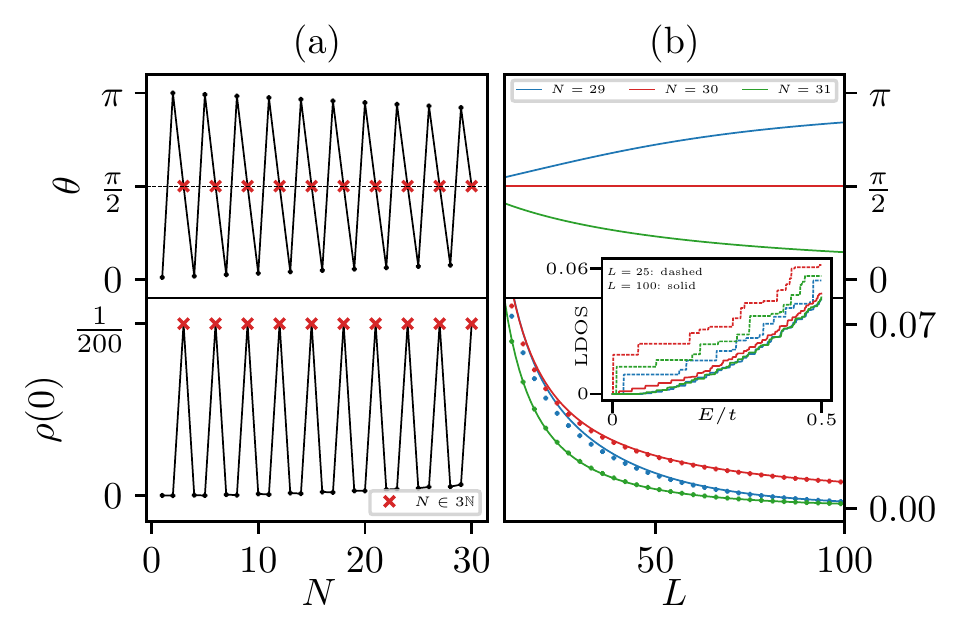}
\caption{(a) The phase shifts $\theta$ of Eq.~(\ref{eq:C1_phase}) and the numerically measured edge densities $\rho(0)$ versus circumference $N$ for the ground state of a system of fixed length $L = 200$. (b) The phase shifts $\theta$ and edge densities $\rho(0)$ versus system length $L$ for the ground state of the gapless systems $N=30$ and its two neighbouring gapped systems $N=29$ and $N=31$. The solid line represents the analytical formulas whilst the points represent numerics. (Inset) The integrated LDOS at the edge $x = 0$ for a nanotube of circumferences $N=29,30,31$ and lengths $L=25$ (dashed lines) and $L=100$ (solid lines).}
\label{fig:phases_densities}
\end{center}
\end{figure}

The stark contrast between gapped and gapless systems is confirmed numerically using the techniques of Appendix~\ref{appendix:numerical_techniques}. We focus on the single-particle eigenstates close to the Fermi energy, as this is where the relativistic description holds. We identify the ground state as the first state above the Fermi energy (which will be the ground state of the Dirac equation), where in order to do this we first identify the band $n$ which the ground state lies on by minimising the gap of Eq.~(\ref{eq:C1_nanotube_gap}). Fig.~\ref{fig:phases_densities}(a) shows the phase shift $\theta_{n,p}$ and edge density $\rho$ of the ground state of a system of length $L = 200$ for varying circumferences $N$. When $N$ is a multiple of three, i.e., when the system is gapless, the edge density spikes to the expected value of $1/L = 0.005$. On the other hand, when $N$ is not a multiple of three, i.e., when the system is gapped, the edge density is small. This behaviour is a consequence of the highly oscillating phase shift $\theta$. When $N$ is a multiple of three, the phase shift is $\pi/2$ exactly, maximising the edge density according to Eq.~(\ref{eq:C1_edge_density}). However, as $N$ increases, all zig-zag nanotubes tend towards gapless systems even if $N$ is not a multiple of three, as there exists a band $n$ such that $n/N \approx \pm 1/3$ when $N$ is large, so the gap of Eq.~(\ref{eq:C1_nanotube_gap}) begins to close. 

The relativistic boundary effects also have a length dependence~\cite{C1_Alberto}. Fig.~\ref{fig:phases_densities}(b) shows the analytical phase shifts $\theta_{n,p}$ and the numerically measured edge density $\rho$ of the ground state of a the gapless system $N=30$ and its two neighbouring gapped systems $N=29$ and $N=31$ for varying system lengths $L$. The edge density of the gapless system $N=30$ goes as $1/L$ whereas the edge density gapped systems $N=29$ and $N=31$ tends to zero quickly, both in accordance with Eq.~(\ref{eq:C1_edge_density}). However, the edge density for \textit{all} systems is prominent if the length is small even if the system is gapped. It is worth noting that, despite the fact that the analytic results have been derived in the large $L$ limit where the continuum approximation holds, the numerics and analytics are in surprisingly good agreement even for very small $L$. This verifies the theoretically predicted relativistic effects of nanotubes with small length $L$ where the violation of the non-relativistic zero edge density is expected.

To explain the system size dependence of the charge density, note that for very small $L$ the allowed momenta $p$ satisfying Eq.~(\ref{eq:C1_transcendental_eq}) become very large. In this case, the imaginary contribution to the phase $\theta_{n,p}  = \arg(\Delta_n + ie_np)$ dominates, giving $\theta_{n,p}\approx \pi/2$ even if the gap is non-zero, as seen in the right-hand column of Fig.~\ref{fig:phases_densities}. Hence, the edge density of Eq.~(\ref{eq:C1_edge_density}) becomes significant for small system sizes. For the gapless case, the phase is exactly equal to $\pi/2$ regardless of the value of $p$ or system size $L$. This yields a uniform charge density throughout the nanotube, resulting in the $1/L$ edge density as observed.

Finally, the inset of Fig.~\ref{fig:phases_densities} shows the integrated local density of states (LDOS) on the edge at $x=0$ given by 
\begin{equation}
N(E,\mathbf{r}) = \sum_m \rho_m(\mathbf{r}) \Theta(E-E_m),
\end{equation}
where $\rho_m(\mathbf{r})$ is the unit cell charge density of the $m$th eigenstate of the $2$D model with eigenvalue $E_m$ and $\Theta$ is the Heaviside step function. We present this for systems $N=29,30,31$ and $L= 25,100$. The edge LDOS is maximised for a fixed $L$ when the system is gapless, so for $N=30$ in this case. Moreover, the LDOS increases as the system size decreases, which provides a clear signature for the observation of the relativistic edge effect which can be measured in the lab using scanning tunnelling microscopy (STM) in the lab~\cite{C1_Andrei, C1_Kim,C1_Venema,C1_Hassanien}.

To summarise, the edge density is prominent if either the system is gapless or the system length $L$ is small. The typical lattice constant of a nanotube is given by $|\mathbf{n}_1| = |\mathbf{n}_2| \approx 2.46$\AA ~\cite{C1_Altland,C1_Saito} (\AA = $10^{-10}$m), so Fig.~\ref{fig:phases_densities} applies to systems on the order of $1$nm in diameter and $10$nm in length. However, the dependence on whether the system is gapless or not is very strong, so we expect these results to hold for much larger systems too.

\subsection{Relativistic spinors from non-relativistic wavefunctions}
\subsubsection{Sublattice wavefunctions}

\begin{figure}[t!]
\begin{center}
\includegraphics[scale=1]{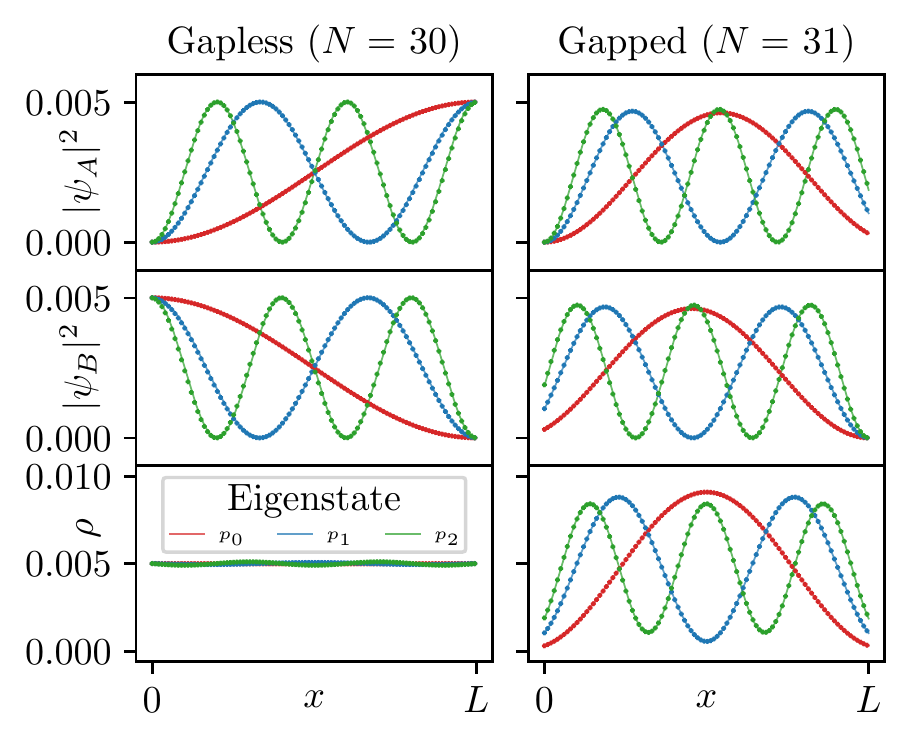}
\caption{A comparison of the analytical wavefunctions $|\psi_A|^2$ and $|\psi_B|^2$ of Eq.~(\ref{eq:C1_nanotube_wavefunctions}) and charge densities $\rho$ of Eq.~(\ref{eq:C1_edge_density}) to the numerical simulation (markers) for the gapless system $(N,L) = (30,200)$ and gapped system $(N,L) = (31,200)$. We present the first three excited states above the Fermi energy $E=0$.}
\label{fig:wavefunctions}
\end{center}
\end{figure}

To explain the emergence of relativistic boundary effects from a non-relativistic model, we focus on the sublattice wavefunctions $\psi_A$ and $\psi_B$. For concreteness, we examine a nanotube of dimension $(N,L) = (30,200)$ and $(N,L) = (31,200)$ which have gapless and gapped spectra, respectively.


In the left-hand column of Fig.~\ref{fig:wavefunctions} we compare the numerical sublattice wavefunctions $\psi_A(x)$, $\psi_B(x)$ and the charge densities $\rho(x)$ to the analytical results of Eq.~(\ref{eq:C1_nanotube_wavefunctions}) and Eq.~(\ref{eq:C1_rho}) respectively, for the first three excited states above the Fermi energy for the gapless system $(N,L) = (30,200)$. We see that the sublattice wavefunctions $\psi_A$ and $\psi_B$ are highly out of phase and maximise the edge support at $x=L$ and $x=0$ respectively, yielding a charge density $\rho(x)$ with minor oscillations about the predicted uniform value of $ 1/L = 0.005$. These oscillations are caused by finite-size effects.

In the right-hand column of Fig.~\ref{fig:wavefunctions}, we present the same information for the gapped system $(N,L) = (31,200)$. Despite $N$ increasing only by $1$, the fact the system now has a gap results in wavefunctions $\psi_A(x)$ and $\psi_B(x)$ that contrast considerably to the gapless case, with a charge density $\rho(x)$ that displays a more Schr\"odinger-like oscillatory profile. As the system size $L$ increases, the relative phase shift $\theta_{n,p}$ modulo $\pi$ between $\psi_A(x)$ and $\psi_B(x)$ decreases, as seen in Fig.~\ref{fig:phases_densities}(b), and the wavefunctions begin to display the Schr\"odinger-like profile that tends to zero on the boundaries. However, this is not the case for gapless systems as the phase shift is always $\pi/2$ regardless of system size, as seen in Fig.~\ref{fig:phases_densities}(b).


\subsubsection{Total lattice wavefunctions}

\begin{figure}[t]
\begin{center}
\includegraphics[scale=1]{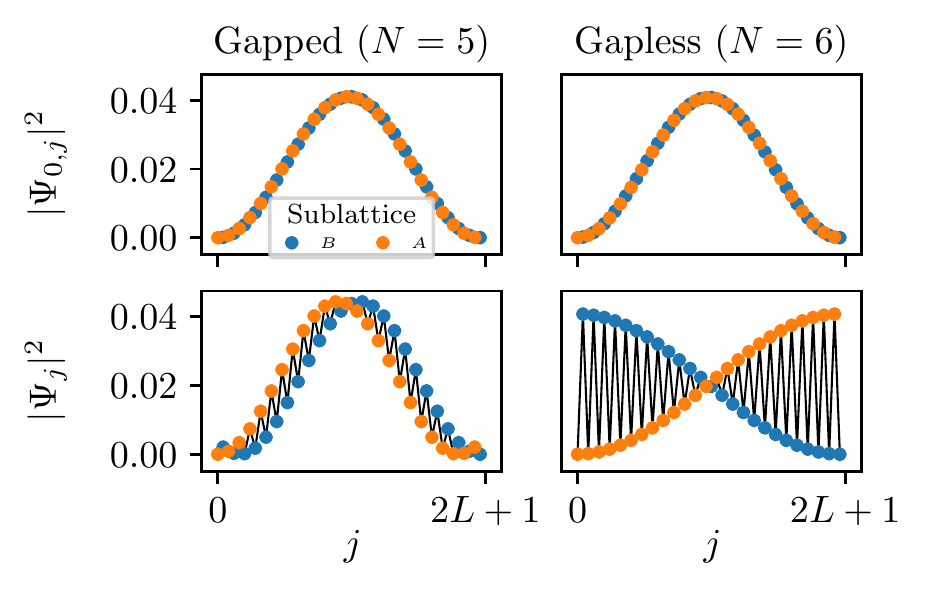}
\caption{The full single-particle ground state $\Psi_{0,j}$ and the first state above the Fermi energy $\Psi_j$ for systems of size $(N,L) = (5,25)$ and $(N,L) = (6,25)$. Comparing this with Fig.~\ref{fig:wavefunctions}, we see how the sublattice wavefunctions $\psi_A$ and $\psi_B$ described by the spinor Eq.~(\ref{eq:C1_nanotube_wavefunctions}) emerge.}
\label{fig:full_wavefunctions}
\end{center}
\end{figure}

We now analyse the total wavefunctions $\Psi_j$ of the lattice fermions, where $j \in \mathbb{N}$ is the real space coordinate of the bipartite lattice, alternating between sublattices $A$ and $B$. This coordinate should be contrasted to the unit cell coordinate $x$ of the spinor $\psi(x)$. Fig.~\ref{fig:full_wavefunctions} shows the wavefunctions of the single-particle eigenstate with the most negative energy below the Fermi energy, $\Psi_{0,j}$, and the first single-particle eigenstate above the Fermi energy, $\Psi_j$, for the gapped system of dimension $(N,L) = (5,25)$ and the gapless system of dimension $(N,L) = (6,25)$. 

The wavefunctions $\Psi_{0,j}$ and $\Psi_j$ are both non-relativistic wavefunctions which vanish on the boundaries. This is to be expected as the microscopic model is non-relativistic. However, due to high frequency oscillations, the support of $\Psi_j$ on each sublattice is highly out of phase if the system is gapped. Comparing with Fig.~\ref{fig:wavefunctions}, we see that these oscillations give the impression of two separate wavefunctions faithfully described by the components of a Dirac spinor.

For the gapped systems of dimension $(N,L) = (5,25)$, the wavefunctions display a Schr\"odinger-like wavefunction for both $\Psi_{0,j}$ and $\Psi_j$, with the edge density tending to zero if the system length $L$ is large. This can be seen clearly in the left-hand column of Fig.~\ref{fig:full_wavefunctions} where the sublattice wavefunctions are almost in phase. 

On the other hand, for the gapless system of dimension $(N,L) = (6,25)$ we see the sublattice wavefunctions are completely out of phase, which is responsible for the maximal edge density. We can describe this wavefunction as a Schr\"odinger wavefunction as $\Psi_j  \propto \sin(pj)$, for momenta $p = (l+1)\pi/2l$, where $l = 2L+1$ is the total length of the bipartite chain and $L+1$ is the number of unit cells in length of the nanotube. The contrasting behaviour of the models when simply increasing $N$ by one is quite remarkable.

In summary, the emergent relativistic physics described by the two-component spinor of Eq.~(\ref{eq:C1_nanotube_wavefunctions}) is a consequence of aliasing from sampling a high-frequency continuous wavefunction at discrete intervals. This effect is independent of length $L$. Such high frequency wavefunctions correspond to eigenstates that lie close to the Fermi energy in the middle of the single-particle spectrum which is where the relativistic physics of graphene emerges~\cite{Wallace,C1_Neto}.

\section{Conclusion \label{nanotube_conclusion}}
Our analysis demonstrates that relativistic effects can dominate certain geometries of Dirac materials, resulting in large edge support. We studied this effect analytically and numerically for zig-zag carbon nanotubes and demonstrated that it is dominant when the system is either gapless or has a small length. This demonstrates that the relativistic effects are strong within system dimensions found within experiments of real carbon nanotubes. In fact, this work has gained interest from our collaborators in China who are attempting to test this effect in the laboratory with real carbon nanotubes.

Despite testing this for carbon nanotubes, this relativistic effect is general and it is expected to be present in 1D, 2D and 3D quantum lattice models with the same qualitative properties presented here. While high edge densities of bulk states should be measurable with STM \cite{C1_Andrei, C1_Kim,C1_Venema,C1_Hassanien}, it is expected to have a significant effect on the conductivity of the material when attaching leads to its boundaries or its response to a magnetic field \cite{C1_Laird,C1_Marganska,C1_Marganska2,C1_Hiroshi} as the boundary LDOS is significant. In addition, determining if such effects will be present in 2D materials containing a finite density of defects which effectively imposes boundary conditions on the wavefunctions within the material will be intriguing \cite{C1_Algharagholy,C1_Araujo,C1_Neto,C1_Dutreix}. We leave these questions for a future work.

\chapter{Kitaev's honeycomb model and Riemann-Cartan geometry \label{chapter:RC_background}}

\section{Introduction}
In this chapter we provide the relevant background material required for chapters~\ref{chapter:kitaev} and \ref{chapter:chiral}. First, in Sec.~\ref{sec:C2_Kitaev's honeycomb model}, we introduce Kitaev's honeycomb model, including the relativistic continuum limit and exact solution which closely follows Refs.~\cite{C2_Kitaev} and \cite{C2_PachosBook}. Then in Sec.~\ref{sec:Riemann_Cartan} we introduce Riemann-Cartan theory which closely follows Ref.~\cite{C2_Nakahara}. We then finish with a brief conclusion in Sec.~\ref{rc_conclusion}.
\section{Kitaev's honeycomb model \label{sec:C2_Kitaev's honeycomb model}}

\subsection{The spin Hamiltonian}

\begin{figure}[t]
\begin{center}
\includegraphics[width=0.75\textwidth]{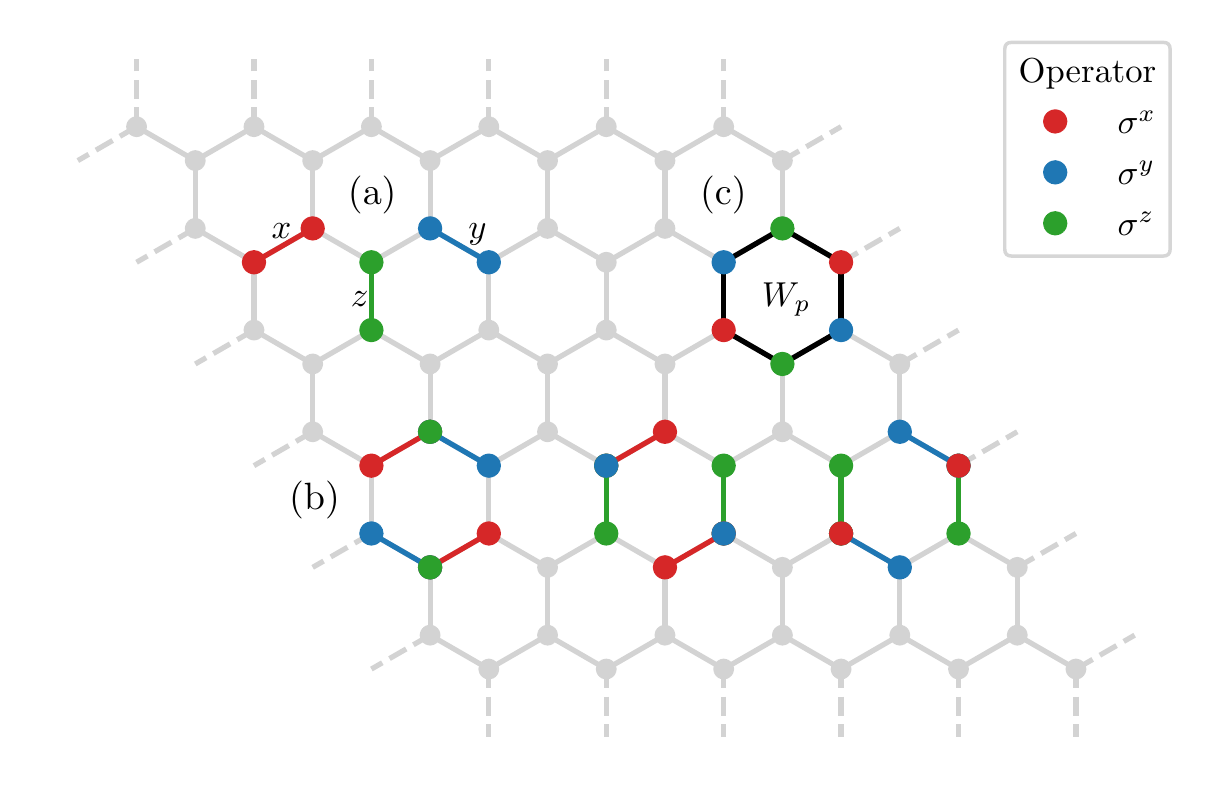}
\end{center}
\caption{(a) The three types of links, $x$, $y$ and $z$, and their corresponding two-body interactions $K_{ij}$. (b) The six types of three-body interactions. (c) A hexagonal plaquette $p$ with its corresponding plaquette operator $W_p$ }
\label{fig:C2_kitaev_honeycomb}
\end{figure}

Kitaev's honeycomb model describes a honeycomb lattice of interacting spin-$1/2$ particles which interact via two- and three-spin interactions. We label the lattice sites with the indices $i,j, \ldots$ and the three types of links with the index $\alpha \in \{ x,y,z\}$ depending on their orientation, as shown in Fig.~\ref{fig:C2_kitaev_honeycomb}(a). We introduce a unit cell as the two sites forming a $z$-link and label these $A$ and $B$.

For every pair of spins $(i,j)$ forming an $\alpha$-link, we define the two-spin interaction $K_{ij} = \sigma^\alpha_i \sigma^\alpha_j,$ as shown in Fig.~\ref{fig:C2_kitaev_honeycomb}(b), where $\sigma^\alpha_i$ are the $\alpha$ Pauli matrices acting only on the $i$th lattice site with an identity acting on all other sites. From the two-spin interactions, we define the three-body interactions $K_{ijk} =-i K_{ij}K_{jk}$ where we perform the product in a clockwise direction $(i,j,k)$ around the each plaquette (to obtain an overall positive sign), as shown in Fig.~\ref{fig:C2_kitaev_honeycomb}. There are six terms in total. The Hamiltonian is given by summing up these interactions across the honeycomb lattice which gives us
\begin{equation}
H = - J_x \sum_{(i,j):x} \sigma^x_i \sigma^x_j -  J_y \sum_{(i,j):y} \sigma^y_i \sigma^y_j - J_z \sum_{(i,j):z} \sigma^z_i \sigma^z_j - K \sum_{(i,j,k)} \sigma^x_i \sigma^y_j \sigma^z_k, \label{eq:C2_kitaev_spin_ham}
\end{equation}
where $(i,j): \alpha$ denotes a sum over lattice sites $i$ and $j$ that form an $\alpha$ link; $(i,j,k)$ denotes a sum over all three-spin interactions; and $J_x,J_y,J_z,K \in \mathbb{R}$ are coupling constants. The three-body interactions of the model arise as a perturbation due to the interaction of the model with an external magnetic field. This breaks the time-reversal symmetry of the Hamiltonian and opens a gap, allowing for a non-trivial Chern number. In this way, the model is topological in nature.

In order to solve this model, we first look for any conserved charges. Consider the plaquette operator $W_p$ defined as the product of all two-spin interactions around each hexagonal plaquette $p$, where
\begin{equation}
W_p = K_{12} K_{23} K_{34} K_{45} K_{56} K_{61} = \sigma^x_1 \sigma^y_2 \sigma^z_3 \sigma^x_4 \sigma^y_5 \sigma^z_6, \label{eq:C2_plaquette_operator}
\end{equation}
as shown in Fig. \ref{fig:C2_kitaev_honeycomb}(c). From its construction, the plaquette operators commute with the Hamiltonian and amongst themselves as $[ H, W_p ] = [W_p,W_q] = 0$ for all $p$ and $q$. Therefore, the Hilbert space of the model $\mathcal{H}$ can be divided up into eigenspaces of each $W_p$ as
\begin{equation}
\mathcal{H} = \bigoplus_{ w } \mathcal{H}_w,
\end{equation}
where $w = \{ w_p \}$ labels the distribution of plaquette eigenvalues across the lattice and $\mathcal{H}_w$ is the corresponding eigenspace. We call each subspace a \textit{vortex sector} which are invariant subspaces of the Hamiltonian.

For a honeycomb lattice with $N$ vertices, the dimension of the Hilbert space is $\mathrm{dim}(\mathcal{H}) = 2^N$ as each spin has a two-dimensional Hilbert space. In addition, the number of plaquettes of the lattice is $M = N/2$, where we assume that $N$ is even. As the plaquette operators obey $W_p^2 = \mathbb{I}$, the eigenvalues are given by $w_p = \pm 1$ which means that there are $2^{M}$ possible distributions of the plaquette eigenvalues and hence $2^{M}$ subspaces $\mathcal{H}_w$. Therefore, we must have $\mathrm{dim}(\mathcal{H}_w) = \mathrm{dim}(\mathcal{H})/2^{M} = 2^{N/2}$. As each subspace is an invariant subspace under the Hamiltonian $H$, we can focus on each subspace individually which significantly reduces the complexity of the problem.

\subsection{Fermionising the model}
The identification of the invariant subspaces significantly helps to diagonalise the Hamiltonian. In fact, the plaquette operators $W_p$ can be interpreted as the ``magnetic flux" of a particular gauge field, more precisely a \textit{Wilson loop}. This gauge theoretic interpretation is the key to solving the model, however as it stands we do not have enough ``room" in our model for it to be described as a gauge theory. This is because a gauge theory contains \textit{redundancy} in the model, where multiple states in the Hilbert space correspond to the same physical state. As it stands, the model does not have this redundancy and the vector potential corresponding to the Wilson loops $W_p$ is hidden. For this reason, we extend the dimension of the Hilbert space of the model by introducing new \textit{unphysical states} to the problem which will be related via gauge transformations.

To reformulate the model as a gauge theory, we rewrite the Hamiltonian in the language of Majorana modes. First, consider a set of fermionic creation and annihilation operators, $f^\dagger_i$ and $f_i$, where $i \in \mathbb{N}$ labels the species. These operators obey the usual fermionic anti-commutation relations $\{ f_j, f_j^\dagger \} = \delta_{ij}$ and $\{ f_i, f_j \} = \{f^\dagger_i, f_j^\dagger \} = 0$. Given a vacuum state $|0\rangle$ obeying $f_i|0\rangle = 0 \ \forall i$, a single fermion generates a two-dimensional Hilbert space spanned by the states $|0\rangle $ and $f^\dagger_i |0\rangle$. From the fermionic modes, we define the \textit{Majorana} modes
\begin{equation}
c_{2i - 1} = f_i + f_i^\dagger, \quad c_{2i} = -i(f_i - f_i^\dagger),
\end{equation}
which are Hermitian, $c_i^\dagger = c_i$, square to the identity, $c_i^2 = \mathbb{I}$, and obey the anti-commutation relations $\{ c_i, c_j \} = 2 \delta_{ij}$. 

For every lattice site $i$ on the honeycomb, we have a spin-$1/2$ particle with a corresponding two-dimensional Hilbert space $\mathcal{H}_i$ which we refer to as a \textit{physical subspace}. We now introduce a pair of fermionic modes $f_{1,i}$ and $f_{2,i}$ to every lattice site, with their corresponding vacuum state $|00\rangle_i$ obeying $f_{1,i} |00\rangle_i = f_{2,i}|00\rangle_i = 0$. These generate a four-dimensional \textit{extended} Hilbert space $\tilde{\mathcal{H}}_i$ spanned by the four states
\begin{equation}
\mathrm{Physical} \  
\begin{cases}
|00\rangle_i  \\
|11\rangle_i = f^\dagger_{1,i} f^\dagger_{2,i} |00\rangle_i 
\end{cases}, \quad  
\mathrm{Unphysical} \  
\begin{cases}
|01\rangle_i = f^\dagger_{2,i} |00\rangle_i \\
|10\rangle_i = f^\dagger_{1,i}  |00\rangle_i
\end{cases},
\end{equation}
which we have arbitrarily divided up into physical and unphysical states by choosing our physical states via the isomorphism $\mathcal{H}_i \cong \mathrm{span}\{ |00\rangle_i , |11\rangle_i \}$. By introducing these two fermionic modes, the additional unphysical states have doubled the degrees of freedom of the model. If we define the operator
\begin{equation}
D_i = (1-2f_{1,i}^\dagger f_{1,i})(1-2f_{2,i}^\dagger f_{2,i}),
\end{equation}
we see that $D_i |\psi \rangle = |\psi \rangle $ for all physical states $|\psi \rangle \in \mathcal{H}_i$. This is an alternative way to define our physical subspace and is a constraint that all physical states must obey. In addition, as $\mathcal{H}_i \subset \tilde{\mathcal{H}}_i$, we can work on the extended Hilbert space as long as we project out any unphysical states that do not live in the physical subspace at the end of our calculation.

From the two fermionic modes, we introduce a set of four Majorana modes 
\begin{equation}
c_i = f_{1,i} + f_{1,i}^\dagger , \quad b_i^x = -i(f_{1,i} - f^\dagger_{1,i}), \quad b_i^y = f_{2,i} + f_{2,i}^\dagger, \quad b_i^z = -i (f_{2,i} - f^\dagger_{2,i}).
\end{equation}
In terms of these Majoranas, we have $D_i = b_i^x b_i^y b^z_i c_i$. From the Majoranas, we additionally define the operators 
\begin{equation}
\tilde{\sigma}_i^\alpha = i b^\alpha_i c_i, \label{eq:pauli_majorana}
\end{equation}
that, when restricted to the physical subspaces, obey the same algebra as the Pauli matrices and commute with $D_i$, so on the physical subspaces they furnish an irreducible representation of the Pauli matrices. We now substitute the Majorana representation of the Pauli matrices of Eq.~(\ref{eq:pauli_majorana}) into the spin Hamiltonian of Eq.~(\ref{eq:C2_kitaev_spin_ham}), so the two- and three-spin interactions are mapped to
\begin{equation}
\sigma^\alpha_i \sigma^\alpha_j = - i \hat{u}_{ij} c_i c_j, \quad \sigma^x_i \sigma^y_j \sigma^z_k =  i \hat{u}_{ik} \hat{u}_{jk} D_k c_i c_j,
\end{equation}
where we have defined the link operators 
\begin{equation}
\hat{u}_{ij} = \begin{cases} ib^\alpha_i b^\alpha_k & \text{if $(i,j)$ forms an $\alpha$-link} \\
0 & \text{otherwise}
\end{cases},
\end{equation}
which obey $\hat{u}_{ij}^\dagger = \hat{u}_{ij}$, $\hat{u}_{ij}^2 = \mathbb{I}$ and $\hat{u}_{ij} = -\hat{u}_{ij}$. Restricting the operators to the physical subspaces by setting $D_i = +1$ for all lattice sites, the original spin Hamiltonian of Eq.~(\ref{eq:C2_kitaev_spin_ham}) takes the form
\begin{equation}
H = \frac{i}{4} \sum_{i,j} \hat{A}_{ij} c_i c_j, \quad \hat{A}_{ij} = 2J_{ij} \hat{u}_{ij} + 2K \sum_k \hat{u}_{ik} \hat{u}_{jk}, \label{eq:C2_kitaev_majorana}
\end{equation}
where $J_{ij}$ is the corresponding nearest-neighbour coupling for the link $(i,j)$. Note that this Hamiltonian is highly interacting still due to the fact the coefficients $\hat{A}_{ij}$ are operators.
\subsection{Emergent $\mathbb{Z}_2$ lattice gauge theory}
Extending the Hilbert space by introducing Majorana modes gives us additional freedom to interpret the model as a $\mathbb{Z}_2$ gauge theory, revealing that $\hat{u}_{ij}$ can be viewed as a \textit{Wilson line} which, roughly speaking, is the vector potential of the gauge theory. The Hamiltonian $H$ given by Eq.~(\ref{eq:C2_kitaev_majorana}) is an operator defined on the physical subspace only. We now extend the domain of this operator to the extended Hilbert space, giving us a new operator $\tilde{H}$ which is identical in form and whose restriction to the physical subspace is the original Hamiltonian $H$. The Hamiltonian $\tilde{H}$ has the symmetry 
\begin{equation}
[\tilde{H},D_i ] = 0,
\end{equation}
so eigenstates of $\tilde{H}$ in the extended subspace will be eigenstates of $H$ in the physical subspace after projecting down. Therefore, we search for states $|\psi \rangle$ which simultaneously obey
\begin{equation}
\tilde{H} |\psi \rangle = E|\psi \rangle, \quad D_i |\psi\rangle = |\psi \rangle. \label{eq:C2_phys_subspace}
\end{equation}
In order to find these states we note that the link operators $\hat{u}_{ij}$ are local symmetries on the extended space as $[\tilde{H},\hat{u}_{ij}]= 0$, so the extended space can be decomposed into eigenspaces as
\begin{equation}
\tilde{\mathcal{H}} = \bigoplus_u \tilde{\mathcal{H}}_u,
\end{equation}
where $u = \{ u_{ij} \}$ represents a specific configuration of eigenvalues of link operators across the lattice. By restricting the problem to one of these subspaces, we can replace the link operators in $\tilde{H}$ with their eigenvalues, giving us an exactly diagonalisable Hamiltonian that is quadratic in Majoranas $c_i$. However, the operators $\hat{u}_{ij}$ and $D_i$ do not commute, but instead anti-commute as $\{ D_i, \hat{u}_{jk} \} = 0$, therefore subspaces of fixed $u$ are unphysical because the constituent states do not obey the second constraint of Eq.~(\ref{eq:C2_phys_subspace}). In fact, applying $D_i$ to a state $| \psi_u \rangle \in \tilde{\mathcal{H}}_u$ flips the sign of all link eigenvalues $u_{ij}$ joining the $i$th site, mapping the state to a different subspace. On the other hand, the plaquette operators in the language of Majoranas on the extended Hilbert space become products of link operators around each plaquette $p$ as
\begin{equation}
\tilde{W}_p = \prod_{i,j \in \partial p} \hat{u}_{ij},
\end{equation}
which importantly commute with both $D_i$ and $\tilde{H}$, and reduce to the original definition of the plaquette operators from Eq.~(\ref{eq:C2_plaquette_operator}) on the physical subspace, so the extended space decomposes into vortex sectors as before. As we have interpreted $\hat{u}_{ij}$ as a $\mathbb{Z}_2$ Wilson line, we can now clearly see that $\tilde{W}_p$ are the Wilson loops which roughly represents the ``magnetic field".

In order to rectify this issue, consider the eigenstate $|\psi_u \rangle \in \tilde{\mathcal{H}}_u$. We generate a physical state from this by projecting down to the physical subspace as
\begin{equation}
|\psi_w \rangle = P |\psi_u \rangle \in \mathcal{H}, \quad P = \prod_i \left(  \frac{\mathbb{I} + D_i}{2} \right), \label{eq:C2_physical_projector}
\end{equation} 
which now obeys the constraints of Eq.~(\ref{eq:C2_phys_subspace}) but is no longer an eigenstate of the link operators $\hat{u}_{ij}$. Due to the form of the projector $P$ in Eq.~(\ref{eq:C2_physical_projector}), the resulting state contains a superposition of all possible eigenvalue distributions of $u_{ij}$ that correspond to a particular vortex sector $w$. This projection is a many-to-one mapping as all states in a particular vortex sector in the extended space are mapped to the same physical state under the projection. As the physical observables of energy $E$ and vortex sector $w$ are unaffected by the projection, we can simply choose a particular distribution of $u_{ij}$ in the extended space that yields the correct vorticity $w$. There are many distributions of $u_{ij}$ that preserve these physical properties which all differ by application of $D_i$, which gives the identification of $u_{ij}$ as a classical $\mathbb{Z}_2$ gauge theory where $D_i$ are the local gauge transformations. Therefore, the original spin Hamiltonian restricted to a particular vortex sector $w$ is equivalent to the Hamiltonian
\begin{equation}
\tilde{H}_w = \frac{i}{4} \sum_{i,j} A^w_{ij} c_i c_j , \quad A^w_{ij} = 2J_{ij} u_{ij} + 2K \sum_k u_{ik} u_{jk} , \label{eq:C2_gauge_fixed_ham_majorana}
\end{equation}
where the matrix $A_{ij}^w$ is a now a  matrix of real numbers so the Hamiltonian is quadratic in Majoranas $c_i$ and can be solved exactly. This is the Hamiltonian we work with and it is an operator defined on the extended Hilbert space. It is worth stressing that even though we do our calculations on the extended Hilbert space by extending the domain of operators, all gauge invariant observables such as the energy spectrum of the Hamiltonian and eigenvalues of the plaquette operators are unaffected by the projection to the physical subspace with $P$, so we do not require its application in the following chapters and hence we can work in the extended Hilbert space for ease of calculation.
\subsection{Solving the model}

\begin{figure}[t]
\begin{center}
\includegraphics[scale=1,valign=t]{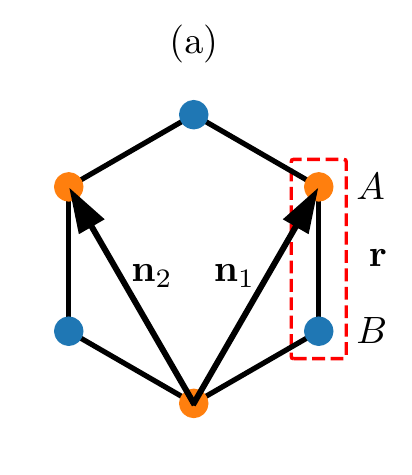}
\hspace{0.5cm}
\includegraphics[width=0.625\textwidth,valign=t]{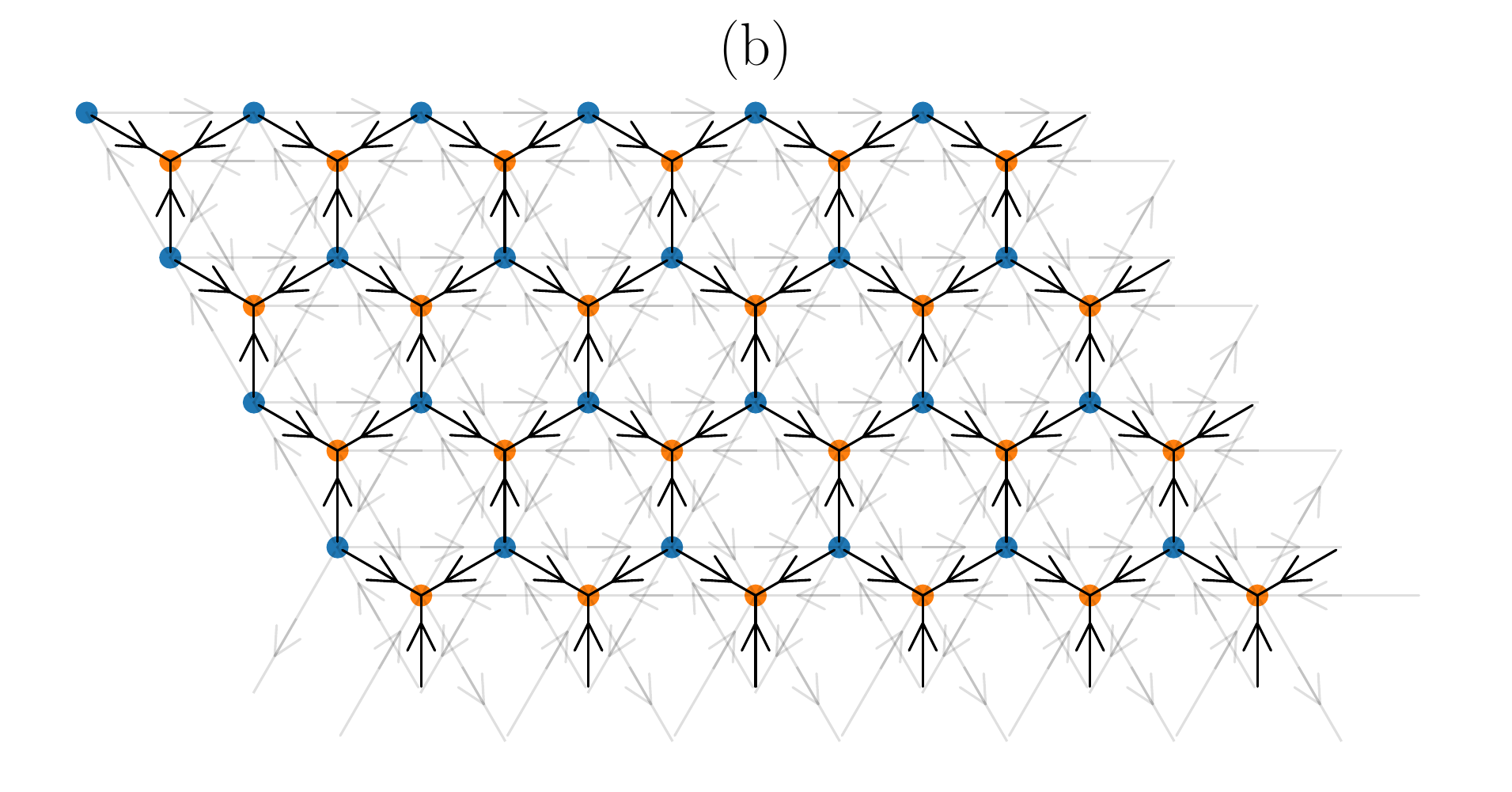}
\end{center}
\caption{(a) Similarly to graphene, we can bi-colour the honeycomb lattice with two sublattices, $A$ and $B$, and choose the unit cell at $\mathbf{r}$ as the pair of sites forming a $z$-link, along with the choice of lattice generators $\mathbf{n}_{1,2} = \frac{a}{2} \left( \pm 1 , \sqrt{3} \right)$, where $|\mathbf{n}_{1,2}| = a$. (b)  A particular gauge choice of the $\mathbb{Z}_2$ gauge field $u_{ij}$ generating the vortex-free sector, $w_p = +1$ for all $p$, where the black arrows indicate the direction for which $u_{ij} = +1$ for the nearest-neighbour links and the grey arrows indicate the direction for which $u_{ij}u_{ik} = +1$ for the next-to-nearest-neighbour links, as given in Eq. (\ref{eq:C2_gauge_fixed_ham_majorana}). }
\label{fig:C2_gauge_fix}
\end{figure}

Just like for graphene, we use the diatomic labelling with the two-site unit cell as shown in Fig~\ref{fig:C2_gauge_fix}(a), and we introduce a set of Majoranas $a_\mathbf{r}$ and $b_\mathbf{r}$ for lattice sites $A$ and $B$ within the unit cell $\mathbf{r} \in \Lambda$ respectively, where the commutation relations read $\{ a_\mathbf{r},a_{\mathbf{r}'} \} = \{ b_\mathbf{r},b_{\mathbf{r}'} \} = 2 \delta_{\mathbf{r},\mathbf{r}'}$, whilst all other anti-commutators vanish.
Let us also consider the isotropic vortex-free sector where $J_x = J_y = J_z = J$ and $w_p = +1$ for all plaquettes $p$. In order to generate the vortex-free sector, we fix the $u_{ij}$ in the pattern shown in Fig.~\ref{fig:C2_gauge_fix}(b), where $u_{ij} = +1$ for all positively oriented links represented by the direction of the arrow. As the matrix $A_{ij}^w$ can be split into two terms, we write the Hamiltonian as $H = H_J + H_K$. With our labelling convention, the Hamiltonian takes the form
\begin{align}
H_J & = \frac{i}{4} \sum_{\mathbf{r} \in \Lambda} 2J a_\mathbf{r} \left( b_\mathbf{r} + b_{\mathbf{r} + \mathbf{n}_1 } + b_{\mathbf{r} + \mathbf{n}_2} \right) + \mathrm{H.c.}, \label{eq:C2_H_J} \\
H_K & = \frac{i}{4} \sum_{\mathbf{r} \in \Lambda} 2K a_\mathbf{r}\left( a_{\mathbf{r} + \mathbf{n}_1} - a_{\mathbf{r} + \mathbf{n}_2} - a_{\mathbf{r} + \mathbf{n}_1 -  \mathbf{n}_2} \right) - (a \leftrightarrow b) + \mathrm{H.c.}, \label{eq:C2_H_K}
\end{align}
where we choose the generators of the honeycomb lattice $\mathbf{n}_{1,2} = \frac{a}{2} \left( \pm 1 , \sqrt{3} \right)$ as shown in Fig.~\ref{fig:C2_gauge_fix}(a), where $|\mathbf{n}_{1,2}| = a $ is the lattice constant.

\begin{figure}
\begin{center}
\begin{subfigure}{0.4\textwidth}
\begin{center}
(a)
\includegraphics[width=\textwidth]{honeycomb_dispersion.pdf}
\end{center}
\end{subfigure}
\hspace{1cm}
\begin{subfigure}{0.4\textwidth}
\begin{center}
(b)
\includegraphics[width=\textwidth]{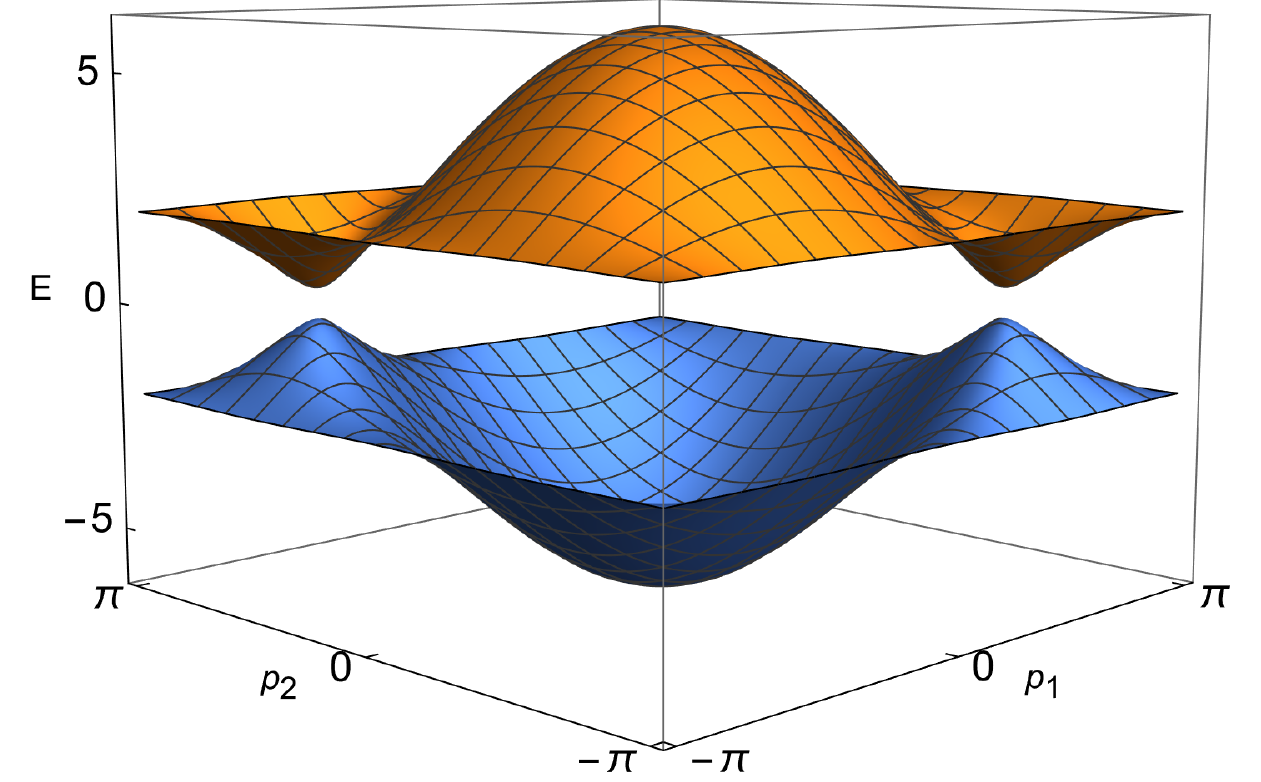}
\end{center}
\end{subfigure}
\end{center}
\caption{The dispersion relation of Eq.~(\ref{eq:C2_dispersion}) for $J = 1$, with $K = 0$ in subfigure (a) and $K = 0.1$ in subfigure (b). We see that the $K$ term opens up a gap in the dispersion relation, where the axes are given by the reciprocal basis coordinates $p_i = \mathbf{p} \cdot \mathbf{n}_i$.}
\label{fig:C2_kitaev_dispersion}
\end{figure}

The Hamiltonian has discrete translational symmetry, so we Fourier transform our Majoranas as
\begin{equation}
a_\mathbf{r} = \frac{1}{\sqrt{N}} \sum_{\mathbf{p} \in \mathrm{B.Z.}} e^{i \mathbf{p} \cdot \mathbf{r}} a_\mathbf{p},
\end{equation}
and similarly for $b_\mathbf{r}$, where $N$ is the number of unit cells in the lattice and the momentum space anti-commutation relations are given by $\{ a_\mathbf{p} , a^\dagger_\mathbf{q} \}  = \{ b_\mathbf{p} , b^\dagger_\mathbf{q} \} = 2\delta_{\mathbf{p},\mathbf{q}} $, whilst all other anti-commutators vanish. Due to the hermiticity of the Majoranas in real space, the momentum space modes obey $a^\dagger_\mathbf{p} = a_{-\mathbf{p}}$ and $b_\mathbf{p}^\dagger = b_{-\mathbf{p}}$. Substituting the Fourier transform into the Hamiltonian yields
\begin{equation}
H  =  \frac{1}{4}  \sum_{\mathbf{p} \in \mathrm{B.Z.}} \left[ i f(\mathbf{p}) a^\dagger_\mathbf{p} b_\mathbf{p} +  \Delta(\mathbf{p}) \left(a^\dagger_\mathbf{p} a_\mathbf{p} - b^\dagger_\mathbf{p} b_\mathbf{p} \right) \right] + \mathrm{H.c.} ,
\end{equation}
which follows a similar route to the calculation for graphene in Eq.~(\ref{eq:C1_fourier_transform_H}), where we have defined
\begin{align}
f(\mathbf{p}) & = 2J \left(1 + e^{i \mathbf{p} \cdot \mathbf{n}_1} + e^{i \mathbf{p} \cdot \mathbf{n}_2} \right), \\
\Delta (\mathbf{p}) & = 4K\left[ - \sin(\mathbf{p} \cdot \mathbf{n}_1) + \sin(\mathbf{p} \cdot \mathbf{n}_2) + \sin(\mathbf{p} \cdot (\mathbf{n}_1 - \mathbf{n}_2) \right].
\end{align}
Pulling everything together, if we define the two-component spinor $\chi_\mathbf{p} = (a_\mathbf{p},ib_\mathbf{p})$, where the $i$ absorbs the $i$ from the off-diagonal terms, we can rewrite the Hamiltonian as
\begin{equation}
H  = \frac{1}{4} \sum_{\mathbf{p} \in \mathrm{B.Z.}} \chi^\dagger_\mathbf{p} h(\mathbf{p}) \chi_\mathbf{p}, \quad 
h(\mathbf{p}) = 
\begin{pmatrix}
\Delta(\mathbf{p}) & f(\mathbf{p}) \\
f^*(\mathbf{p}) & -\Delta(\mathbf{p}) 
\end{pmatrix}, \label{eq:C2_diagonal_hamiltonian}
\end{equation}
where $h(\mathbf{p})$ is the single-particle Hamiltonian. Just like for graphene, the dispersion relation of the model is obtained from the eigenvalues of $h(\mathbf{p})$, giving us
\begin{equation}
E(\mathbf{p}) = \pm \sqrt{ |f(\mathbf{p})|^2 + \Delta^2(\mathbf{p})}  , \label{eq:C2_dispersion}
\end{equation}
which is shown in Fig.~\ref{fig:C2_kitaev_dispersion} for various values of $K$. We see that the dispersion relation is a gapped version of graphene's dispersion relation where $K$ controls the gap. 

\subsection{Continuum limit (two-dimensional representation) \label{sec:C2_2x2_cont_lim}}
\subsubsection{The Hamiltonian}
We are only interested in the low-energy properties properties of the model close to the ground state. As the model is fermionic when diagonalised in the form of Eq.~(\ref{eq:C2_diagonal_hamiltonian}), the ground state consists of filling up the Fermi sea of negative energy single-particle modes. For the case where $K = 0$, the dispersion relation is gapless and the upper and lower bands touch at two Fermi points $\mathbf{P}_\pm$, where $E(\mathbf{P}_\pm) = 0$, as seen in Fig.~\ref{fig:C2_kitaev_dispersion}. These points are given by
\begin{equation}
\mathbf{P}_\pm = \pm \left( \frac{4 \pi}{3 a} , 0 \right),
\end{equation}
which is expressed in Cartesian coordinates.

We first project our Hamiltonian onto a small neighbourhood of the Fermi points by Taylor expanding the Hamiltonian to first order about these points. We have
\begin{align}
f(\mathbf{P}_\pm + \mathbf{p}) & = \sqrt{3} aJ(\mp p_x - ip_y) + O(\mathbf{p}^2), \\
\Delta(\mathbf{P}_\pm + \mathbf{p}) & = \mp 6 \sqrt{3}K + O(\mathbf{p}^2),
\end{align}
where we assume that $|\mathbf{p}| $ is smaller than some $O(1/a)$ cut-off. We now take the continuum limit $a \rightarrow 0$ and thermodynamic limit $N \rightarrow \infty$ so the momentum $\mathbf{p}$ becomes a continuous variable on $\mathbb{R}^2$. In addition, we take the limit in such a way to keep the Fermi velocity $v_\mathrm{F} = |\nabla E(\mathbf{P}_\pm)| = \sqrt{3} aJ$ fixed. In order to do this, we renormalise the couplings as $aJ \rightarrow J$ and ensure that this quantity remains finite and constant as we take the continuum limit. Substituting this back into the single-particle Hamiltonian of Eq.~(\ref{eq:C2_diagonal_hamiltonian}) yields a Hamiltonian about each Fermi point defined as $h_\pm(\mathbf{p}) = h(\mathbf{P}_\pm + \mathbf{p})$, where
\begin{equation}
h_\pm(\mathbf{p}) = v_\mathrm{F} ( \mp \sigma^x p_x + \sigma^y p_y)  \mp \Delta \sigma^z ,
\end{equation}
where $v_\mathrm{F} = \sqrt{3} J$ is the Fermi velocity and $\Delta = 6 \sqrt{3}K$ is the gap. This takes the form of a momentum space Dirac Hamiltonian for a particle of mass $\Delta$ on a $(2+1)$D Minkowski spacetime with an effective speed of light given by the Fermi velocity $v_\mathrm{F}$.

In addition to taking the continuum limit of the single-particle Hamiltonian, we must address how to correctly take the continuum limit of the many-body Hamiltonian. In the language of second quantisation, we define the continuum limit fields about each Fermi point as
\begin{equation}
c_\mu(\mathbf{p}) = \lim_{\Delta p_x \rightarrow 0} \lim_{\Delta p_y \rightarrow 0} \frac{c_{\mathbf{P}_\mu + \mathbf{p}}}{\sqrt{2 \Delta p_x \Delta p_y}}, 
\end{equation}
where $\mu = \pm $ labels the Fermi points, $c = a,b$ labels the sublattice, and $\Delta p_x \Delta p_y$ is the Cartesian area element in momentum space. The factor of $1/\sqrt{2}$ is required to achieve standard commutation relations. With this, the continuum limit of the many-body Hamiltonian takes the form
\begin{equation}
H = \frac{1}{2} \sum_{\mu = \pm} \int \mathrm{d}^2 p \chi_\mu^\dagger(\mathbf{p}) h_\mu(\mathbf{p}) \chi_\mu(\mathbf{p}), \quad  \chi_\mu(\mathbf{p}) \equiv \begin{pmatrix}a_\mu(\mathbf{p}) \\ b_\mu(\mathbf{p}) \end{pmatrix}, \label{eq:C2_cont_ham_momentum}
\end{equation}
which is the Hamiltonian of a Dirac field on a $(2+1)$D Minkowski spacetime. 

Ultimately, we are interested in the continuum limit in real space as it will allow us to generalise to curved spacetimes later. Therefore, we return to position space with an inverse Fourier transform given by
\begin{equation}
c_\mu(\mathbf{x}) = \frac{1}{2 \pi} \int \mathrm{d}^2 x e^{i \mathbf{p} \cdot \mathbf{x}} c_\mu(\mathbf{p}), \label{eq:C2_fourier_transform_field}
\end{equation}
where now we see that, despite the original Majorana modes in real space being hermitian, the continuum fields are not as they obey $c^\dagger_\pm (\mathbf{x}) = c_\mp(\mathbf{x})$. With this, we find after Fourier transforming Eq.~(\ref{eq:C2_cont_ham_momentum}) the continuum limit in real space is given by
\begin{equation}
H = \frac{1}{2} \sum_{\mu = \pm} \int \mathrm{d}^2 x \chi^\dagger_\mu(\mathbf{x}) h_\mu(\mathbf{x}) \chi_\mu(\mathbf{x}), \quad  \chi_\mu(\mathbf{x}) \equiv \begin{pmatrix}a_\mu(\mathbf{x}) \\ b_\mu(\mathbf{x}) \end{pmatrix}, \label{eq:C2_2x2_field_ham}
\end{equation}
where the single-particle Hamiltonian now takes the form of the differential operator
\begin{equation}
h_\pm(\mathbf{x}) = -i v_\mathrm{F} \left( \mp \sigma^x \overset{\leftrightarrow}{\partial_x}  + \sigma^y \overset{\leftrightarrow}{\partial_y}  \right) \mp \Delta \sigma^z,
\end{equation}
and where we have defined the differential operator $A\overset{\leftrightarrow}{\partial_i}B \equiv \frac{1}{2}( A \partial_i B - (\partial_i A) B )$ which acts on spinor fields only.
\subsubsection{Commutation relations}
We now calculate the commutation relations of the continuum fields. Consider the field $a_\mu(\mathbf{p})$: we immediately see that
\begin{equation}
\begin{aligned}
\{ a_\mu(\mathbf{p}) , a^\dagger_\nu(\mathbf{q}) \} & = \lim_{\Delta p_x \rightarrow 0 } \lim_{\Delta p_y \rightarrow 0 } \frac{ \{  a_{\mathbf{P}_\mu + \mathbf{p}} , a^\dagger_{\mathbf{P}_\nu + \mathbf{p}} \} }{2 \Delta p_x \Delta p_y} \\
& = \lim_{\Delta p_x \rightarrow 0 } \lim_{\Delta p_y \rightarrow 0 } \frac{ \delta_{\mathbf{P}_\mu + \mathbf{p}, \mathbf{P}_\nu + \mathbf{q} }}{\Delta p_x \Delta p_y} \\
& = \delta(\mathbf{P}_\mu - \mathbf{P}_\nu + \mathbf{p} - \mathbf{q}) \\
& = \delta_{\mu \nu} \delta(\mathbf{p} - \mathbf{q} ),
\end{aligned}
\end{equation}
where in the final equality we have used the fact that for $\mu \neq \nu$ the delta function is $\delta( 2\mathbf{P}_\mu + \mathbf{p} - \mathbf{q})$ as $\mathbf{P}_\mu = -\mathbf{P}_\nu$, but as we are assuming that $|\mathbf{p} - \mathbf{q}| \ll |\mathbf{P}_\mu| = O(1/a)$ due to the momentum cut-off, then the argument of the delta function will never be zero. Through a similar reasoning, we also have
\begin{equation}
\{ a_\mu(\mathbf{p}), a_\nu(\mathbf{q}) \} = \{ a^\dagger_\mu(\mathbf{p}), a^\dagger_\nu(\mathbf{q}) \}  = 0.
\end{equation}
The field $b_\mu(\mathbf{p})$ obeys the same commutation relations as above, whilst all mixed anti-commutators between the $a_\mu(\mathbf{p})$ and $b_\nu(\mathbf{q})$ fields vanish.  Pulling everything together, we see the field $\chi_\mu(\mathbf{p})$ obeys the commutation relations
\begin{equation}
\{ \chi_\mu(\mathbf{p}),\chi_\nu^\dagger(\mathbf{q}) \} = \delta_{\mu \nu} \delta(\mathbf{p}-\mathbf{q}) \mathbb{I}_2, 
\quad \{ \chi_\mu(\mathbf{p}),\chi_\nu(\mathbf{q}) \} =  \{ \chi^\dagger_\mu(\mathbf{p}),\chi^\dagger_\nu(\mathbf{q}) \} = 0.
\end{equation}
Using the definition of the position space field in Eq.~(\ref{eq:C2_fourier_transform_field}), we can deduce the commutation relations in position space take the form
\begin{equation}
\{ \chi_\mu(\mathbf{x}), \chi_\nu^\dagger(\mathbf{y}) \} = \delta_{\mu \nu} \delta(\mathbf{x} - \mathbf{y}) \mathbb{I}_2, \quad \{ \chi_\mu(\mathbf{x}),\chi_\nu(\mathbf{y}) \} =  \{ \chi^\dagger_\mu(\mathbf{x}),\chi^\dagger_\nu(\mathbf{y}) \} = 0, \label{eq:C2_continuum_commutation}
\end{equation}
which are the standard commutation relations for a quantum field theory of a pair of two-component spinor fields $\chi_\mu(\mathbf{x})$~\cite{Srednicki2}. 
\subsection{Continuum limit (four-dimensional representation) \label{sec:C2_4x4_cont_lim}}
\subsubsection{The Hamiltonian}
The continuum limit yields a two-dimensional single-particle Hamiltonian $h_\mu(\mathbf{x})$ about each Fermi point. It is convenient to consider each Fermi point simultaneously and combine these Hamiltonians into a single four-dimensional object as
\begin{equation} 
\begin{aligned}
h_\mathrm{total}(\mathbf{x}) & = h_+(\mathbf{x}) \oplus \sigma^x h_-(\mathbf{x}) \sigma^x \\
& = -i v_\mathrm{F} \left( - \sigma^z \otimes \sigma^x \overset{\leftrightarrow}{\partial_x} + \sigma^z \otimes \sigma^y \overset{\leftrightarrow}{\partial_y} \right) + \Delta \mathbb{I} \otimes \sigma^z  ,
\end{aligned}
\end{equation}
where we have done a $\sigma^x$ rotation on $h_-(\mathbf{x})$ in order to bring the Hamiltonian into a standard representation. This is a Dirac Hamiltonian using the four-dimensional representation of the Dirac alpha and beta matrices $\alpha^A = -\sigma^z \otimes \sigma^A, \quad \beta = \sigma^x \otimes \mathbb{I}_2$, which obey $(\alpha^A)^2 = \beta^2 = \mathbb{I}$, where $A \in \{1,2,3 \}$. Defining the Dirac gamma matrices with the definitions $\gamma^0 = \beta$ and $\gamma^A = \beta^{-1} \alpha^A$, we find
\begin{equation}
\gamma^0 = \sigma^x \otimes \mathbb{I}_2, \quad \gamma^A = i \sigma^y \otimes \sigma^A,  \label{eq:C2_4x4_gammas}
\end{equation}
which is the Chiral representation of the gamma matrices~\cite{Peskin,C2_Maggiore}, which obey the Clifford algebra $\{ \gamma^a, \gamma^b \} = 2 \eta^{ab}$, where $\eta^{ab} = \mathrm{diag}(1,-1,-1,-1)$ is the $(3+1)$-dimensional Minkowski metric and $a \in \{0,1,2,3 \}$. Note that any representation of the gamma matrices is suitable, however we choose the chiral representation for convenience as it treats each Fermi point as an effective chirality, which is why we do not choose $\mathbb{I} \otimes \sigma^z$ as $\beta$. In addition, despite working in $(2+1)$D spacetime, we can still define a $\gamma^z$ due to the additional freedom of working with four-dimensional matrices. With this, the many-body Hamiltonian reads
\begin{equation}
H = \frac{1}{2} \int \mathrm{d}^2 x \chi^\dagger(\mathbf{x}) h(\mathbf{x}) \chi(\mathbf{x}), \quad \chi(\mathbf{x}) = \begin{pmatrix}\chi_+(\mathbf{x}) \\ \sigma^x \chi_-(\mathbf{x}) \end{pmatrix},
\end{equation}
where the fields $\chi_\pm(\mathbf{p})$ are the same fields used in the two-dimensional representation of Eq.~(\ref{eq:C2_2x2_field_ham}).
\subsubsection{Majorana spinors and commutation relations}
From the gamma matrices, we can define a \textit{Majorana spinor}. A Majorana spinor $\psi$ is a spinor field that is equal to its charge conjugate, $\psi^{(c)} = \psi$, where charge conjugation is defined as
\begin{equation}
\psi^{(c)}(\mathbf{x}) = C \psi^*(\mathbf{x}), \label{eq:charge_conjugation}
\end{equation}
where $C$ is the unitary charge conjugation operator which obeys $C^\dagger \gamma^a C = - (\gamma^a)^*$ for all gamma matrices and the notation $\psi^*$ here represents taking the hermitian conjugate of each component of $\psi$ without taking the transpose, i.e., $\psi^* = (\psi^\dagger)^\mathrm{T}$~\cite{C2_Maggiore,srednicki_2007}. In our representation of Eq.~(\ref{eq:C2_4x4_gammas}), this is given by $C = - \sigma^y \otimes \sigma^y$. Consider our spinor $\chi(\mathbf{x})$: we have
\begin{equation}
\chi^{(c)}  =  
\begin{pmatrix}
0 & 0 & 0 & 1 \\ 0 & 0 & -1 & 0 \\ 0 & -1 & 0 & 0 \\ 1 & 0 & 0 & 0
\end{pmatrix}
\begin{pmatrix}
a_+ \\ 
ib_+ \\
i b_-  \\
a_- 
\end{pmatrix}^* = 
\begin{pmatrix}
a_-^\dagger \\
-(-ib_-^\dagger) \\
-(-ib_+^\dagger) \\
a_+^\dagger \end{pmatrix} 
= 
\begin{pmatrix}
a_+ \\ 
ib_+ \\
i b_-  \\
a_-
\end{pmatrix}
 = \chi, \label{eq:C2_majorana_spinor}
\end{equation}
where in the second equality we used the fact that the position space fields obey $a_\pm^\dagger(\mathbf{x}) = a_\mp(\mathbf{x})$, and similarly for $b_\pm(\mathbf{x})$, so $\chi(\mathbf{x})$ is a Majorana spinor. Hence, the continuum limit of Kitaev's honeycomb model is described by a Majorana spinor on a $(2+1)$-dimensional Minkowski spacetime with an effective speed of light $v_\mathrm{F}$.

Due to the Majorana property, in the four-dimensional language the commutation relations are slightly different. From before we have
\begin{equation}
\{ \chi(\mathbf{x}) , \chi^\dagger(\mathbf{y}) \} = \delta(\mathbf{x} - \mathbf{y}) \mathbb{I}_4,
\end{equation}
but now the Majorana constraint gives us
\begin{equation}
\{ \chi(\mathbf{x}) , \chi(\mathbf{y}) \}  = \{ \chi (\mathbf{x}), C \chi^\dagger(\mathbf{y}) \}  = C \delta(\mathbf{x} - \mathbf{y}), \label{eq:C2_majorana_commutation}
\end{equation}
which are the standard commutation relations of a Majorana spinor on a Minkowski spacetime~\cite{srednicki_2007} and contrast to the standard Dirac commutation relations in the two-dimensional language of Eq.~(\ref{eq:C2_continuum_commutation}).

\section{Riemann-Cartan Geometry \label{sec:Riemann_Cartan}}
In the next chapter of this thesis we are interested in generalising the continuum limit description of the previous section to curved spacetimes using the language of general relativity. The usual treatment of general relativity uses \textit{Riemannian geometry}, where the fundamental object is the metric $g_{\mu \nu}$ which describes the geometry of the spacetime. In particular, the metric uniquely determines the torsion-free Levi-Civita connection, or Christoffel symbols $\Gamma^\mu_{\nu \sigma}$, which in turn describes the curvature via the Riemann tensor $R^\mu_{\ \nu \rho \sigma}$. On the other hand, in order to describe the continuum limit we require the second-order formalism of general relativity. This is where the metric and connection are considered independent objects. Contrary to my original paper of Ref.~\cite{C2_Farjami}, the language of the second formalism is most elegantly and efficiently presented using the language of differential forms and Cartan's structure equations, which we shall discuss in this section. This section closely follows Ref.~\cite{C2_Nakahara}.

\subsection{The metric and the veilbein}

Consider an $n$-dimensional differentiable manifold $M$ equipped with a Lorentzian metric $g$. For every point $p \in M$ we have a tangent space $T_p(M)$ defined as the vector space of tangent vectors to that point, where $g : T_p(M) \times T_p(M) \rightarrow \mathbb{R}$ plays the role of an inner product allowing us to measure lengths and angles on the manifold. Consider a local coordinate system $x^\mu$ in a neighbourhood of $p$, where $x^\mu \in \{ t,x,y,\ldots\}$ labels the axes, then the standard basis of $T_p(M$) is given by the coordinate basis $\{ e_\mu = \partial_\mu \}$ which roughly speaking ``point" along the coordinate axes. In addition, the cotagent spaces $T^*_p(M)$ dual to every tangent space $T_p(M)$ are spanned by the basis $\{ e^\mu = \mathrm{d} x^\mu \}$ defined such that $e^\mu(e_\nu) = \delta^\mu_\nu$. From the coordinate basis, we can define the basis of all tensors by taking tensor products.

In order to work with spinors on $M$ we need to introduce an orthonormal basis called the veilbein basis.

\begin{figure}[t]
\begin{center}
\includegraphics[scale=1]{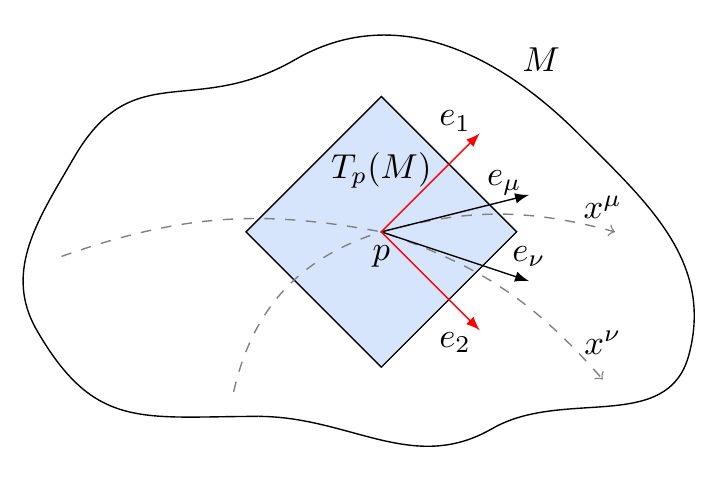}
\end{center}
\caption{The tangent space $T_p(M)$ to the point $p \in M$. The coordinate basis vectors $e_\mu$ and $e_\nu$ are parallel to the coordinate axes $x^\mu$ and $x^\nu$ respectively. The veilbein basis $e_1$ and $e_2$ is orthonormal with respect to the metric and is not aligned with any coordinate system.}
\label{fig:C2_veilbein}
\end{figure}

\begin{definition}[Veilbein]
The veilbein basis is an orthonormal basis of the tangent spaces given by $\{ e_a = e_a^{\ \mu} e_\mu \}$, as shown in Fig.~\ref{fig:C2_veilbein}, with the corresponding dual basis $\{ e^a = e^a_{\ \mu} e^\mu \}$ such that $g(e_a , e_b ) = \eta_{ab}$ and $e^a(e_b) = \delta^a_b$, where $\eta_{ab} = \mathrm{diag}(1,-1,\ldots,-1)$ is the Minkowski metric. In components this reads
\begin{equation}
g_{\mu \nu} e_a^{\ \mu} e_b^{\ \nu} = \eta_{ab}, \quad e^a_{\ \mu} e_b^{\ \mu} = \delta^a_b, 
\end{equation}
where $g_{\mu \nu} = g(e_\mu , e_\nu )$ are the components of the metric with respect to the coordinate basis. Given a veilbein basis we can uniquely define the corresponding metric as 
\begin{equation}
g_{\mu \nu} = e^a_{\ \mu} e^b_{\ \nu} \eta_{ab},
\end{equation}
so we see that the veilbein are only defined up to an orthogonal transformation. 
\end{definition}
The name ``veilbein'' means ``many legs" in German and is sometime renamed depending on what dimension we are working in, e.g., for two dimensions they are called \textit{zweibein} and for three dimensions \textit{dreibein}.

We use Greek indices $\mu , \nu \in \{t,x,y,\ldots\}$ to represent components with respect to the coordinate basis and Latin indices $a ,b \in \{ 0 , 1,2,\ldots \}$ equivalently for the veilbein basis. The components of the veilbein are typically called the veilbein too, so we shall refer to them as such when it is not ambiguous. They allow us to transform tensor components between the coordinate and veilbein basis, e.g. for a vector $X$, the components in each basis are related via $X^\mu = e^{\ \mu}_a X^a$ and $X^a = e^a_{\ \mu} X^\mu$. We can also use the metric to raise and lower indices, e.g., $X_\mu = g_{\mu \nu } X^\nu$ and $X^\mu = g^{\mu \nu} X_\nu$ and similarly for veilbein indices as $X_a = \eta_{ab} X^b$ and $X^a = \eta^{ab} X_b$. These properties generalise to higher order tensors too.

\subsection{The spin connection, curvature and torsion \label{sec:spin_connection}}
In this section we first briefly introduce the language of differential forms as it provides an extremely powerful tool to tackle geometric quantities to be defined in this section.

\begin{definition}[$p$-form]
A differential $p$-form $A$ is a completely anti-symmetric rank $(0,p)$ tensor with the components $A_{\mu_1\mu_2\ldots \mu_p} = A_{[\mu_1\mu_2 \ldots \mu_p]}$ where the square brackets denote anti-symmetrising over the indices.
\end{definition}

\begin{definition}[Wedge product]
The wedge product between a $p$-form $A$ and a $q$-form $B$ is given by the $(p+q)$-form $A \wedge B$ with components
\begin{equation}
(A \wedge B)_{\mu_1 \ldots \mu_p \nu_1 \ldots \nu_q} = \frac{(p+q)!}{p! q!} A_{[\mu_1 \ldots \mu_p} B_{\nu_1 \ldots \nu_q]} .
\end{equation}
\end{definition}

\begin{definition}[Exterior derivative]
The exterior derivative of a $p$-form $A$ is the $(p+1)$-form $\mathrm{d}A$, with components
\begin{equation}
(\mathrm{d} A)_{\mu_1 \ldots \mu_{p+1}} = (p+1) \partial_{[\mu_1} A_{\mu_2 \ldots \mu_{p+1}]},
\end{equation}
which obeys $\mathrm{d}^2 = 0$.
\end{definition}

With these definitions, we can now define the spin connection. When taking the derivative of a tensor, we must compare it at infinitesimally separated points. On a general manifold, there is no canonical way to compare an object at different points as they reside in different tangent spaces. For this reason, we must introduce a connection which allows us to move tensors around the manifold in such a way that they remain parallel. This is called \textit{parallel transport}. Parallel transport allows us to compare objects defined at different locations and we can write down a covariant derivative. We denote $\nabla_\mu$ as the covariant derivative with respect to the $e_\mu$ direction which preserves the rank of any tensor it acts on, e.g., for a vector field $X$ then $\nabla_\mu X$ is also a vector field and similarly for higher order tensors, so it can be expanded with respect to a basis of our choice. From this, we \textit{define} the action of the covariant derivative on the veilbein basis as $\nabla_\mu e_a = \Omega^b_{\ a  \mu} e_b $ which defines the set of coefficients $\Omega^a_{\ b \mu} \in \mathbb{R}$. From this, we can define the spin connection.

\begin{definition}[Spin connection]
Let $ e_a  $ be a veilbein basis. The spin connection is defined as the set of one-forms 
\begin{equation}
\Omega^a_{\ b} = \Omega^a_{\ b \mu} \mathrm{d}x^\mu,
\end{equation}
where $\nabla_\mu e_a = \Omega^a_{\ b \mu} e_b$ defines the components. These one-forms are labelled by the Latin veilbein indices which are not to be confused with the components. The spin connection obeys $\Omega_{ab} = - \Omega_{ab}$, where $\Omega_{ab} = \eta_{ac} \Omega^c_{\ b}$. 
\end{definition}
The spin connection must obey certain transformation rules, however these are not stringent enough to uniquely define the spin connection. Therefore, a manifold contains many different notions of ``parallel". However, a standard definition to take is a \textit{metric compatible connection} defined by $\nabla g= 0$, which ensures the lengths of a pair of vectors and the angle between them is fixed while parallel transporting them together. 

\begin{figure}
\begin{center}
\hspace{1cm} (a) \hfill (b) \hspace{6cm} \par\medskip
\includegraphics[valign=t,scale=1]{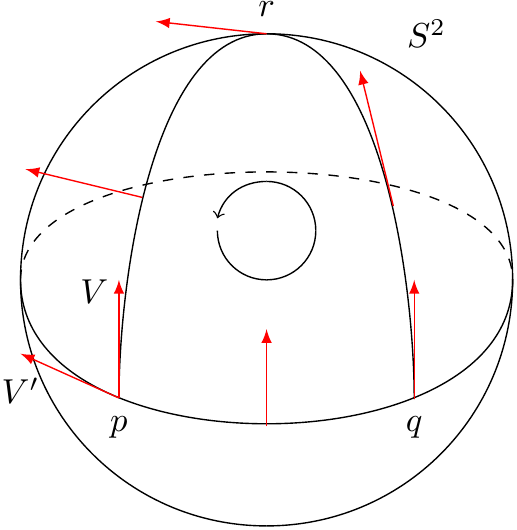}
\hspace{1cm}
\includegraphics[valign=t,scale=1]{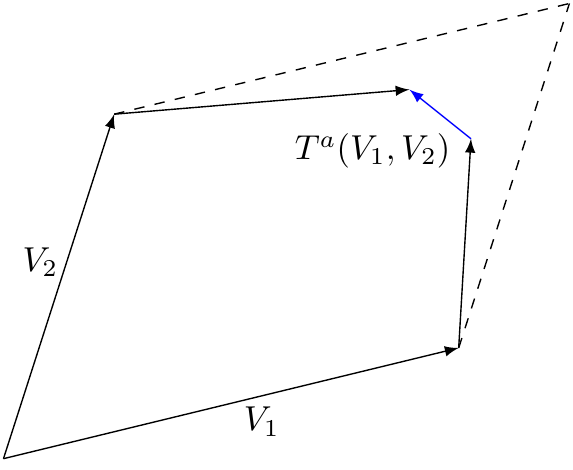}
\end{center}
\caption{(a) Parallel transport of a vector $V$ along a closed loop $pqrp$ on the sphere $S^2$. After returning to its original position at $p$, the vector has transformed into a new vector $V'$ which has rotated relative to $V$ due to the presence of curvature. (b) The presence of torsion results in the failure of a pair of vectors $V_1$ and $V_2$ to produce a parallelogram (represented by the dashed lines) when parallel transported along each other. The failure is measured by the torsion tensor $T^a$ acting upon the two vectors $V_1$ and $V_2$, to yield the blue vector $T^a(V_1,V_2)$.}
\label{fig:C2_parallel_transport} 
\end{figure}

When parallel transporting a vector with respect to some connection, it can so happen that when the vector traverses a closed loop, it has a different orientation to how it started. The standard example is that of a sphere, whereby a parallelly transported vector changes its angle when it return to the same point, as shown in Fig.~\ref{fig:C2_parallel_transport}(a). This effect is a consequence of curvature.
\begin{definition}[Curvature]\label{def:curvature}
Given a spin connection $\{ \Omega^a_{\ b} \}$, the curvature is defined as the set of $2$-forms
\begin{equation}
R^a_{\ b}(\Omega) = \mathrm{d}\Omega^a_{\ b} + \Omega^a_{\ c} \wedge \Omega^c_{\ b} \equiv \frac{1}{2}R^a_{\ b \mu \nu } \mathrm{d} x^\mu \wedge \mathrm{d}x^\nu,
\end{equation}
where $R^a_{\ b \mu \nu }$ is the Riemann tensor.
\end{definition}
In addition to curvature, there is a slightly less intuitive concept of torsion. Roughly speaking, the torsion measures the failure of a pair of vectors to create a parallelogram when parallel transported along each other, as shown in Fig.~\ref{fig:C2_parallel_transport}(b). We now introduce its definition.

\newpage 
\begin{definition}[Torsion]\label{def:torsion}
Given a veilbein basis $\{ e^a \}$ spin connection $\{ \Omega^a_{\ b}\}$, the torsion is defined as the set of $2$-forms
\begin{equation}
T^a(e,\Omega) = \mathrm{d}e^a + \Omega^a_{\ b} \wedge e^b \equiv \frac{1}{2} T^a_{\ \mu \nu} \mathrm{d}x^\mu \wedge \mathrm{d}x^\nu,
\end{equation}
where $T^a_{\ \mu \nu}$ is the torsion tensor. This, together with Def.~\ref{def:curvature}, are known as \textit{Cartan's structure equations}.
\end{definition}

It can be shown~\cite{C2_Nakahara} that a general metric compatible connection can decomposed into two pieces as
\begin{equation}
\Omega^a_{\ b} = \omega^a_{\ b} + C^a_{\ b}, \label{eq:C2_connection_decomposition}
\end{equation}
where $\omega^a_{\ b}$ is the unique \textit{torsion-free} metric compatible connection, called the Levi-Civita connection, which is uniquely defined by the the veilbein $e^a$, whilst $C^a_{\ b}$ is the contortion tensor which contains all of the torsional information. In other words, given a veilbein basis $ e_a $, we can uniquely define the Levi-Civita connection $\omega^a_{\ b}$ via the torsion free condition $T^a(\omega) = 0$ which implies
\begin{equation}
\mathrm{d}e^a = -\omega^a_{\ b} \wedge e^b,
\end{equation}
whilst the contortion gives us
\begin{equation}
T^a = C^a_{\ b} \wedge e^b.
\end{equation}

\subsection{The absence of torsion in general relativity \label{sec:torsion_in_GR}}

In general relativity, we typically ignore torsion and set $T^a = 0$. This can be seen for the following reasons. The Einstein-Hilbert action describing the interaction of matter fields $\psi$ with geometry, described by a veilbein $e_a$ and spin connection $\Omega^a_{\ b}$, is given by
\begin{equation}
S[e,\Omega,\psi] = \frac{1}{16 \pi} \int \mathrm{d}^{n+1}x \sqrt{|g|} R + S_\mathrm{matter}[e,\Omega,\psi],
\end{equation}
where $R \equiv R^{ab}_{\ \ \mu \nu} e_a^{\ \mu} e_b^{\ \nu}$ is the Ricci scalar corresponding to the connection $\Omega^a_{\ b}$ which is defined via a contraction of the Riemann tensor $R^a_{\ b \mu \nu}$, see Def.~\ref{def:curvature}, and $S_\mathrm{matter}$ is the action describing matter fields $\psi$~\cite{Gravity_book}. 

On a Riemann-Cartan spacetime the veilbein $e_a$ and spin connection $\Omega^a_{\ b}$ are considered independent objects, so we have an equation of motion for both of these objects. The variation with respect to the veilbein yields the equation of motion
\begin{equation}
\mathrm{Ric}_{\ \mu}^a - \frac{1}{2} e_{\ \mu}^a R = 8 \pi t^a_{\ \mu} , \quad t^a_{\ \mu} =  - \frac{2}{|e|} \frac{\delta S_\mathrm{matter}}{\delta e_a^{\ \mu}},
\end{equation}
where $\mathrm{Ric}^a_{\ \mu} \equiv R^{a b}_{\ \ \alpha \mu} e^{\ \alpha}_b$ is the Ricci tensor and $t_a^{\ \mu}$ is the energy-momentum tensor of the matter fields. This is the usual Einstein equation of general relativity which describes the response of geometry to matter. On the other hand, variation with respect to the spin connection yields
\begin{equation}
\frac{1}{2} T^a_{\ \mu \nu} e^b_{\ \rho} \epsilon_{abcd} \epsilon^{\mu \nu \rho \sigma} =  8\pi S^{\ \ \sigma}_{cd}, \quad S^{ab \sigma} = - \frac{2}{|e|} \frac{\delta S_\mathrm{matter}}{\delta \Omega_{ab \sigma}}, \label{eq:einstein_cartan_2} 
\end{equation}
where $S^{ab \sigma}$ is the \textit{spin current} which is the Noether current corresponding to the internal rotational symmetry of the action. For scalar fields this is zero.

This combination of Riemann-Cartan geometry with general relativity is known as \textit{Einstein-Cartan} theory~\cite{Gravity_book}. The first equation of motion is the usual Einstein equation for the metric. As this is a differential equation, there are solutions which permit curvature in the vacuum, where the energy-momentum tensor is zero. This explains the existence of gravitational fields and gravitational waves in the vacuum. On the other hand, the second equation of motion relating the torsion to the spin tensor is an algebraic equation. This means that the torsion is zero if the spin tensor is zero, therefore the torsion will be zero in the vacuum. As we are usually only interested in general relativity in the vacuum, the torsion tensor will always be zero so we can safely ignore torsion in general relativity.

In addition to this, torsion has some unintuitive features which we do not detect in the world around us. The first is related to geodesics: we can define a geodesic as either the shortest path between two points, or the path which parallel transports its own tangent vector which roughly speaking is the ``straightest path". In a space with torsion, these two definitions do not align and these geodesics will differ~\cite{C2_Nakahara}. Another unusual feature is that torsion is related to how frames \textit{twist} about geodesics. This has the unusual effect that if one was to throw a rugby ball without spinning it about its axis, the effect of torsion would cause the ball to spin. 

It is important to stress that the emergent spacetimes in this thesis are not spacetimes generated from the response to a distribution of matter and spin, but are instead simply emergent effective \textit{background} geometries generated by the couplings of the model with zero back-reaction of matter on the geometry. However, the important feature is that we \textit{can} simulate solutions to the Einstein equation by suitable choice of the couplings.

\subsection{Spinor fields on Riemann-Cartan geometry}
Relativistic spinors are defined as objects which transform in the spinor representation of the Lorentz algebra $\mathfrak{so}(1,2)$, however there is no spinor representation with respect to the diffeomorphism symmetry of a general manifold so we must construct a locally inertial frame at each location, which the veilbein defined previously provide us with.

The action for a Majorana spinor field $\psi$ of mass $m$ defined on a $(2+1)$-dimensional Riemann-Cartan spacetime $M$ with dreibein $ e_a $, corresponding to a metric $g_{\mu \nu}$, and spin connection $ \Omega^a_{\ b} $ is given by~\cite{C2_Nakahara}
\begin{equation}
S_\mathrm{RC} = \frac{i}{2} \int_M \mathrm{d}^{2+1}x |e| \left( \bar{\psi} \gamma^\mu D_\mu \psi - \overline{D_\mu \psi} \gamma^\mu \psi + 2im \bar{\psi} \psi \right) \equiv \int_M \mathrm{d}^{2+1}x \mathcal{L}_\mathrm{RC}, \label{eq:C2_RC_action}
\end{equation}
where the fields $\psi$ obey the Majorana constraint $\psi = C\psi^*$ and have anti-commuting Grassmann numbers~\cite{C2_Nakahara} as components. The matrices $\{ \gamma^\mu = e_a^{\ \mu} \gamma^a \}$ are the curved space gamma matrices which obey the Clifford algebra $\{ \gamma^\mu , \gamma^\nu \} = 2 g^{\mu \nu}$ and are related to the local flat gamma matrices $\{ \gamma^a \}$ which obey the flat space Clifford algebra $\{ \gamma^a , \gamma^b \} = 2 \eta^{ab}$. The Dirac adjoint is defined as $\bar{\psi} = \psi^\dagger \gamma^0$, where $\gamma^0$ is the flat space gamma matrix, and $|e| = \det [e_\mu^a ] = \sqrt{-g}$. The covariant derivative of spinors $D_\mu$ is defined as
\begin{equation}
D_\mu \psi = \partial_\mu \psi + \Omega_\mu \psi, \quad \overline{D_\mu \psi} = \partial_\mu \bar{\psi} - \bar{\psi} \Omega_\mu,
\end{equation}
where $\Omega_\mu$ plays the role of the connection for spinor field which is related to the spin connection of $M$ via
\begin{equation}
\Omega_\mu = \frac{i}{2} \Omega_{a b \mu } \Sigma^{ab}, \quad \Sigma^{ab} = \frac{i}{4} [\gamma^a , \gamma^b],
\end{equation}
where we lower the indices as $\Omega_{ a b \mu} = \eta_{ac} \Omega^c_{\ b \mu}$ and $\{ \Sigma^{ab}\}$ form a representation of the Lorentz algebra $\mathfrak{so}(1,2)$. 

We are only interested in $(2+1)$-dimensional theories on a manifold of the form $M = \mathbb{R} \times \Sigma$, where $\mathbb{R}$ represents the time dimension and $\Sigma$ is a two-dimensional spatial hypersurface, which simplifies the action considerably. First of all, we can always find a coordinate system such that the metric and dreibein takes the block-diagonal form
\begin{equation}
g_{\mu \nu} = 
\begin{pmatrix} 
1 & 0 \\ 
0 & G_{ij}
\end{pmatrix} , \quad e^a_{\ \mu} = \begin{pmatrix} 
1 & 0 \\
0 & E^A_{\ i}
\end{pmatrix}, \quad e_a^{\ \mu} = \begin{pmatrix} 
1 & 0 \\
0 & E_A^{\ i}
\end{pmatrix}
\label{eq:2+1_space}
\end{equation}
where $G_{ij}$ is a 2D metric defined on the spatial hypersurface $\Sigma$ with the corresponding spatial-only \textit{zweibein} $E^A_{\ i}$ and $E_A^{\ i}$, where the indices $i$ and $j$ represent spatial coordinate only and $A$ represents spatial only dreibein indices. We define the canonical momentum as
\begin{equation}
\pi  := \frac{\partial_L \mathcal{L}_\mathrm{RC}}{\partial (\partial_t\psi)},
\end{equation}
where we have defined it using the left derivative $\frac{\partial_L }{\partial (\partial_t\psi)}$ as we are dealing with Grassmann numbers. In order to take this derivative, we explicitly substitute in the covariant derivative to separate the partial derivatives from the connection and use the Majorana condition to rewrite $\bar{\psi} = \psi^\mathrm{T} C \gamma^0$ to get
\begin{equation}
\mathcal{L}_\mathrm{RC}  = \frac{i}{4} |e| \left( \bar{\psi} \gamma^\mu \partial_\mu \psi - \partial_\mu \psi^\mathrm{T} C \gamma^0 \gamma^\mu \psi + \bar{\psi} \{ \gamma^\mu, \Omega_\mu \} \psi \right),
\end{equation}
so we see that the second term is no longer independent when varying with respect to $\dot{\psi}$. Therefore, we have
\begin{equation}
\pi  = - \frac{i}{2}|E| \psi^\dagger,
\end{equation}
where to derive this we have used the rules for differentiating Grassmann numbers from the left~\cite{C2_Nakahara}, used the fact that with the metric above we have $|e| = |E| = \det [E^A_{\ i}]$, $\gamma^t = e_0^{\ t} \gamma^0 = \gamma^0$ and $(\gamma^0)^2 = \mathbb{I}$, and applied the Majorana condition $C \psi = \psi^*$. 

As the field obeys the Majorana condition $\psi = C\psi^*$ the canonical momentum defines a constraint on the fields as $\psi^\dagger$ is no longer treated as an independent object to $\psi$, so the machinery of constrained systems must be employed to derive the anti-commutation relations, which give~\cite{srednicki_2007}
\begin{equation}
\{ \psi_\alpha(\mathbf{x}) , \psi^\dagger_\beta(\mathbf{y}) \} = \delta_{\alpha \beta} \frac{\delta(\mathbf{x}-\mathbf{y})}{|E|}, \quad \{ \psi_\alpha(\mathbf{x}) , \psi_\beta(\mathbf{y}) \} = C_{\alpha \beta} \frac{\delta(\mathbf{x}-\mathbf{y})}{|E|}. \label{eq:C2_RC_commutators}
\end{equation}
We can also unambiguously define the Hamiltonian via a Legendre transform as
\begin{equation}
\begin{aligned}
H &  = \int_\Sigma \mathrm{d}^2 x \left( \dot{\psi}_\alpha \pi_\alpha  - \mathcal{L}_\mathrm{RC} \right) \\
& = \frac{1}{2} \int_\Sigma \mathrm{d}^2 x |E| \bar{\psi} \left( -\frac{i}{2} \gamma^i \overset{\leftrightarrow}{\partial_i}  - \frac{i}{2}  \{ \gamma^\mu , \Omega_\mu \}  + m \right) \psi. 
\end{aligned}
\end{equation}
The anti-commutator is given by
\begin{equation}
\{ \gamma^\mu ,\Omega_\mu \} = \frac{1}{8} e^{\ \mu}_a \Omega_{ b c \mu} \{ \gamma^a , [\gamma^b , \gamma^c ]\} = \frac{1}{2} \Omega_{abc} \epsilon^{abc} \gamma^0 \gamma^1 \gamma^2,
\end{equation}
where we used the $(2+1)$D gamma matrix identity $\{ \gamma^a , [\gamma^c , \gamma^c ]\} = 4 \epsilon^{abc} \gamma^0 \gamma^1 \gamma^2$ and we have defined $\Omega_{abc} = e^{\ \mu}_c \Omega_{ab\mu} $. We see that the spinor field $\psi$ only couples to the completely anti-symmetric portion of the spin connection. Moreover, the Levi-Civita connection will have no time components either because the metric has a trivial temporal part and does not mix up space and time directions. For this reason, using the fact that we are working with a metric compatible connection so can decompose the spin connection into its Levi-Civita and contortion components as in Eq.~(\ref{eq:C2_connection_decomposition}), we find
\begin{equation}
 \Omega_{abc} \epsilon^{abc} = \omega_{abc}\epsilon^{abc} + C_{abc} \epsilon^{abc} =  C_{abc} \epsilon^{abc} \equiv  c.  
\end{equation}
With this identification, the Hamiltonian takes the form
\begin{equation}
H_\mathrm{RC}  = \frac{1}{2} \int_\Sigma \mathrm{d}^2 x |E| \bar{\psi} \left( -\frac{i}{2}  \gamma^i \overset{\leftrightarrow}{\partial_i}  - \frac{i c}{4}  \gamma^0 \gamma^1 \gamma^2 + m \right) \psi.
\end{equation}
This Hamiltonian is currently in the language of the Dirac gamma matrices, which is a convenient form to use when keeping track of Lorentz invariance, however the Hamiltonians we shall obtain from the continuum limit of Kitaev's honeycomb model will be in a slightly different form. We define the Dirac alpha and beta matrices via $\alpha^i = \gamma^0 \gamma^i$ and $\beta = \gamma^0$, which brings the Hamiltonian into the final form
\begin{equation}
H_\mathrm{RC} = \frac{1}{2} \int_\Sigma \mathrm{d}^2 x |E| \psi^\dagger \left(- \frac{i}{2} \alpha^i \overset{\leftrightarrow}{\partial_i} + \frac{ i c}{4}\alpha^x \alpha^y + m \beta \right) \psi, \label{eq:C2_RC_ham}
\end{equation}
which is our final result.

\section{Conclusion \label{rc_conclusion}}
This chapter introduced the relevant background material required for chapters~\ref{chapter:kitaev} and \ref{chapter:chiral}. We first introduced Kitaev's honeycomb lattice model as an model describing interacting spin-$1/2$ particles arranged onto a honeycomb lattice and showed its relativistic continuum limit in terms of Majorana fermions on a Minkowski spacetime. We then discussed Riemann-Cartan geometry, which is a generalised of Riemannian geometry used in general relativity, whereby the spacetime additionally contains torsion.

\chapter{Emergent geometry in Kitaev's honeycomb model \label{chapter:kitaev}}

\section{Introduction}

In recent years there has been a surge of interest in the geometrical degrees of freedom that characterise the response of topologically-ordered phases of matter beyond the well-known limit governed by an effective topological quantum field theory. An important class of such systems, the fractional quantum Hall (FQH) states, have been understood to exhibit a universal response to the variations of ambient geometry. This response leads to many fruitful investigations of an interplay between topology and geometry in these strongly-correlated systems~\cite{C2_wen1992shift, C2_avron1995viscosity, C2_read2009non, C2_HaldaneViscosity, C2_abanov2014electromagnetic, C2_gromov2015framing, C2_BradlynRead, C2_CanLaskinWiegmann, C2_klevtsov2015geometric, C2_hughes2011torsional, C2_gromov2014density}. In particular, the neutral collective mode of FQH systems~\cite{C2_GMP85} has been described as a fluctuating spacetime metric~\cite{C2_HaldaneGeometry, C2_GromovSon,C2_wiegmann2017nonlinear}. 
On the other hand, a recent study~\cite{C2_Golan} has shown that by minimally coupling a spinless $p$-wave superconductor on a square lattice to an electromagnetic field, the continuum limit takes the form of a Dirac Hamiltonian defined on a spacetime with both curvature and torsion. Such curved spaces with torsion are called \textit{Riemann-Cartan} spacetimes~\cite{C2_Hehl} as introduced in Sec.~\ref{sec:Riemann_Cartan}. Riemann-Cartan geometry also naturally arises in the theory of defects in lattices, whereby disclinations and dislocations in the continuum limit are described by curvature and torsion, respectively~\cite{C2_Katanaev, C2_deJuan}, which has been investigated in strained graphene~\cite{C2_Wagner,C2_deJuan,C2_deJuan2}. 

Other techniques from quantum gravity have also been employed in condensed matter. The holographic correspondence or AdS/CFT correspondence has been used to model gapless modes living on the defect lines of class D topological superconductors~\cite{C2_Palumbo} and to determine the specific heat of a two-dimensional interacting gapless Majorana system~\cite{C2_Maraner}. Moreover, the emergence of gravitational anomalies has been considered in topological superconductors~ \cite{C2_Jaakko}. Building upon Luttinger's proposal~\cite{C2_Luttinger}, gravitational techniques applied to the thermal Hall effect have also attracted many theoretical~\cite{C2_Nakai, C2_Cooper, C2_ReadGreen, C2_Wang2, C2_Ryu, C2_Qin, C2_Shitade} and experimental~\cite{C2_Jezouin,C2_Banerjee2} investigations.

Nevertheless, despite much progress in the investigation of geometric effects in the continuum field theory description, the study of Riemann-Cartan geometry in microscopic, solvable lattice models has received less attention. In this paper, we investigate geometric description of the Kitaev's honeycomb model~\cite{C2_Kitaev}, the well-known two-dimensional (2D) model of interacting spin-$1/2$ particles that gives rise to a quantum spin liquid phase with topological order. A salient feature of the Kitaev's honeycomb model is that it can support non-Abelian anyons in the form of Majorana zero modes trapped at its vortices~\cite{C2_Kitaev,C2_Kitaev2,C2_Otten,C2_Vidal3,C2_Vidal4}. 
Similar to the FQH effect~\cite{C2_Moore}, Kitaev's honeycomb model is both topologically ordered in the sense that it can support anyonic excitations and it is a topological phase categorised by a non-trivial Chern number~\cite{C2_Kitaev,C2_Vidal2}. Unlike the FQH effect, Kitaev's honeycomb model is exactly solvable, which has provided unique opportunities to analytically probe its anyonic properties~\cite{C2_Ville1,C2_Vidal3, C2_Vidal4}, its topological edge currents~\cite{C2_Chris1}, its finite temperature behaviour~\cite{C2_Chris2,C2_Ville3,C2_Ville4,C2_Nasu,C2_Nasu2,C2_Nasu3} and to investigate dimer limits of the model~ \cite{C2_Vidal,C2_Vidal5}. Moreover, many features of the Kitaev's honeycomb are recognised in experimentally realisable materials, such as complex iridium oxides~\cite{C2_Chaloupka, C2_Choi, C2_Jackeli} or ruthenium chloride~\cite{C2_Banerjee}. This makes Kitaev's honeycomb model of interest to numerous theoretical and experimental investigations.

In this chapter we address the following question: can we use Kitaev's honeycomb model to simulate Majorana fermions embedded in a Riemann-Cartan spacetime? To answer this question, we allow the couplings of Kitaev's honeycomb model to take general configurations that are anisotropic and inhomogeneous, while leaving the lattice configuration of the model unaffected. We demonstrate that in this case the low energy limit of the model can be effectively described by massless Majorana spinors obeying the Dirac equation embedded in a Riemann-Cartan spacetime which is locally Lorentz invariant. Moreover, the Majorana spinors are coupled to a non-trivial torsion. It is important to stress that this geometry emerges purely from distortions in the couplings of the system and \textit{not} from the geometry of the lattice itself.
 
This chapter is based off the study done in Ref.~\cite{C2_Farjami} and is structured in the following way. First in Sec.~\ref{sec:kitaev_dilation} we will introduce the most general anisotropic model, where the couplings of the model are unequal, which requires us to briefly re-derive the continuum limit from first principles. We will then focus on the special case for $J_x = J_y = 1 $ and $J_z \in [0,2]$ and see that this model describes a dilated spacetime in the continuum limit, the evidence of which is seen in the dilation of lattice Majorana correlation functions and Majorana zero-mode wavefunctions. Then in Sec.~\ref{sec:RC_kitaev} we will study the most general inhomogeneous model and identify its geometric quantities, such as the spin connection, curvature, and torsion. In addition, we discuss the Kekul\'e distortion~\cite{C2_Hou,C2_Yang} to generate mass, which will introduce a new topological phase to the model with a corresponding topological phase transition. Finally, in Sec.~\ref{sec:spin_densities} we discuss the unpublished work on how to observe the effects of curvature on the lattice by studying the spin densities of the model. We then finish with a conclusion in Sec.~\ref{kitaev_conclusion}.

\section{Emergent spacetime dilation \label{sec:kitaev_dilation}}
\subsection{The Hamiltonian}
In this section we consider the simplest non-trivial modification of Kitaev's honeycomb model introduced in Sec. \ref{sec:C2_Kitaev's honeycomb model}. We saw in Secs. \ref{sec:C2_2x2_cont_lim} and \ref{sec:C2_4x4_cont_lim} that the continuum limit of the model is relativistic and described by the Dirac equation on a flat Minkowski spacetime, which hints at a more general geometric description. In Sec.~\ref{sec:RC_kitaev} we shall employ the full machinery of Riemann-Cartan geometry to describe the most general models which have curvature, but for now we shall focus on a simple flat space generalisation of the previous analysis which still has enough to detect the presence of an underlying metric.

Let us consider the vortex-free sector where $J_x$, $J_y$ and $J_z$ are positive. In this case, the Hamiltonian that we start from takes the slightly modified form, given by $H = H_J + H_K$ where
\begin{align}
H_J & = \frac{i}{4} \sum_{\mathbf{r} \in \Lambda} 2 a_\mathbf{r} \left( J_z b_\mathbf{r} + J_x b_{\mathbf{r} +  \mathbf{n}_1 } +J_y b_{\mathbf{r} + \mathbf{n}_2} \right) + \mathrm{H.c.}, \\
H_K & = \frac{i}{4} \sum_{\mathbf{r} \in \Lambda} 2 a_\mathbf{r}\left( K_1 a_{\mathbf{r} + \mathbf{n}_1} - K_2 a_{\mathbf{r} + \mathbf{n}_2} - K_3 a_{\mathbf{r} + \mathbf{n}_1 -  \mathbf{n}_2} \right) - (a \leftrightarrow b) + \mathrm{H.c.},
\end{align}
which we call the \textit{anisotropic model}. Notice that we have generalised $H_K$ by introducing a set of three couplings $\{ K_i \}$. This is required because, as we shall see, $H_K$ may shift the location of the Fermi points. In the introduction to the model in Sec.~\ref{sec:C2_Kitaev's honeycomb model}, we saw that $H_K$ simply opened up a gap in the isotropic model---we would like to retain this feature here in the anisotropic case.

The anisotropic model retains its translational symmetry so can be diagonalised with a discrete Fourier transform as before. This brings the Hamiltonian into the same form as before in Eq.~(\ref{eq:C2_diagonal_hamiltonian}), except now the components of the single-particle Hamiltonian are slightly different, given by
\begin{equation}
f(\mathbf{p}) = 2\left( J_x e^{i \mathbf{p} \cdot \mathbf{n}_1 } + J_y e^{i \mathbf{p} \cdot \mathbf{n}_2} + J_z \right),
\end{equation}
and
\begin{equation}
\Delta(\mathbf{p})  = 4\big\{ - K_1\sin(\mathbf{p} \cdot \mathbf{n}_1) + K_2 \sin(\mathbf{p} \cdot \mathbf{n}_2) + K_3 \sin(\mathbf{p} \cdot [\mathbf{n}_1 - \mathbf{n}_2]) \big\}.
\end{equation}
Again, the dispersion relation of the model is obtained from the eigenvalues of the single-particle Hamiltonian $h(\mathbf{p})$ which gives us
\begin{equation}
E(\mathbf{p}) = \pm \sqrt{|f(\mathbf{p})|^2 + \Delta^2(\mathbf{p})},
\end{equation} 
which takes the same form as before, however the Fermi points of the model will be dependent upon the couplings.
\subsection{The Fermi points}
First, consider the case where $K_1 = K_2 = K_3 = 0$ which corresponds to a gapless system. We search for the Fermi points where $E(\mathbf{p}) = 0$. As before, there are two Fermi points, which are now given by 
\begin{equation}
\mathbf{P}_\pm = \pm \frac{1}{a} \begin{pmatrix}
 \cos^{-1}(\alpha) +  \cos^{-1}(\beta)\\
\frac{1}{\sqrt{3}} \left( \cos^{-1}(\alpha) - \cos^{-1}(\beta)\right)
\end{pmatrix}, \quad \alpha = \frac{J_y^2 - J_x^2 - J_z^2}{2J_x J_z}, \quad \beta = \frac{J_x^2 - J_y^2 - J_z^2}{2J_y J_z}. \label{eq:C2_general_fermi_points} 
\end{equation}
In the isotropic case discussed before in Sec.~\ref{sec:C2_Kitaev's honeycomb model}, we found that the $K$ term simply opened up a gap. We would like to retain this feature for the anisotropic model, however this will not be true in general because the $K$ term may shift the Fermi points as we change the couplings. In order to ensure this is not the case, we require that the Fermi points of Eq.~(\ref{eq:C2_general_fermi_points}) are the minima of $\Delta(\mathbf{p})$, in other words, we require 
\begin{equation}
\begin{aligned}
\boldsymbol{\nabla} \Delta(\mathbf{P}_\pm)  = 4 \bigg[& - K_1 \alpha \mathbf{n}_1 + K_2 \beta \mathbf{n}_2 \\
& + K_3 \left( \alpha \beta - \sqrt{(1-\alpha^2)(1-\beta^2)} \right) (\mathbf{n}_1 - \mathbf{n}_2 ) \bigg]  = 0 .
\end{aligned}
\end{equation}
This implies that we must choose the couplings
\begin{align}
K_1 & = 4K \beta \left( \alpha \beta - \sqrt{(1-\alpha^2)(1-\beta^2)} \right), \\
K_2 & = 4K \alpha \left( \alpha \beta - \sqrt{(1-\alpha^2)(1-\beta^2)} \right), \\
K_3 & = 4K\alpha \beta,
\end{align}
in which case the gap at the Fermi points is given by $\Delta(\mathbf{P}_\pm) = \mp \Delta$, where
\begin{equation}
\Delta  = 16K \left(\alpha \sqrt{1-\beta^2} + \beta \sqrt{1-\alpha^2} \right) \sqrt{(1-\alpha^2)(1- \beta^2)},
\end{equation}
which reduces to the original gap of $\Delta = 6\sqrt{3} K$ in the isotropic case for $J_x = J_y = J_z$.
\subsection{The continuum limit}
We now take the continuum limit of this model by playing the same game as before in section \ref{sec:C2_2x2_cont_lim} and \ref{sec:C2_4x4_cont_lim}. 

\subsubsection*{Step 1 - Taylor expand about the Fermi points}
First, we Taylor expand the single-particle Hamiltonian $h(\mathbf{p})$ about the Fermi points $\mathbf{P}_\pm$ to first order in momentum. We have
\begin{equation}
\begin{aligned}
f(\mathbf{P}_\pm + \mathbf{p}) = \frac{2ia}{\sqrt{3}} \bigg[& J_x \left(\alpha \pm i \sqrt{1-\alpha^2}\right) \mathbf{n}_1  
\\
& + J_y \left( \beta \mp i \sqrt{1 - \beta^2} \right) \mathbf{n}_2 \bigg] \cdot \mathbf{p} + O(\mathbf{p}^2),
\end{aligned} 
\end{equation}
and
\begin{equation}
\Delta(\mathbf{P}_\pm + \mathbf{p}) = \mp \Delta + O(\mathbf{p}^2),
\end{equation}
therefore when substituting this back into the single-particle Hamiltonian we arrive at a pair of single-particle Hamiltonians, one about each Fermi point as $h_\pm(\mathbf{p}) = h(\mathbf{P}_\pm + \mathbf{p})$, where
\begin{equation}
h_\pm(\mathbf{p}) = \left( \mp A \sigma^x + C \sigma^y \right)p_x + B \sigma^y p_y \mp \Delta \sigma^z + O(\mathbf{p}^2),
\end{equation}
where the coefficients are given by
\begin{align}
A&=  \sqrt{ 4J^2_x - \frac{(J_y^2 - J_x^2 - J_z^2)^2}{J_z^2}}, \\
B & = -\sqrt{3} J_z, \\
C & =  \frac{J_y^2 - J_x^2}{J_z}.
\end{align}
During this process, we take the continuum limit $a \rightarrow 0 $ and thermodynamic limit $N \rightarrow \infty$ in such a way that the Fermi velocity $v_\mathrm{F}$ of the model remains fixed. To do this, we simply have to renormalise the couplings as $aJ_i \rightarrow J_i$ whilst keeping $J_i$ finite and constant.

\subsubsection*{Step 2 - Combine Hamiltonians about each Fermi point}

We now consider each Fermi point simultaneously by combining the two Hamiltonians into a single Hamiltonian in the same way as done in Sec. \ref{sec:C2_4x4_cont_lim}. We have
\begin{equation}
\begin{aligned}
h(\mathbf{p}) & = h_+(\mathbf{p}) \oplus \sigma^x h_-(\mathbf{p}) \sigma^x \\
& =  \left(A \sigma^z \otimes \sigma^x + C \sigma^z \otimes \sigma^y  \right) p_x + B \sigma^z \otimes \sigma^y p_y + \Delta \mathbb{I} \otimes \sigma^z \\
& \equiv E_A^{\ i} \alpha^A p_i  - i \Delta \alpha^x \alpha^y,
\end{aligned}
\end{equation} 
where in the last equality we have defined the coefficients $E_A^{\ i}$ and the Dirac alpha and beta matrices given by
\begin{equation}
E_A^{\ i} = \begin{pmatrix}
A & 0 \\
C & B
\end{pmatrix}, \quad \alpha^A = \sigma^z \otimes \sigma^A, \quad \beta = \sigma^x \otimes \mathbb{I}_2,
\end{equation}
where $A = 1,2$ and $\sigma^1 \equiv \sigma^x$ and $\sigma^2 \equiv \sigma^y$. As expected, this Hamiltonian takes the form of a momentum space Dirac Hamiltonian using the standard chiral representation of the alpha and beta matrices, albeit with generalised coefficients now. In the isotropic case for $J_x = J_y = J_z$ as discussed in Sec.~\ref{sec:C2_4x4_cont_lim}, the continuum limit Hamiltonian had an overall factor of the Fermi velocity $v_\mathrm{F}$ which we were able to interpret as an effective speed of light. Now, due to the unequal coefficients, we choose not to do this and instead assume we are working on a spacetime with the speed of light set to unity.

\subsubsection*{Step 3 - Write down the many-body Hamiltonian}
Pulling everything together and transforming back to position space gives us the many-body continuum limit
\begin{equation}
H = \frac{1}{2} \int_\Sigma \mathrm{d}^2 x \chi^\dagger(\mathbf{x})\left(-i E_A^{\ i} \alpha^A \overset{\leftrightarrow}{\partial_i} -i \Delta \alpha^x \alpha^y \right) \chi(\mathbf{x}), \label{eq:C2_many_body_hamiltonian_general} 
\end{equation}
where the fields $\chi$ are Majorana spinors, i.e., $\chi = C \chi^*$ where $C = - \sigma^y \otimes \sigma^y$ is the charge conjugation matrix, as derived before in Eq.~(\ref{eq:C2_majorana_spinor}), and $\Sigma$ is the two-dimensional spatial hypersurface representing the honeycomb. In addition, the commutation relations are given by
\begin{equation}
\{ \chi(\mathbf{x}), \chi^\dagger(\mathbf{y}) \} = \delta(\mathbf{x} - \mathbf{y}) \mathbb{I}_4, \quad \{ \chi(\mathbf{x}) , \chi(\mathbf{y}) \} = C \delta(\mathbf{x} - \mathbf{y}), \label{eq:C2_majorana_commutation_2}
\end{equation}
which are derived in the same way as before in Eq.~(\ref{eq:C2_majorana_commutation}). 

\subsection{Emergent metric}

As it stands, this Hamiltonian is a generalised Dirac Hamiltonian on a flat Minkowski spacetime as the fields obey flat space commutation relations and the integration measure is the standard flat one. However, if we compare it to the general Riemann-Cartan Hamiltonian of Eq.~(\ref{eq:C2_RC_ham}), we see that the Hamiltonian can be viewed as describing a Majorana spinor field $\psi$, which is related to the non-relativistic continuum field $\chi$ via 
\begin{equation}
\psi(\mathbf{x}) = \frac{1}{\sqrt{|E|}} \chi(\mathbf{x}), \label{eq:C2_spinor_E}
\end{equation}
propagating on a Riemann-Cartan spacetime of the form $M = \mathbb{R} \times \Sigma$, where we \textit{interpret} the coefficients $E_A^{\ i}$ as the spatial components of the dreibein corresponding to the total dreibein
\begin{equation}
e_a^{\ \mu} =
\begin{pmatrix}
1 & 0 & 0 \\
0 & A & 0 \\
0 & C & B 
\end{pmatrix}, \quad e^a_{\ \mu} = 
\begin{pmatrix}
1 & 0 & 0 \\
0 & \frac{1}{A} & 0 \\
0 & - \frac{C}{AB} & \frac{1}{B}
\end{pmatrix}, \label{eq:C2_general_dreibein}
\end{equation} 
where the field $\psi(\mathbf{x})$, using the commutation relations of Eq.~(\ref{eq:C2_majorana_commutation_2}), obeys the correct commutation relations of a curved space Majorana fermion derived before in Eq.~(\ref{eq:C2_RC_commutators}). The corresponding metric is
\begin{equation}
g_{\mu \nu} = e^a_{\ \mu} e^b_{\ \nu} \eta_{ab} = \begin{pmatrix}
1 & 0 & 0 \\
0 & - \frac{1}{A^2} \left( 1 + \frac{C^2}{B^2} \right) & \frac{C}{AB^2} \\
0 & \frac{C}{AB^2} & - \frac{1}{B^2}
\end{pmatrix}. \label{eq:C2_general_metric}
\end{equation}
We call this metric an \textit{internal metric} as it does not represent the geometry of the underlying lattice, which of course remains fixed and situated in flat, non-relativistic Euclidean space, but it is instead an effective metric induced by the couplings of the model. It is convenient to interpret the model geometrically, as it provides us with a useful dictionary of geometric quantities to describe lattice observables with, provides us with an explanation of many phenomena observed and also motivates us to hunt for other geometric observables that would not have an obvious interpretation at the lattice level.

\begin{figure}[t]
\begin{center}
\includegraphics[scale=1]{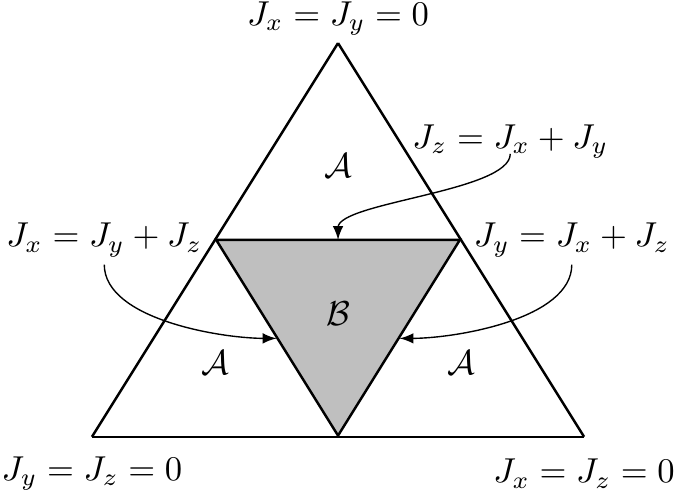}
\end{center}
\caption{The phase diagram of the vortex-free sector of Kitaev's honeycomb model, where the dependence on $J_x$, $J_y$ and $J_z$ for a small $K$ is shown. The centre of the diagram is where $J_x = J_y = J_z$ and all corners are limiting cases where two of the $J$'s are zero. The diagram contains two distinct phases $\mathcal{A}$ and $\mathcal{B}$, where $\mathcal{A}$ is the gapped toric code phase and $\mathcal{B}$ is the gapless non-abelian phase.}
\label{fig:C3_phase_diagram}
\end{figure}

This metric is rather complicated but we can immediately extract information from this pertaining to the phase properties of the underlying lattice model. We see that the metric has singularities when either $A = 0$ or $B = 0$. The condition $A = 0$ is equivalent to $2J_x J_z = \pm (J_y^2 - J_x^2 - J_z^2)$ which has the solutions
\begin{align}
J_x - J_y - J_z & = 0, \\
J_x - J_y + J_z & = 0, \\
J_x + J_y - J_z & = 0,
\end{align}
where we have assumed that the couplings are positive. These three conditions are nothing but the conditions for the critical points of the underlying lattice model. They define a triangular phase boundary between the gapped toric code phase $\mathcal{A}$ and the gapless non-abelian phase $\mathcal{B}$, as shown in Fig.~\ref{fig:C3_phase_diagram}. On the other hand, the second condition $B = 0$ implies $J_z = 0$. In this case, the model behaves as a set of gapless, disentangled, one-dimensional chains. We see that there is an intimate relationship between the phase diagram of the model and the singular points of the geometric model, providing a quick route to the conditions for criticality.

The reason for this relationship between the phase diagram of the model and the metric is because the metric encodes the dispersion relation $E(\mathbf{p})$ via the equation $g^{\mu \nu} p_\mu p_\nu = 0$, where $p_\mu = (E(\mathbf{p}),\mathbf{p})$ and we have taken $\Delta = 0$. The \textit{inverse} metric $g^{\mu \nu}$ measures the slope of the Dirac cones in the dispersion, so the metric $g_{\mu \nu}$ becomes singular if the dispersion becomes flat at the Fermi points, which is a signature of a phase transition as we will see later in Chap.~\ref{chapter:black_hole}.

\subsection{Metric stretching \label{sec:metric_stretching}}
Let us now consider a simple case. We consider the model for which $J_x = J_y = 1$ and $J_z \in [0,2]$. In this case, from Eq.~(\ref{eq:C2_general_metric}), the metric takes the form
\begin{equation}
g_{\mu \nu} = \begin{pmatrix} 
1 & 0 & 0 \\
0 & -\frac{1}{4 - J_z^2} & 0 \\
0 & 0 & - \frac{1}{3J_z^2}
\end{pmatrix}. \label{eq:C2_anisotropic_metric}
\end{equation}
At this stage, the couplings are constant so this model describes no curvature yet, however the effect of the metric can still be detected as the diagonal elements are unequal in magnitude. This indicates that the length scales in the $x$- and $y$-directions are unequal which is the signature of a \textit{dilation} of spacetime. To be more precise, consider two points $p, q \in \Sigma$ at a fixed time $t$. The effective distance between these two points is given by
\begin{equation}
d = \sqrt{- g_{ij} \Delta x^i \Delta x^j}, \label{eq:C3_internal_metric_distance}
\end{equation}
where $\Delta x^i$ is the spatial coordinate separation of the two points and $g_{ij}$ is the spatial portion of the metric only. We can visualise the effect of changing $J_z$ on the geometry of the model by considering a circle of constant \textit{coordinate} radius of $\Delta x^i = 1$. As we change $J_z$, the length scales of the $x$- and $y$-directions will change, stretching the axes of spacetime and deforming the circle into an ellipse. Let $d_x$ and $d_y$ be the $x$- and $y$-dimensions of the ellipse, we have
\begin{equation}
\frac{d_x}{d_y} = \frac{\sqrt{-g_{xx}}}{ \sqrt{-g_{yy}}} = \frac{\sqrt{3}J_z}{\sqrt{4 - J_z^2}}. \label{eq:C2_metric_dilation}
\end{equation}
We must be careful however: the ``distances" measured using the metric are \textit{effective} distances. The actual physical model is a non-relativistic model situated in Euclidean space described by the trivial Euclidean metric---the lattice is completely flat and undistorted as we change $J_z$. We only interpret the couplings of the model as an effective \textit{internal} metric for our convenience. Numerically, we use the flat Euclidean metric $\delta_{ij}$ to measure our spatial lengths, however we observe that any observable with the units of length with respect to the Euclidean metric will scale \textit{inversely} to how the internal metric scales in Eq.~(\ref{eq:C2_metric_dilation}). In other words, the Euclidean distance of observables measured on the lattice using the numerical simulation will scale in such a way to ensure the quantity $d$ of Eq.~(\ref{eq:C3_internal_metric_distance}) is held constant. We observe this effect by studying two-point Majorana correlators and Majorana zero-mode wavefunctions at the lattice level.

\subsubsection{Two-point Majorana correlators}

\begin{figure}[t]
\begin{center}
\includegraphics[scale=0.35]{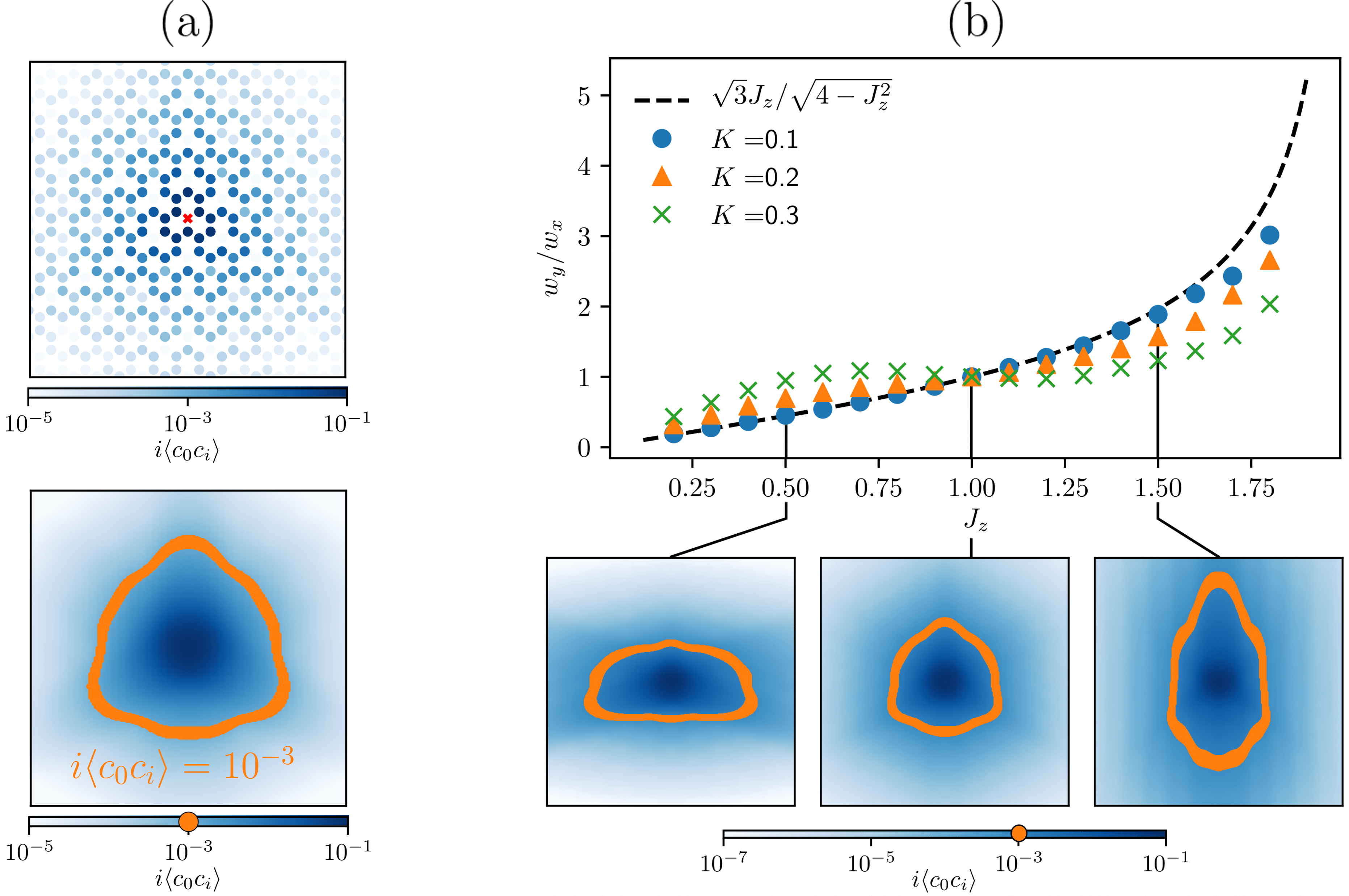}
\end{center}
\caption{(a) The two-point correlations $i\langle c_0c_i\rangle$ between each site $i$ and a central reference site $0$ denoted with a red cross and the continuous approximation as described by Eq.~(\ref{eq:C2_gaussian_majorana}). The size and shape of the correlations are characterised by finding the set of points where $i\langle c_0c_i\rangle = 10^{-3}$, as illustrated. Here we used $J_x=J_y=J_z=1$, system size $36\times36$, $K=0.1$ and $\epsilon = 1$. (b) At the top is the ratio of the height and width of the ``boundary'' $w_y/w_x$ for $J_x=J_y=1$, $\epsilon =1$, system size $36 \times 36$ for a range of $K$. The dashed line is the theoretical ratio from Eq.~(\ref{eq:C2_metric_dilation}).  At the bottom are examples of the boundaries we find for various $J_z$ and $K=0.1$ using Eq.~(\ref{eq:C2_gaussian_majorana}).}
\label{fig:C2_correlation_stretching}
\end{figure}

The two-point Majorana correlators are defined as
\begin{equation}
C_{ij} = i \langle \Omega | c_i c_j  |\Omega\rangle,
\end{equation}
where $c_i$ are the original Majorana modes of the lattice Hamiltonian in Eq.~(\ref{eq:C2_kitaev_majorana}), $|\Omega\rangle$ is the ground state of the model and the factor of $i$ is to ensure the correlation matrix is hermitian. These correlators can be calculated using exact diagonalisation of the Hamiltonian, as discussed in Appendix~\ref{appendix:numerical_techniques}. The correlation length $\xi$ of this model goes as $\xi \propto 1/\Delta$, therefore the smaller the gap, the further the correlations extend and the better the continuum limit approximation becomes.

As we are comparing a continuum limit result with a lattice simulation, we need to approximate the continuum by spreading out the correlators with a Gaussian which gives us a smooth distribution of correlators. Consider a fixed lattice site $i$, we define the scalar field
\begin{equation}
C(\mathbf{r}) =  i \sum_i \langle c_i c_j \rangle \delta(\mathbf{r}_j - \mathbf{r}) \rightarrow i \sum_i \frac{ \langle c_i c_j \rangle}{2 \pi \epsilon} e^{-\frac{|\mathbf{r} - \mathbf{r}_j|^2}{2 \epsilon}} \label{eq:C2_gaussian_majorana},
\end{equation}
where $\epsilon$ is taken to be around the lattice spacing to ensure that Gaussians of neighbouring lattice sites overlap, giving us a smooth plot as shown in Fig.~\ref{fig:C2_correlation_stretching}(a). This allows us to reduce the lattice effects, enabling a comparison with the continuum limit calculations.


In the 2D plots of Fig.~\ref{fig:C2_correlation_stretching}(b) we plot $C(\mathbf{r})$ for set of points centred on a reference site $i$ located at the origin. We plot the contours of this function and see that when $J_z = 1$, corresponding to the isotropic model, the function is approximately circular and isotropic. However, when $J_z$ moves away from the isotropic point, the correlators begin to deform in shape, forming an ellipse as expected. We measure the ratio of the height and width of one of the ellipses, $w_y/w_x$, as we change the value of $J_z$. We see that this ratio closely follows the \textit{inverse} of Eq.~(\ref{eq:C2_metric_dilation}) as expected. In addition, the agreement is stronger as we reduce $K$. This is because as $K$ gets smaller, the gap decreases and the continuum approximation is more accurate, giving us smoother and more elliptical contours.

\subsubsection{Zero mode wavefunctions}

\begin{figure}[t]
\begin{center}
\includegraphics[scale=0.35]{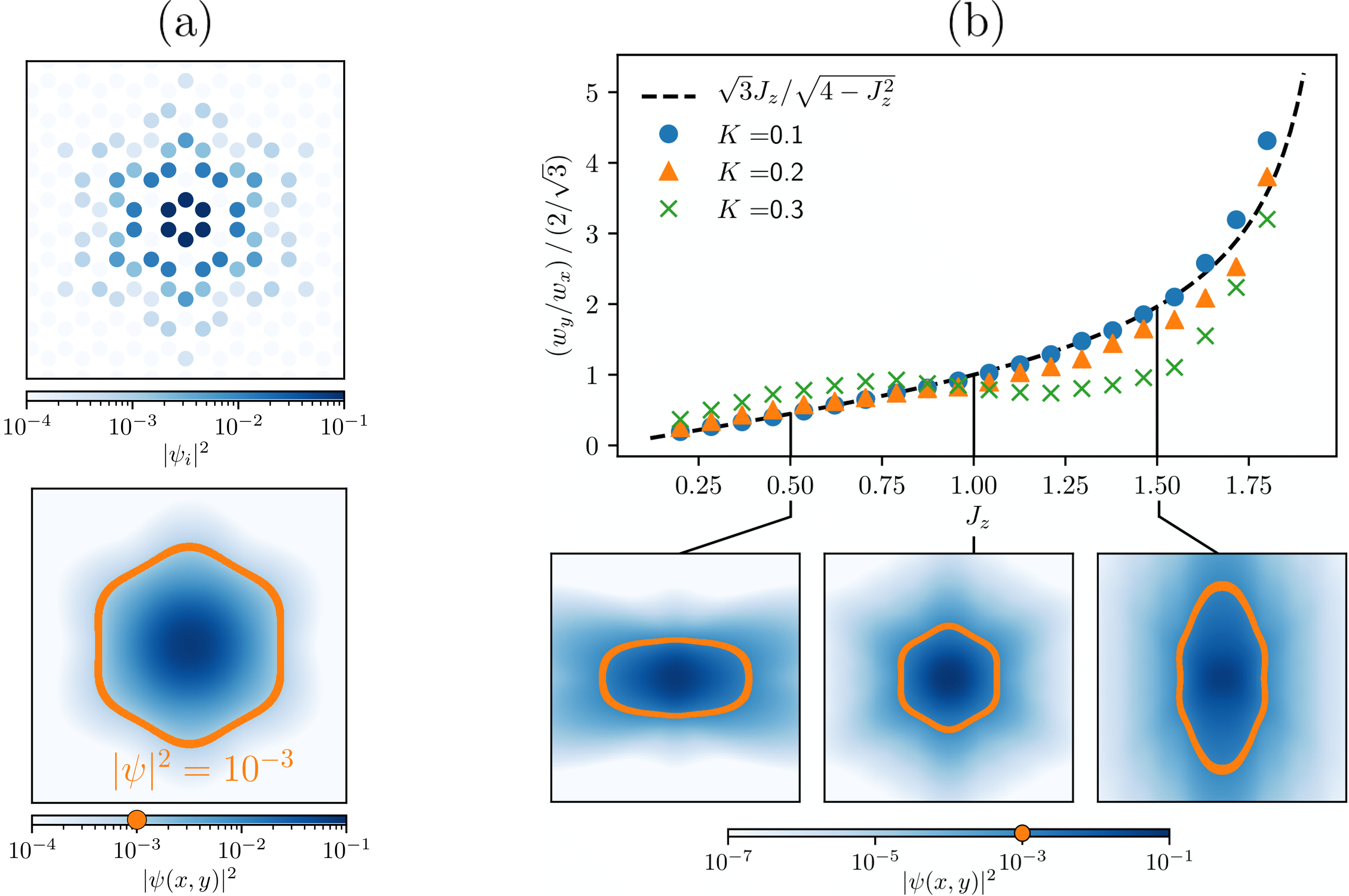}
\end{center}
\caption{(a) The discrete lattice probability density $|\psi_i|^2$ of the wave function for a vortex, located on the central plaquette, and its continuous approximation as given by Eq.~(\ref{eq:C2_smooth_wavefunction}). The size and shape of the vortex are characterised by finding the set of points where $|\psi(\boldsymbol{r})|^2 = 10^{-3}$, as illustrated. Here we used $J_x=J_y=J_z=1$, system size $36\times36$, $K=0.125$ and $\epsilon = 1$. (b) At the top is the ratio of the height and width of the ``boundary'' $w_y/w_x$ for $J_x=J_y=1$, $\epsilon =1$, system size $36 \times 36$ for a range of $K$. The dashed line is the theoretical ratio from Eq.~(\ref{eq:C2_metric_dilation}). Due to the hexagonal geometry of the lattice, we have divided the ratio $w_y/w_x$ by the ratio of the height and width of a regular hexagon $2/\sqrt{3}$. At the bottom are examples of the boundaries we find for various $J_z$ and $K=0.1$ using Eq.~(\ref{eq:C2_smooth_wavefunction}).}
\label{fig:C2_zero_mode_stretching}
\end{figure}

Vortex excitations can be introduced in pairs by inserting $\pi$-fluxes into the $\mathbb{Z}_2$ gauge field $u_{ij}$ that couples to the Majorana modes $c_i$ on the lattice, c.f. Eq.~(\ref{eq:C2_gauge_fixed_ham_majorana}). In the spectrum of the model, the vortex pair manifests as a zero-energy fermionic mode called a \textit{Majorana zero mode}. We study the spatial wave function $\psi_i$ of a single vortex sufficiently separated from its partner which can be done via exact diagonalisation using the methods of Appendix~\ref{appendix:numerical_techniques}.

To analyse the geometric profile of the zero modes we
adopt the same procedure we used for the Majorana correlations. We approximate the discrete wavefunction profile with a continuous distribution by replacing the probability density $|\psi_i|^2$ at each lattice point with a two-dimensional Gaussian centred at the site as
\begin{equation}
|\psi(\mathbf{r})|^2 = \sum_i |\psi_i|^2 \delta(\mathbf{r}_j - \mathbf{r}) \rightarrow \sum_i \frac{|\psi_i|^2}{2 \pi \epsilon} e^{-\frac{|\mathbf{r} - \mathbf{r}_j|^2}{2 \epsilon}}, \label{eq:C2_smooth_wavefunction}
\end{equation}
where $\epsilon$ is taken to be around the lattice spacing to ensure that Gaussians of neighbouring lattice sites overlap, giving us a smooth plot, as seen in Fig~\ref{fig:C2_zero_mode_stretching}(a).

In Fig.~\ref{fig:C2_zero_mode_stretching}(b), we see that for the isotropic case when $J_z = 1$, the wavefunctions are circular and isotropic, just like for the correlators. As we increase $J_z$, the wavefunction stretches in the same way as described by the inverse of Eq.~(\ref{eq:C2_metric_dilation}).

\section{Emergent Riemann-Cartan geometry \label{sec:RC_kitaev}}
We now consider the conditions required to generate curvature in the model. Our starting point is the anisotropic model discussed in the previous section, along with the dreibein and metric of Eq.~(\ref{eq:C2_general_dreibein}) and Eq.~(\ref{eq:C2_general_metric}) respectively. This continuum limit rested upon the assumption of translational invariance in the model, where the couplings were constant. We now relax this condition and allow the couplings to be functions of space (but not time). The translational invariance will now be broken, however if we assume that the couplings vary slowly with respect to the lattice spacing $a$, i.e., $|\nabla J_i | \ll 1/a$, then simply upgrading the couplings in the Hamiltonian of Eq.~(\ref{eq:C2_many_body_hamiltonian_general}) to functions of space gives us a faithful approximation to the true continuum limit of the inhomogeneous system. 

In the following we shall investigate the most general model and derive the spin connection, curvature, and torsion. In my original work of Ref.~\cite{C2_Farjami}, the geometric quantities were calculated using a \textit{brute force} approach by calculating the connection coefficients individually. In this section, we will instead take the more direct route of using the language of differential forms as introduced in Sec.~\ref{sec:Riemann_Cartan}, which allows us to calculate the curvature fully.

\subsection{The spin connection}
From the dreibein derived in Eq.~(\ref{eq:C2_general_dreibein}) we can uniquely determine the Levi-Civita spin connection $\omega^a_{\ b}$ using the useful torsion-free identity $\mathrm{d} e^a = - \omega^a_{\ b} \wedge e^b$, where the dreibein are expressed as one-forms $e^a = e^a_{\ \mu} \mathrm{d}x^\mu$. In order to take the exterior derivative we use the fact that it obeys a (generalised) Leibniz rule~\cite{C2_Nakahara}, so for the dreibein we have 
\begin{equation}
\mathrm{d}e^a = (\mathrm{d} e^a_{\ \mu}) \wedge \mathrm{d}x^\mu + e^a_{\ \mu}  \mathrm{d}^2 x^\mu  = \partial_\nu e^a_{\ \mu} \mathrm{d}x^\nu \wedge \mathrm{d}x^\mu,
\end{equation}
where in the second equality we used the identity $\mathrm{d}^2 = 0$. From Eq. (\ref{eq:C2_general_dreibein}), the dreibein are given by
\begin{equation}
e^0 = \mathrm{d}t , \quad e^1 = \frac{1}{A} \mathrm{d}x, \quad e^2 = - \frac{C}{AB} \mathrm{d}x + \frac{1}{B} \mathrm{d}y, \label{eq:C2_general_dreibein_one_forms}
\end{equation}
where now $A$, $B$ and $C$ are functions of space, which gives us the derivatives
\begin{align}
\mathrm{d}e^0 &  = 0, \label{eq:C2_de^0} \\
\mathrm{d}e^1 & = -\frac{\partial_y A}{A^2} \mathrm{d}x \wedge \mathrm{d}y \equiv F \mathrm{d}x \wedge \mathrm{d}y , \\
\mathrm{d}e^2 & = \left( \frac{\partial_y C}{AB} - \frac{C \partial_y A}{A^2B} - \frac{C \partial_y B}{AB^2} - \frac{\partial_x B}{B^2} \right) \mathrm{d}x \wedge \mathrm{d}y \equiv G \mathrm{d}x \wedge \mathrm{d}y,
\end{align}
where used the fact that the wedge products are anti-symmetric $\mathrm{d} x \wedge \mathrm{d}y = - \mathrm{d} y \wedge \mathrm{d}x$ and obey $\mathrm{d}x \wedge \mathrm{d}x = \mathrm{d} y \wedge \mathrm{d}y = 0$ to simplify the expressions.

The first result of Eq.~(\ref{eq:C2_de^0}), combined with the facts that the spatial dreibein in Eq.~(\ref{eq:C2_general_dreibein_one_forms}) have no time component, do not mix up space and time directions, and do not have any time-dependent components, tells us that the temporal dreibein are $\omega^0_{\ a} = \omega^a_{\ 0} = 0$, whilst the remaining non-zero dreibein $\omega^1_{\ 2} = -\omega^2_{\ 1}$ must take the form
\begin{equation}
\omega^1_{\ 2} = u \mathrm{d}x + v \mathrm{d}y ,
\end{equation}
for some scalar fields $u$ and $v$ which greatly simplifies the calculation. From the torsion-free relation $\mathrm{d}e^a = -\omega^a_{\ b} \wedge e^b$, the right hand side gives us
\begin{align}
\mathrm{d}e^1 & = -(u \mathrm{d} x + v \mathrm{d} y ) \wedge \left( - \frac{C}{AB} \mathrm{d}x + \frac{1}{B} \mathrm{d}y \right) = - \frac{1}{B} \left(u + \frac{vC}{A} \right) \mathrm{d}x \wedge \mathrm{d}y \\
\mathrm{d}e^2 & = (u \mathrm{d}x + v \mathrm{d}y ) \wedge \frac{1}{A} \mathrm{d}x = -\frac{v}{A} \mathrm{d}x \wedge \mathrm{d}y.
\end{align}
Comparing coefficients of the two expressions for $\mathrm{d}e^a$, we can read off $u$ and $v$ to give us the final result
\begin{equation}
\omega^1_{\ 2} = - \omega^2_{\ 1} = (CG - BF) \mathrm{d}x - AG \mathrm{d}y. \label{eq:C3_levi_civita_connection}
\end{equation}
\subsection{Curvature}
In this chapter, we are only interested in the curvature of the Levi-Civita connection $\omega^a_{\ b}$ as later we will see its application to an observable we can measure at the lattice level. From Def.~\ref{def:curvature}, the curvature of $\omega^a_{\ b}$ is given by
\begin{equation}
R^a_{\ b}(\omega) = \mathrm{d} \omega^a_{\ b} + \omega^a_{\ c} \wedge \omega^c_{\ b}.
\end{equation}
However, due to the form of our spin connection in Eq.~(\ref{eq:C3_levi_civita_connection}), the only non-zero component is given by
\begin{equation}
\begin{aligned}
R^1_{\ 2}(\omega) = -R^2_{\ 1}(\omega) & = \mathrm{d}\omega^1_{\ 2} + \omega^1_{\ c} \wedge \omega^c_{\ 2}  \\
& = \mathrm{d} \omega^1_{\ 2} + \omega^1_{\ 2} \wedge \omega^2_{\ 1} \\
& = \mathrm{d} \omega^1_{\ 2},
\end{aligned}
\end{equation}
where in the last equality we used the fact that $\omega^1_{\ 2} = -\omega^2_{\ 1}$ and the wedge product of a one-form with itself is zero. Therefore, the curvature is given by
\begin{equation}
R^1_{\ 2}(\omega)  = - R^2_{\ 1}(\omega) =  - \left[ \partial_y(C G - BF) + \partial_x(AG) \right] \mathrm{d}x \wedge \mathrm{d}y,
\end{equation}
while all other $R^a_{\ b}$ vanish. This is an extremely complicated expression for the curvature, however the upshot is that we require space-dependent couplings in the model to generate curvature in the continuum.

A simpler quantity that is of interest to us is the \textit{Ricci scalar} which is defined from a full contraction of the Riemann tensor, c.f. Def.~\ref{def:curvature}, as
\begin{equation}
R = R^{ab}_{\ \ \mu \mu} e_a^{\ \mu} e_b^{\ \nu}.
\end{equation}
As we are working on a spacetimes of the form $M = \mathbb{R} \times \Sigma$ with a metric given by Eq.~(\ref{eq:2+1_space}), time essentially plays the role of a parameter so the non-trivial geometry is contained on the two-dimensional hypersurface $\Sigma$. In two-dimensional spaces, the Riemann tensor only has one independent component~\cite{C2_Carroll} which simplifies things considerably. The Ricci scalar $R$ does not contain all of the information of the Riemann tensor, however in two-dimensional spaces it vanishes if and only if the Riemann tensor vanishes, so it provides us with a useful measure of the curvature. In addition, it will provide us with an observable which we can directly measure at the lattice level, as we shall see in the next section. 

To form the Ricci scalar $R$, we can use any coordinate system we like, including the dreibein basis, however we must always ensure that if any two indices are contracted they are with respect to the same coordinate system. For us, the Riemann tensor has mixed indices as $R^a_{\ b \mu \nu}$, c.f. Def.~\ref{def:curvature}, which is due to working with the second-order formalism, so we must use the dreibein to switch between the coordinate and dreibein bases. Explicitly, we have
\begin{equation}
R = g^{\mu \nu} R^{\alpha}_{\ \mu \alpha \nu} = g^{\mu \nu} e^{\ \alpha}_{a} e^b_{\ \mu} R^a_{\ b \alpha \nu}.
\end{equation}
Let us look at the original isotropic model for which $J_x=  J_y = J_z \equiv J$. In this case, the dreibein and metric are diagonal, given by Eq.~(\ref{eq:C2_general_dreibein}) and Eq.~(\ref{eq:C2_general_metric}) respectively, vastly simplifying our calculation, to give us
\begin{equation}
R(\omega) = 2 \partial^2 \ln J, \label{eq:isotropic_curvature}
\end{equation}
where $\partial^2 = g^{\mu \nu} \partial_\mu \partial_\nu$. Note that the curvature becomes singular if $J = 0$. This is to be expected, as if $J = 0$ the Hamiltonian is identically zero.
\subsection{Torsion}
The torsion of the model is much simpler. As we are working with a metric compatible connection, we showed that the torsion is related to the contorsion via $T^a = K^a_{\ b} \wedge e^b$. In components this reads
\begin{equation}
T^a_{\ \mu \nu} = 2 K^a_{\ [\mu \nu]}.
\end{equation}
In Eq. (\ref{eq:C2_RC_ham}) we saw that the spinor field only couples to the completely anti-symmetric portion of the connection $\Omega^a_{\ b}$ which then boiled down to the torsion pseudoscalar $c$. As this is all we can possibly deduce about torsion from the action, without any loss of generality we can take the contorsion to be completely anti-symmetric as $K_{abc} = \frac{c}{3!} \epsilon_{abc} $, therefore the torsion is given by
\begin{equation}
T^a_{\ \mu \nu} = \frac{c}{3} \epsilon^a_{\ \mu \nu},
\end{equation}
where $\epsilon^a_{\ \mu \nu} = \eta^{ab} e^c_{\ \mu} e^d_{\ \nu} \epsilon_{bcd}$. A comparison of the general continuum limit of Eq.~(\ref{eq:C2_many_body_hamiltonian_general}) with the general Hamiltonian from Riemann-Cartan theory of Eq.~(\ref{eq:C2_RC_ham}) reveals that the torsion pseudoscalar is given by $c = - 4\Delta$, therefore the $K$ term of the Hamiltonian generates torsion.
\subsection{Kekul\'{e} distortion}

\begin{figure}
\begin{center}
\includegraphics[align=c,scale=1]{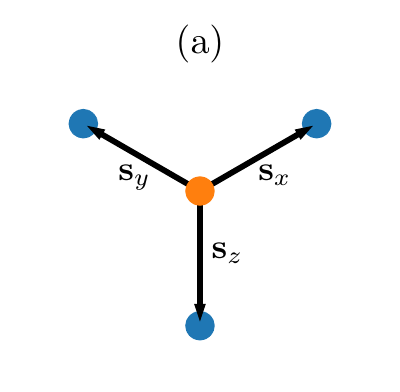}
\includegraphics[align=c,scale=1]{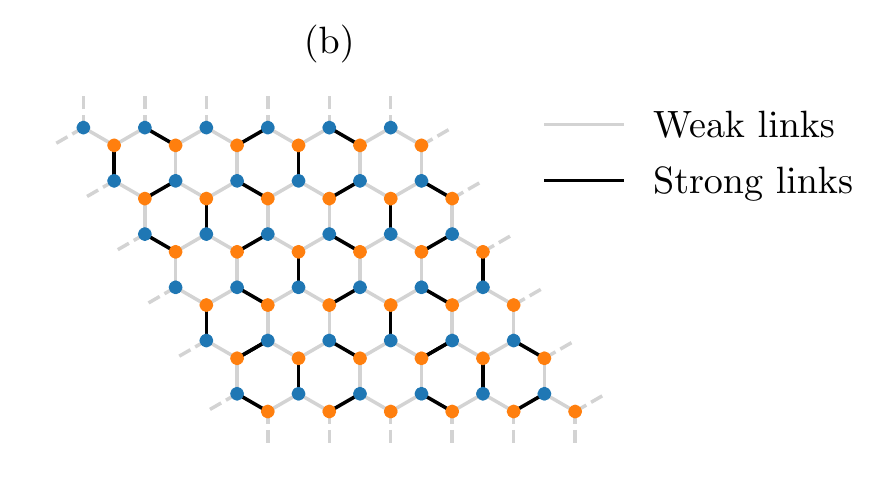}
\end{center}
\caption{(a) The nearest-neighbour link vectors $\mathbf{s}_x = \frac{b}{2} \left( \sqrt{3} , 1 \right) $, $\mathbf{s}_y = \frac{b}{2} \left( -\sqrt{3} , 1 \right)$ and $\mathbf{s}_z = b\left(0,-1\right)$, where $|\mathbf{s}_i| = b$ which is related to the lattice spacing $a$ via $b = a /\sqrt{3}$. (b) The Kekul\'e distortion of the nearest-neighbour couplings as given by Eq. (\ref{eq:C3_kekule}), where the strong and weak links are indicated by the black and grey links respectively. These couplings generate a mass in the continuum limit.}
\label{fig:C2_kekule}
\end{figure}

The $K$ term generates a gap which is interpreted as torsion in the continuum limit. To generate a mass term in the continuum limit, we need to introduce a Kekul\'e distortion to the nearest-neighbour couplings. This method is similar to the one employed in graphene to theoretically generate a mass gap~\cite{C2_Hou} and adjusted further for use with Majorana modes instead~\cite{C2_Yang}.

Consider the isotropic model where $J_x = J_y = J_z = J$. We modify the nearest-neighbour lattice couplings of Eq.~(\ref{eq:C2_gauge_fixed_ham_majorana}) as $J_{ij} \rightarrow J_{ij} + \delta J_{ij}$ for
\begin{equation}
\delta J_{ij} = \begin{cases} \frac{m}{3} e^{i \mathbf{P}_+ \cdot \mathbf{s}_\alpha} e^{i(\mathbf{P}_+- \mathbf{P}_-)\cdot \mathbf{r}_i } +\text {c.c.} & \text{if $(i,j)$ forms an $\alpha$-link} \\
0 & \text{otherwise}
\end{cases},
\label{eq:C3_kekule}
\end{equation}
where $\{ \mathbf{s}_\alpha \}$ are the three nearest-neighbour vectors as shown in Fig.~\ref{fig:C2_kekule}(a) and $m \in \mathbb{R}$. This additional term causes a Kekul\'e distortion in the couplings of the honeycomb lattice that has the form shown in Fig.~\ref{fig:C2_kekule}(b) and does not shift the Fermi points. 

The Kekul\'e distortion changes the unit cell of the honeycomb lattice to include six sites, causing the Brillouin zone to fold three times compared to the undisturbed case. Subsequently, we Fourier transform and restrict ourselves to the low-energy neighbourhood of the Fermi points. Up to first order in momentum, the additional term generated by the couplings gives the following contribution around the Fermi points~\cite{C2_Yang}
\begin{equation}
\delta H =  \int_\Sigma \mathrm{d}^2 x m \chi^\dagger \beta \chi,
\end{equation}
where $\beta \equiv \gamma^0 = \sigma^x \otimes \mathbb{I}$, which is quoted in the four-dimensional language of Sec.~\ref{sec:C2_4x4_cont_lim}. This is the mass term given in the general Riemann-Cartan Hamiltonian of Eq.~(\ref{eq:C2_RC_ham}), hence a Kekul\'e distortion generates a mass $m$.

When $K=0$ the Kekul\'e distortion creates an energy gap due to a non-zero mass $m$. In this situation, vertices of sub-lattice $A$ are coupled exclusively to vertices of sub-lattice $B$, so the model has chiral symmetry under relabelling of the sublattices $A \leftrightarrow B$ and the phase of the system belongs in the BDI class which has trivial Chern number using the Altland-Zirnbauer classification of topological systems \cite{AZ,C2_Chiu}. Nevertheless, zero-dimensional defects, such as vortices, can trap chiral Majorana zero modes~\cite{C2_Jackiw,C2_Yang,C2_Chiu}. This should be contrasted with the case where only the $K$ term is present and the model is equivalent to the $p+ip$ superconductor belonging to class D~\cite{C2_Chiu}. 

\begin{figure}[t]
\begin{center}
\includegraphics[scale=0.8]{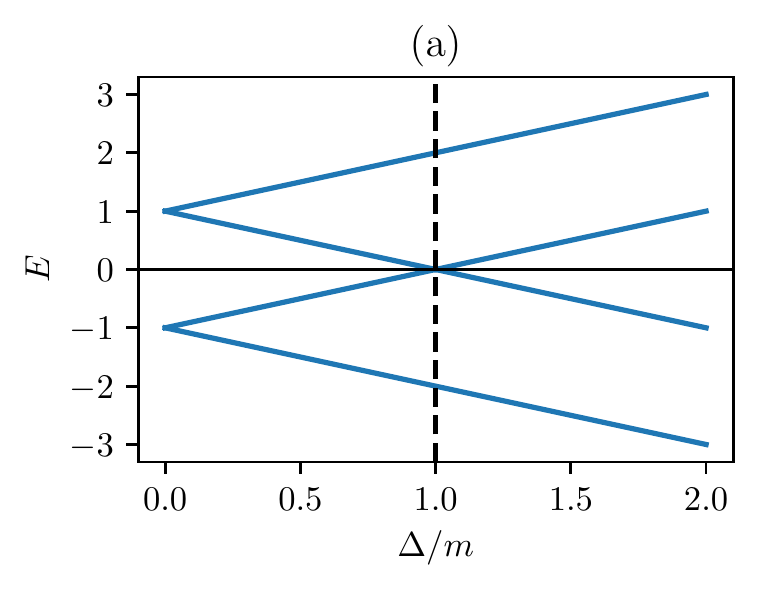}
\includegraphics[scale=0.8]{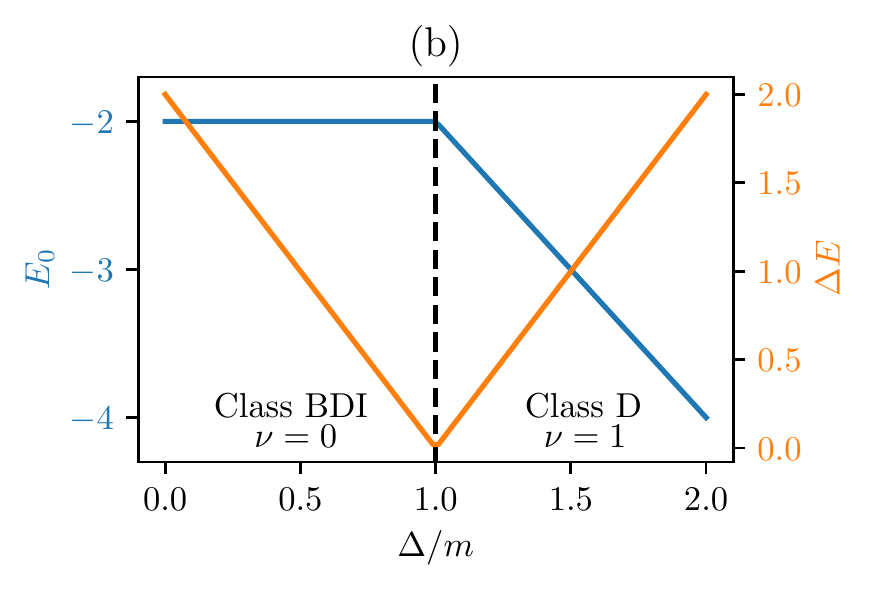}
\end{center}
\caption{(a) The single-particle energy levels of $h'(\mathbf{0})$ of Eq.~(\ref{eq:C2_mass_kekule_ham}) as a function of the ratio $\Delta/m$. We see that at $\Delta/m = 1$ there is a level crossing which is the signature of a quantum phase transition. (b) The ground state energy $E_0$ and energy gap $\Delta E$ for the many-body Hamiltonian. We see that at the location of the level crossing, $\Delta/m = 1$, there is a discontinuity in the the first derivative of $E_0$ and the gap closes, which signifies a topological phase transition between a Class BDI model with Chern number $\nu = 0$ and a Class D model with Chern number $\nu = 1$.}
\label{fig:C3_kitaev_kelule_phase}
\end{figure} 

By varying the couplings of the extended Kitaev model with the Kekul\'{e} distortion, it is possible to induce a topological phase transition between the BDI and D phases, whereby the Chern number $\nu$ of the model changes from $0$ to $1$. We investigate this by studying the single-particle Hamiltonians obtained from the quantum field theory. With the Kekul\'{e} distortion, the single-particle Hamiltonian takes the form
\begin{equation}
h'(\mathbf{p}) = E_A^{\ i} \alpha^A p_i - i\Delta \alpha^x \alpha^y + \beta m . \label{eq:C2_mass_kekule_ham}
\end{equation}
We are interested in the gap of this Hamiltonian, so we can safely set all $p_i = 0$ as the gap is located at the Fermi points (recall that $p_i$ measures the distance from the Fermi points). From this we can deduce that there must be a first-order phase transition exhibited by this model as the two remaining terms in the Hamiltonian, $\alpha^x \alpha^y$ and $\beta$, commute, where the location of the critical point is when the magnitude of their coefficients are equal. In Fig.~\ref{fig:C3_kitaev_kelule_phase}, we present the single-particle energy levels of $h'(\mathbf{0})$ as we change the ratio $\Delta/m$. We see that there is a level crossing at $\Delta/m = 1$ in the single-particle spectrum. In the many-body picture, this corresponds to a discontinuity in the first derivative of the ground state energy $E_0$ and the closing of the gap $\Delta E$ which are the signatures of the topological phase transition, in this case between the BDI and D classes.

In Eq.~(\ref{eq:C2_RC_ham}) we showed that in $(2+1)$D the torsion pseudoscalar arises in the Hamiltonian as the coefficient of $\alpha^x \alpha^y$, whilst the mass term arises as the coefficient of $\beta$. Any representation of the alpha and beta matrices would suffice and for certain choices, $\alpha^x \alpha^y \propto \beta$, in which case the torsion pseudoscalar $c$ is simply absorbed into the mass $m$ so the two are indistinguishable. Nevertheless, this section has demonstrated that the Kekul\'e distortion and $K$ term have distinguishable topological properties, which was revealed by the continuum limit with our choice of gamma matrices. For this reason, we choose to interpret $\Delta$ as generating a  torsion $c$, whilst $m$ generates a mass. In the next section we will relax this when we are forced to work with the two-dimensional representation, in which case we have no choice but to interpret $\Delta $ as a mass.

\section{Spin densities \label{sec:spin_densities}}
\subsection{Background theory of spin currents}
In Sec.~\ref{sec:metric_stretching} we were able to deduce the underlying metric of the model by studying how correlators and zero mode wavefunctions reacted to the scaling of couplings, however we now ask: what observable can we measure that detects the curvature of the model? In order to do this, it is convenient to work with the two-dimensional language of Sec.~\ref{sec:C2_2x2_cont_lim}. In this case, we have a pair of two-dimensional Hamiltonians $h_\pm(\mathbf{p})$, one about each Fermi point. In this language, the two-component spinors $\chi_\mu$ are \textit{not} Majorana and are just standard Dirac spinors. However, we can still interpret the the action for these as a Riemann-Cartan action given by Eq.~(\ref{eq:C2_RC_action}) but without the factor of $1/2$ in front which is required for a Majorana spinor only. The following section is unpublished work. 

A useful observable to use that responds to the presence of curvature is the spin current which was defined previously in Sec.~\ref{sec:torsion_in_GR}. To remind the reader, given an action $S$ of a quantum field theory, we define the spin current as the functional derivative of the action with respect to the spin connection as
\begin{equation}
S^{ a b \mu} := \frac{1}{|e|} \frac{\delta S}{\delta \Omega_{ ab \mu}},
\end{equation}
where we assume that the dreibein $e^a$ and spin connection $\Omega^a_{\ b}$ vary independently. This object is found on the right hand side of the equation of motion for the connection in Einstein-Cartan theory, see Eq.~(\ref{eq:einstein_cartan_2}). Here we have ignored the factor of $-2$ in the definition of the spin current as it is not important. This observable is obtained from Noether's theorem applied to the internal SU(2) rotational symmetry of the Dirac action in $(2+1)$D. In our case, the ``spin" components of our spinors represent the two sublattice degrees of freedom, either sublattice $A$ or $B$. Therefore, the SU(2) symmetry of the continuum limit corresponds to rotational symmetry in the internal sublattice space.

In order to evaluate the ground state expectation value of this operator, we resort to using path integral techniques. We consider the dreibein $e_a$ and spin connection $\Omega^a_{\ b}$ as static background fields as they are fixed by the couplings which do not fluctuate. The ground state expectation value is given by the path integral
\begin{equation}
\langle S^{ab \mu} \rangle = \frac{1}{Z} \int \mathcal{D} \psi \mathcal{D} \bar{\psi} S^{ab \mu} e^{i S[\psi,\bar{\psi},e,\Omega]},
\end{equation}
where $Z = \int D \psi D\bar{\psi} e^{iS} \equiv e^{iW}$ is the partition function and $W = W[e,\Omega]$ is the \textit{effective action} obtained by integrating out the fermionic fields which depends only upon the dreibein and spin connection. Due to the definition of the spin current, we have
\begin{equation}
\langle S^{ab \mu} \rangle  = \frac{1}{Z} \int \mathcal{D} \psi \mathcal{D} \bar{\psi} \left( \frac{1}{|e|} \frac{\delta S}{\delta \Omega_{ ab \mu}} \right) e^{i S}  = \frac{1}{|e|} \frac{\delta W}{\delta \Omega_{ab \mu}},
\end{equation}
where in the last line we used the fact that $Z = e^{iW}$. This is a useful result which allows us to evaluate the spin current. For the case of a Riemann-Cartan action $S_\mathrm{RC}$ of Eq.~(\ref{eq:C2_RC_action}), the effective action $W_\mathrm{RC}$ after integrating out the fermions is given perturbatively by~\cite{Hughes,Parrikar,C2_Golan}
\begin{equation}
\begin{aligned}
W_\mathrm{RC}[e,\Omega] & = \frac{k_H}{2} \int Q_3(\omega) - \frac{k_H}{2} \int R(\omega) e^a \wedge De_a  + \frac{\zeta_H}{2} \int e^a \wedge D e_a \\
& + \frac{1}{2k_N} \int \left( R(\omega) - 2\Lambda + \frac{3}{2} c^2 \right) |e| \mathrm{d}^3 x + \ldots , \label{eq:C2_effective_action}
\end{aligned}
\end{equation}
where 
\begin{equation}
Q_3(\omega) =  \omega^a_{\ b} \wedge \mathrm{d} \omega^b_{\ a} + \frac{2}{3} \omega^a_{\ b} \wedge \omega^b_{\ c} \wedge \omega^c_{\ a},
\end{equation}
is the Chern-Simons $3$-form, $c = C_{abc} \epsilon^{abc}$ is the torsion pseudoscalar, $R(\omega)$ is the Ricci scalar of the Levi-Civia connection and $De_a = \mathrm{d}e_a + \Omega_{ab} \wedge e^b$. The coefficients $k_\mathrm{H}$, $\zeta_H$ take the form
\begin{equation}
k_H = |k_H| \mathrm{sgn}(m)o ,\quad \zeta_H = |\zeta_H| \mathrm{sgn}(m) o,
\end{equation}
where $o = \mathrm{sgn}\left(\det\left[e^a_{\ \mu}\right] \right)$ is the orientation of the dreibein, whilst $k_N$ is unimportant for us~\cite{C2_Golan}. This perturbative result is UV insensitive, so is insensitive to the microscopic lattice details.

Let us call these four terms $W_\mathrm{RC} = W_1 + W_2 + W_3 + W_4$. Immediately, we see that $W_1$ will not contribute to the spin density as it does not depend upon the connection $\Omega^a_{\ b}$. This is because $W_1$ is a functional of the Levi-Civita connection $\omega^a_{\ b}$ only, which in turn is a functional of the dreibein, i.e., $\omega^a_{\ b} = \omega^a_{\ b}(e)$. When we take the derivative with respect to the spin connection we keep the dreibein $e_a$ fixed by definition, so $W_1$ does not vary. 

Now let us mould $W_2$ into something more useful. We have
\begin{equation}
\begin{aligned}
W_2 & = - \frac{k_H}{2} \int R(\omega) e^a \wedge D e_a \\
& = - \frac{k_H}{2} \int R(\omega) e^a \wedge \left( \mathrm{d} e_a + \Omega_{ab} \wedge e^b \right) \\
& = + \frac{k_H}{2} \int R(\omega) \Omega_{ a b \mu} e_c^{\ \mu} e^a \wedge e^b \wedge e^c + \ldots ,
\end{aligned}
\end{equation}
where we have explicitly expanded out the spin connection into the dreibein basis using $\Omega_{ab} = \Omega_{ab \mu} \mathrm{d}x^\mu = \Omega_{ab \mu} e^{\ \mu}_c e^c$ and changed the sign by reordering the wedge products. The ellipsis here represent the part that does not depend on the spin connection so is unimportant. If we expand out the dreibein explicitly as $e^a = e^a_{\ \mu} \mathrm{d}x^\mu$, we can rewrite the double wedge product in a more useful form as
\begin{equation}
e^a \wedge e^b \wedge e^c  = e^a_{\ \alpha}e^b_{\ \beta} e^c_{\ \gamma} \mathrm{d}x^\alpha \wedge \mathrm{d}x^\beta \wedge \mathrm{d}x^\gamma 
  = e^a_{\ \alpha}e^b_{\ \beta} e^c_{\ \gamma} \epsilon^{\alpha \beta \gamma} \mathrm{d}^3 x   = o |e| \epsilon^{abc} \mathrm{d}^3 x,
\end{equation}
where we used the determinant identity $e^a_{\ \alpha} e^b_{\ \beta} e^c_{\ \gamma} \epsilon^{\alpha \beta \gamma} = \det\left[e^a_{\ \alpha}\right] \epsilon^{abc}$, $o = \mathrm{sgn}\left(\det \left[e^a_{\ \alpha} \right] \right)$ is the orientation of the basis, $e \equiv \det \left[ e^a_{\ \alpha} \right]$ and $\mathrm{d}^3x = \mathrm{d}t \wedge \mathrm{d}x \wedge \mathrm{d}y$ is the standard integration measure, so
\begin{equation}
W_2 =  \frac{k_H}{2} \int R(\omega) \Omega_{ ab \mu} e_c^{\ \mu} o |e| \epsilon^{abc} \mathrm{d}^3 x + \ldots.
\end{equation}
This is now in a convenient form to take the derivative, so we just peel off the coefficient of the spin connection to yield
\begin{equation}
\frac{1}{|e|}\frac{\delta W_2}{\delta \Omega_{ a b \mu}} =  \frac{k_H}{2} R(\omega) e_c^{\ \mu} o  \epsilon^{abc}.
\end{equation}
The form of $W_3$ is the same as $W_2$, except it does not have a factor of $R(\omega)$ in the integrand, so following the same steps as above we end up with
\begin{equation}
\frac{1}{|e|} \frac{\delta W_3}{\delta \Omega_{\mu a b}} = - \frac{\zeta_H}{2}e_c^{\ \mu} o \epsilon^{abc}.
\end{equation}
Finally, the only part of $W_4$ that depends on the connection is the torsion pseudoscalar $c = C_{abc}\epsilon^{abc}$. As the spin connection can be split up into the Levi-Civita connection and contorsion as $\Omega_{ a b \mu} = \omega_{ a b \mu} + C_{ a b \mu}$, the variation is given by $\delta \Omega_{\mu a b} = \delta C_{\mu a b}$ as the Levi-Civita connection $\omega_{a b \mu}$ is fixed. Performing a variation to first order gives
\begin{equation}
\delta(c^2) = 2 c \delta c = 2 c \delta C_{ ab \mu} e_c^{\ \mu} \epsilon^{cab}  = 2c \delta \Omega_{ab \mu} e^{\ \mu}_c \epsilon^{cab},
\end{equation}
so the variation of $W_4$ w.r.t. the spin connection is
\begin{equation}
\delta W_4 = \frac{1}{2k_N} \int \frac{3}{2} \delta (c^2) |e| \mathrm{d}^3 x = \frac{3}{2 k_N} \int c \delta \Omega_{ a b \mu} e_c^{\ \mu} \epsilon^{cab} |e| \mathrm{d}^3 x ,
\end{equation}
which allows us to read off the derivative as
\begin{equation}
\frac{1}{|e|} \frac{\delta W_4}{\delta \Omega_{ab \mu}} = \frac{3}{2k_N} c e_c^{\ \mu} \epsilon^{cab}.
\end{equation}

Now let us focus on the temporal component of the spin current $ \rho \equiv S^{12t}$, namely the spin density. Pulling everything together from above, the ground state expectation value of the spin density is given by
\begin{equation}
\langle \rho \rangle  = \frac{k_H}{2} R(\omega) o - \frac{\zeta_H}{2} o + \frac{3}{2k_N} c + \ldots , \label{eq:spin_density_gs}
\end{equation}
where we have assumed we are working on the spacetime $M = \mathbb{R} \times \Sigma$ so the only non-zero temporal dreibein is $e_0^{\ t} = 1$. This observable depends upon the curvature via the Ricci scalar of the Levi-Civita connection $R(\omega)$, where the ellipsis represents higher order terms in the mass $m$. As this observable was derived using a continuum limit field theory, we now ask what is the corresponding observable in Kitaev's honeycomb model at the lattice level.

\subsection{Spin densities in Kitaev's honeycomb model}
The effective action $W_\mathrm{RC}$ is derived under the assumption that we are working with a two-dimensional representation of the gamma matrices, so we must ensure we use the two-dimensional language of Sec.~\ref{sec:C2_2x2_cont_lim} instead of the four-dimensional representation we used in this chapter so far. To remind the reader, we took the continuum limit of Kitaev's honeycomb model by Taylor expanding the single-particle Hamiltonian about each Fermi point. This process yielded two two-dimensional Hamiltonians $h_\pm(\mathbf{p})$---one about each Fermi point---given by
\begin{equation}
h_\pm(\mathbf{p}) = \left( \mp A \sigma^x + C \sigma^y \right)p_x + B \sigma^y  \mp \Delta \sigma^z \equiv (E_\pm)^{\ i}_A \hat{\alpha}^A p_i \mp \Delta \hat{\beta},
\end{equation}
where it employs a two-dimensional representation of the Dirac alpha and beta matrices $\hat{\alpha}^i = \sigma^i$ and $\hat{\beta} = \sigma^z$. Unlike in the four-dimensional representation, we have less flexibility and must make this choice. This yields the gamma matrix representation $\hat{\gamma}^0 = \hat{\beta}$ and $\hat{\gamma}^A = \hat{\beta}^{-1} \hat{\alpha}^A$, where
\begin{equation}
\hat{\gamma}^0 = \sigma^z, \quad \hat{\gamma}^1 = i \sigma^y, \quad \hat{\gamma}^2 = -i \sigma^x, \label{eq:C2_2D_gamma}
\end{equation}
which indeed obeys the $(2+1)$D Clifford algebra $\{ \hat{\gamma}^a , \hat{\gamma}^b \} = 2\eta^{ab}$. A consequence of this, however, is that torsion and mass are indistinguishable as in this representation $\hat{\gamma}^0 \hat{\gamma}^1 \hat{\gamma}^2 = i$, so the torsion pseudoscalar of Eq.~(\ref{eq:C2_RC_ham}) is absorbed into the mass. As the effective action of Eq.~(\ref{eq:C2_effective_action}) is derived under the assumption of a non-zero mass,  we choose to interpret the quantity $\Delta$ as a mass.

As we did before in the four-dimensional language in this chapter, we are free to interpret these Hamiltonians as a pair of Riemann-Cartan Hamiltonians, except each Fermi point now has its own set of dreibein which differ by an \textit{orientation}, i.e., the determinant $\det \left[e^a_{\ \mu}\right]$ about each point differs in sign. Both Fermi points will yield the same metric as before as an orientation change can be viewed as a gauge transformation, so basis independent quantities remain unchanged. Another important feature to point out is that in the four-dimensional language, the four-component field $\chi$ is a Majorana spinor, however in the two-dimensional language, the fields $\chi_\mu$ are not Majorana, therefore, we must use the language of standard Dirac fermions.

From Eq.~(\ref{eq:C2_RC_action}), the Riemann-Cartan action for a Dirac spinor $\psi$ can be written in the convenient form
\begin{equation}
\begin{aligned}
S_\mathrm{RC} &  =  \int_M \mathrm{d}^2x |e|  \left( \ldots + \frac{i}{2}  \bar{\psi} \{ \gamma^\mu , \Omega_\mu \}  \psi \right) \\
& =  \int_M \mathrm{d}^2 x |e| \left( \ldots + \frac{i}{16} \Omega_{ab\mu} \bar{\psi} \{ \gamma^\mu , [\gamma^a , \gamma^b ]\} \psi \right),
\end{aligned}
\end{equation}
where we have removed the factor of $1/2$, and the ellipsis represents the kinetic term and mass term of the action which are both unimportant here.

Taking the derivative is now simple: we just peel off the coefficient of $\Omega_{ab \mu}$ to yield the spin current
\begin{equation}
S^{ab\mu} = \frac{i}{16} \bar{\psi} \{ \gamma^\mu, [\gamma^a , \gamma^b] \} \psi = \frac{i}{4} \epsilon^{\ \mu}_c \epsilon^{cab} \psi^\dagger \gamma^1 \gamma^2 \psi,
\end{equation}
where we used the $(2+1)$D gamma matrix identity $\{ \gamma^a, [\gamma^b ,\gamma^b] \} = 4 \epsilon^{abc} \gamma^0 \gamma^1 \gamma^2$ and $\bar{\psi} \gamma^0 = \psi^\dagger$. However, as we are now working with a two-dimensional representation of the gamma matrices in Eq.~(\ref{eq:C2_2D_gamma}), we have less flexibility with our gamma matrices, therefore the spin density is given by
\begin{equation}
\rho = S^{12t} = -\frac{1}{4} \psi^\dagger \sigma^z \psi, \label{eq:spin_density}
\end{equation}
where again we used the fact that we are working on a spacetime of the form $M = \mathbb{R} \times \Sigma$ so the only non-zero temporal component of the dreibein is $e_0^{\ t} = 1$. 

In Eq.~(\ref{eq:C2_spinor_E}) we showed that we can interpret the continuum limit of Kitaev's honeycomb model if we assume the spinor fields $\chi$ of the continuum limit and spinor fields of the effective Riemann-Cartan theory $\psi$ are related via a factor of $\sqrt{|E|}$. This remains in the two-dimensional language too, where each Fermi point contributes a field
\begin{equation}
\psi_\mu = \frac{1}{\sqrt{|E_\mu|}} \chi_\mu ,
\end{equation}
where $\mu = \pm$. Using this fact and combining with Eqs.~(\ref{eq:spin_density_gs}) and (\ref{eq:spin_density}), we find the spin density is related to our lattice observables with the useful result
\begin{equation}
\langle \chi^\dagger_\mu \sigma^z \chi_\mu \rangle  = - 2 |E| \left( k_H R(\omega) o_\mu - \zeta_H o_\mu  + \ldots \right),
\end{equation}
where we have set the torsion pseudoscalar $c = 0$ as we have chosen to interpret $\Delta$ as a mass now.

What observable does this correspond to on the lattice? Terms that go as $\sigma^z$ generate a gap which the the $K$ term does. Referring back to Eq.~(\ref{eq:C2_2x2_field_ham}), we found in the two-dimensional language that the $K$ term of the Hamiltonian in the continuum limit is
\begin{equation}
H_K  = \Delta \int_\Sigma  \mathrm{d}^2 x \left( -\chi^\dagger_+ \sigma^z \chi_+ + \chi^\dagger_- \sigma^z \chi_- \right) \equiv \int_\Sigma \mathrm{d}^2 x \mathcal{H}_K ,
\end{equation}
where $\mathcal{H}_K$ is the Hamiltonian density corresponding to the $K$ term. Using the results above, the ground state expectation value of this quantity is given by
\begin{equation}
\langle \mathcal{H}_K \rangle  = \Delta  \left(- \langle \chi_+^\dagger \sigma^z \chi_+ \rangle + \langle \chi^\dagger_- \sigma^z \chi_- \rangle \right)  = 4\Delta |E| \left( |k_H| R(\omega) - |\zeta_H| \right) \label{eq:GS_spin_density},
\end{equation}
where we have used the fact that the masses about each Fermi point are given by $m_\pm = \mp \Delta$. We see that the density of the $K$ term is proportional to the curvature of the model. As the $K$ term is easily accessible at the lattice level, we have an observable that we can measure to detect the presence of curvature.  

For the isotropic model, using the result of Eq.~(\ref{eq:isotropic_curvature}) for the curvature, we have $R(\omega) \propto g^{\mu \nu} \partial_\mu \partial_\nu \ln J$. We see that the curvature vanishes only if $J$ is a constant. Therefore, applying the result of Eq.~(\ref{eq:GS_spin_density}), we see that the ground state expectation value of the density $\langle \mathcal{H}_K \rangle$ is only space-dependent if the model contains curvature which allows us to detect its presence, where we assume that $K$, and hence $\Delta$, is a constant.
\section{Conclusion \label{kitaev_conclusion}}
In this chapter we expanded upon the known result that the low-energy limit, or continuum limit, of Kitaev's honeycomb model is described by massless Majorana fermions obeying the Dirac equation on a Minkowski spacetime. We took this idea further by investigating whether the continuum limit could possibly yield non-trivial \textit{curved} geometries. A suitable generalisation of Minkowski spacetime, namely a Riemann-Cartan geometry which contains both curvature and torsion, manifests itself in Kitaev's honeycomb via non-trivial dreibein $ e_a $ and spin connection $\Omega^a_{\ b}$. It was shown that if the couplings of the model take a general space-dependent form, a Riemann-Cartan continuum limit can indeed be obtained.

We first showed theoretically that the continuum limit of Kitaev's honeycomb with general space-dependent couplings can be identified as a Riemann-Cartan theory, where the nearest-neighbour couplings $\{ J_i \}$ determine the dreibein $e_a$, whilst the next-to-nearest-neighbour couplings $\{ K_i \}$ determine the torsion pseudoscalar. We noted that, quite remarkably, the singularities of the emergent metric coincided with the location of the critical points of the model.

As an initial investigation of the geometric interpretation, we showed that a homogeneous model with couplings $J_x = J_y = 1$ and $J_z \in [0,2]$ generated a continuum limit whose metric described a \textit{dilation} of spacetime, where $J_z$ controls the stretching. Using the numerics of Ref.~\cite{C2_Farjami}, we saw this emergent metric faithfully predicted how Majorana correlation functions $i \langle c_i c_j \rangle$ and zero-mode wavefunctions $|\psi_i|^2$ warp and stretch under scaling of $J_z$, with the geometric description improving as we reduced $K$ (reducing the gap). This provides us with strong evidence of the emergent metric of the model.

We then introduced a Kekul\'e distortion to the model, generating a mass $m$ in the continuum limit, which should be contrasted to the gap $\Delta$ which is actually interpreted as the torsion pseudoscalar $c$. By tuning the magnitude of $m$ and $\Delta$, the continuum limit reveals a first-order phase transition between two topological phases due to the level crossing in the single-particle spectrum or equivalently the closing of the many-body gap. We showed that for $\Delta > m$ the system is in a topological superconducting phase of class D, whereas for $m > \Delta$ the system is found in the class BDI phase, with the gapless critical point between these two phases where $m = \Delta$. 

Finally, we upgraded the couplings to space-dependent functions, which generated a curvature in the continuum limit. In order to probe how the model reacts to the presence of curvature, we studied the spin densities and showed they are related to the curvature of the Levi-Civita connection $\omega^a_{\ b}$. At the lattice level, the spin density corresponds to the density of the $K$ term of the Hamiltonian, which is an easily accessible observable to measure, giving us a way to measure the curvature at the lattice level.

\chapter{Chiral Gauge fields in Kitaev's honeycomb model \label{chapter:chiral}}
\section{Introduction}
We have seen in the previous chapters that the continuum limit provides a powerful tool to interpret many lattice effects using the language of geometry. In this chapter, we propose to build upon these studies by considering \textit{chirality} and chiral gauge fields, which is a rather exotic concept of high energy physics that permeates to condensed matter systems. 

Massless fermions in $(3+1)$D can be described by spinors which are reducible into a pair of Weyl fermions of opposite chirality. This chirality, either left-handed or right-handed, signals how these objects transform under Lorentz transformations~\cite{C2_Maggiore,Peskin,srednicki_2007}. The weak interaction of the Standard Model is chiral in nature as its interactions treat left- and right-handed particles differently~\cite{C3_Maggiore}. Chirality also arises naturally in lattice gauge theories~\cite{C3_Creutz} and condensed matter systems such as Weyl semimetals, whose low-energy excitations are described by Weyl fermions. There is an intimate relationship between chiral gauge fields and torsion in the continuum limit which allows one to produce strain-induced gauge fields by inserting deformations to the lattice~\cite{C3_laurila2020torsional,C3_Cortijo,C3_Sumiyoshi,C3_Landsteiner,C3_Grushin_2016,C3_Pikulin_2016,C3_Gorbar_2017,C3_Ferreiros_2019}. Upon coupling to gauge fields, these systems can exhibit the chiral anomaly~\cite{C3_Gian,C3_Bertlmann,C3_laurila2020torsional,C3_Landsteiner,C3_Pikulin_2016}, where chiral symmetry is broken resulting in a non-conserved current and a generalised quantum Hall effect~\cite{C3_Liu}. Chirality has also been discussed in the context of graphene~\cite{C3_Jackiw2}, phase transitions~\cite{C3_Ville3}, and Landau levels~\cite{C3_Rachel,C3_laurila2020torsional,C3_Grushin_2016}. 

We saw in Sec.~\ref{sec:C2_Kitaev's honeycomb model} that the continuum limit of Kitaev's honeycomb is described by \textit{Majorana spinors}, where each Fermi point behaves like a chiral degree of freedom in the high-energy sense. As Majorana fermions are charge-neutral they cannot couple to a U(1) gauge field, however they can interact with a U(1)\textsubscript{A} \textit{chiral} gauge field, where the A stands for \textit{axial}. These chiral gauge fields naturally generalise the $\mathbb{Z}_2$ gauge field that is present only at the lattice level of Kitaev's honeycomb to the continuum limit. Indeed, we apply techniques from lattice gauge theory to demonstrate the equivalence between $\mathbb{Z}_2$ gauge fields on the lattice and U(1)\textsubscript{A} chiral gauge fields in the continuum, generalising the results of the U(1) lattice gauge theory description of graphene~\cite{C3_Giuliani,C3_Neto,C3_Gusynin}. Moreover, we show these chiral gauge fields also provide a faithful encoding of lattice deformations such as dislocations and twists in the continuum level, while preserving the relativistic description of the model. Hence, we are able to demonstrate that in the continuum limit of the model the lattice twists are equivalent to $\mathbb{Z}_2$ gauge transformations. 

This chapter is structured as follows. We first define chiral symmetry in Sec.~\ref{sec:chiral_gauge_fields} and discuss how one gauges this symmetry by introducing chiral gauge fields. In Sec.~\ref{section:shifting_fermi_points} we discuss how one can interpret gauge fields as a shifting of the Fermi points in the lattice model, resulting in a gauge field in the continuum. In Sec.~\ref{section:adiabatic_equivalence} we then demonstrate that the $\mathbb{Z}_2$ gauge fields and lattice twists yield the same continuum chiral gauge fields and are adiabatically connected, demonstrating their equivalence. Finally, in Sec.~\ref{section:pi_fluxes} we demonstrate that the emergent chiral gauge field encodes the same information as the $\mathbb{Z}_2$ gauge field, specifically encoding $\pi$-fluxes. We close the chapter with a conclusion in Sec.~\ref{chiral_conclusion}.
\section{Chiral gauge fields \label{sec:chiral_gauge_fields}}
We saw in Sec.~\ref{sec:C2_Kitaev's honeycomb model} that the spinor field $\chi$ describing the continuum limit of Kitaev's honeycomb model is a Majorana spinor. These spinors have no electric charge as, unlike the Dirac action, the Majorana action of Eq.~(\ref{eq:C2_RC_action}) has no U(1) symmetry. This is because if $\chi$ is a Majorana spinor, then under the U(1) transformation $\chi \rightarrow \tilde{\chi} = e^{i \theta} \chi$, the transformed spinor is no longer a Majorana spinor because $\tilde{\chi} \neq \tilde{\chi}^{(c)}$, where $\chi^{(c)}$ is the charge conjugate spinor defined in Eq.~(\ref{eq:charge_conjugation}). Additionally, the U(1) current $j^\mu = \bar{\psi} \gamma^\mu \psi$ is identically zero for Majorana spinors.

On the other hand, the Majorana action possesses the chiral U(1)\textsubscript{A} \textit{chiral} symmetry~\cite{C2_Maggiore,Peskin,srednicki_2007}, where the A stands for \textit{axial}. For a spinor field in the chiral representation of the gamma matrices taking the form $\chi = (\chi_\mathrm{L}, \chi_\mathrm{R})^\mathrm{T}$, then a $\mathrm{U}(1)_A$ transformation is defined as
\begin{equation}
\chi \rightarrow \tilde{\chi} = e^{i \theta \gamma^5} \chi \equiv 
\begin{pmatrix} 
e^{i \theta} \chi_\mathrm{L} \\ 
e^{-i \theta } \chi_\mathrm{R} 
\end{pmatrix}, 
\end{equation}
where $\gamma^5$ is the fifth gamma matrix defined as
\begin{equation}
\gamma^5 = i\gamma^0 \gamma^1 \gamma^2 \gamma^3 = \begin{pmatrix} \mathbb{I} & 0 \\
0 & -\mathbb{I} \end{pmatrix},
\end{equation}
which anti-commutes with all other gamma matrices. We see that the U(1)\textsubscript{A} transformation is essentially a pair of opposite U(1) transformation of each chiral component. 

We can gauge this symmetry by upgrading it to a local symmetry $\theta \rightarrow \theta(x)$. As it stands, this will not be a symmetry of the Dirac action because partial derivatives $\partial_\mu$ will not transform correctly. We introduce a covariant derivative $D_\mu^A$ such that, under a local chiral transformation $\psi \rightarrow e^{i \theta(x) \gamma^5} \psi$, the derivatives transform the same way too
\begin{equation}
D^A_\mu \psi \rightarrow D^{A'}_\mu \left(  e^{i \theta(x) \gamma^5} \psi \right) = e^{i \theta(x) \gamma^5 } D^A_\mu \psi.
\end{equation}
In order to achieve this, we must take
\begin{equation}
D^A_\mu = \partial_\mu + i A_\mu \gamma^5,
\end{equation}
where $A_\mu \gamma^5$ is a U(1)\textsubscript{A} gauge field, such that under a U(1)\textsubscript{A} transformation, the gauge field transforms as $A_\mu \rightarrow A'_\mu = A_\mu - \partial_\mu \theta(x)$. Replacing all partial derivatives with the covariant derivative will make the theory locally gauge invariant. For the case of the Riemann-Cartan action, after a Fourier transform to momentum space this will give us the single-particle Hamiltonians
\begin{equation}
h(\mathbf{p}) = e_a^{\ i} \alpha^a ( p_i + A_i \gamma^5) + \ldots ,
\end{equation}
where the ellipsis represents the non-kinetic terms such as mass and torsion that we are not interested in here. We also assume that $A_0 = 0$ but we can generate it by modifying the $K$ term of the Kitaev Hamiltonian, see Ref.~\cite{chiral_paper}.
\section{Gauge fields from shifting Fermi points \label{section:shifting_fermi_points}}

We obtain the quantum field theory description of a lattice model by taking the continuum limit, therefore we now discuss how one can generate a gauge field in Kitaev's honeycomb model.

When taking the continuum limit of Kitaev's honeycomb model, we Taylor expand about the Fermi points of the model. In lattice models these Fermi points always come in pairs~\cite{C4_Nielsen1,C4_Nielsen2}, which is seen explicitly in Kitaev's honeycomb model as we have two inequivalent Fermi points $\mathbf{P}_\pm$ in the Brillouin zone. Given a momentum space single-particle Hamiltonian $h(\mathbf{p})$, we define the continuum limit Hamiltonians about each Fermi point as
\begin{equation}
h_\pm(\mathbf{p}) \equiv h(\mathbf{P}_\pm + \mathbf{p}) = \mathbf{p} \cdot \boldsymbol{\nabla} h(\mathbf{P}_\pm ) + O(\mathbf{p}^2).  \label{eq:h_taylor}
\end{equation}
Modifications to the model such as varying the strength of the couplings $\{ J_i \}$, inserting a $\mathbb{Z}_2$ gauge field or adding in extra couplings, will have the effect of modifying the single-particle Hamiltonian as $h(\mathbf{p}) \rightarrow {h'}(\mathbf{p})$. In general, this Hamiltonian will have new Fermi points $\mathbf{P}'_\pm$, giving rise to a shift relative to the old ones:
\begin{equation}
\Delta \mathbf{P}_\pm = \mathbf{P}'_\pm - \mathbf{P}_\pm.
\end{equation}
The continuum limit Hamiltonians about the new points are given by
\begin{equation}
h'_\pm(\mathbf{p}') \equiv h'(\mathbf{P}'_\pm + \mathbf{p}') = \mathbf{p}' \cdot \boldsymbol{\nabla} h'(\mathbf{P}'_\pm) + O(\mathbf{p}'^2). \label{eq:h_taylor_prime}
\end{equation}
In general, $ \mathbf{p} \neq \mathbf{p}'$ as they are defined relative to different Fermi points, so direct comparison of the continuum limits Eqs. (\ref{eq:h_taylor}) and (\ref{eq:h_taylor_prime}) cannot be done. Nevertheless, employing the relation $\mathbf{p}' = \mathbf{p} - \Delta \mathbf{P}_\pm$ the expansion Eq. (\ref{eq:h_taylor_prime}) becomes
\begin{equation}
{h}'_\pm(\mathbf{p}) = (\mathbf{p} - \Delta \mathbf{P}_\pm ) \cdot \boldsymbol{\nabla} {h}'(\mathbf{P}'_\pm) + O(\mathbf{p}'^2). 
\label{eqn:trans}
\end{equation}
Now that both Hamiltonians Eqs. (\ref{eq:h_taylor}) and (\ref{eqn:trans}) are written down using the same definition of $p$, one can compare them. We see that the shift in the Fermi points appears in the Hamiltonian in the same way that a gauge field would appear if we were to apply the minimal coupling prescription, therefore we interpret the presence of a gauge field $\mathbf{A}_\pm = \Delta \mathbf{P}_\pm$~\cite{C4_Volovik_helium_droplet}. 

\subsection{$\mathbb{Z}_2$ gauge field}

We first briefly remind the reader of zero modes required for the following section. Kitaev's honeycomb model, c.f. Eq.~(\ref{eq:C2_gauge_fixed_ham_majorana}), contains a $\mathbb{Z}_2$ gauge field $u_{ij}$. If a particular plaquette operator $W_p$ has an eigenvalue of $w_p = -1$, then we say there is a \textit{flux} through the plaquette $p$. If this is the case, the Hamiltonian contains a zero-energy eigenstate called a Majorana zero-mode whose wavefunction is localised on the flux.

As the $\mathbb{Z}_2$ gauge field $u_{ij}$ is always multiplied with the coupling constants $J_{ij}$ in the Majorana Hamiltonian of Eq.~(\ref{eq:C2_gauge_fixed_ham_majorana}), we can encode the profile of the gauge field by simply choosing the sign of the coupling constants across the lattice and fixing $u_{ij} = +1$ for every link. In particular, this allows us to smoothly interpolate between different gauge configurations as the couplings are allowed to change continuously, whilst $u_{ij}$ are always discrete. Consider the isotropic model for $J_x = J_y = 1$, $J \in [-1,1]$ and $K= 0$. The corresponding Hamiltonian of this model is given by
\begin{equation}
H  = \frac{i}{4} \sum_{\mathbf{r} \in \Lambda} 2 a_\mathbf{r} \left(J_z b_\mathbf{r} +b_{\mathbf{r} +  \mathbf{n}_1 } + b_{\mathbf{r} + \mathbf{n}_2} \right) + \mathrm{H.c.}.
\end{equation}
Following the same procedure by Fourier transforming as before, the Fermi points of this model are given by 
\begin{equation}
\mathbf{P}_\pm(J_z) = \pm \frac{2}{a} \left[ \cos^{-1}\left(- \frac{J_z}{2} \right) , 0\right]. \label{eq:C3_z2_fermi_points}
\end{equation}
Transitioning from the vortex-free sector, where $u_{ij} = +1$ for all links, to the vortex-full sector, where $u_{ij} = -1$ for all $z$-links, is equivalent to swapping $J_z = +1$ with $J_z = -1$ in this case. Under this transformation, we see that the Fermi points of the model change, giving rise to a gauge field
\begin{equation}
\mathbf{A}_\pm = \mp \left( \frac{2 \pi}{3 a} , 0 \right). \label{eq:C3_z2_gauge_field}
\end{equation}
We see that the shift of each Fermi point is chiral, that is, the Fermi points have shifted oppositely, so the Hamiltonian of each Fermi point has a different gauge field which collectively is nothing but a chiral gauge field as discussed earlier. The corresponding single-particle Hamiltonian in the four-dimensional language of Sec.~\ref{sec:C2_4x4_cont_lim} is given by
\begin{equation}
h(\mathbf{p}) = v_\mathrm{F} \left[ -\alpha^x \left( p_x + \frac{ 2 \pi}{3a} \gamma^5 \right) + \alpha^y p_y \right]. \label{eq:C3_z2_cont_lim}
\end{equation}

\subsection{Lattice twists}
We now modify the isotropic model by removing \textit{all} $z$-links and adding two diagonal links across each plaquette of the honeycomb lattice. The corresponding Hamiltonian for $K=0$ is given by
\begin{equation}
H' = \frac{i}{4} \sum_{\mathbf{r} \in \Lambda}  2 a_{\mathbf{r}} \left(b_{\mathbf{r} + \mathbf{n}_1} + b_{\mathbf{r} +\mathbf{n}_2} + b_{\mathbf{r} + \mathbf{n}_1 - \mathbf{n}_2} \right) + 2a_{\mathbf{r} + \mathbf{n}_1 - \mathbf{n}_2 } b_{\mathbf{r}}  + \text{H.c.},
\label{eq:two_cross_ham}
\end{equation}
which is seen in Fig.~\ref{fig:zero-mode-comparison-A1-z}(b) (albeit implemented globally instead). This lattice modification does not break translational symmetry or increase the size of the unit cell of the lattice, therefore the Brillouin zone is unchanged. The Fermi points of this model are given by
\begin{equation}
\mathbf{P}'_\pm = \pm \left( \frac{2 \pi}{3 a} , 0 \right), \label{eq:twist_fermi_point}
\end{equation}
which are the same Fermi points as the ones obtained from a global $J_z$ sign change given by Eq. (\ref{eq:C3_z2_fermi_points}). We again interpret the shift in the Fermi points relative to the isotropic case as a chiral gauge field, which yields the same chiral gauge field Eq. (\ref{eq:C3_z2_gauge_field}). The corresponding Hamiltonian is given by
\begin{equation}
h(\mathbf{p}) = v_\mathrm{F} \left[ 3 \alpha^x \left(p_x + \frac{2 \pi}{3 a} \gamma^5 \right)  p_x  + \alpha^y p_y  \right]. \label{eq:cross_ham_cont}
\end{equation}
If we compare Eq. (\ref{eq:cross_ham_cont}) to Eq. (\ref{eq:C3_z2_cont_lim}), we see that the continuum limits look identical, apart from a factor of $3$ in front of the $x$-component kinetic term. The emergent chiral gauge fields are the same as the Fermi points of both models have shifted by the same amount relative to the isotropic case. The factor of $3$ is the result of the additional next-to-next-to-nearest-neighbour couplings that changed the geometry of the lattice. Its effect is to scale the $x$-direction of the continuum limit and can be absorbed in the dreibein of the continuum limit using the geometric language of Chap.~\ref{chapter:kitaev}. For this reason, we conclude that both lattice models are equivalent as they yield the same continuum limits up to a smooth deformation of the dreibein, so correspond to the same phase.

\section{Adiabatic equivalence between vortices and twists \label{section:adiabatic_equivalence}}
\subsection{Majorana zero modes}

\begin{figure}[t]
\begin{center}
\begin{subfigure}{0.4\textwidth}
\begin{center}
\ \ \ (a)
\includegraphics[width=\textwidth]{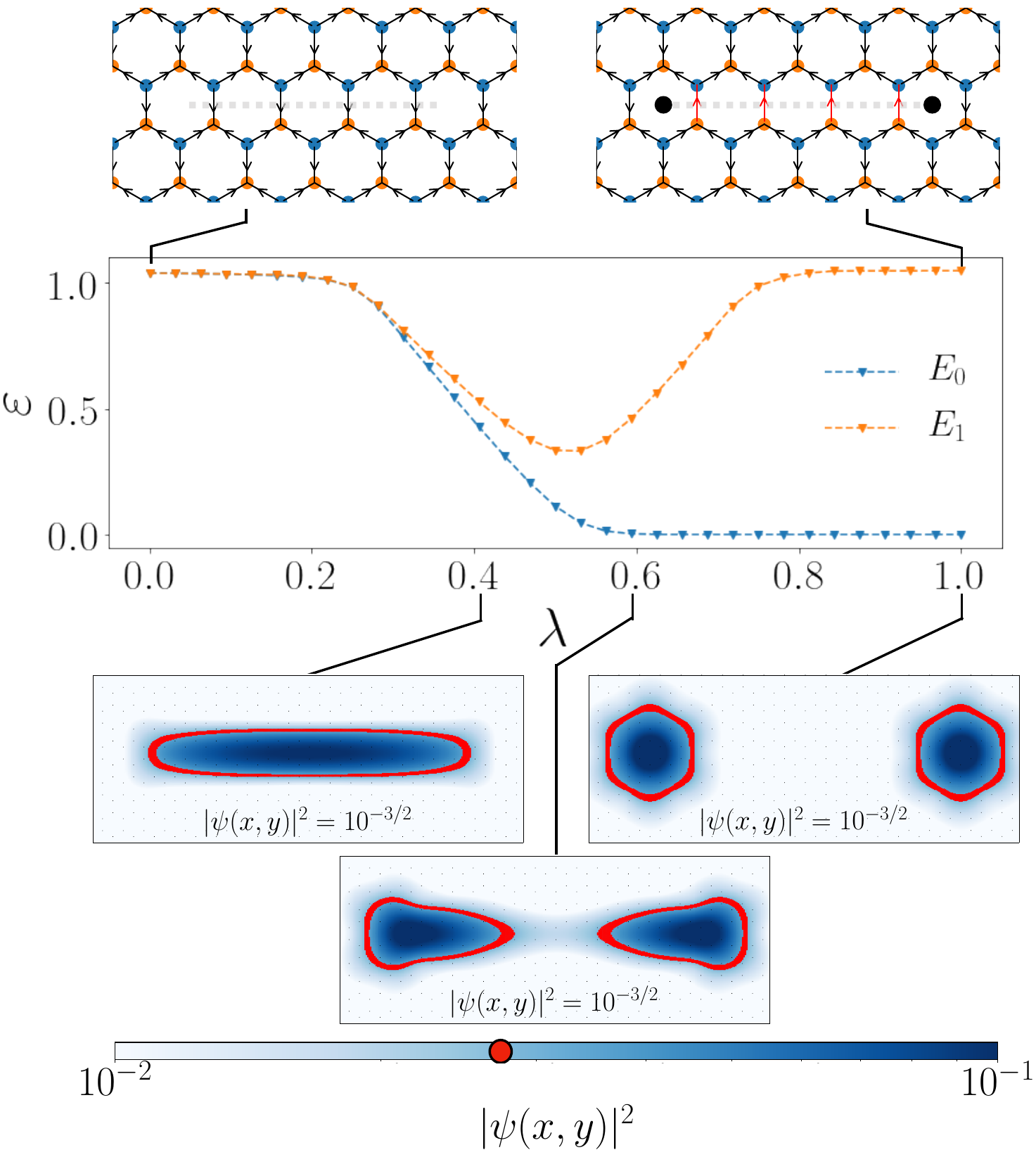}
\end{center}
\end{subfigure}
\hspace{1cm}
\begin{subfigure}{0.4\textwidth}
\begin{center}
\ \ \ (b)
\includegraphics[width=\textwidth]{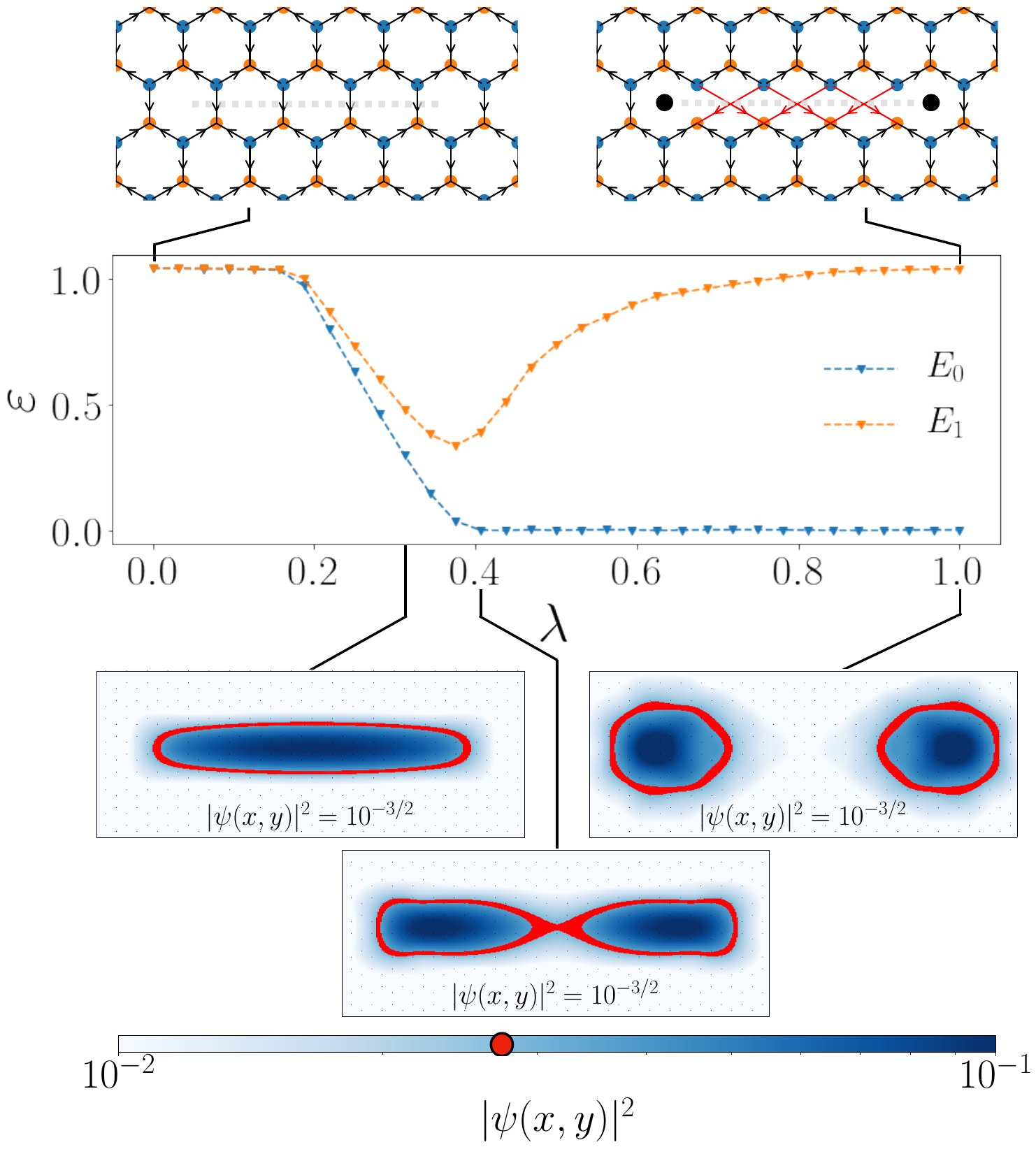}
\end{center}
\end{subfigure}
\end{center}
\caption{ (a) Top: A sketch of $H_{\mathbb{Z}_2}(\lambda)$ of Eq.~(\ref{eq:Z_2_adiabatic}) at $\lambda=0$ and $\lambda=1$ The path $P$, indicated by a dashed grey line, runs perpendicular to the $z$-links of the lattice. The red links are the links for which the $\mathbb{Z}_2$ gauge field takes the value $u_{ij} = -1$, and where the vortices, for $W_p = -1$, are marked by the black circles. Middle: The energy gap of $H_{\mathbb{Z}_2}(\lambda)$ as a function of $\lambda$ for a system with linear dimension $L=30$, isotropic $J$, and $K=0.1$. Bottom: The continuous profile of the wavefunction $|\psi(\mathbf{r})|^2$ of the zero modes at $\lambda\approx 0.4,0.6,1$. (b) The same information but for $H_\mathbf{A}(\lambda)$ of Eq.~(\ref{eq:H_A_adiabatic}), where the red lines represent the twisted links.}
\label{fig:zero-mode-comparison-A1-z}
\end{figure}

While the $\mathbb{Z}_2$ gauge field $u_{ij}$ can only change through a discrete process, it is possible to change them in a continuous way by encoding the gauge field in the couplings as we discussed in the previous section. We observe the formation of zero modes throughout this continuous process by studying the behaviour of the energy spectrum and wave functions. For example, consider adiabatically transitioning between the vortex-free sector, where $u_{ij} = +1$ for all links, and a gauge sector where $u_{ij} = -1$ for a horizontal line of $z$-links, as shown in Fig.~\ref{fig:zero-mode-comparison-A1-z}(a). To shift from one Hamiltonian to the other, we introduce the interpolating Hamiltonian
\begin{equation}
H_{\mathbb{Z}_2}({\lambda})=(1-\lambda)H_0+\lambda H_{\mathbb{Z}_2}, \quad \lambda \in [0,1]. \label{eq:Z_2_adiabatic}
\end{equation}
The result is a continuous change in the value of $u_{ij}$ between $+1$ and $-1$ for the $z$-links along the path. Thus, we expect to see Majorana zero modes appearing at the end points of $P$ as $\lambda$ approaches $1$. All numerical simulations presented in this section are from Ref.~\cite{chiral_paper} for models with periodic boundary conditions, system size $L=30$, isotropic $J=1$ and $K=0.1$.

The generation of localised Majorana zero modes is shown in Fig.~\ref{fig:zero-mode-comparison-A1-z}(a) as $\lambda$ increases in discrete steps demonstrating that the local $\mathbb{Z}_2$ gauge field creates $\pi$-vortices. The single particle Hamiltonian $H_{\mathbb{Z}_2}(\lambda)$ is diagonalised for each discrete value of $\lambda$ and the energies $E_0$ and $E_1$ of the two lowest eigenstates are plotted in Fig.~\ref{fig:zero-mode-comparison-A1-z}(a). At $\lambda=0$ the model is clearly gapped with no zero energy modes, while at $\lambda=1$ there is a clear zero energy mode with a gap above it. The gap between $E_0$ and $E_1$ forms at a transition point around $\lambda\approx 0.5$. From the diagonalisation of $H_{\mathbb{Z}_2}(\lambda)$, we also obtain the probability density at each lattice site $|\psi_i|^2$ for the lowest energy eigenstate. We call this the spatial wave function of the vortices. To visualise the shape of the zero modes, we approximate them with a continuous function as done before in Eq. (\ref{eq:C2_smooth_wavefunction}). As we approach the transition point $\lambda\approx 0.5$ a single fermion mode appears over the length of the path $P$. This mode splits into two Majorana zero modes as $\lambda$ increases, becoming exponentially localised at the end points of $P$ as we approach $\lambda=1$.

We now consider the isotropic vortex-free Hamiltonian $H_0$ and we create a non-zero chiral gauge field $\mathbf{A}$ by introducing lattice deformations with the Hamiltonian of Eq.~(\ref{eq:two_cross_ham}). We consider these deformations locally along the same horizontal path $P$ that result in the creation of twists at the endpoints of the path, as shown in Fig.~\ref{fig:zero-mode-comparison-A1-z}(b). We denote the resulting Hamiltonian as $H_\mathbf{A}$. We use the same method as above to continuously shift between these two Hamiltonians:
\begin{equation}
H_\mathbf{A}({\lambda})=(1-\lambda)H_0+\lambda H_\mathbf{A}, \quad \lambda \in [0,1]. \label{eq:H_A_adiabatic}
\end{equation}
Fig.~\ref{fig:zero-mode-comparison-A1-z}(b) shows the energies of the two lowest eigenstates of the single particle Hamiltonian produced by varying $\lambda$ as well as the continuous approximations of the spatial wave function as vortices are produced. Similar to the vortex creation, we observe that the formation of twists give rise of stable Majorana zero modes as $\lambda$ increases and the gap begins to open. Hence, twists bound Majorana zero modes much like the $\mathbb{Z}_2$ vortices do. In addition, we see the zero modes for the twists are slightly stretched relative to the $\mathbb{Z}_2$ zero modes due to the non-trivial dreibein here using the language of Sec.~\ref{chapter:kitaev}.

\subsection{Adiabatic equivalence}

\begin{figure}
\center
\includegraphics[width=0.4\linewidth ]{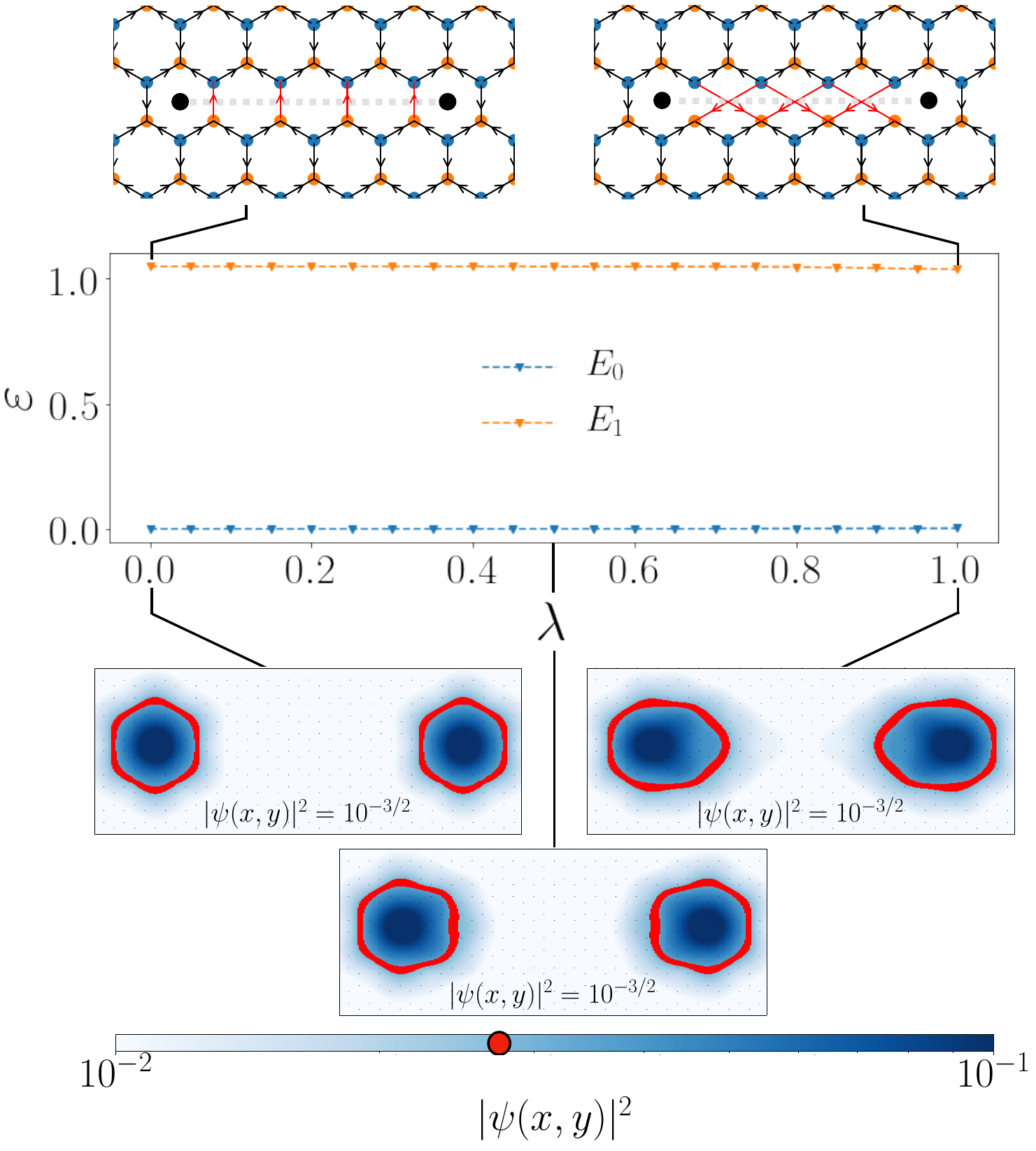}
\caption{Top: A sketch of $H'(\lambda)$ of Eq.~(\ref{eqn:adia}) at $\lambda=0$ and $\lambda=1$. The path $P$, indicated by a dashed grey line, remains constant, runs perpendicular to the $z$-links of the lattice. The modified links along the path $P$ are highlighted in red. Middle: The energy gap of $H'(\lambda)$ as a function of $\lambda$ for a system with linear dimension $L=30$, isotropic $J=1$, and $K=0.1$ Bottom: The continuous profile zero modes at $\lambda\approx 0,0.5,1$ shows they remain fixed in place and well-localised throughout the adiabatic transition.}
\label{fig:adiabatic1}
\end{figure}

We established in the previous section that string-like configurations of twists in the lattice give rise to Majorana zero modes at the end points of the string. This is very similar to the zero modes trapped by string-like configurations of the $\mathbb{Z}_2$ gauge field that creates $\pi$ flux vortices at its end-points. Here we demonstrate that these two apparently different ways of realising Majorana zero modes, i.e., by changing the sign of certain links or by modifying the connectivity of the lattice, are actually physically equivalent. We demonstrate this by adiabatically transforming between these two configurations and considering both the behaviour of the energy spectrum as well as the wave function of the zero modes. 

We take the Hamiltonians $H_{\mathbb{Z}_2}$ and $H_\mathbf{A}$, defined in the previous section and depicted in the top left and right of Fig.~\ref{fig:adiabatic1} respectively. We define the Hamiltonian
\begin{equation}
H'(\lambda) = (1-\lambda)H_{\mathbb{Z}_2} + \lambda H_\mathbf{A}, \quad \lambda \in [0,1].
\label{eqn:adia}
\end{equation}
This allows us to adiabatically transition between the two Hamiltonians by varying $\lambda$. The path $P$ remains fixed throughout this transition. Fig.~\ref{fig:adiabatic1} shows the energy gap of the system and the continuous approximation of the wave function of a pair of zero modes as we adiabatically transition between $H_{\mathbb{Z}_2}$ and $H_\mathbf{A}$. We observe that  the zero modes remain energetically separated from the rest of the states for all $\lambda$ with an energy gap that remains more or less constant throughout the process. Moreover, the zero modes of the model remain fixed in place and well-localised throughout the adiabatic transition. Hence, the two ways of generating vortices are physically equivalent. The shape of the zero modes of $H_\mathbf{A}$ appear stretched in the $x$-direction compared to $H_{\mathbb{Z}_2}$. This is due to the change in the dreibein in Eq.~(\ref{eq:cross_ham_cont}). This adiabatic process also demonstrates that there is a continuous family of lattice configurations given by $H'(\lambda)$ for $\lambda\in [0,1]$ that give rise to the same localised Majorana zero modes. 

\section{$\pi$-fluxes in the continuum \label{section:pi_fluxes}}

We know that at the lattice level, the $\mathbb{Z}_2$ gauge field encodes fluxes $\phi$ through the plaquette operators $W_p$ as we can write their eigenvalues as $w_p = e^{i \phi} = \pm 1$. This is inherited from the continuum gauge theory, where the \textit{Wilson loops} are exponentials of the fluxes enclosed by the loop. At the lattice level, the plaquette operators $W_p$ are the discrete versions of the Wilson loops so we can relate them to a flux in the same way. If the plaquette has a flux through it, so if $W_p = -1$, then we would say the flux is $\phi = \pi$. 

A natural question to ask is whether the $\mathbb{Z}_2$ gauge field on the lattice and chiral gauge field $\mathbf{A}$ generated in the continuum encode the same information. In particular, we ask whether the gauge field $\mathbf{A}$ encodes the $\pi$-fluxes of the $\mathbb{Z}_2$ gauge field. In order to see this, we need to slightly modify the analysis of the previous section and introduce the $\mathbb{Z}_2$ gauge field on the $x$- or $y$-links. The reason for this is the following: in the continuum limit, we use the coordinate system $\mathbf{x}$ which labels the \textit{unit cells} of the lattice and not the lattice sites themselves, which is why we were forced to introduce two sublattices $A$ and $B$ and a two-component spinor to account for the lattice sites. Using this language, the model actually looks like a rhombic lattice generated by $\mathbf{n}_2$ and $\mathbf{n}_2$, where the $z$-links have ``disappeared" and their information has been absorbed into the spinor where the sublattice degree of freedom behaves as the ``spin" degree of freedom of the spinor. To avoid these subtleties about where the $z$-link gauge field lives, we work with $x$- and $y$-links instead.

In this section it is convenient to work with the coordinate system adapted to the lattice. This contrasts to my paper of Ref.~\cite{chiral_paper}, where I used the Cartesian coordinate system, however the analysis of this paper required me to map from the honeycomb lattice to the brick wall lattice which likely raises more questions than it answers. The lattice coordinate system provides the most elegant solution and avoids this mapping. 

\begin{figure}[t]
\hspace{4cm} (a) \hspace{6.5cm} (b) \\
\begin{center}
\includegraphics[scale=1,valign=c]{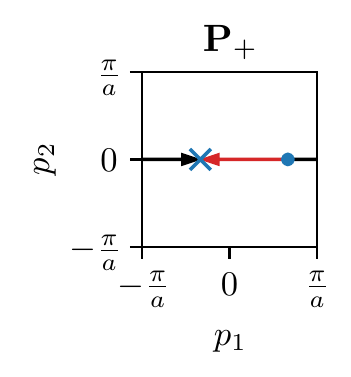}
\includegraphics[scale=1,valign=c]{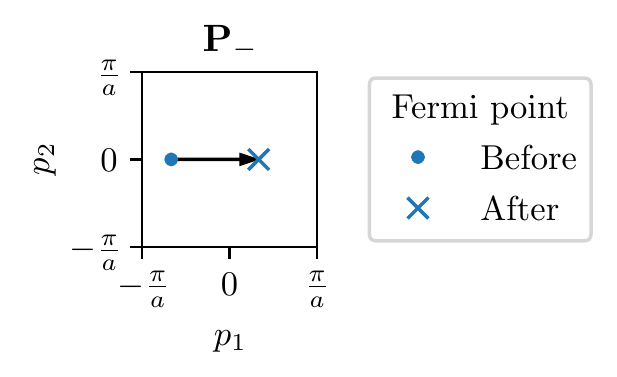}
\includegraphics[scale=0.3,valign=c]{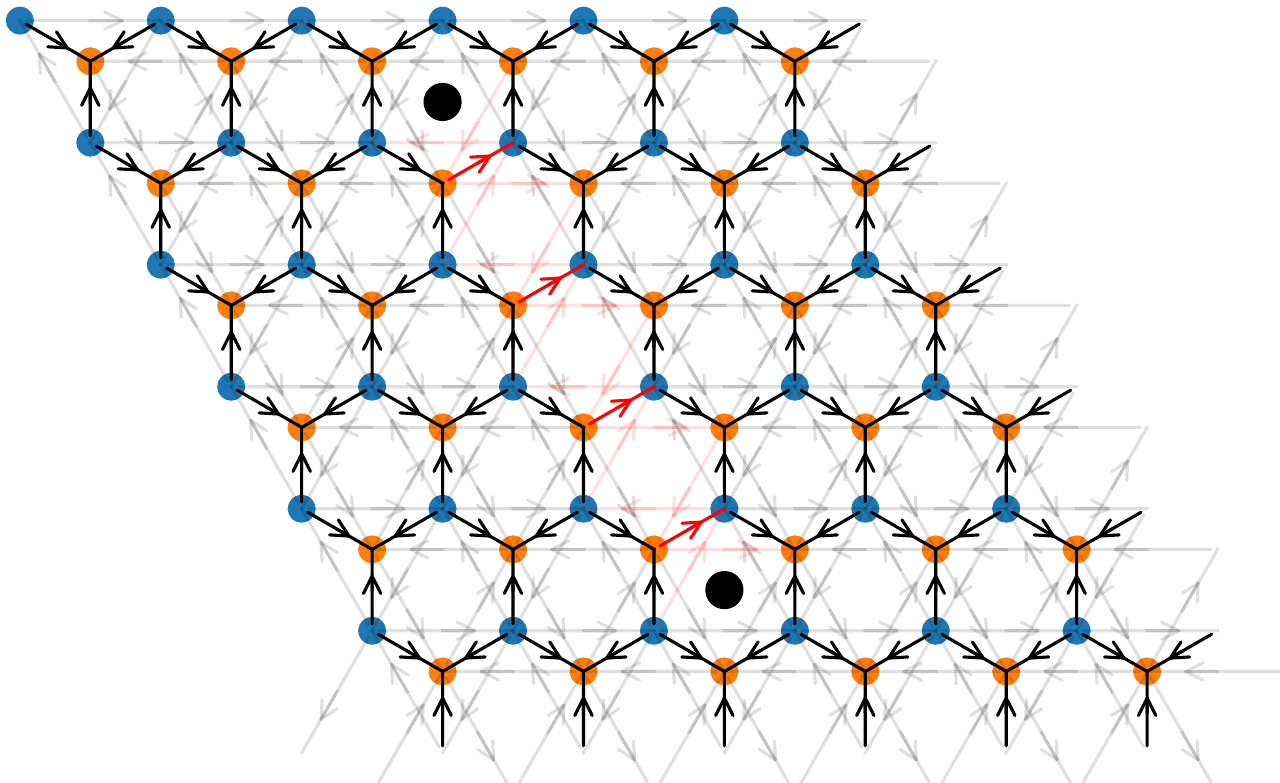}
\end{center}
\caption{(a) The shift of the two Fermi points $\mathbf{P}_\pm$ in the Brillouin zone with respect to the reciprocal basis, where $p_i = \mathbf{p} \cdot \mathbf{n}_i /a $. We see that switching on a global $\mathbb{Z}_2$ gauge field on the $x$-links shifts the Fermi points in the $p_1$ direction by $\pi/a$, represented by the black arrows. Due to the $2\pi/a$ periodicity of the Brillouin zone, the shift can be viewed chirally, where $\mathbf{P}_+$ actually shifts by $-\pi /a$ instead, show by the red arrow. (b) The path $P$ of $x$-links in the $\mathbf{n}_2$ direction with $u_{ij} = -1$, shown by the red links. At each end we have a $\pi$-flux located on the black circles, as $w_p = -1$ here. These trap zero-energy eigenstates called Majorana zero modes.}
\label{fig:C3_z2_gauge_field_fermi_point}
\end{figure}

We briefly remind the reader of the important aspects of Bravais lattices introduced in Sec.~\ref{sec:bravais_lattice}. Given a Bravais lattice $\Lambda$ with generators $\{ \mathbf{n}_i\} $ and its corresponding dual lattice $\Lambda^*$ with generators $\{ \mathbf{G}_i \}$ which obey $\mathbf{n}_i \cdot \mathbf{G}_j = 2 \pi \delta_{ij}$, the position $\mathbf{r} \in \Lambda$ and momenta $\mathbf{p} \in \Lambda^*$ can be expressed with respect to the these bases as 
\begin{equation}
\mathbf{r} = \sum_i x_i \mathbf{n}_i, \quad \mathbf{p} = \frac{a}{2 \pi} \sum_i p_i \mathbf{G}_i,
\end{equation}
where $x_i \in \mathbb{Z}$ are the position coordinates and $a = |\mathbf{n}_i|$ is the unit cell spacing, whilst $p_i \in [-\pi/a,\pi/a)$ are the momentum coordinates which obey the useful relationship $a p_i = \mathbf{p} \cdot \mathbf{n}_i$. In this language, we can write down the Hamiltonian with a $\mathbb{Z}_2$ gauge field on the $x$-links as
\begin{equation}
H_\theta  = \frac{iJ}{4} \sum_{\mathbf{r} \in \Lambda} 2 a_\mathbf{r} \left( b_\mathbf{r} + e^{i \theta} b_{\mathbf{r} +  \mathbf{n}_1 } +  b_{\mathbf{r} + \mathbf{n}_2} \right) + \mathrm{H.c.},
\end{equation}
where the phase on the $x$-links represent the $\mathbb{Z}_2$ gauge field written in a convenient form called a \textit{Peierls substitution}, or a \textit{Wilson line} in lattice gauge theory, where $\theta \in \{0, \pi\}$. If we Fourier transform following the same procedure as before, we arrive at the same form of Hamiltonian in Eq.~(\ref{eq:C2_diagonal_hamiltonian}), except now with $f(\mathbf{p}) \rightarrow f_\theta(\mathbf{p})$, where
\begin{equation}
f_\theta(\mathbf{p}) = 2J \left( 1 + e^{i \theta} e^{i \mathbf{p} \cdot \mathbf{n}_1} +  e^{i \mathbf{p} \cdot \mathbf{n}_2} \right) = 2J\left(1 + e^{i a \left(p_1 + \frac{\theta}{a} \right)} + e^{i a p_2} \right) ,
\end{equation}
where in the second equality we expressed our momenta with respect to the reciprocal basis. The dispersion relation is given by $E_\theta(\mathbf{p}) = \pm |f_\theta(\mathbf{p})|$ so we see that the phase has shifted the dispersion by $-\theta/a$ in the $p_1$ direction. We could interpret this shift chirally too if we wanted to, as due to the $2\pi/a$ periodicity of the Brillouin zone in the reciprocal basis $p_i$, a shift of $-\pi/a$ in the $p_1$ direction is equivalent to a shift of $+\pi/a$. Therefore, if we shifted one Fermi point by $+\pi/a$ and the other by $- \pi/a$, we would achieve the same effect, as seen in Fig.~\ref{fig:C3_z2_gauge_field_fermi_point}(a). As the continuum limit fields are Majorana, we prefer to interpret it as chiral shift of the Fermi points, yielding a chiral gauge field instead which is compatible with Majorana spinors.

Let us consider the case where the $\mathbb{Z}_2$ gauge field is inserted along a semi-infinite path $P$ of $x$-links starting at the origin going in the $\mathbf{n}_2$ direction, as seen in Fig.~\ref{fig:C3_z2_gauge_field_fermi_point}(b). Let us also focus on just one of the Fermi points (as the other will be the opposite due to chirality). The Fermi point shifts only on the path $P$ so this would be represented by the gauge field
\begin{equation}
\mathbf{A} = \frac{1}{a} \begin{pmatrix} -\pi \\ 0 \end{pmatrix} \delta_{x_1, 0} \Theta_{x_2}  \xrightarrow[a \to 0]{} \begin{pmatrix} -\pi \\ 0 \end{pmatrix} \delta(x_1) \Theta(x_2),
\end{equation}
where we have used the fact that $\lim_{a \rightarrow 0} \delta_{x_1,0}/a = \delta(x_1)$, and $\Theta$ is the Heaviside step function. We can calculate the flux of this vector field using Stokes' theorem. As Stokes' theorem can be used in any coordinate system we like, we choose the lattice coordinate system instead of the Cartesian coordinate system as this is the simplest to work with here. However, this coordinate system is not orthogonal so we must resort to the general definition of Stokes' theorem written in the language of differential forms $\int_{\partial S} A = \int_S \mathrm{d}A$~\cite{C2_Nakahara}. In components reads
\begin{equation}
\int_{\partial S} A_i \mathrm{d}x^i = \int_S \partial_i A_j \epsilon^{ij} \mathrm{d}x^1 \mathrm{d}x^2 ,
\end{equation}
where the indices $i$ and $j$ refer to the lattice coordinate indices. For any path $\partial S$ that encloses a surface $S$ containing the origin, the gauge field $\mathbf{A}$ has the flux 
\begin{equation}
\phi = \int_{\partial S} A_i \mathrm{d}x^i  = \int_S (\partial_1 A_2 - \partial_2 A_1) \mathrm{d}x^1 \mathrm{d}x^2  = \int_S \pi \delta(x_1)\delta(x_2) \mathrm{d}x^1 \mathrm{d}x^2  = \pi ,
\end{equation}
where we use the fact that $\partial_x \Theta(x) = \delta(x)$, so the continuum limit gauge field also encodes the $\pi$-flux as we hoped. This analysis would work for $y$-link gauge fields too, but not for $z$-links as discussed earlier due to how we took the continuum limit. If we wish to use $z$-links, then we would have to choose a different unit cell for the lattice to ensure the $z$-links are not absorbed into the definition of the spinors.

\section{Conclusion \label{chiral_conclusion}}
The generation and manipulation Majorana fermions is one of the central problems in the current effort to understand the physics of non-Abelian anyons and employ them for quantum technologies. Here we demonstrated that two of the leading ways of trapping Majorana zero modes, employing vortices and employing lattice twists, are physically equivalent. We demonstrated this equivalence by finding the appropriate representation of these lattice defects in the continuum limit in terms of chiral gauge fields. We showed analytically that both $\mathbb{Z}_2$ gauge fields and lattice deformations have an equivalent representation in the low-energy spectrum of the system in terms of chiral gauge field coupled to the Majorana version of the Dirac equation. As the two continuum limits differed only by a smooth transformation of the dreibein, this suggested that the lattice level Hamiltonians must also be equivalent which was investigated numerically in Ref.~\cite{chiral_paper} by simulation local configurations of the $\mathbb{Z}_2$ gauge field and lattice deformations. 

We observed numerically that local configurations of this chiral gauge field can create $\pi$ flux vortices. Motivated by this equivalence we investigated the possibility of Majorana bounding twists being physically equivalent to Majorana bounding vortices. We performed an adiabatic transformation between Hamiltonians that encode twists and vortices and showed that both the structure of the energy spectrum as well as the localisation properties of the Majorana zero modes remain invariant during the adiabatic transformation.

Our investigation demonstrates that Majorana bounding twists are physically equivalent to vortices even though they do not have a gauge field representation in the lattice level. Nevertheless, they give rise to a chiral gauge field in the continuum limit equivalent to the gauge field that a $\mathbb{Z}_2$ gives rise to. This opens up a variety of possible investigations. First, it is possible to realise gauge theories that do not necessarily have a traditional interpretation in the lattice level in terms of Wilson lines. This can give wider flexibility for the realisation of gauge theories in the laboratory, e.g., with optical lattices~\cite{C3_Alba}. Second, the adiabatic transformation between vortices and twists created a continuous spectrum of defects that can support Majorana zero modes beyond the two limiting cases. The possibility of having a wider range of Majorana bounding defects can facilitate their experimental generation and detection.

\chapter{Chiral spin chain interfaces as event horizons \label{chapter:black_hole}}
\begin{chapquote}{Freeman Dyson (1967)}
However, my personal reason for
working on one-dimensional problems is
merely that they are fun. A man grows
stale if he works all the time on the
insoluble and a trip to the beautiful work
of one dimension will refresh his
imagination better than a dose of LSD.
\end{chapquote}
\section{Introduction}
Interfaces of quantum systems offer a fertile environment for rich and exotic physics to emerge that often cannot be met without the support of bulk systems. For example, domain walls between fractional quantum Hall states can give rise to parafermions, anyons with non-Abelian statistics \cite{C4_Clarke}, while domain walls between 2D Heisenberg models can give rise to deconfined fractional excitations \cite{C4_Batista}. Moreover, higher order topological phases support gapless edge states at boundary defects \cite{C4_Benalcazar,C4_Langbehn,C4_Ezawa} and non-unitary conformal field theories can emerge at the boundary of interacting field theories \cite{C4_Soderberg}. The complexity of interfaces, especially for interacting systems, is so high that simple and effective modelling is instrumental to obtaining a qualitative and quantitative understanding of their behaviour.

Here we consider the 1D XY model supplemented by a three-spin chirality operator making the system intrinsically \textit{interacting}, unlike the previous chapters which were all non-interacting models. Such chiral systems are of interest as they exhibit a rich spectrum of quantum correlations \cite{C4_Pachos1} and can give rise to skyrmionic configurations \cite{C4_Tikhonov}. Using mean field theory, we show an interface between a chiral and non-chiral phase is effectively modelled by a Dirac fermion on a black hole background, where the event horizon is positioned at the interface. The chiral phase is identified with the interior of the black hole, whilst the non-chiral phase is identified with the exterior of the black hole. 

In this chapter, we first establish the validity of the mean field theory that gives rise to the back hole description and further understand the behaviour of the interacting chiral system by comparing it to the Matrix Product State (MPS) numerical analysis conducted in Ref.~\cite{black_hole_paper} and analytically by bosonising the system and using the theory of Luttinger liquids. In this way, we show that the horizon description is quantitatively and quantitatively faithful to the one provided by mean field theory. We further demonstrate that the interface is between two conformal field theories with central charge $c=2$ inside the black hole and $c=1$ outside, thus identifying the change in the fermionic degrees of freedom across the interface. 

To test the faithfulness of the black hole description, we use the mean field description to investigate the time evolution following a quench that propagates through the horizon. We prepare an interface between two opposite chiralities that models a black hole-white hole interface and demonstrate numerically that a pulse in one chiral phase is thermalised as it passes through the interface. This pulse is interpreted as thermal radiation with a temperature well approximated by the Hawking temperature for a wide range of coupling profiles and initial conditions. Hence, the black hole description accurately models the evolution of the interacting chiral phases across the boundary. We envision that gravity at extreme curvatures can provide an elegant formalism that can efficiently model several strongly interacting systems and their interfaces in higher dimensions.

This chapter is structured as follows. In Sec.~\ref{section:mean_field} we introduce the spin model and its fermionic mean field description. In Sec.~\ref{section:phase_transitions} we study the phase diagram predicted by the mean field theory and compare it to the results obtained through matrix product state (MPS) techniques applied to the full spin model from Ref.~\cite{black_hole_paper}. In Sec.~\ref{section:luttinger}, we apply the theory of Luttinger liquids to study the model analytically beyond mean field theory. Finally, in Sec.~\ref{section:hawking}, we study the continuum limit of the model and reveal an emergent black hole description and its simulation of the Hawking effect. We close the chapter with a conclusion in Sec.~\ref{black_hole_conclusion}.
\section{The lattice model \label{section:mean_field}}
\subsection{The Hamiltonian}

\begin{figure}[t]
\begin{center}
\includegraphics[scale=1]{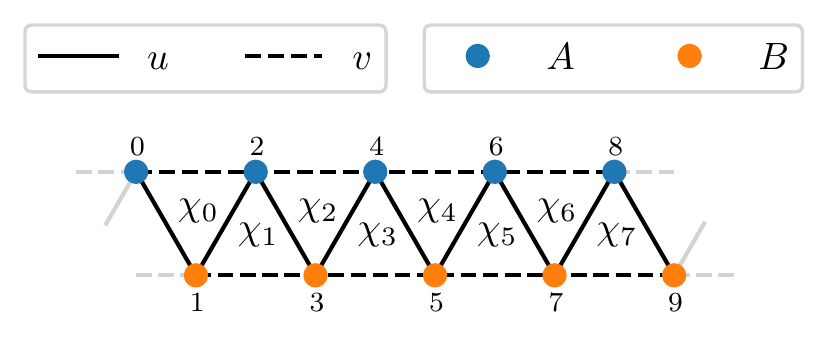}
\end{center}
\caption{The geometry of the lattice of interactions described by the Hamiltonian of Eq.~(\ref{eq:C4_spin_ham}). We label the lattice sites with the index $n \in \{ 0,1,\ldots,N\}$ and impose periodic boundary conditions. As $H_\mathrm{XY}$ contains nearest-neighbour interactions with coupling $u$, whilst $H_\chi$ contains next-to-nearest-neighbour interactions with coupling $v$, the lattice has the geometry of a zig-zag ladder. For this reason, it is convenient to label the lattice sites $A$ and $B$, and the chirality operator $\chi_n$ is an interaction of the three spins forming the $n$th triangular plaquette, i.e., a plaquette operator.}
\label{fig:C4_zig_zag_lattice}
\end{figure}

In this work we study a modification of the 1D spin-$1/2$ XY model. For a system containing $N$ spins, the Hilbert space of the model is given by $\mathcal{H} = \mathcal{H}_{1/2}^{\otimes N} $, where $\mathcal{H}_{1/2}$ is the two-dimensional Hilbert space of a spin-$1/2$ particle. The Hamiltonian is given by
\begin{equation}
H = \frac{1}{2} \sum_{n=0}^{N-1} \left[- \frac{u}{2} \left(\sigma^x_n \sigma^x_{n+1} + \sigma^y_n \sigma^y_{n+1} \right) + \frac{v}{4} \chi_n \right] \equiv H_\mathrm{XY} + H_\chi, \label{eq:C4_spin_ham}
\end{equation}
where $u,v \in \mathbb{R}$, $\{ \sigma^x_n , \sigma^y_n , \sigma^z_n \}$ are the Pauli matrices that act on the $n$th spin only, given by the tensor product
\begin{equation}
\sigma^\alpha_n = \underbrace{\mathbb{I} \otimes \mathbb{I} \otimes \ldots \otimes \mathbb{I} \otimes \sigma^\alpha}_\text{$n +1$ factors} \otimes \mathbb{I} \otimes \ldots \otimes \mathbb{I} \otimes \mathbb{I},
\end{equation}
where $\alpha = x,y,z$ and the Pauli matrix $\sigma^\alpha$ is in the $n$th position, where we count from $0$. The operator $\chi_n$ is the spin chirality given by the three-spin interaction \cite{C4_Pachos1,C4_Pachos2}
\begin{equation}
\chi_n \equiv \boldsymbol{\sigma}_n \cdot (\boldsymbol{\sigma}_{n+1} \times \boldsymbol{\sigma}_{n+2} ),
\end{equation}
where $\boldsymbol{\sigma}_n = (\sigma^x,\sigma^y,\sigma^z)$ is the vector of Pauli matrices of the $n$th spin. We apply periodic boundary conditions $\boldsymbol{\sigma}_{n+N} \equiv \boldsymbol{\sigma}_n$ throughout, however we always have the thermodynamic limit $N \rightarrow \infty$ in mind. Due to the form of the interactions in the Hamiltonian, the geometry of the lattice is given by a zig-zag ladder as shown in Fig.~\ref{fig:C4_zig_zag_lattice}.

In order to make analytic progress with this model, we map from the language of spins to the language of fermions by applying a Jordan-Wigner transformation defined as~\cite{coleman_2015}
\begin{equation}
\sigma_n^+   = \exp\left(-i \pi \sum_{m=0}^{n-1} c^\dagger_m c_m \right) c^\dagger_n, \quad \sigma_n^-   = \exp\left(i \pi \sum_{m =0}^{n-1} c^\dagger_m c_m \right) c_n, \quad \sigma^z_n  = 1 - 2c_n^\dagger c_n,
\end{equation}
where $\sigma_n^\pm = (\sigma^x_n \pm i\sigma^y_n)/2$ and $c_n$ are fermionic operators obeying the commutation relations $\{ c_n, c_m^\dagger \} = \delta_{nm}$ and $\{ c_n, c_m \} = \{c_n^\dagger,c_m^\dagger\} = 0$. Using the definition of $\sigma^\pm_n$, we have the useful identities
\begin{align}
\sigma^x_n \sigma^x_{n+1} + \sigma^y_n \sigma^y_{n+1} & = 2 \sigma^+_n\sigma^-_{n+1} + \mathrm{H.c.}, \\
\sigma^x_n \sigma^x_{n+1} - \sigma^y_n \sigma^y_{n+1} & = 2i \sigma^+_n\sigma^-_{n+1} + \mathrm{H.c.}.
\end{align}
The first identity allows us to rewrite $H_\mathrm{XY}$ straight away, whilst the second identity allows us rewrite the chirality $\chi_n$ as
\begin{equation}
\begin{aligned}
\chi_n &  = \epsilon_{abc} \sigma^a_n \sigma^b_{n+1} \sigma^c_{n+2} \\
& = \left(\sigma^x_n \sigma^y_{n+1} - \sigma^y_n \sigma^x_{n+1} \right) \sigma^z_{n+2} + \left(\sigma^x_{n+1} \sigma^y_{n+2} - \sigma^y_{n+1} \sigma^x_{n+2} \right) \sigma^z_n \\ 
& \ \ \ \ + \left(\sigma^x_{n+2} \sigma^y_n  -  \sigma^y_{n+2} \sigma^x_n\right) \sigma^z_{n+1} \\
& = 2i \left( \sigma^+_n \sigma^-_{n+1} \sigma^z_{n+2} + \sigma^+_{n+1} \sigma^-_{n+2} \sigma^z_n + \sigma^+_{n+2} \sigma^-_n \sigma^z_{n+1}  \right) + \mathrm{H.c.}.
\end{aligned}
\end{equation}
With this, the Hamiltonian of Eq.~(\ref{eq:C4_spin_ham}) takes the form
\begin{equation}
H =   \sum_n \left[ -u \sigma^+_n \sigma^-_{n+1} + \frac{iv}{2} \left( \sigma^+_n \sigma^-_{n+1} \sigma^z_{n+2} + \sigma^+_{n+1} \sigma^-_{n+2} \sigma^z_n +  \sigma^+_{n+2} \sigma^-_n \sigma^z_{n+1} \right) \right] + \mathrm{H.c.}.
\end{equation}
Now the Hamiltonian is in a convenient form, we apply a Jordan-Wigner transformation, which gives us
\begin{align}
\sigma^+_n \sigma^-_{n+1} & = c^\dagger_n c_{n+1}, \\
\sigma^+_{n+2} \sigma^-_n & = c^\dagger_{n+2} \exp ( -i \pi c^\dagger_{n+1} c_{n+1} ) c_n, \label{eq:app_JW_2}
\end{align}
where the second term gains a phase as the exponentials from the Jordan-Wigner transformation do not fully cancel. Using these results, the Hamiltonian transforms to
\begin{equation}
H  =  \sum_n \left[ -u c_n^\dagger c_{n+1} +\frac{iv}{2} (c_n^\dagger c_{n+1} \sigma^z_{n+2} + c^\dagger_{n+1} c_{n+2} \sigma^z_n -  c^\dagger_n c_{n+2})\right]  + \text{H.c.}, \label{eq:app_h_interacting}
\end{equation} 
where the final term loses its $\sigma^z_{n+1}$ because $\sigma^z_{n+1} = \exp( i \pi c^\dagger_{n+1}c_{n+1})$ which cancels with the exponential obtained from the Jordan-Wigner transformation in Eq. (\ref{eq:app_JW_2}). We also swap the final term for its Hermitian conjugate which picks up a minus sign. 

For a system with periodic boundary conditions, after applying the Jordan-Wigner transformation we would pick up boundary terms which couple the last lattice sites $n =N-1$ and $n = N-2$ to the first lattice sites $n = 0$ and $ n =1$, however these terms contribute an order $O(1/N)$ correction to the Hamiltonian which can we can safely ignore as we assume we work in the thermodynamic limit for large $N$ \cite{C4_DePasquale}.

\subsection{Mean field theory \label{sec:mean_field_theory}}

The Hamiltonian we arrived at in Eq.~(\ref{eq:app_h_interacting}) after a Jordan-Wigner transformation is an exact result---it is just the original Hamiltonian expressed in a different basis. However, this is an \textit{interacting} fermionic Hamiltonian due to the four-fermion interaction terms such as $c_n^\dagger c_{n+1} \sigma^z_{n+2}$, therefore this Hamiltonian cannot be diagonalised using the method of Appendix~\ref{appendix:numerical_techniques} which is only suited to quadratic Hamiltonians. In order to make progress, we apply mean field theory to transform this Hamiltonian into a non-interacting quadratic Hamiltonian. We define the fluctuation of an operator $A$ as $\delta A = A - \langle A \rangle $, where $\langle A \rangle$ is the expectation value of the operator $A$ with respect to the mean field ground state. Therefore, for a product of two operators we have 
\begin{equation}
AB   = \langle A \rangle B + A \langle B \rangle - \langle A \rangle \langle B \rangle + \delta A \delta B,
\end{equation}
where we ignore the last term that is second order in fluctuations. Applying this to the interacting Hamiltonian of Eq.~(\ref{eq:app_h_interacting}), we can tackle the two interacting terms by choosing $A = c_n^\dagger c_{n+1}$ and $B = \sigma^z_{n+2}$ for the first interacting term, with a similar choice for the second interacting term, so the Hamiltonian maps to
\begin{equation}
H_\mathrm{MF}(\alpha,Z) = \sum_n \left( -(u - iv Z) c^\dagger_n c_{n+1} - \frac{iv}{2} c^\dagger_n c_{n+2} \right)  + \mu \sum_n c^\dagger_n c_n + E_0 + \mathrm{H.c.},
\end{equation}
where $\mu =  2 v \mathrm{Im}(\alpha)$ is an effective chemical potential controlling the number of particles in the ground state, $E_0 =  v ( Z - 1) \mathrm{Im}(\alpha) $ is a constant energy shift, and  $\langle \sigma_n^z \rangle = Z$, $\langle c^\dagger_n c_{n+1} \rangle = \alpha$, where the expectation value is done with respect to the ground state of the mean field Hamiltonian. This gives us a quadratic Hamiltonian as a function of $\alpha$ and $Z$, however in order for this to be self-consistent, we require 
\begin{equation}
\langle \Omega(\alpha , Z)|\sigma^z_n | \Omega(\alpha, Z)\rangle =Z, \quad \langle \Omega(\alpha , Z)|c^\dagger_n c_{n+1} | \Omega(\alpha,Z)\rangle =\alpha,\label{eq:self_con}
\end{equation}
where $| \Omega(\alpha,Z)\rangle$ is the ground state of $H_\text{MF}(\alpha, Z)$. 

There are many solutions to these equations, however we can single one out on physical grounds. As observed via \textit{exact} diagonalisation of the spin Hamiltonian of Eq.~(\ref{eq:C4_spin_ham}), the $z$-component of spin is given by $\langle \Omega | \sigma^z_n |\Omega \rangle = 0$, where $|\Omega\rangle$ is the ground state of the total spin Hamiltonian. From the definition of $\sigma^z_n = 1 - 2c^\dagger_n c_n$ this implies we must have half-filling with $\langle c^\dagger_n c_n \rangle = 1/2$ which in turn implies we must have a chemical potential of $\mu = 0$. This can also be deduced analytically from the particle-hole symmetry of the model, see Eq.~(\ref{eq:half_filling}). Therefore, we take the solution $Z = \mathrm{Im}(\alpha) = 0$ and our mean field Hamiltonian is given by
\begin{equation}
H_\text{MF} = \sum_n \left( -u c^\dagger_n c_{n+1} - \frac{iv}{2} c^\dagger_n c_{n+2} \right) + \text{H.c.}. \label{eq:mf_ham}
\end{equation}
This Hamiltonian is now in a form that can be diagonalised exactly using the method of Appendix~\ref{appendix:numerical_techniques}, including for inhomogeneous generalisations for when we upgrade $u$ and $v$ to space-dependent parameters. In this chapter, all mean field (MF) numerics are done using the methods of Appendix \ref{appendix:numerical_techniques} with this Hamiltonian.

\subsection{Diagonalising $H_\mathrm{MF}$}

\begin{figure}[t]
\begin{center}
\includegraphics[scale=1]{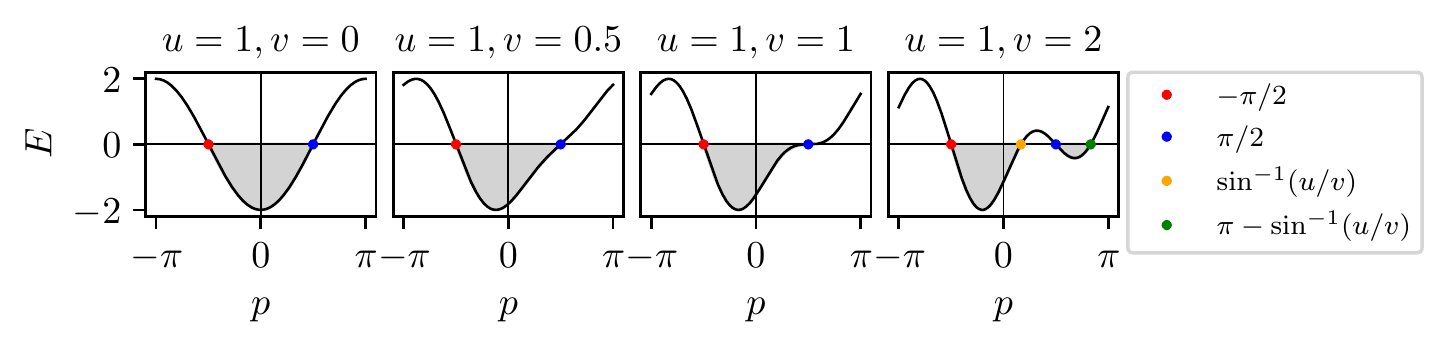}
\end{center}
\caption{The dispersion relation of Eq.~(\ref{eq:C4_mono_dispersion}) for various values of $v$, for $a = 1$. We see that two additional Fermi points appear if $v > u$ which divides the negative-energy portion of the Brillouin zone into two disconnected regions.}
\label{fig:C4_mono_dispersion}
\end{figure}

For the case of the homogeneous model (constant $u$ and $v$) the model has translational symmetry so it can be diagonalised exactly with a Fourier transform given by
\begin{equation}
c_n = \frac{1}{\sqrt{N}} \sum_{p \in \mathrm{B.Z}} e^{ipan} c_p, \label{eq:C4_fourier_transform}
\end{equation}
where $\mathrm{B.Z.} = [-\pi/a,\pi/a)$ is the Brillouin zone, $p = \frac{2n\pi}{N} \in \mathrm{B.Z.}$ for $n \in \mathbb{Z}$, and $a $ is the lattice spacing. Substituting this into the Hamiltonian gives us
\begin{equation}
\begin{aligned}
H_\mathrm{MF} &  = \frac{1}{N}  \sum_{n,p,q} \left( -u e^{-ipan} e^{iqa(n+1)} - \frac{iv}{2} e^{-ipan} e^{iqa(n+2)} \right)c^\dagger_p c_q + \mathrm{H.c.} \\
& = \sum_{p,q} \left( -u e^{iqa} - \frac{iv}{2} e^{2iaq} \right) \left( \frac{1}{N} \sum_n e^{-i(p-q)an} \right) c^\dagger_p c_q + \mathrm{H.c.} \\
& = \sum_p \left( -u e^{ipa} - \frac{iv}{2} e^{2ipa} \right) c^\dagger_p c_p + \text{H.c.} \\& = \sum_p E(p) c^\dagger_p c_p, 
\end{aligned}
\end{equation}
where $E(p)$ is the dispersion relation given by
\begin{equation}
E(p) = -2u \cos(ap) + v \sin(2ap), \label{eq:C4_mono_dispersion}
\end{equation}
which is plotted in Fig.~\ref{fig:C4_mono_dispersion} for different values of $v$. The ground state is therefore the state for which all negative-energy states are occupied as
\begin{equation}
|\Omega_\mathrm{MF} \rangle = \prod_{p:E(p) < 0 } c^\dagger_p |0\rangle,
\end{equation}
where $|0\rangle$ is the vacuum state annihilated by all annihilation operators as $c_p |0\rangle = 0$ for all $p$. For the rest of this section we simply refer to the ground state as $|\Omega \rangle$.

The Fermi points of this model are the points $p$ for which $E(p) = 0$, which implies
\begin{equation}
\left[ u - v\sin(ap) \right] \cos(ap) = 0 ,
\end{equation}
where we used the trigonometric identity $\sin(2x) = 2 \cos(x)\sin(x)$. First, suppose that $|v| < |u|$. As $v\sin(ap) < u$ for all $p$, the first bracket will never equal zero and the only solution to this is when $\cos( ap) = 0$, which gives us two solutions $p_\pm = \pm \frac{\pi}{2a}$. However, for $|v| > |u|$, we still have the usual Fermi points at $p_\pm = \pm \frac{\pi}{2a}$, but now the first bracket can vanish too, so we get the additional two solutions at
\begin{equation}
p_1 = \frac{1}{a} \sin^{-1} \left( \frac{u}{v} \right) , \quad p_2 = \frac{\pi}{a} - p_1 .
\end{equation}
These additional zero energy crossings are a result of the Nielsen-Ninomiya theorem which states that the number of left-movers and right-movers in a lattice model must be equal~\cite{C4_Nielsen1,C4_Nielsen2}, where the direction of motion is given by the sign of the Fermi velocity $v_i = E'(p_i)$ at the Fermi points $p_i$. These additional crossings also change the topology of the negative energy portion of the Brillouin zone by splitting it into two disconnected regions as we see in Fig. \ref{fig:C4_mono_dispersion}.

\section{Phase transitions \label{section:phase_transitions}}
To investigate the nature of quantum phases supported by Eq.~(\ref{eq:C4_spin_ham}), and the transitions between them, we consider the case of homogeneous couplings $u$ and $v$. In this section, we focus on the predictions of the mean field Hamiltonian of Eq.~(\ref{eq:mf_ham}) and compare it with the results obtained using matrix product state (MPS) analysis of the spin Hamiltonian of Eq.~(\ref{eq:C4_spin_ham}) published in Ref.~\cite{black_hole_paper}. All analytic calculations of this section are done using the mean field theory.
\subsection{Correlations}

The correlation matrix is defined as $C_{nm} = \langle \Omega | c^\dagger_n c_m |\Omega \rangle$, where $|\Omega \rangle$ is the ground state of the Hamiltonian. Mapping to momentum space with a discrete Fourier transform as in Eq. (\ref{eq:C4_fourier_transform}), we can write
\begin{equation}
\begin{aligned}
C_{nm} & = \frac{1}{N} \sum_{p,q \in \text{BZ}} e^{-i p n} e^{i q m} \langle \Omega | c^\dagger_p c_q |\Omega \rangle \\
&  = \frac{1}{2 \pi} \sum_{p:E(p)<0} \Delta p e^{-ip(n-m)}  \\
& \rightarrow  \frac{1}{2 \pi} \int_{p:E(p)< 0} \mathrm{d}p e^{-ip(n-m)},
\end{aligned} \label{eq:C4_correlation_def}
\end{equation} 
where in the second equality we used the fact that the ground state $|\Omega\rangle$ has all negative energy states occupied, so $\langle \Omega | c^\dagger_p c_q |\Omega \rangle = \delta_{pq} \theta(-E(p))$ and used the fact that eigenstates are separated in momentum space by $\Delta p = 2 \pi/N$ for a lattice spacing $a = 1$ to rewrite the sum as a Riemann sum. In the third equality we took the thermodynamic limit $N\rightarrow \infty$ mapping the Riemann sum to an integral which can now be solved analytically given the dispersion relation $E(p)$ from Eq. (\ref{eq:C4_mono_dispersion}).

For $|v| < |u|$ the correlation function is given by
\begin{equation}
C_{nm} = \frac{1}{2\pi} \int_{-\pi/2}^{\pi/2} \mathrm{d}p e^{-ip(n-m)}  = \frac{\sin\left[ \frac{\pi}{2} (n-m) \right] }{\pi (n-m)}, \label{eq:C4_correlation_v<u}
\end{equation}
which is an elementary result. For $|v| > |u|$, the negative energy portion of the Brillouin zone splits into two disconnected regions so the integral splits into two as
\begin{equation}
\begin{aligned}
C_{nm} & =\frac{1}{2\pi} \left( \int_{-\frac{\pi}{2}}^{p_1} \mathrm{d}p + \int_{\frac{\pi}{2}}^{\pi - p_1} \mathrm{d}p \right) e^{-ipa(n-m)} \\
& = \frac{i}{2 \pi (n-m)} \left\{ -2 \cos\left[ (n-m)\frac{\pi}{2} \right] + (-1)^{n-m} e^{i p_1 (n-m)} + e^{-ip_1(n-m)}\right\}, \label{eq:C4_correlation_v>u}
\end{aligned} 
\end{equation}
which is now a function of $v$. 

The change in the topology of the dispersion relation is the root cause of the phase transition exhibited by the model to be seen in the next three sections. This is because the change in topology of the dispersion relation has resulted in correlation functions that are \textit{not} smooth functions. As all observables can be expressed in terms of correlations, they will not be smooth in general too.
\subsection{Energy density}

The ground state energy density is given by
\begin{equation}
\rho_0  =\lim_{N \rightarrow \infty} \frac{1}{N} \sum_{p:E(p) < 0} E(p) = \frac{1}{2 \pi} \int_{p:E(p)<0} \mathrm{d}p E(p),
\end{equation}
where we took the thermodynamic limit $N \rightarrow \infty$ by using the standard trick of moulding the sum into a Riemann sum and taking the limit. For $|v| \leq |u|$, we have
\begin{equation}
\rho_0 = \frac{1}{2 \pi} \int_{-\frac{\pi}{2}}^{\frac{\pi}{2}} \mathrm{d}p \left[  -2u   \cos(p) + v\sin(2p) \right]  = - \frac{2u}{\pi },
\end{equation}
whilst for $|v| > |u|$ we have
\begin{equation}
\rho_0 = \frac{1}{2 \pi}\left( \int_{-\frac{\pi}{2}}^{p_1} \mathrm{d}p + \int_{\frac{\pi}{2}}^{\pi - p_1} \mathrm{d}p \right) \left[  -2u   \cos(p) + v\sin(2p) \right]  = -\frac{1}{\pi } \left( \frac{u^2}{v} + v \right).
\end{equation}

If we look at the derivatives of the energy density, we see that the model exhibits a second order phase transition as we change $v$:
\begin{equation}
\frac{\partial \rho_0}{\partial v}  = \begin{cases}
0 & |v| \leq |u| \\
\frac{1}{\pi } \left(  1 - \frac{u^2}{v^2} \right) & |v| > |u|
\end{cases}, \quad \frac{\partial^2 \rho_0}{\partial v^2}  = \begin{cases}
0 & |v| \leq |u| \\
- \frac{1}{\pi} \frac{u^2}{v^3} & |v| > |u|
\end{cases},
\end{equation}
so we see that the first derivative is continuous, but the second derivative contains a discontinuity at $|u| = |v|$ which therefore corresponds to the critical point of a second order phase transition.

\begin{figure}[t]
\begin{center}
\includegraphics[scale=1,valign=t]{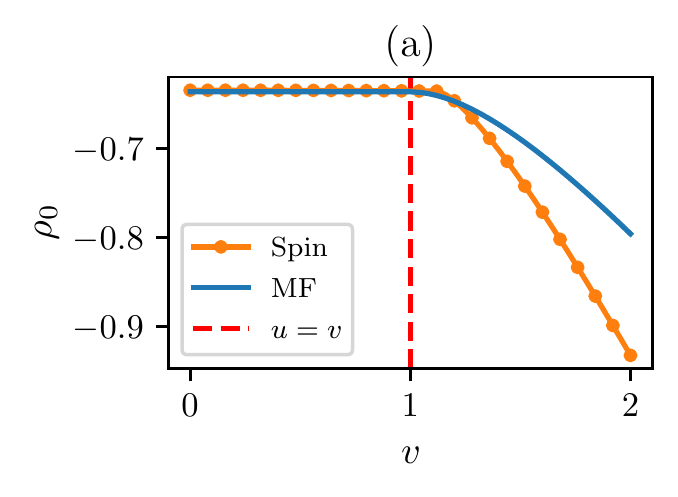}
\includegraphics[scale=1,valign=t]{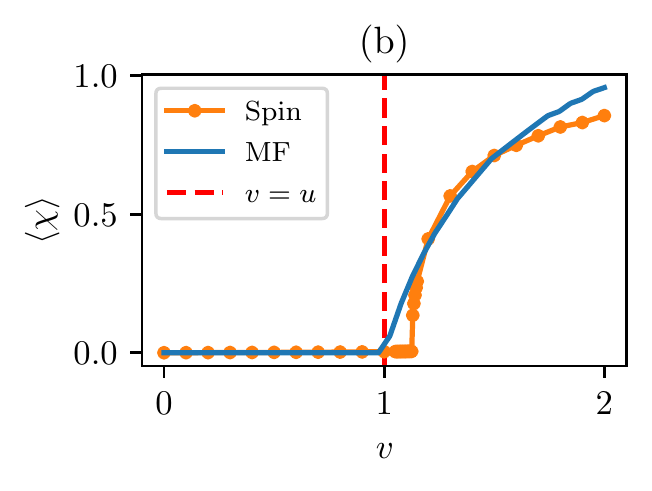}
\end{center}
\caption{(a) A comparison of the ground state energy density vs. $v$ obtained from MPS simulation of the spin model from Ref.~\cite{black_hole_paper} and the mean field (MF) approximation for $N= 200$ spins. (b) A comparison of the total ground state chirality $\langle \chi \rangle = \sum_n \langle \chi_n \rangle$ density vs. $v$ obtained from the MPS simulation of the spin model from Ref.~\cite{black_hole_paper} and the mean field (MF) approximation for $N= 200$ spins. }
\label{fig:E&chi_vs_v}
\end{figure}

In Fig.~\ref{fig:E&chi_vs_v}(a), we compare the ground state energy density vs. $v$ for the MPS numerics of the spin model taken from Ref.~\cite{black_hole_paper} and the mean field approximation for a system of $N = 200$. We see that the mean field agrees extremely well with the spin model, accurately predicting the location of the critical point. Below the critical point, the two models agree exactly, which suggests that the interactions induced by the chiral term are irrelevant in the ground state.

\subsection{Chirality}
We now study the chirality of the ground state. Using mean field theory, we showed in Eq.~(\ref{eq:mf_ham}) that the chirality is given by
\begin{equation}
\chi_n = -2i c^\dagger_n c_{n+2} + \mathrm{H.c.}.
\end{equation}
Using the expression for the correlation matrix derived in Eqs.~(\ref{eq:C4_correlation_v<u}) and (\ref{eq:C4_correlation_v>u}), the ground state chirality is given by the simple expression
\begin{equation}
\langle \chi_n \rangle = 4\mathrm{Im}\left( C_{n,n+2} \right) = 
\begin{cases}
0 & |v| < |u| \\
\frac{4}{\pi} \left(1 - \frac{u^2}{v^2}\right) & |v| \geq |u|
\end{cases}. \label{eq:C4_MF_chi}
\end{equation}
By Taylor expanding just above the critical point, we find the chirality goes as
\begin{equation}
\langle \chi_n \rangle \sim (v-v_\mathrm{c})^\gamma,
\end{equation}
where $v_\mathrm{c} = u$ is the critical point and $\gamma = 1$ is the critical exponent.

On the other hand, it was shown in Ref.~\cite{black_hole_paper} by studying the full spin model of Eq.~(\ref{eq:C4_spin_ham}) using finite DMRG \cite{C4_schollwock2011density} that the phase transition of the full model is located at $v_\mathrm{c} \approx 1.12u$ with a critical exponent of  $\gamma \approx 0.39$. A comparison between the chirality of this MPS spin model simulation and the mean field approximation can be seen in Fig.~\ref{fig:E&chi_vs_v}(b). The mean field faithfully captures the important information of a non-chiral to chiral phase transition. In particular, just like for the energy density, the two models agree exactly below the critical point where the chirality is zero. The behaviour suggests the chirality is an order parameter for the model and emphases again that, below the critical point, the interactions are irrelevant in the ground state.

From Eq.~(\ref{eq:C4_MF_chi}), we see that the chirality is non-zero if and only if we have complex next-to-nearest-neighbour correlations $C_{n,n+2}$. We ask under what conditions is this the case. Consider a general tight-binding model with discrete translational symmetry and periodic boundary conditions. From above in Eq.~(\ref{eq:C4_correlation_def}) the correlation matrix is given by
\begin{equation}
C_{nm} = \frac{1}{2 \pi} \int_{E(p) < 0} \mathrm{d}p e^{-ip(n - m)}.
\end{equation}
Suppose we had a model with inversion symmetry under the transformation $c_n \rightarrow  c_{-n}$. This implies that the dispersion relation is an even function obeying $E(p) = E(-p)$, so our Fermi points come in $\pm$ pairs. First consider the simple case where the model has two Fermi points $\pm p_0$ such that for $|p| < p_0$ we have $E(p) < 0$, then we have
\begin{equation}
C_{nm} = \frac{1}{2\pi} \int_{-p_0}^{p_0} \mathrm{d}p e^{-ip(n-m)}  = \frac{\sin\left[p_0(n-m)\right]}{\pi (n-m)} ,
\end{equation}
which is real. For two pairs where, we label the Fermi points at $\pm p_1$ and $\pm p_2$ for $p_1 < p_2$, where now if $p_1 < |p| < p_2$ then $E(p) < 0$, we have
\begin{equation}
C_{nm} = \frac{1}{2\pi}\left( \int_{-p_2}^{-p_1} \mathrm{d}p + \int_{p_1}^{p_2} \mathrm{d}p \right) e^{-ip(n-m)}  = \frac{\sin\left[p_2(n-m)\right] - \sin\left[p_1(n-m)\right]}{\pi(n-m)}, 
\end{equation}
which, again, is real. This generalises to models with higher numbers of Fermi points. We see that that models with inversion symmetry have zero chirality.

Let us now break inversion symmetry. A simple model to look at that breaks inversion symmetry is a model with nearest-neighbour hoppings and complex couplings, with Hamiltonian
\begin{equation}
H = -ue^{-i \theta}\sum_n c^\dagger_n c_{n+1} + \text{H.c.},
\end{equation}
where $u \in \mathbb{R}$ and $\theta \in [0,2\pi)$. The breaking of inversion symmetry is apparent from the dispersion relation $E(p) = -2u \cos(p - \theta)$ as it is no longer an even function. The Fermi points of this model are at $p_0 = \theta \pm \pi/2$, therefore the correlations of this model are given by
\begin{equation}
C_{nm} = \frac{1}{2 \pi} \int_{\theta - \frac{\pi}{2}}^{\theta + \frac{\pi}{2}} \mathrm{d}p e^{-ip(n - m)} = \frac{\sin\left[(n-m)\frac{\pi}{2}\right]}{\pi(n-m)}e^{-i \theta(n-m)},
\end{equation}
which are complex, but notice that correlations between next-to-nearest-neighbours, where $|n -m| = 2$, are zero, therefore the chirality of this model will be zero too.

The simplest way to achieve complex next-to-nearest neighbour correlations is to include a term in the Hamiltonian which couples next-to-nearest neighbour sites and breaks inversion symmetry. A simple example of this is nothing but our mean field Hamiltonian of Eq. (\ref{eq:mf_ham}). The interesting feature of this model is that for $|v| < |u|$, the dispersion relation retains its symmetric Fermi points at $p_\pm = \pm \pi /2$ despite the dispersion not being symmetric. Therefore, all correlators in this phase will be real as seen in Eq.~(\ref{eq:C4_correlation_v<u}) and hence the chirality will be zero. On the other hand, for $|v| > |u|$ the topology of the dispersion relation changes resulting in complex correlations which yields a non-zero chirality, giving the chirality its order parameter behaviour. 

\begin{figure}[t]
\begin{center}
\includegraphics[scale=1]{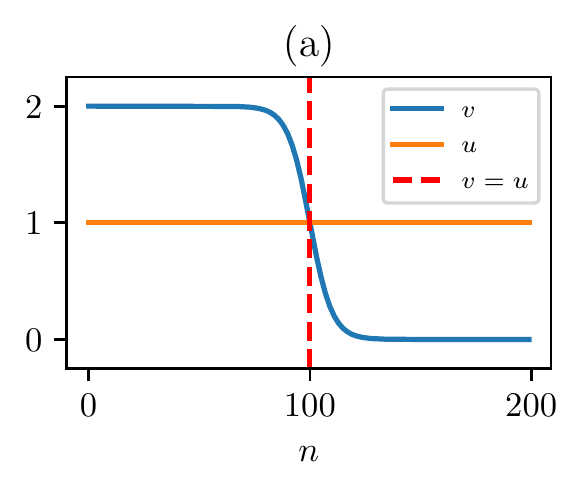}
\includegraphics[scale=1]{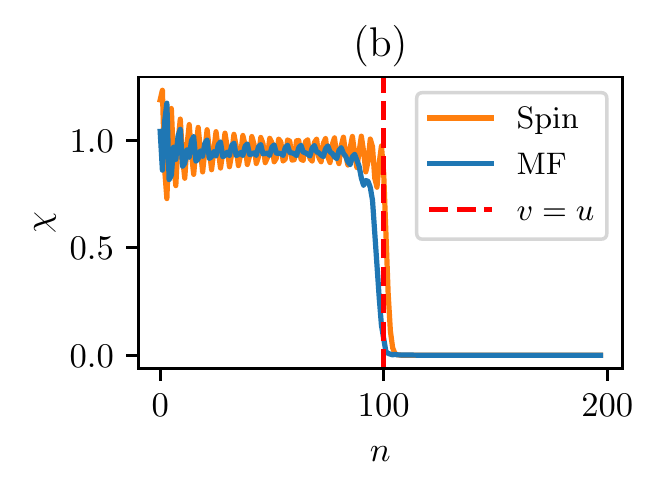}
\caption{(a) An example of an inhomogeneous distribution for the couplings $v$. (b) The corresponding chirality obtained from the spin model MPS from Ref.~\cite{black_hole_paper} and the mean field model. We see that the distribution of $v$ describes a phase boundary between a chiral ($v > u$) and non-chiral ($v <u$) phase.}
\label{fig:C4_inhomogenous_v}
\end{center}
\end{figure}

The above analysis was conducted for homogeneous systems where $u$ and $v$ are constants, however we still expect this to hold when we upgrade $v$ to a slowly varying function. We consider profiles where $v(x)$ changes slowly in space such that $v = u$. In Fig.~\ref{fig:C4_inhomogenous_v} we present the chirality distribution across the system for a given $v(x)$. We observe the result that the system is  chiral where $|v| > |u|$, whereas for $|v| < |u|$ the system is non-chiral, therefore we have an interface between two phases.

\subsection{Central charge}

\begin{figure}[t]
\begin{center}
\includegraphics[scale=1]{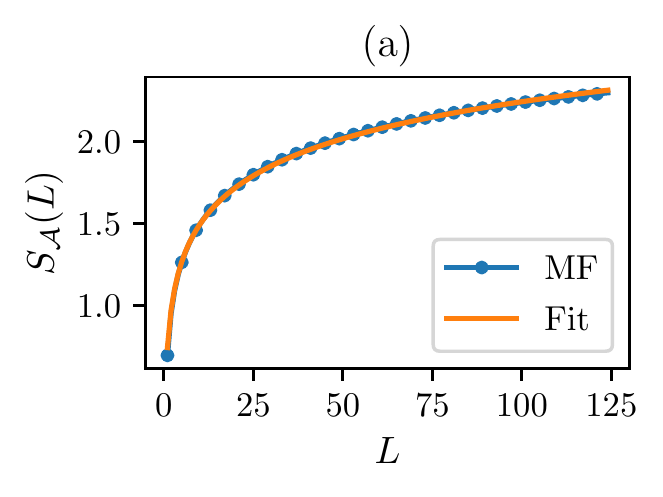}
\includegraphics[scale=1]{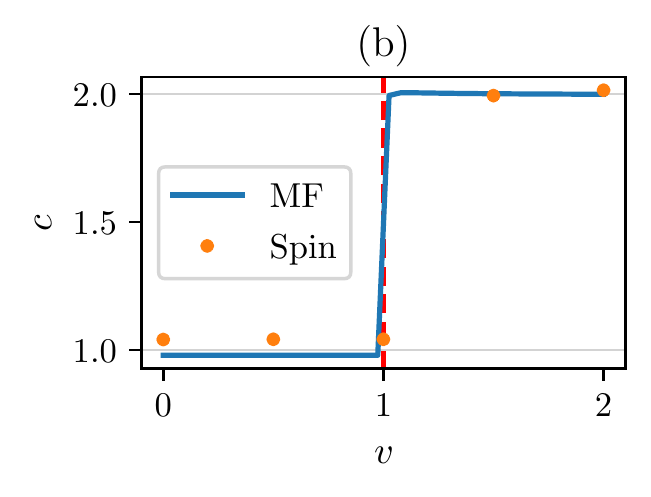}
\end{center}
\caption{(a) The entanglement entropy $S_L$ of the mean field (MF) model vs. $L$ for a system of size $N = 200$. We see the entanglement entropy follows Eq. (\ref{eq:C4_entanglement_entropy}), allowing us to extract the central charge. (b) A comparison of the central charge $c$ of the mean field model and spin model vs. $v$ for the same system. We see that the central charge jumps from $c = 1$ to $c = 2$ across the phase transition for the mean field, suggesting that the degrees of freedom of the model have changed.}
\label{fig:C4_entanglement_entropy}
\end{figure}

To gain further insight into the nature of the chiral phase transition, we consider the behaviour of the ground state bipartite entanglement entropy as a function of $v$. Consider partitioning the system into two subsystems, $\mathcal{A}$ and $\mathcal{B}$, where $\mathcal{A}$ contains $L \ll N$ adjacent spins. We define the reduced density matrix of $\mathcal{A}$ as the partial trace over the remaining $N - L$ spins of $\mathcal{B}$ as $\rho_\mathcal{A} = \mathrm{Tr}_\mathcal{B}(\rho )$, where $\rho$ is the state of the whole system. As we are interested in the ground state only, we have $\rho = |\Omega \rangle \langle \Omega |$, where $|\Omega \rangle$ is the (pure) ground state of the total system. The entanglement entropy is defined as $S_\mathcal{A} = -\mathrm{Tr}(\rho_\mathcal{A} \ln \rho_\mathcal{A})$. As discussed above, the model is gapless for all $v$ so it can be described by a conformal field theory (CFT)~\cite{CFT}. In this case we expect the ground state entanglement entropy of a partition of  spins to obey the Cardy formula
\begin{equation}
S_\mathcal{A}(L) = \frac{c}{3} \ln L + S_0, \label{eq:C4_entanglement_entropy}
\end{equation}
where $c$ is the central charge of the CFT and $S_0$ is a constant \cite{C4_Calabrese,PhysRevLett.90.227902}, which applies to both the original spin model and the mean field approximation. Using the numerical techniques of Appendix~\ref{appendix:numerical_techniques}, we can measure the entanglement entropy of the mean field model quite simply by using the correlation matrix. We find that scaling behaviour of the entanglement entropy follows this formula, as shown in Fig.~\ref{fig:C4_entanglement_entropy}(a), allowing us to extract the central charge $c$ for various values of $v$.

Using the MPS results from Ref.~\cite{black_hole_paper}, we compare the spin model and the mean field approximation. In Fig.~\ref{fig:C4_entanglement_entropy}(b) we see that $ c \approx 1 $ in the XY phase which jumps to $ c \approx 2$ in the chiral phase, with good agreement between the spin and mean field results. We can clearly interpret this in the mean field model: the additional Fermi points appearing when $|v| > |u|$ cause the model to transition from a $c=1$ CFT with a single Dirac fermion to a $c=2=1+1$ CFT with two Dirac fermions, as seen by the additional Fermi points of the dispersion in Fig. \ref{fig:C4_mono_dispersion}. This can also be understood from the lattice structure of the MF model, as seen in Fig.~\ref{fig:C4_zig_zag_lattice}(a), where for $|v|\ll |u|$ a single zig-zag fermionic chain dominates ($c=1$) while for $|v|\gg |u|$ two fermionic chains dominate, corresponding to the edges of the ladder, thus effectively doubling the degrees of freedom ($c=2$).

\section{Luttinger model \label{section:luttinger}}
The previous sections have demonstrated that in the phase $|v| < |u|$ the ground state of the spin model and mean field approximation agree extremely well. In fact, the ground state observables of energy density, chirality and central charge are independent of $v$ in this phase, behaving exactly the same as the XY model where $v = 0$. We investigate why this is the case by employing the machinery of Luttinger liquids and bosonisation. The analytical techniques in this section follow closely the review of Ref.~\cite{C4_Miranda}.

\subsection{Particle-hole symmetry}

Let us return to the full spin model of Eq.~(\ref{eq:C4_spin_ham}). After a Jordan-Wigner transformation, we arrived at the interacting Hamiltonian $H = H_0 + H_\mathrm{int}$, where 
\begin{align}
H_0  & = \sum_n \left( - u  c_n^\dagger c_{n+1} - \frac{iv}{2} c^\dagger_n c_{n+2} \right) + \mathrm{H.c.}, \\
H_\mathrm{int} & = \frac{iv}{2} \sum_n  \left(c_n^\dagger c_{n+1} \sigma^z_{n+2} + c^\dagger_{n+1} c_{n+2} \sigma^z_n  \right)  + \text{H.c.}, \label{eq:app_H_full_fermions}
\end{align}
where $\sigma^z_n = 1 - 2c^\dagger_n c_n$. This fully interacting Hamiltonian has particle-hole symmetry under the transformation
\begin{equation}
c_n \rightarrow U^\dagger c_n U = (-1)^n c_n^\dagger , \quad c_n^\dagger \rightarrow U^\dagger c_n^\dagger U = (-1)^n c_n.
\end{equation}

\begin{figure}[t]
\begin{center}
\includegraphics[scale=1]{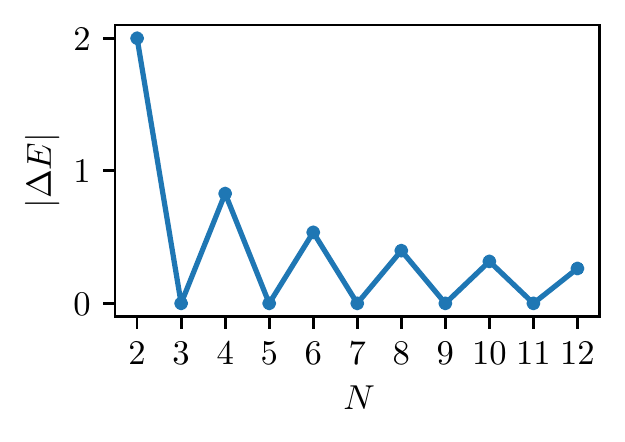}
\end{center}
\caption{The energy difference $\Delta E$ between the lowest two energy levels of the spin Hamiltonian of Eq.~(\ref{eq:C4_spin_ham}) obtained through exact diagonalisation.}
\label{fig:spin_gap}
\end{figure}

Let us look at the consequences of this symmetry. Exact diagonalisation of the spin Hamiltonian reveals that the ground state is non-degenerate for an even number of lattice sites, as seen in Fig.~\ref{fig:spin_gap}, so we fix $N \in 2 \mathbb{N}$ to avoid any subtleties due to degeneracy, therefore our ground state will be an eigenstate of $U$ with an eigenvalue of either $+1$ or $-1$ as $U^2 = \mathbb{I}$. Suppose we calculated the ground state density, we have
\begin{equation}
\langle \Omega | c^\dagger_n c_n | \Omega \rangle  = \langle \Omega | U^\dagger c^\dagger_n c_n U|\Omega \rangle 
 = (-1)^{2n} \langle \Omega | c_n c_n^\dagger | \Omega \rangle  
 = 1 - \langle \Omega | c^\dagger_n c_n |\Omega \rangle , \label{eq:half_filling}
\end{equation}
implying $\langle\Omega |c^\dagger_n c_n |\Omega \rangle  = \frac{1}{2}$, which is our usual half-filling result, where in the last equality we used the fermionic anti-commutation relations. Now, for the nearest-neighbour correlations we have
\begin{equation}
\langle \Omega | c^\dagger_n c_{n+1} |\Omega \rangle  = \langle \Omega|U^\dagger c^\dagger_n c_{n+1} U |\Omega \rangle 
= (-1)^{2n + 1} \langle \Omega | c_n c^\dagger_{n+1} |\Omega \rangle 
 =  \langle \Omega | c^\dagger_n c_{n+1} |\Omega \rangle^*,
\end{equation}
where in the last equality we used the fermionic anti-commutation relations and the fact that for any operator $A$ that $\langle A^\dagger \rangle = \langle A \rangle^*$. Therefore, the nearest-neighbour correlators are real. We use these results in the following calculation.

For a product of two operators, normal ordering amounts to subtracting off the ground state expectation value as $: A : \ = A - \langle \Omega | A | \Omega \rangle$. We can use this to simplify the interaction term of Eq. (\ref{eq:app_H_full_fermions}) which prepares us for bosonisation later. We have
\begin{equation}
c^\dagger_n c_{n+1} = \ : c^\dagger_n c_{n+1} : + \langle \Omega| c^\dagger_n c_{n+1} |\Omega\rangle \equiv \ :c^\dagger_n c_{n+1} : + \alpha,
\end{equation}
where we have defined the correlation $\alpha = \langle \Omega |c^\dagger_n c_{n+1} |\Omega \rangle$. Similarly, we have
\begin{equation}
\sigma^z_n = 1 - 2 c^\dagger_n c_n =  1 - 2( : c^\dagger_n c_n : + \langle \Omega | c^\dagger_n c_n| \Omega \rangle ) = -2 :c^\dagger_n c_n :,
\end{equation}
where we used the half filling result $\langle \Omega|  c_n^\dagger c_n |\Omega \rangle = \frac{1}{2}$. From this, we can substitute this into the interaction Hamiltonian of Eq. (\ref{eq:app_H_full_fermions}) to give
\begin{equation}
\begin{aligned}
H_\mathrm{int} & = -iv \sum_n \left[ \left( :c^\dagger_n c_{n+1}: + \alpha \right) :c^\dagger_{n+2} c_{n+2}: + \left( :c^\dagger_{n+1} c_{n+2} :+ \alpha \right) :c^\dagger_n c_n: \right] + \mathrm{H.c.} \\
& = -iv \sum_n \left( :c^\dagger_n c_{n+1} : : c^\dagger_{n+2} c_{n+2} : + :c^\dagger_{n+1} c_{n+2} : : c^\dagger_n c_n : \right) + \text{H.c.},
\end{aligned}
\end{equation}
where we used the fact that $\alpha$ is real and $:c^\dagger_n c_n:$ is Hermitian to get rid of $\alpha$.

\subsection{Continuum limit}
For the phase $|v| < |u|$, the mean field theory agrees extremely well with the total spin model and demonstrates that the additional chirality term interaction is irrelevant for ground state properties, whereby the model behaves as if it is the XY model ($v = 0$). We focus on this phase in the following. 

In order to bosonise this model, we must take the continuum limit. In previous chapters, we took the continuum limit by mapping to momentum space, defining a two-component spinor and expanding about the Fermi points. However, our Hamiltonian is interacting as it is not quadratic, therefore we cannot diagonalise with a Fourier transform. In order to make progress, we have to take the continuum limit in real space with an approximation. In the non-chiral phase for $|v| < |u|$, the mean field of Sec.~\ref{sec:mean_field_theory} suggested that the model has two Fermi points at $p_\mathrm{R,L} = \pm \pi/2$, so we expand our fields as
\begin{equation}
\frac{c_n}{\sqrt{a}} = \sum_{\mu = \mathrm{R,L}} e^{ip_\mu an} \psi_\mu(x_n), \label{eq:expansion}
\end{equation}
where the sum is over the Fermi points, $\psi_\mu(x_n)$ is a slowly-varying continuous field sampled at discrete lattice sites $x_n = na$ and we have reinstated the lattice spacing $a$. 

First, we substitute the expansion of Eq. (\ref{eq:expansion}) into $H_0$ of Eq. (\ref{eq:app_H_full_fermions}) to give
\begin{equation}
H_0  = \sum_{\mu,\nu}\sum_n a e^{-i(p_\mu - p_\nu) an} \left( -u  e^{ip_\nu a} \psi^\dagger_\mu(x_n)\psi_\nu(x_{n+1}) - \frac{iv}{2}  e^{2ip_\nu a} \psi^\dagger_\mu(x_n) \psi_\nu(x_{n+2}) \right) + \text{H.c.}. 
\end{equation}
We now discard any oscillating term in the Hamiltonian as these integrate to zero, so we requires $p_\mu = p_\nu$ in the first phase. This yields
\begin{equation}
\begin{aligned}
H_0 & = \sum_{\mu} \sum_n a \left( -u e^{ip_\mu a}  \psi^\dagger_\mu \left( \psi_\mu + a \partial_x \psi_\mu \right) -\frac{iv}{2} e^{2ip_\mu a}  \psi^\dagger_\mu \left( \psi_\mu + 2a \partial_x \psi_\mu\right)  \right) +O(a^3)  + \text{H.c.} \\
& = -i \sum_{\mu} \sum_na^2 \left( \pm u \psi^\dagger_\mu \partial_x \psi_\mu  
- v \psi^\dagger_\mu \partial_x \psi_\mu \right) +O(a^3) + \text{H.c.}  \\
&  \rightarrow -2i\sum_{\mu} \int  \mathrm{d}x  v_\mu \psi_\mu^\dagger \partial_x \psi_\mu , 
\end{aligned}
\end{equation}
where in the second line $\pm$ corresponds to $\mu = \mathrm{R,L}$ and we have renormalised the couplings as $au \rightarrow u$ and $av \rightarrow v$. We have defined 
\begin{equation}
v_\mathrm{R,L} = 2(\pm u - v), \label{eq:MF_fermi_v}
\end{equation}
which are nothing but the Fermi velocities $v_\mu = E'(p_\mu)$ obtained from the mean field approximation with the dispersion relation of Eq.~(\ref{eq:C4_mono_dispersion}).

We now repeat the procedure for the interaction term $H_\mathrm{int}$ of Eq.~(\ref{eq:app_H_full_fermions}). We substitute in the expansion of Eq. (\ref{eq:expansion}) into $H_\mathrm{int}$ to give
\begin{equation}
\begin{aligned}
H_\text{int} = -iv   \sum_{\mu,\nu,\alpha,\beta} \sum_n  & ae^{-i(p_\mu - p_\nu + p_\alpha - p_\beta)an} \left( e^{i(p_\nu - 2(p_\alpha - p_\beta))a}+ e^{-i(p_\mu - 2p_\nu)a}\right)  \\
& \times :\psi^\dagger_\mu \psi_\nu: : \psi^\dagger_\alpha \psi_\beta : +O(a^3) + \text{H.c.}, \label{eq:int}
\end{aligned}
\end{equation}
where we have expanded all fields to zeroth order in $a$ to ensure the Hamiltonian retains order $a^2$ and renormalised the couplings as $av \rightarrow v$. We discard any term that oscillates which requires $p_\mu - p_\nu + p_\alpha - p_\beta = 2n \pi/a$ for $n \in \mathbb{Z}$. With this we find only four terms survive giving us
\begin{equation}
H_\text{int}  = 2v \int \mathrm{d}x \left( \rho_\mathrm{R}^2 + \rho_\mathrm{R} \rho_\mathrm{L}  - \rho_\mathrm{L} \rho_\mathrm{R} - \rho_\mathrm{L}^2 \right) + \text{H.c.} = 4v  \int \mathrm{d}x  \left( \rho^2_\mathrm{R} - \rho^2_\mathrm{L} \right),
\end{equation}
where we have defined the normal-ordered densities $\rho_{\mathrm{R,L}} = \ :\psi^\dagger_{\mathrm{R,L}} \psi_{\mathrm{R,L}}:$.
\subsection{Bosonising the Hamiltonian}
If we pull everything together, the normal-ordered Hamiltonian is given by
\begin{equation}
:H: \ = \ :H_0 + H_\text{int}: \ = -i\sum_{\mu = \mathrm{R,L}} \int  \mathrm{d}x  \left( v_\mu : \psi_\mu^\dagger \partial_x \psi_\mu :   \pm 4v  : \rho_\mu^2 : \right),
\end{equation}
where the $\pm$ corresponds to R and L respectively. Following Ref. \cite{C4_Miranda}, we map the fermionic fields $\psi_\mu$ to bosonic fields $\phi_\mu$ with the mapping
\begin{equation}
\psi_\mathrm{R,L} = F_\mathrm{R,L} \frac{1}{\sqrt{2\pi \alpha}} e^{\pm i \frac{2 \pi \hat{N}_\mathrm{R,L} }{L}x} e^{-i \sqrt{2\pi} \phi_\mathrm{R,L}}, \quad \rho_\mathrm{R,L} = \frac{\hat{N}_\mathrm{R,L}}{L} \mp \frac{1}{\sqrt{2\pi}} \partial_x \phi_\mathrm{R,L},
\end{equation}
where $\hat{N}_\mathrm{R,L}$ are defined as the normal ordered number operators for the right- and left-moving excitations respectively, $L = Na$ is the system's length, $F_\mathrm{R,L}$ are a pair of Klein factors and $\alpha$ is a cut-off. The bosonic fields obey the commutation relations
\begin{equation}
[ \phi_\mathrm{R,L}(x),\phi_\mathrm{R,L}(y)] = \pm\frac{i}{2} \mathrm{sgn}(x-y),
\end{equation} 
whilst pairs of fields about different Fermi points commute. The fermionic fields and densities obey the useful identities
\begin{equation}
:\psi^\dagger_\mathrm{R,L} \partial_x \psi_\mathrm{R,L}: = \pm \frac{i}{2}  \partial_x \phi_\mathrm{R,L} , \quad \rho_\mathrm{R,L}  = \mp \frac{1}{\sqrt{2\pi}}\partial_x \phi_\mathrm{R,L},
\end{equation}
where we have taken $L \rightarrow \infty$. With this, the Hamiltonian is mapped to
\begin{equation}
\begin{aligned}
:H: \ &  =  \int \mathrm{d}x \left( \frac{1}{2} \left[ |v_\mathrm{R}| : (\partial_x \phi_\mathrm{R})^2 : + |v_\mathrm{L}| :(\partial_x \phi_\mathrm{L})^2 : \right] + \frac{2v}{\pi} \left[ (\partial_x \phi_\mathrm{R})^2 - (\partial_x \phi_\mathrm{L})^2 \right]  \right) \\
& = \frac{1}{2} \int \mathrm{d}x \left( |v'_\mathrm{R}| :(\partial_x \phi_\mathrm{R})^2 :+ |v_\mathrm{L}'| : (\partial_x \phi_\mathrm{L})^2 : \right), \label{eq:h_boson}
\end{aligned}
\end{equation}
where the renormalised Fermi velocities are given by
\begin{equation}
v_\mathrm{R,L}' = 2 \left[ \pm u - v\left(1 - \frac{2}{\pi}\right) \right]. \label{eq:C4_renorm_fermi_velocities}
\end{equation}
As the Fermi velocities of the model are not equal, we must generalise the bosonisation procedure of Ref. \cite{C4_Miranda}. We define the canonical transformation
\begin{equation}
\Phi = \sqrt{\frac{\mathcal{N}}{2}} \left( \sqrt{|v_\mathrm{L}'|} \phi_\mathrm{L} - \sqrt{|v_\mathrm{R}'|} \phi_\mathrm{R} \right), \quad \Theta = \sqrt{\frac{\mathcal{N}}{2}} \left( \sqrt{|v_\mathrm{L}'|}\phi_\mathrm{L} + \sqrt{|v_\mathrm{R}'|}\phi_\mathrm{R} \right), \label{eq:canonical}
\end{equation}
where $\mathcal{N}$ is a constant to ensure the fields obey the correct commutation relations. Just as for the case of equal Fermi velocities in Ref. \cite{C4_Miranda}, we require the fields $\Phi$ and $\Theta$ to obey the commutation relations 
\begin{equation}
[ \Phi(x), \Theta(y) ] = -\frac{i}{2} \mathrm{sgn}(x-y).
\end{equation}
In terms of our canonical transformation, we have
\begin{equation}
\begin{aligned}
[\Phi(x),\Theta(y)] & = \frac{\mathcal{N}}{2} \left( |v'_\mathrm{L}|[\phi_\mathrm{L}(x),\phi_\mathrm{L}(y)] - |v_\mathrm{R}'| [\phi_\mathrm{R}(x),\phi_\mathrm{R}(y)] \right) \\
& = \frac{\mathcal{N}}{2}\left( -\frac{i|v_\mathrm{L}'|}{2} \mathrm{sgn}(x-y) - \frac{i|v_\mathrm{R}'|}{2} \mathrm{sgn}(x-y) \right) \\
& = - \frac{i\mathcal{N}}{4} (|v_\mathrm{L}'| + |v_\mathrm{R}'|) \mathrm{sgn}(x-y),
\end{aligned}
\end{equation}
therefore we require
\begin{equation}
\mathcal{N} = \frac{2}{|v'_\mathrm{L}| + |v_\mathrm{R}'|} = \frac{1}{2u}.
\end{equation}
Inverting the canonical transformation of Eq. (\ref{eq:canonical}), we have
\begin{equation}
\sqrt{|v_\mathrm{L}'|}\phi_- = \sqrt{u} \left( \Theta + \Phi \right) , \quad \sqrt{|v_\mathrm{R}'|}\phi_+ = \sqrt{u} \left( \Theta - \Phi \right).
\end{equation}
Substituting this back into the bosonised Hamiltonian of Eq. (\ref{eq:h_boson}) gives
\begin{equation}
:H: \ = u \int \mathrm{d}x \left[ :(\partial_x \Theta)^2: + :(\partial_x \Phi)^2: \right].
\end{equation}
Differentiating the commutator $[ \Phi(x),\Theta(y)]$ with respect to $y$, we find $[\Phi(x), \partial_y \Theta(y)] = i \delta(x-y)$, so we can identify the canonical momentum as $\Pi(x) = \partial_x \Theta(x)$. Therefore, the bosonised Hamiltonian takes the form of the free boson
\begin{equation}
:H: \ = u \int \mathrm{d}x \left[ :\Pi^2: + :(\partial_x \Phi)^2: \right],
\end{equation}
which is exactly the same result obtained from bosonising the XY model ($v = 0$). According to the theory of Luttinger liquids, this implies that $K = 1$ which is the sign of non-interacting fermions \cite{C4_Giamarchi}, demonstrating that the interactions for $|v| < |u|$ are irrelevant in the ground state.

\begin{figure}[t]
\begin{center}
\includegraphics[scale=1]{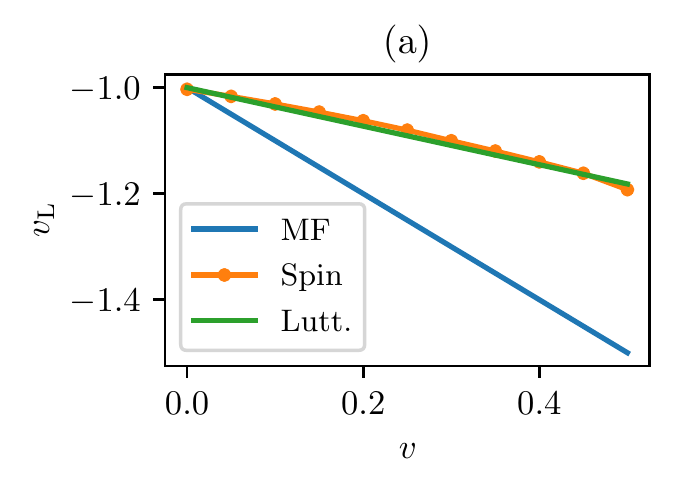}
\includegraphics[scale=1]{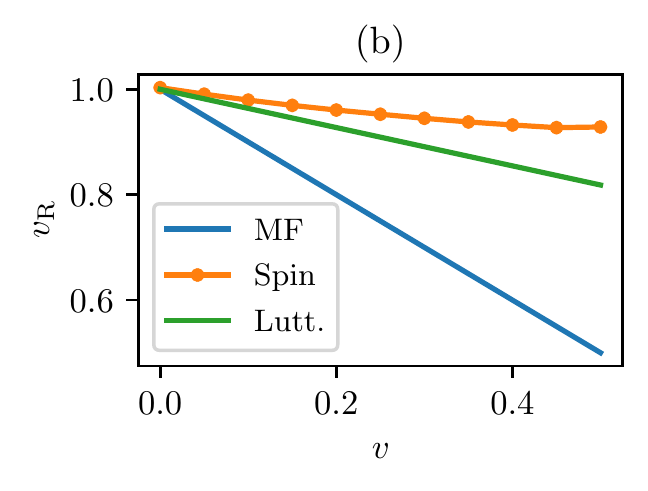}
\caption{The Fermi velocities, $v_\mathrm{L}$ and $v_\mathrm{R}$ respectively ($u=1$) derived from the mean field (MF) of Eq.~(\ref{eq:MF_fermi_v}) and Luttinger liquid description of Eq.~(\ref{eq:C4_renorm_fermi_velocities}) compared to the numerical results of the MPS excitation ansatz from Ref.~\cite{black_hole_paper} for the spin model at bond dimension $D=36$ in the thermodynamic limit.}
\label{fig:C4_Fermi_velocities}
\end{center}
\end{figure}

The dispersion of the full spin model as a function of $v$ for $|v| < |u|$ can be calculated using the MPS excitation ansatz working in the thermodynamic limit \cite{C4_haegeman2013post}, as done in Ref.~\cite{black_hole_paper}. The full dispersion features unequal left- and right-moving Fermi velocities whose magnitudes change oppositely with $v$---the signature of tilting of the cones---and the appearance of additional Fermi points, similar to the mean field model. In Fig.~\ref{fig:C4_Fermi_velocities} the Fermi velocities $v_\mathrm{L,R}$ obtained from the mean field approximation of Eq.~(\ref{eq:MF_fermi_v}), the Luttinger liquid model of Eq.~(\ref{eq:C4_renorm_fermi_velocities}), and the MPS numerics of the spin Hamiltonian are compared. We see that the Luttinger liquid model is much more accurate than the mean field approximation. We expect that the quantitative disagreement and the observed asymmetry in the change to the left and right velocities to be lifted at higher order in perturbation theory. The upshot is that the prediction of unequal left- and right-moving Fermi velocities from mean field theory is accurate. As this is the signature of a \textit{tilting} Dirac cone it suggests that the model may behave like a black hole. We investigate this further in the next section.

\section{Emergent black hole \label{section:hawking}}
\subsection{Dispersion relation and Fermi points}
In order to make the link with relativity, we now label the lattice sites as alternating between sub-lattices $A$ and $B$ by introducing a two-site unit cell, as shown earlier in Fig.~\ref{fig:C4_zig_zag_lattice}. We can rewrite the mean field Hamiltonian of Eq. (\ref{eq:mf_ham}) as 
\begin{equation}
H_\mathrm{MF} =  \sum_n \left[-ua^\dagger_n(b_n + b_{n-1}) - \frac{iv}{2} (a_n^\dagger a_{n+1} + b^\dagger_n b_{n+1}) \right] + \text{H.c.}, \quad u,v \in \mathbb{R} \label{eq:lattice_ham},
\end{equation}
where the fermionic modes $a^\dagger_n$ and $b^\dagger_n$ are creation operators for sublattice $A$ and $B$, respectively, of the unit cell located at site $n$. These modes obey the commutation relations $
\{ a_n,a^\dagger_m \}  = \{ b_n, b_m^\dagger\} = \delta_{nm}$, while all mixed anti-commutators vanish. The index $n$ now labels the unit cells. We Fourier transform the fermions with the definition 
\begin{equation}
a_n = \frac{1}{\sqrt{N_c}} \sum_{p \in \mathrm{B.Z.}} e^{i pa_\mathrm{c}n } a_p,
\end{equation}
and similarly for $b_n$, where $N_c = N/2$ is the number of unit cells in the system, $a_c = 2a$ is the unit cell spacing for a given lattice spacing $a$, and $\mathrm{B.Z.} = [0,2 \pi/a)$ is the Brillouin zone. Applying this to the Hamiltonian, we arrive at
\begin{equation}
H_\mathrm{MF} = \sum_{p \in \mathrm{B.Z.}} \chi^\dagger_p h(p) \chi_p, \quad h(p) = \begin{pmatrix} g(p)  & f(p) \\ f^*(p) & g(p)  \end{pmatrix},
\end{equation}
where we have defined the two-component spinor $\chi_p = (a_p, b_p)^\mathrm{T}$ and the components
\begin{equation}
f(p) = -u(1+e^{-ia_\mathrm{c}p}), \quad g(p) = v \sin(a_\mathrm{c}p).
\end{equation}

\begin{figure}
\begin{center}
\includegraphics[scale=1]{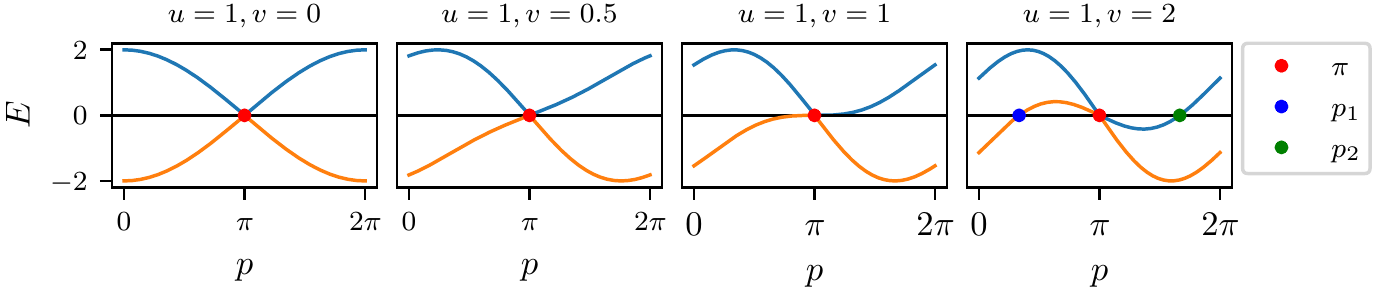}
\end{center}
\caption{The dispersion relation Eq.~(\ref{eq:app_dispersion_2}) for various values of $v$ and a fixed $u = 1$. We see that the parameter $v$ tilts the cones in a similar way to a black hole and $v = u$ is the critical value corresponding to the event horizon. Due to the Nielsen-Ninomiya theorem, additional zero-energy crossings appear when the cones over-tilt.} 
\label{fig:appendix_dispersion_2}
\end{figure}

As usual, the dispersion relation is given by the eigenvalues of the single-particle Hamiltonian $h(p)$ which yields 
\begin{equation}
E(p) = g(p) \pm |f(p)| =  v \sin(a_cp) \pm u \sqrt{2 + 2 \cos(a_\mathrm{c} p) } \label{eq:app_dispersion_2}.
\end{equation}
In Fig. \ref{fig:appendix_dispersion_2}, we see that the parameter $v$ has the effect of tilting the cones.
The Fermi point $p_0$ satisfies $E(p_0) = 0$ which implies, after some simple algebra, that the Fermi point satisfies
\begin{equation}
v^2 \cos^2(pa_\mathrm{c}) + 2u^2 \cos(pa_\mathrm{c}) + (2u^2 - v^2) =0,
\end{equation}
which is quadratic in the variable $ \cos(pa_\mathrm{c})$. Solving this yields three roots in the Brillouin zone
\begin{equation}
p_0 =  \frac{\pi}{a_\mathrm{c}}, \quad p_\pm  = \pm \frac{1}{a_\mathrm{c}} \arccos \left( 1 -  \frac{2u^2}{v^2} \right) .
\end{equation}
The roots $p_\pm$ only exist if the argument of $\arccos$ is in the range $[-1,1]$. As $2u^2/v^2$ is always positive, we require
\begin{equation}
\frac{2u^2}{v^2} \leq 2, \quad \Rightarrow \quad  |u| \leq |v|.
\end{equation}
Therefore, if $|v| \leq |u|$, the only Fermi point is located at $p_0 = \frac{\pi}{a_\mathrm{c}}$ which is where the Dirac cone is located. When the cone over tilts, so when $|u| < |v|$, additional zero-energy crossings at $p_\pm$ appear which is due to the Nielsen-Ninomiya theorem which states that the number of left- and right-movers must be equal~\cite{C4_Nielsen1,C4_Nielsen2}.
\subsection{Continuum limit}
Let us focus on the cone at $p_0$: we take the continuum limit by Taylor expanding the single-particle Hamiltonian $h(p)$ about the Fermi point $p_0$. We have
\begin{align}
f(p_0 + p) &  = -ia_\mathrm{c} u p + O(p^2), \\
g(p_0 + p) & = -a_\mathrm{c} v p + O(p^2),
\end{align}
which gives us 
\begin{equation}
h(p_0 + p)  = u\sigma^y p -  v \mathbb{I} p + O(p^2) \equiv e_a^{\ i} \alpha^a p_i +O(p^2),
\end{equation}
where we have defined the coefficients $e^{\ x}_0 = - v,e^{\ x}_1 = u $ and the Dirac matrices $\alpha^0 = \mathbb{I},\alpha^1 = \sigma^y $. In addition, we have also absorbed a factor of $a_\mathrm{c}$ into the couplings as $a_\mathrm{c}u \rightarrow u$ and $a_\mathrm{c}v \rightarrow v$ to keep the Fermi velocity fixed to ensure the continuum limit $a \rightarrow 0$ is well-defined. In addition, we define the continuum limit coordinate $x = na_c$, which, due to the bipartite labelling of the lattice, labels the unit cells. Therefore, the continuum limit Hamiltonian after an inverse Fourier transform to real space is given by
\begin{equation}
H  =  \int_\mathbb{R} \mathrm{d}x \chi^\dagger(x) \left( -ie_a^{\ i} \alpha^a \overset{\leftrightarrow}{\partial_i} \right) \chi(x), \label{eq:cont_ham}
\end{equation}
where we have defined $A\overset{\leftrightarrow}{\partial_\mu}B =\frac{1}{2} \left( A \partial_\mu B - (\partial_\mu A)B \right)$ which only acts on spinors, the Dirac alpha and beta matrices $\alpha^a = (\mathbb{I},\sigma^y)$ and $\beta = \sigma^z$. 

Comparing this Hamiltonian to the general one of Eq.~(\ref{eq:C2_RC_ham}), we can interpret the continuum limit of the lattice model as a curved space field theory with zweibein
\begin{equation}
e_a^{\ \mu} = \begin{pmatrix} 1 & -v \\ 0 & u \end{pmatrix}, \quad e^a_{\ \mu} = \begin{pmatrix} 1 & v/u \\ 0 & 1/u \end{pmatrix} \label{eq:tetrad}
\end{equation}
and Dirac gamma matrices $\gamma^0  = \sigma^z$ and $\gamma^1 = -i \sigma^x$
which obey the anti-commutation relations $\{ \gamma^a, \gamma^b \} = 2\eta^{ab}$, where $\eta^{ab} = \mathrm{diag}(1,-1)$. The zweibein corresponds to the metric
\begin{equation}
g_{\mu \nu} = e^a_{\ \mu}e^b_{\ \nu} \eta_{ab} =  \begin{pmatrix}   1 - v^2/u^2  &  -v/u^2 \\ -v/u^2 & -1/u^2 \end{pmatrix},
\end{equation}
or equivalently in terms of differentials
\begin{equation}
\mathrm{d}s^2 =  \left( 1 -\frac{v^2}{u^2} \right) \mathrm{d}t^2 -  \frac{2v}{u^2} \mathrm{d}t \mathrm{d}x - \frac{1}{u^2}\mathrm{d}x^2  \label{eq:app_GP_metric}. 
\end{equation}
This is the Schwarzschild metric expressed in Gullstrand-Painlev\'e coordinates \cite{C4_Volovik_helium_droplet} which is sometimes know as the \textit{acoustic metric}. We refer to this metric as an \textit{internal metric} of the model as it depends upon the internal couplings of the Hamiltonian and not the physical geometry of the lattice. In addition, this is a fixed classical background metric and the quantum fields have no back-reaction on the metric. 

In order to bring the metric Eq. (\ref{eq:app_GP_metric}) into standard form, we employ the coordinate transformation $(t,x) \mapsto (\tau,x)$ via
\begin{equation}
\tau(t,x)  = t - \int_{x_0}^x \mathrm{d} z  \frac{v(z)}{u^2 - v^2(z)} ,
\end{equation}
which maps the metric to
\begin{equation}
\mathrm{d}s^2 =  \left( 1 - \frac{v^2}{u^2} \right) \mathrm{d}\tau^2 -  \frac{1}{u^2 \left( 1 - \frac{v^2}{u^2} \right)} \mathrm{d}x^2,
\end{equation}
which is the Schwarzschild metric. If we upgrade $u$ and $v$ to slowly-varying functions of space, then the preceding calculation is still valid and the event horizon is therefore located at the point $x_\mathrm{h}$, where $|v(x_\mathrm{h})| = |u(x_\mathrm{h})|$. In this project, we fix $u(x) = 1$ so it aligns with the standard Schwarzschild metric in natural units. If we refer back to the phase diagram of Fig.~\ref{fig:E&chi_vs_v}, we see that the location of the event horizon coincides with the boundary between two chiral phases. In this way, we see an intimate link between chiral models and event horizons.

Using the Hawking formula for the temperature of a black hole, the temperature is given by \cite{C4_Volovik3}
\begin{equation}
T_\text{H} = \frac{1}{2\pi} | v'(x_\mathrm{h})|. \label{eq:C4_Hawking_temperature}
\end{equation}
which we shall investigate in the next section.
\subsection{Hawking radiation}
\begin{figure}[t]
\begin{center}
\includegraphics[scale=1]{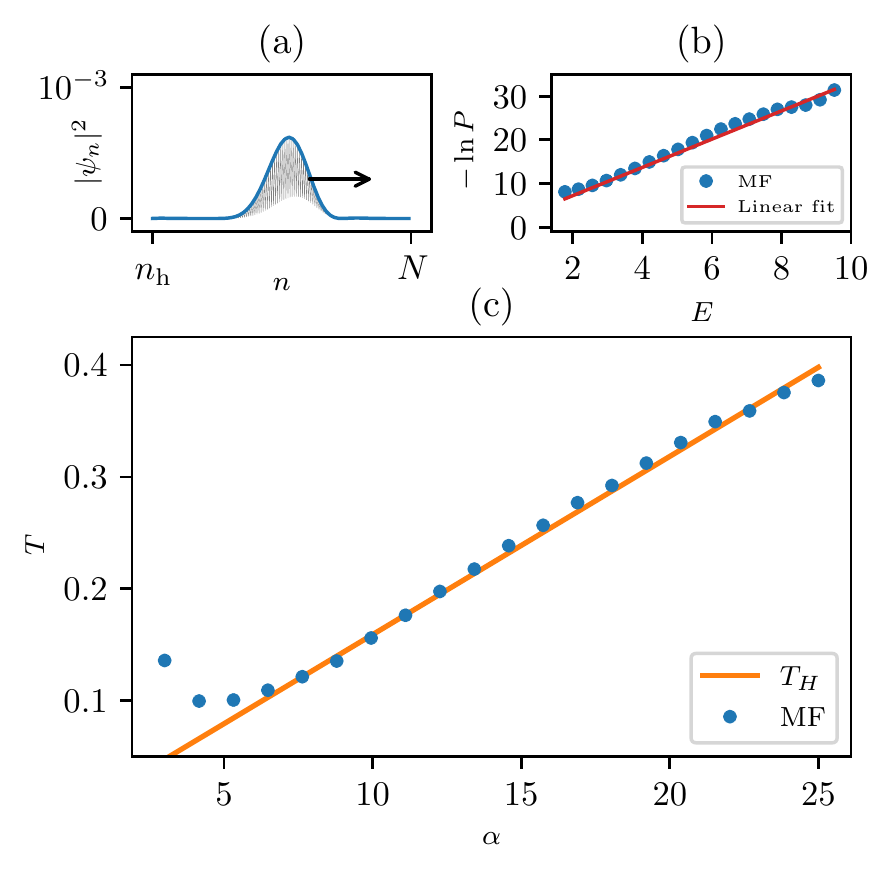}
\end{center}
\caption{(a) The lattice wavefunction $\psi_n$ on the right half of the system ($n \in [n_\text{h},N]$) transmitted through the horizon, for the couplings $u=1$ and $v$ given by Eq. \eqref{en:couplings} with $\alpha = 20$, $\beta = 0.1$, and the horizon at $n_\text{h} = N/2$ with $N=500$. The particle tunnels across at $t \approx 2$ and a small wavepacket escapes into the other half, which we interpret as Hawking radiation. (b) A snapshot of the overlap $-\ln P$ vs. the energy of the state $E$ at time $t = 4.5$. The system thermalises shortly after the particle passes through the interface, displaying a linear dependence on $E$, where the gradient is given by $1/T$. (c) The extracted temperature $T$ of the radiation vs. $\alpha$ extracted after a short time $t=4.5$. $T$ grows linearly with $\alpha$, very close to the predictions of the Hawking formula $T_\text{H} \approx \alpha \beta/2 \pi$.}
\label{fig_3}
\end{figure}


It has been shown in the literature that many analogue gravitational systems in condensed matter will exhibit a Hawking-like effect \cite{C4_Volovik1,C4_Volovik2,C4_Volovik3,C4_Yang,C4_Sabsovich,C4_Maertens,C4_Huang,C4_Hang,C4_Guan,C4_Retzker,C4_Rodriguez,C4_Roldan-Molina,C4_Kosior,C4_Steinhauer_2016,C4_Stone_2013}. Reversing the argument, we would like to see if the Hawking radiation can effectively describe quenched time evolutions across the chiral interface. To be able to simulate large system sizes and long evolution times, we resort to the mean field description of Eq. (\ref{eq:mf_ham}) rather than the full spin model. Consider an open, inhomogeneous system with couplings $u(x) = 1$ and
\begin{equation}
v(x) = \alpha \tanh[\beta(x-x_\mathrm{h})],
\label{en:couplings}
\end{equation}
where $\alpha ,\beta \in \mathbb{R}$ and $x_\text{h}$ is in the centre of the system. Here, we take $x$ as the unit cell coordinate in order to align with our continuum limit conventions (see Supplementary Material). This produces a positive and negative chiral region separated by a small zero-chirality region in the centre of the system. For large enough $\alpha$ and $\beta$, the zero-chirality region has arbitrarily small size, so the system effectively models an interface between two oppositely polarized chiral phases, corresponding to a black hole-white hole interface respectively in the continuum.

Following the method of Ref. \cite{C4_Yang}, we initialise a single-particle state $|n_0 \rangle = c^\dagger_{n_0}|0\rangle$ on the $n_0$th lattice site inside the left half of the system, and let the wavefunction evolve freely across the boundary into the other half with the Hamiltonian $H_\mathrm{MF}$, as shown in Fig. \ref{fig_3}(a). We then measure the overlap of the wavefunction with localised energy modes that exist only on the other side of the boundary as
\begin{equation}
P(k,t) = |\langle k | e^{-iHt} |n_0\rangle|^2,
\end{equation}
where $|k \rangle$ are the single-particle eigenstates of the Hamiltonian $H_\mathrm{out}$, where $H_\mathrm{out}$ is the mean field Hamiltonian of Eq. (\ref{eq:mf_ham}) truncated to the outside region only. This method utilises the result that Hawking radiation can be viewed as quantum tunnelling \cite{C4_Wilczek}. This differs from the standard treatment of Hawking radiation, see Refs.~\cite{Hawking,Page_2005}, whereby the radiation is produced via vacuum fluctuations of a quantum field by the horizon, which additionally causes the black hole to evaporate. In this study, we do not model these features.

We find numerically that the interface between the two chiral phases thermalises the wavefunction: once the wavefunction evolves across the interface, the distribution takes the form $P(k,t) \propto e^{-E(k)/T}$, where $T$ is some effective temperature. Fig.~\ref{fig_3}(b) shows the distribution $P(k,t)$ at time $t = 4.5$ for a system with parameters $N= 500$, $n_h=250$, $\alpha = 20$ and $\beta = 0.1$, where we prepared the particle at $n_0 = 230$. It is clear that $P(k,t)$ follows a Boltzmann distribution at some temperature $T$, where the gradient of the line is given by $1/T$. We observe the system strongly thermalises. The value $\beta = 0.1$ is taken to suppress the effects from having finite lattice spacing and finite system size. In Fig.~\ref{fig_3}(c), we present the dependence of the measured temperature $T$ on the magnitude $\alpha$ for $\beta = 0.1$. We see it closely follows the predictions of the Hawking formula $T_\mathrm{H} = \alpha \beta/2 \pi$, obtained from Eq.~(\ref{eq:C4_Hawking_temperature}) and Eq.~(\ref{en:couplings}), for a wide range of couplings, $\alpha$, thus accurately modelling the physics of the chiral interface. The thermalisation to the Hawking temperature breaks down when $\alpha < 4$ as the couplings will not be sharp enough to provide a sufficient interface, whereas for large $\alpha$ the couplings vary too fast for the continuum approximation to be valid, which is where the black hole physics should emerge.

\begin{figure}[t]
\begin{center}
\includegraphics[scale=1]{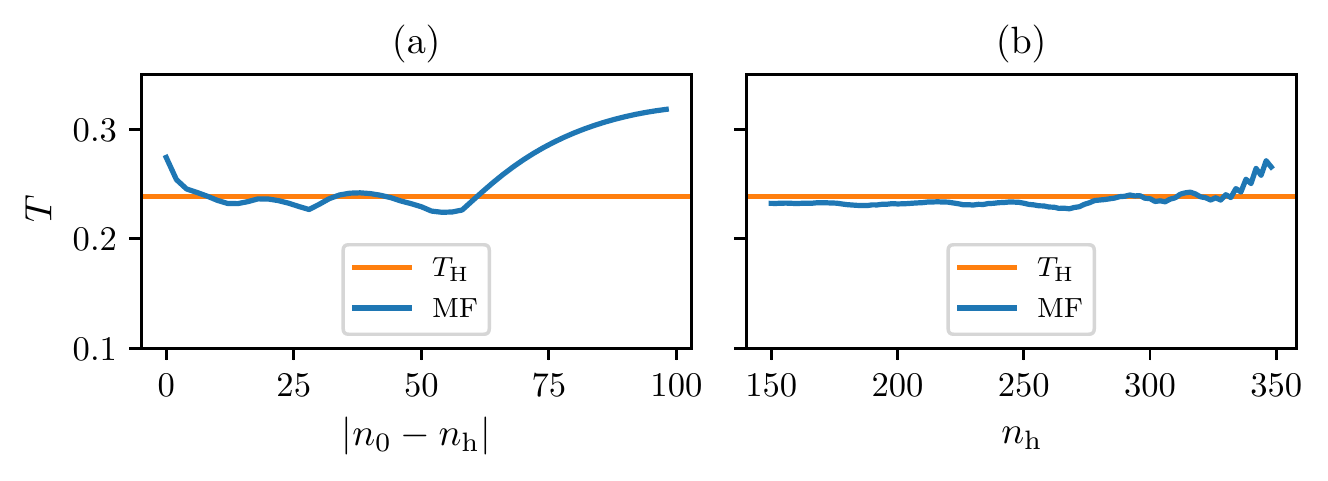}
\end{center}
\caption{(a) The measured temperature $T$ vs. the distance $|n_0 - n_\mathrm{h}|$ from the event horizon that the particle is released for the mean field (MF) system of size $N = 500$, $\alpha = 15$, $\beta = 0.1$ and $n_\mathrm{h} = N/2$.  (b) The measured temperature $T$ vs. the position of the horizon $n_\mathrm{h}$ for the same system, where $n_0 = n_\mathrm{h} - 25$.}
\label{fig_4}
\end{figure}

The Hawking temperature $T_\text{H}= \alpha \beta/2 \pi$ of our system is a very simple analytic formula that describes a complex thermalisation process. In particular, it does not depend on the initial value $n_0$, i.e., the position where quench starts, nor the horizon location $n_\text{h}$ which effectively gives the size ratio between the two chiral phases in our system. To verify these properties, we numerically determine the dependence of $T$ on $n_0$ and $n_\mathrm{h}$ in Fig.~\ref{fig_4}(a) and Fig.~\ref{fig_4}(b) respectively. We see that the measured temperature $T$ is largely insensitive to the initial position of the particle $n_0$. The black hole description only fails if $n_0$ is initially too close or too far away from the interface or when the interface $n_\mathrm{h}$ is too close the edges of the system. In all these cases boundary effects start to contribute as the exterior region which the overlap $P(k,t)$ is measured in becomes too small. These two observations show that the thermalisation across the interface is robust and will aid in any potential experimental realisation of the model.

Sec.~\ref{section:phase_transitions} revealed how well the mean field description agrees with the spin model. For this reason, we expect the thermalisation properties of the full spin model to agree with the mean field approximation. We leave a study of this to further work.


\subsection{Generalised Gibbs ensemble}

It is important to note that, as has been discussed elsewhere, e.g. \cite{C4_Sabsovich}, the Hawking temperature obtained in this study is observed through \textit{scattering} processes rather than in the thermal equilibration of observables. The effective thermalisation observed with $H_\mathrm{MF}$ takes place at very short time-scales after release of the particle. If we allow the system to evolve for a long time, it will not equilibrate to a thermal state at the Hawking temperature which one may have expected (or hoped for), but instead it will equilibrate to a generalised Gibbs ensemble \cite{C4_Perarnau_Llobet,C4_Vidmar}---this is because the mean field Hamiltonian $H_\mathrm{MF}$ is integrable. 

The eigenstate thermalisation hypothesis (ETH)~\cite{Srednicki2,Deutsch,Rigol} states that some quantum systems prepared out of equilibrium evolve to a thermal state after a long time described by the Gibbs ensemble
\begin{equation}
\rho_\text{GE} = \frac{1}{Z_\text{GE}} e^{- \beta (H - \mu N)}, \quad Z_\text{GE} = \mathrm{Tr}\left(e^{- \beta (H - \mu N)}\right),
\end{equation}
where $\beta = 1/T$ is the inverse temperature and $\mu$ is the chemical potential. In other words, expectation values of observables $\mathcal{O}$ are expected to evolve to a thermal equilibrium value given by $\langle \mathcal{O} \rangle_{t \rightarrow \infty} = \mathrm{Tr}(\rho_\text{GE} \mathcal{O})$. The constants $\beta$ and $\mu$ are fixed by the requirements that the energy and particle number are conserved, therefore if the system is initialised in the pure state $|\psi\rangle$, we expect $\langle \psi | H | \psi\rangle = \mathrm{Tr}(\rho_\text{GE} H)$ and $\langle \psi |\hat{N} | \psi \rangle = \mathrm{Tr}(\rho_\text{GE} N)$.

On the other hand, integrable systems are not expected to equilibrate to the predictions of the Gibbs ensemble. Instead, they relax to the predictions of the generalised Gibbs ensemble (GGE), which is defined as
\begin{equation}
\rho_\text{GGE} = \frac{1}{Z_\text{GGE}}e^{- \sum_k \lambda_k Q_k},\quad Z_\text{GGE} = \mathrm{Tr}\left(e^{- \sum_k \lambda_k Q_k}\right),
\end{equation}
where $\{ Q_k \}$ is a set of conserved charges which commute with the Hamiltonian, $[H,Q_k] = 0$ for all $k$, and $\{ \lambda_k \}$ are their corresponding Lagrange multipliers. The Lagrange multipliers are fixed by ensuring that the charges are conserved, i.e. for an initial pure state $|\psi\rangle$, we require $\langle \psi |Q_k | \psi\rangle = \mathrm{Tr}( \rho_\text{GGE} Q_k )$ which constrains the $\{ \lambda_k \}$. Typically, the charges are given by $Q_k = c^\dagger_k c_k$ which are constructed from the modes which diagonalise the Hamiltonian. 

\begin{figure}[t]
\begin{center}
\includegraphics[scale=1]{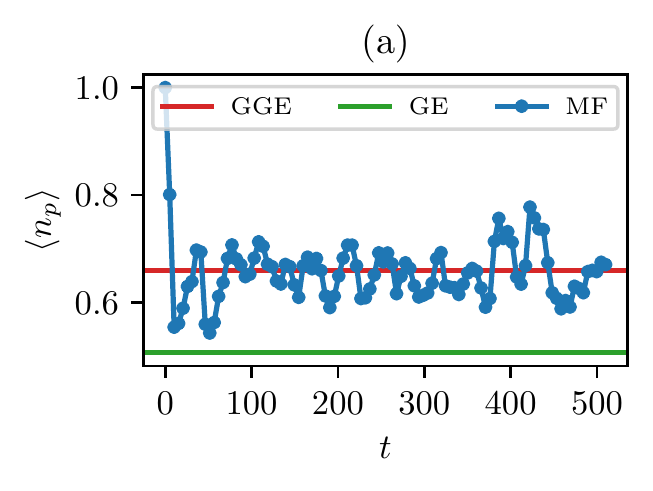}
\includegraphics[scale=1]{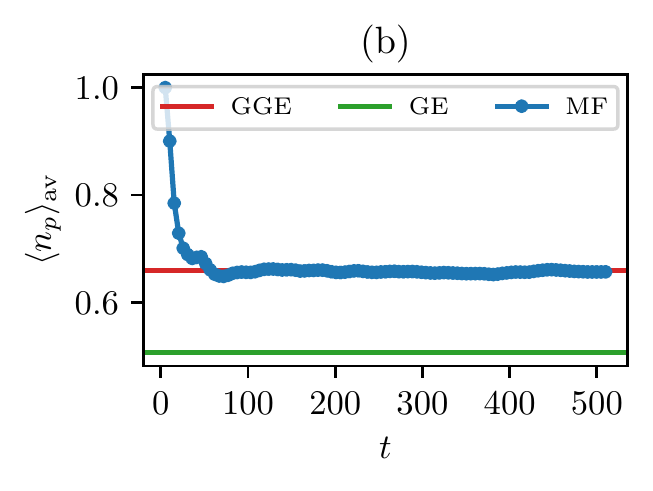}
\end{center}
\caption{(a) The mode occupation of the modes $c_{0,k}$ after quenching from the $v = 0$ ground state (the XY model) for a system of size $N = 200$ for the black hole couplings $v(x)$. We see the system equilibrates to a GGE instead of a GE, therefore we cannot assign a temperature to the system. (b) The time-averaged data $\langle n_p \rangle_\mathrm{av}(t) = \frac{1}{t} \int^t_0 \langle n_p \rangle(t') \mathrm{d}t'$.}
\label{fig:GGE}
\end{figure}

Suppose we looked at the mean field Hamiltonian $H_\mathrm{MF}$. It appears to thermalise the wavefunction of a particle if it passes through the phase boundary between two different chiral regions. It is natural to ask whether this system thermalises in the sense of a Gibbs ensemble. However, the thermalisation observed in the analogue black hole system is an effective thermalisation, related to the scattering process across the horizon and only takes place shortly after release of the particle. If we let the system evolve for a long time, it will equilibrate to a GGE. Consider preparing the system in the ground state $|\Omega_0\rangle$ of the Hamiltonian $H_\text{MF}$ when $v = 0$. In other words, we prepare the system in the ground state of the XY model $H_\mathrm{XY}$ in the absence of any black hole. The initial Hamiltonian is diagonalised to $H_\mathrm{XY} = \sum_k E_0(k)c^\dagger_{0,k}c_{0,k}$. Then we quench by instantaneously switching on the black hole profile $v(x)$ and letting the system evolve with the full Hamiltonian $H_\text{MF}$. We then measure the occupancy of the modes $c_{0,k}$ as time evolves:
\begin{equation}
n_0(k;t) = \langle \Omega_0 | e^{iH_\mathrm{MF}t} c^\dagger_{0,k}c_{0,k} e^{-iH_\mathrm{MF} t} |\Omega_0\rangle.
\end{equation}
In Fig.~\ref{fig:GGE} we find the system equilibrates to the predictions of the GGE and \textit{not} the GE as expected. 

This section has studied the thermalisation from the mean field approximation only. An analysis of the thermalisation properties of the full spin Hamiltonian of Eq.~(\ref{eq:C4_spin_ham}) will be far more involved due to the interacting nature of the model. However, it is important to investigate this as the thermalisation predicted from the mean field theory and the spin model may behave significantly differently. An initial analysis of the energy-level statistics of the full spin Hamiltonian of Eq.~(\ref{eq:C4_spin_ham}) suggests to us that the model may be integrable too as it displays Poisson level statistics~\cite{Kramer2002}, which is the signature of integrability, further demonstrating the accuracy of the mean field description (as this is integrable too), however we leave a systematic study of this to future work.
\section{Conclusion \label{black_hole_conclusion}}
In this chapter, we considered a modified 1D XY model with additional three-spin interactions and demonstrated that that low-energy behaviour can be described by Dirac fermions on a black hole background. We tackled this from both the condensed matter and high-energy perspective.

Using mean field theory, we first analysed the physics of the model by studying its phase diagram, comparing it to the matrix product state (MPS) results of Ref.~\cite{black_hole_paper}. By studying the energy density, chirality and entanglement entropy of the model, we saw that the mean field description accurately predicted a first-order phase transition between a chiral and non-chiral phase, in strong agreement with the MPS techniques. We noted that in the non-chiral phase, the ground state is unaffected by the additional interactions. We made this precise by bosonising the model by using the theory of Luttinger liquids. This Luttinger approach also revealed the tilting Dirac cones in the dispersion relation in agreement with the mean field theory, suggesting a deeper interpretation in terms of a black hole.

By projecting onto the low-energy sector of the mean field model, we obtained a Dirac equation on a black hole background. We saw that the interface of a chiral phase with a non-chiral phase is the location of the event horizon. Using the theory of Hawking radiation, we were able to predict the time evolution of a quenched system with a chiral interface and saw it thermalised to the Hawking temperature $T_\mathrm{H}$ through a scattering process. We demonstrated that this is indeed the case for a wide variety of quenches, positions of the interface and coupling parameters, thus providing a faithful high level description of chiral interfaces. We then demonstrated that the observed thermalisation is strictly not described by a Gibbs ensemble in the traditional sense, but a generalised Gibbs ensemble. 

We envision that this bridge between chiral systems and black holes can facilitate the quantum simulation of Hawking radiation in the laboratory with, for example, cold atom technology \cite{C4_Jaksch,C4_Rodriguez,C4_Kosior}. Moreover, our investigation opens the way for modelling certain strongly correlated systems by effective geometric theories with extreme curvature, thus providing an intuitive tool for their analytical investigation. However, the analysis of the thermalisation of this model used mean field theory only. An initial study demonstrated that the total spin model displays Poisson level statistics, suggesting it may be integrable. As the mean field analysis agreed extremely well with the MPS simulation of the total spin model, we suspect that the spin model may be integrable and the thermalisation may follow a similar behaviour to the mean field model, which only reinforces the black hole analogy. A systematic study of this is left to future work. Additionally, we only simulated semi-classical gravity. A simulation of quantum gravity could be accessible by modifying the model to realise a fluctuating metric in the continuum limit~\cite{Patricio}, which we leave to future work as well.

\chapter{Conclusion \label{chapter:conclusion}}
In this thesis, we investigated emergent spacetime in quantum lattice models. We built upon the known result that the low-energy limit, or continuum limit, of many lattice models such as graphene~\cite{Wallace,C1_Neto}, Kitaev's honeycomb model~\cite{C2_Kitaev} and the XY model~\cite{C4_DePasquale} have an emergent relativistic description. We attempted to generalise this to \textit{curved} spacetimes which gave us a dictionary of geometric observables to explain observed phenomena at the lattice level, whilst also motivating us to hunt for geometric observables which at first glance do not correspond to an obvious observable at the lattice level. This also allows one to simulate relativistic effects in the lab that would otherwise be out of reach of experiment, such as Hawking radiation. This thesis contained four chapters of original research which we shall conclude now and discuss any open problems.


In Chapter~\ref{chapter:nanotube}, we investigated zig-zag carbon nanotubes. These are constructed from a sheet of graphene rolled up in a particular direction to yield a cylindrical geometry with a zig-zag boundary at each end. Like graphene, the carbon nanotube admits a low-energy relativistic description in terms of the Dirac equation for states near the Fermi energy. One of the interesting features of the Dirac equation is that is admits bulk solutions $\psi$ whose support is non-zero on the edges of an open system. This contrasts to the Schr\"odinger equation which demands all wavefunctions vanish on the edges. 

We investigated the low-energy properties of a zig-zag carbon nanotube by studying the eigenstates $\psi$ of the Dirac equation with zig-zag nanotube boundary conditions. We identified the two components of the eigenstates, $\psi_A$ and $\psi_B$, as the wavefunctions of the two sublattices $A$ and $B$ of the honeycomb lattice respectively. We showed that the relativistic effects generate a phase shift of $\theta_{n,p}$ between the sublattice wavefunctions, which is related to the density at the edges of the nanotube by $\rho \equiv \psi^\dagger \psi \propto \sin^2(\theta_{n,p})$. 

For the special case of \textit{gapless} zig-zag nanotubes, which are nanotubes whose circumference contains a multiple of three unit cells, we showed that the phase shift is maximal with $\theta_{n,p} = \pi/2$, which in turn maximises the edge density $\rho$. In addition, the density across the lattice is also completely uniform for all eigenstates taking a value of $1/L$, where $L$ is the height of the tube. This contrasted significantly with \textit{gapped} nanotubes, whose circumference does \textit{not} contain a multiple of three unit cells. In this case,  the phase shift $\theta_{n,p} \approx 0$ so the density across the lattice followed a more Schro\"odinger like profile, going to zero at the edges. This contrast between gapped and gapless nanotubes was striking as the behaviour of the eigenstates changes dramatically if we simply change the circumference by one cell, as seen in Fig.~\ref{fig:wavefunctions}. 

We also investigated the total lattice wavefunctions of the model. In this case, we see that the emergent relativistic behaviour near the Fermi energy is a result of high frequency \textit{aliasing}, as seen in Fig.~\ref{fig:full_wavefunctions}. The underlying wavefunction of the model is always non-relativistic, being described by a Schr\"odinger equation. However states near the Fermi energy, which is where the relativistic description holds, have such high frequency oscillations that the sublattice wavefunctions $\psi_A$ and $\psi_B$ are out of phase, giving the impression of two separate wavefunctions for each sublattice which are nothing but the two components of the Dirac equation. This identification gives us an easily accessible observable to measure in the lab with STM to detect the presence of the emergent relativistic behaviour. In fact, this effect holds for nanotubes of dimensions accessible to experiment, so this work has gained interest from our collaborators in China who are attempting to test this in the lab with real nanotubes.

Of course, many systems in condensed matter have non-trivial edge effects which are of great interest, such as localised zero-mode solutions in the SSH model for example, however these solutions die off exponentially into the bulk of the system. The Dirac equation allows one to find \textit{bulk} solutions which have edge support instead, which could have a significant effect on the conductivity of the material when attaching leads to its boundaries or its response to a magnetic field. An investigation into this is left as an open problem. 


In Chapter~\ref{chapter:kitaev} we investigated the relativistic description of Kitaev's honeycomb model. Just like graphene, the low-energy degrees of freedom of Kitaev's honeycomb model are described by a Dirac equation. The standard continuum limit is taken for the case where all of the couplings $\{ J_i \}$ are equal and constant, which results in a Dirac equation on a \textit{Minkowski} spacetime. In this chapter we asked how one could generate more exotic spacetimes in the continuum limit which contain curvature.

The continuum limit of Kitaev's honeycomb model relied heavily upon the translational invariance of the model. This was because we take the continuum limit by mapping to momentum space and projecting the Hamiltonian onto a small neighbourhood of the Fermi points where the dispersion relation is relativistic. When upgrading the couplings of the model to space-dependent functions we break translational symmetry so this procedure does not work exactly, however we assume that the couplings vary slowly on the scale of the lattice spacing. This means we can, to a good approximation, take the continuum limit derived for the homogeneous case and replace the couplings with functions. 

When doing this, we find that the most general model is described by a Dirac equation on a \textit{Riemann-Cartan} which is a generalisation of Riemannian geometry used in general relativity as it contains both curvature \textit{and} torsion. We identify the metric $g_{\mu \nu}$ which depends on the couplings of the underlying lattice model. This metric is an \text{emergent} metric that does not describe the physical geometry of the lattice, but instead describes an effective internal metric of the model. With this, we were able to identify the geometric observables of the model, such as the spin connection, curvature and torsion. Immediately, we were able to identify the torsion of the model as the gap $\Delta$ arising from the $K$ term of the Hamiltonian, as the effect of torsion in $(2+1)$D reduces to a scalar. Interestingly, we found that the singular points of the emergent metric correspond with the critical points of the phase diagram of the model.

In order to observe the metric from the lattice model, we first studied the simply example for which $J_x = J_y = 1$ and $J_z \in [0,2]$. Analytically, this resulted in a continuum metric $g_{\mu \nu}$ describing a \textit{dilation} of spacetime. That is, the internal metric of the model described a space whose $x$- and $y$-directions stretch unequally, where $J_z$ controls the stretching. We observed this stretching by studying Majorana correlators $i \langle c_i c_j \rangle$ and zero-mode wavefunctions $\psi_i$. We saw using the results from Ref.~\cite{C2_Farjami} that for both cases, these lattice observables warped and stretched in the same way predicted by the emergent metric, where the agreement improved as we reduced the gap of the model as the continuum limit is more accurate here.

We also studied the Kekul\'e distortion. The $K$ term of Kitaev's honeycomb model generates a gap, however we interpreted this as torsion instead of mass because the gap couples \textit{chirally} to the Fermi points. Instead, a mass should couple the same way to each Fermi point. We showed that a Kekul\'e distortion can be used in Kitaev's honeycomb model to generate a mass $m$. With both the $K$ term and the Kekul\'e distortion, we have a competition between the parameters $K$ and $m$, which we see introduces a first-order phase transition to the model, transitioning between the topological phases of class D and BDI.

Finally, we investigated what lattice observable one would have to measure to observe \textit{curvature}. We showed that at the continuum level, the \textit{spin density} is dependent on the Ricci scalar of the Levi-Civita connection of the model. This provides us with an observable that we were able to translate back to the lattice level, whereby we identified the spin density as the density of the $K$ term. 


In Chapter~\ref{chapter:chiral} we investigated emergent chiral gauge fields in Kitaev's honeycomb model. As the model has two Fermi points, the low-energy degrees of freedom are described by two \textit{chiralities}. In high-energy physics, this chirality corresponds to how spinors transform under representations of the Lorentz group. For Kitaev's honeycomb model, the chirality is a pseudo-chirality corresponding to which Fermi point the low-energy excitations are close to. However, the machinery of chirality from high-energy physics can still be employed.

At the lattice level, Kitaev's honeycomb lattice model is coupled to a $\mathbb{Z}_2$ gauge field. If we introduce a constant gauge field globally, we find that this has the effect of shifting the Fermi points in momentum space. This shift appears in the continuum limit Hamiltonian the same way a gauge field would via minimal coupling. For this reason, we interpret the shift of the Fermi point as a gauge field.

As the continuum limit of Kitaev's honeycomb model is described by Majorana spinors, they cannot couple to a U(1) gauge field in the continuum. On the other hand, they \textit{can} couple to a chiral U(1)\textsubscript{A} gauge field, which is where the Fermi points of the model shift oppositely. We demonstrated that if we insert a $\mathbb{Z}_2$ gauge field locally on the lattice, this corresponds to a chiral gauge field in the continuum limit. In addition, we introduce defects to the model, which shifts the Fermi points in exactly the same manner, yielding the same continuum limit. This suggested that $\mathbb{Z}_2$ gauge fields and lattice defects can behave in similar ways. We demonstrated this by showing an adiabatic equivalence between the $\mathbb{Z}_2$ gauge field and a special type of defect. This method allows one to generate gauge fields that do not necessarily have a traditional interpretation in terms of Wilson lines on the lattice.

Finally, we discussed how one can extract lattice information about the $\mathbb{Z}_2$ gauge field from the continuum gauge field. If a $\mathbb{Z}_2$ gauge field is inserted locally along a path, at the end of the path we find $\pi$-fluxes of the gauge field which we find in the continuum limit as well. 


In Chapter~\ref{chapter:black_hole} we investigate how to modify the XY model to yield an effective black hole in the continuum limit. The XY model is a simple one-dimensional chain of spin-$1/2$ particles that interact via nearest-neighbour interactions and has a relativistic continuum limit description. By introducing additional three-spin interactions to the model, we can introduce \textit{spin chirality} to the model and an additional chiral phase. We studied this model from both the condensed matter perspective and the high-energy physics limit.

First, due to the additional three-spin interactions, the model is intrinsically interacting in the fermionic picture after performing a Jordan-Wigner transformation. For this reason, the model cannot be solved exactly as it is not quadratic. However, we employ mean field theory to map the interacting Hamiltonian to a non-interacting quadratic Hamiltonian that can be tackled analytically. We verify the validity of the mean field description by comparing it to matrix product state techniques of Ref.~\cite{black_hole_paper}. We study the energy density, chirality and entanglement entropy of the model and see that there is a good agreement between the mean field and spin model. We see that there exists a second-order phase transition in the model between a chiral and non-chiral phase. For this reason, the spin chiraltiy of the model behaves as an order parameter. We also note that the non-chiral phase has a central charge of $c= 1$ whilst the chiral phase has a central charge of $c=2$, which can be seen as the number of Fermi points, and hence the number of effective Dirac fermions in the continuum limit, doubles across this phase transition.

One interesting feature of the model is that in the non-chiral phase, the ground state properties of the model behave the same as the XY model demonstrating that the interactions are irrelevant in this phase. We investigate this further by bosonising the model and applying the theory of Luttinger liquids. To first order, we see that the interactions disappear when we bosonise, explaining the observed irrelevance of the interactions. The bosonisation also renormalises the Fermi velocities compared to what was obtained from the mean field description and demonstrated the presence of a tilting Dirac cone in the dispersion, hinting at a deeper black hole interpretation of the model.

Finally, we study the high-energy continuum limit. The condensed matter analysis suggested the model has two phases, unequal left- and right-moving Fermi velocities and showed that the mean field description was a good approximation. The unequal left- and right-moving Fermi velocities is the signature of a \textit{tilting} Dirac cone in the continuum limit. This is made precise by deriving the continuum limit of the mean field model which yields a Dirac Hamiltonian on a curved spacetime with a Gullstrand-Painlev\'e metric, which is the Schwarzschild metric in a different coordinate system. In this coordinate system, the Dirac cones tilt as we change the parameters of the model. For this reason, we see the low-energy limit of the model behaves as a black hole.

We investigate this analogy by studying the thermal properties of the model. Hawking showed that black holes radiate particles at the Hawking temperature. As our model is static, we do not expect to see any spontaneous radiation in eigenstates as they are in equilibrium, yet radiation is a non-equilibrium process. For this reason, we induce Hawking radiation by quenching the system. We place a single particle inside the black hole and allow it to quantum tunnel across the horizon. We observe that when the particle crosses the horizon, the transmitted portion of the wavefunction is thermalised to the Hawking temperature. This allows us to simulate Hawking radiation in the laboratory. However, it is important to stress that the observed thermalisation is a scattering event \textit{interpreted} as a thermalisation. In the long time limit, the model will not equilibrate to a temperature but instead will equilibrate to a generalised Gibbs ensemble.

We envision that this bridge between chiral systems and black holes can facilitate the quantum simulation of Hawking radiation, e.g. with cold atom technology \cite{C4_Jaksch,C4_Rodriguez,C4_Kosior}. Moreover, our investigation opens the way for modelling certain strongly correlated systems by effective geometric theories with extreme curvature, thus providing an intuitive tool for their analytical investigation. However, this model describes semi-classical gravity. An open problem remains of how one could modify this model to that simulates a fluctuating metric in the continuum limit~\cite{Patricio}.

To conclude, this thesis studied and generalised the known relativistic continuum limit of many quantum lattice models. This allowed us to solve models analytically, explain observed phenomena using geometric quantities and provided us with an arena to simulate curved spacetimes in the laboratory. Open problems for future work include studying dynamic spacetimes such as expanding universes~\cite{C2_Carroll}, dynamically evaporating black holes~\cite{Wald} and observing the Page curve of entanglement entropy~\cite{Page1,Page2}, to more exotic ideas such as fluctuating metrics for a simulation of quantum gravity~\cite{Patricio} which should provide a means to understand these ideas better theoretically too.

\begin{appendix}
\chapter{Numerical Techniques \label{appendix:numerical_techniques}}
The lattice models of this thesis can be solved exactly only if we assume they exhibit \textit{translational invariance}. If we introduce inhomogeneity to these models, translational invariance is broken and we cannot solve the model exactly using analytic methods. For this reason, we must resort to numerical techniques to find exact solutions. In this appendix, we discuss the numerical techniques required to diagonalise the fermionic and Majorana Hamiltonians of this thesis. 
\section{Fermionic Hamiltonians}
\subsection{Exact diagonalisation}
Consider a set of fermionic creation and annihilation operators, $c^\dagger_n$ and $c_n$, where $n$ is an index which labels the possible single-particle quantum states, such as position, spin, and particle species. These fermionic modes obey the anti-commutation relations
\begin{equation}
\{ c_m, c^\dagger_n \} = \delta_{mn}, \quad \{ c_m,c_n\} = \{ c^\dagger_m , c^\dagger_n \} = 0.
\end{equation}
In the language of second quantisation, we focus on non-interacting fermionic Hamiltonians of the form
\begin{equation}
H = \sum_{m,n} h_{mn} c^\dagger_m c_n, \label{eq:appendix_fermionic_H}
\end{equation}
where $h_{mn}$ is an Hermitian matrix called the single-particle Hamiltonian. This defines a subset of Hamiltonians that many models of interest fall into, such as graphene in Chap.~\ref{chapter:nanotube} and the spin chains after a Jordan-Wigner transformation in Chap.~\ref{chapter:black_hole}. As the matrix $h$ is hermitian, there exists a unitary matrix $U$ whose columns are the eigenvectors of $h$ such that
\begin{equation}
h = UDU^\dagger, \label{eq:appendix_diagonal_h}
\end{equation}
where $D$ is a diagonal matrix of corresponding eigenvalues: $D_{pq} = \delta_{pq} E(p) $. Substituting Eq.~(\ref{eq:appendix_diagonal_h}) into the Hamiltonian of Eq.~(\ref{eq:appendix_fermionic_H}), we arrive at
\begin{equation}
\begin{aligned}
H & = \sum_{m,n} \sum_{p,q} (UDU^\dagger)_{mn} c^\dagger_m c_n \\
& = \sum_{m,n} \sum_{p,q} U_{mp} D_{pq} U^\dagger_{qn} c^\dagger_m c_n \\
&= \sum_{p,q} \delta_{pq} E(p) \left( \sum_m U_{mp} c^\dagger_m \right)\left( \sum_n U_{nq}^* c_n \right)\\
& \equiv \sum_p E(p) c^\dagger_p c_p,
\end{aligned}
\end{equation}
where we defined the new fermionic modes via a canonical transformation 
\begin{equation}
c^\dagger_p = \sum_n U_{np} c^\dagger_n , \quad c_p = \sum_n U^*_{np} c_n ,
\end{equation}
which obey the fermionic anti-commutation relations 
\begin{equation}
\{ c_p, c^\dagger_q \} = \delta_{pq}, \quad \{ c_p,c_q\} = \{ c^\dagger_p , c^\dagger_q \} = 0.
\end{equation}
We say that the Hamiltonian $H$ in this basis is now \textit{diagonalised} as it takes the form of a sum of simple harmonic oscillators and we can simply read off the eigenstates as the states generated by repeated action of the modes $c_k^\dagger$ on the vacuum $|0\rangle$. In particular, the ground state $|\Omega \rangle$, defined as the state with the lowest energy eigenvalue, is the state where all negative-energy modes are occupied 
\begin{equation}
|\Omega \rangle = \prod_{p:E(p) < 0 } c^\dagger_p |0\rangle,
\end{equation}
where the notation ``$p:E(p) < 0 $'' denotes a product over all values of $p$ such that the energy $E(p)$ is less than zero.
\subsection{Correlation matrix}
We define the correlation matrix as
\begin{equation}
C_{mn} = \langle \Omega | c^\dagger_n c_m| \Omega \rangle.
\end{equation}
From the definition of the diagonal modes, we have
\begin{equation}
\begin{aligned}
C_{mn} & = \sum_{p,q} U^*_{mp} U_{nq} \langle \Omega | c^\dagger_p c_q |\Omega \rangle  \\
& = \sum_{p,q} U^*_{mp} U_{nq} \theta(-E(p)) \delta_{pq} \\
& = \sum_{p,q:E < 0} U^*_{mp} U_{np},
\end{aligned}
\end{equation}
where in the second equality we used the fact that, as the ground state $|\Omega \rangle$ has all negative-energy states occupied, then $\langle \Omega | c^\dagger_p c_q |\Omega \rangle = \theta(-E(p)) \delta_{pq}$. This is because if $p$ or $q$ correspond to a positive energy state, then $c_q$ annihilates the ground state as all positive-energy modes are unoccupied; whilst if $p$ or $q$ correspond to negative-energy modes then $c_p |\Omega \rangle$ will be orthogonal to $c_q |\Omega\rangle$ as these states have different holes, unless $p = q$. 

Armed with this formula, we can calculate the correlation matrix $C_{mn}$ numerically which gives us access to all possible observables we could measure.
\subsection{Entanglement Entropy}
This section closely follows Refs.~\cite{Latorre_2009,PhysRevLett.90.227902,Peschel_2009}. We now wish to calculate the ground state entanglement entropy of a non-interacting fermionic system with a Hamiltonian of the form Eq.~(\ref{eq:appendix_fermionic_H}).  In this, we consider a simple fermionic system with modes $c_n$, where $n$ labels the lattice sites, which is found in the pure state $\rho = |\Omega \rangle \langle \Omega |$, where $|\Omega\rangle$ is the ground state. We consider partitioning the system into subsystem $\mathcal{A}$ and subsystem $\mathcal{B}$. The reduced density matrix of $\mathcal{A}$ is defined as $\rho_\mathcal{A} = \mathrm{Tr}_\mathcal{B} \rho$ which contains all of the information about subsystem $\mathcal{A}$, where $\mathrm{Tr}_\mathcal{B}$ denotes the partial trace over subsystem $\mathcal{B}$ only. 

The correlation matrix for the total system is given by $C_{nm} = \mathrm{Tr}(\rho c^\dagger_n c_m ) = \langle \Omega |c^\dagger_m c_n |\Omega \rangle $. On the other hand, the correlation matrix of subsystem $\mathcal{A}$ is given by $C^\mathcal{A}_{nm} = \mathrm{Tr}_\mathcal{A}(\rho_\mathcal{A} c^\dagger_n c_m)$. In order for this to be consistent, we require $C^\mathcal{A}_{mn} = C_{mn}$ for $m,n \in \mathcal{A}$, which reads
\begin{equation}
C^\mathcal{A}_{mn} \equiv \mathrm{Tr}_\mathcal{A} (\rho_\mathcal{A} c_m^\dagger c_n ) = \langle \Omega | c^\dagger_m c_n |\Omega  \rangle , \quad \forall m,n \in \mathcal{A}.
\end{equation}
As $C^\mathcal{A}_{nm}$ is an Hermitian matrix, it can be diagonalised with a suitable unitary as $D^\mathcal{A} = U^\dagger C^\mathcal{A} U$, where $D^\mathcal{A}$ is diagonal, which in index notation gives
\begin{equation}
\begin{aligned}
D^\mathcal{A}_{pq}&   = \sum_{m,n \in \mathcal{A}} U^*_{mp} U_{nq} C^A_{mn} \\
&  = \sum_{m,n \in \mathcal{A}} U^*_{mp} U_{nq} \mathrm{Tr}_\mathcal{A}(\rho_\mathcal{A} c_m^\dagger c_n ) \\
&  = \mathrm{Tr}_A ( \rho_\mathcal{A} f^\dagger_p f_q ) \\
& = \lambda_p \delta_{pq},
\end{aligned}
\end{equation}
where we have defined the fermionic modes
\begin{equation}
f_p = \sum_{i \in \mathcal{A}} U_{ip} c_i,
\end{equation}
and to arrive at the last line we used the fact that $D_{pq}^\mathcal{A}$ must be diagonal so can be written in terms of the eigenvalues as $D^\mathcal{A}_{pq} = \lambda_p \delta_{pq}$. The Fock space can be decomposed with respect to these modes as 
\begin{equation}
\mathcal{H}_\mathcal{A} = \bigotimes_p \mathcal{H}_p,
\end{equation}
where $\mathcal{H}_p = \mathrm{span}\{|0\rangle_p , |1\rangle_p = f^\dagger_p |0\rangle_p \}$ is a two-dimensional Hilbert space acted upon by the fermionic modes $f^\dagger_p$ and $f_p$ only, where $|0\rangle_p$ is the vacuum of the $p$th mode and annihilated by $f_p$. With this identification, the fact there exists modes such that $\mathrm{Tr}_\mathcal{A}(\rho_\mathcal{A} f^\dagger_p f_q ) = \lambda_p \delta_{pq}$ implies the reduced density matrix must decompose with respect to these modes too as  
\begin{equation}
\rho_\mathcal{A} = \bigotimes_p \rho_p,
\end{equation}
where $\rho_p$ is a two-dimensional density matrix acting on the Hilbert space $\mathcal{H}_p$ only. This identification relies upon the result that the reduced density matrix must be Gaussian. Using this identification, the entanglement entropy is given by
\begin{equation}
S_\mathcal{A} = - \mathrm{Tr}(\rho_\mathcal{A} \ln \rho_\mathcal{A})  = - \sum_p  \mathrm{Tr}(\rho_p \ln \rho_p ) \equiv \sum_p S_p,
\end{equation}
where $S_p$ is the entropy of the $p$th mode. For the $p$th mode, we can explicitly take the matrix representation of our operators on $\mathcal{H}_p$ to be
\begin{equation}
\rho_p =
\begin{pmatrix} 
\alpha_p & \beta_p \\ \beta_p^* & 1 - \alpha_p
\end{pmatrix}, \quad 
f_p = 
\begin{pmatrix} 
0 & 0 \\ 
1 & 0 
\end{pmatrix}
, \quad 
f^\dagger_p = 
\begin{pmatrix}
0 & 1 \\
0 & 0 
\end{pmatrix},
\end{equation}
where the expression for $\rho_p$ gives us the most general form of $\rho_p$ satisfying $\mathrm{Tr}(\rho_p) = 1$ and $\rho_p = \rho_p^\dagger$; whilst the representation for $f_p$ and $f_p^\dagger$ indeed obeys the commutation relations $\{ f_p, f_p^\dagger \} = 1$ on $\mathcal{H}_p$.

Using this representation, we can use this to calculate $\alpha_p$ and $\beta_p$. First, it is easy to see that 
\begin{equation}
\mathrm{Tr}_A ( \rho_A f_p)   = \mathrm{Tr}(\rho f_p ) = \langle \Omega |f_p | \Omega \rangle =0,
\end{equation}
where we have used the fact that $\langle \Omega | f_p |\Omega \rangle = 0$ as $f_p |\Omega \rangle$ has a different number of particles to $|\Omega \rangle$. From the explicit matrix representation of these operators, we also have
\begin{equation}
\begin{aligned}
\mathrm{Tr}_\mathcal{A}(\rho_\mathcal{A} f_p) & = \mathrm{Tr} \left[ \begin{pmatrix} 
\alpha_p & \beta_p \\ \beta_p^* & 1 - \alpha_p
\end{pmatrix} 
\begin{pmatrix} 
0 & 0 \\ 
1 & 0 
\end{pmatrix} \right] \\
&  = \mathrm{Tr} \begin{pmatrix} \beta_p & 0 \\ 1- \alpha_p & 0 \end{pmatrix}\\
&  = \beta_p ,
\end{aligned} 
\end{equation}
so comparing with the previous result implies that $\beta_p = 0$. We also have
\begin{equation}
\mathrm{Tr}_\mathcal{A} (\rho_\mathcal{A} f_p^\dagger f_p) = \mathrm{Tr} \left[ \begin{pmatrix} 0 & 1 \\ 0 & 0 \end{pmatrix} \begin{pmatrix} 0 & 0 \\ 1 & 0 \end{pmatrix} \begin{pmatrix} \alpha_p & 0 \\ 0 & 1 - \alpha_p \end{pmatrix} \right] = \alpha_p,
\end{equation}
which, combing with the result $\mathrm{Tr}_\mathcal{A}(\rho_\mathcal{A} f^\dagger_p f_q ) = \lambda_p \delta_{pq}$, implies $\alpha_p = \lambda_p$. Pulling everything together, the entropy of the $p$th mode is given by
\begin{equation}
S_p = -\mathrm{Tr}(\rho_p \ln \rho_p) = - \lambda_p \ln \lambda_p - (1-\lambda_p) \ln (1- \lambda_p),
\end{equation}
and the total entanglement entropy $S_\mathcal{A}$ is therefore given by
\begin{equation}
S_\mathcal{A} = \sum_p S^\mathcal{A}_p = - \sum_{p} \left[ \lambda_p \ln \lambda_p + (1-\lambda_p) \ln (1 - \lambda_p) \right],
\end{equation}
where $\{ \lambda_p \} $ are the eigenvalues of $C^\mathcal{A}_{nm}$.

\section{Majorana Hamiltonians}
\subsection{Exact diagonalisation}
Consider a set of Majorana operators $c_n$, where $c^\dagger_n = c_n$, $c_n^2 = \mathbb{I}$ and $n$ is an index which labels the possible single-particle quantum states, such as position, spin, and particle species. These modes obey the anti-commutation relations
\begin{equation}
\{ c_m, c_n \} = 2\delta_{mn}.
\end{equation}
In the language of second quantisation, we focus on Majorana Hamiltonians of the form
\begin{equation}
H = \frac{1}{4} \sum_{m,n} h_{nm} c_m c_n,
\end{equation}
where $h_{nm}$ are the components of an hermitian matrix $h$ and the factor of $1/4$ is for convenience later. This is the general category of Majorana Hamiltonians that many models fall into, such as Kitaev's honeycomb introduced in Sec.~\ref{sec:C2_Kitaev's honeycomb model} and the Kitaev chain~\cite{Kitaev_Chain}. 

In contrast to the fermionic case in Eq.~(\ref{eq:appendix_fermionic_H}), the anti-commutation property of Majorana modes constrains $h$ to be an anti-symmetric matrix as well, where $h^\mathrm{T} = -h$. These two properties imply that the eigenvectors of $h$ come in complex conjugate pairs as $u_p$ and $u^*_p$ with eigenvalues of $\pm E(p)$ respectively, which is because $h u_p = E(p) u_p$ implies $h u_p^* = -E(p) u_p^*$. For this reason, we choose to label our eigenstates with the index $p$ such that the complex conjugate pairs are labelled by $u_{-p} = u_{p}^*$ and $E(-p) = -E(p)$ and $E(p) > 0$. 

Again, we can diagonalise the matrix $h$ with a unitary as $h = UDU^\dagger$, where $D$ is a diagonal matrix of eigenvalues of $h$, bringing the Hamiltonian into the form
\begin{equation}
H = \frac{1}{2} \sum_p E(p) f_p^\dagger f_p , \label{eq:appendix_majorana_H_diagonal_1}
\end{equation}
where the operators $f_p$ are given by
\begin{equation}
f_p^\dagger = \frac{1}{\sqrt{2}} \sum_n U_{np} c_n, \quad f_p = \frac{1}{\sqrt{2}} \sum_n U^*_{np} c_n, \label{eq:appendix_majorana_mode_expansion}
\end{equation}
where the $1/\sqrt{2}$ is to ensure fermionic anti-commutation relations later. The Hamiltonian in Eq.~(\ref{eq:appendix_majorana_H_diagonal_1}) is not diagonal yet as the modes $f_p$ do not obey the fermionic algebra for \textit{all} $p$. However, due to the complex conjugate pairing of the eigenvectors, we have $U_{np}^* = U_{n,-p}$ as the columns of $U$ are the eigenvectors of $h$, therefore the operators obey $f^\dagger_p = f_{-p}$. With this identification, the operators obey the fermionic algebra
\begin{equation}
\{ f_p, f^\dagger_q \} = \delta_{pq}, \quad \{ f_p,f_q\} = \{ f^\dagger_p , f^\dagger_q \} = 0, \quad p,q > 0,
\end{equation}
so we must restrict to $p > 0$. Rewriting the Hamiltonian of Eq.~(\ref{eq:appendix_majorana_H_diagonal_1}) with sums over $p > 0$ only brings it into the form
\begin{equation}
\begin{aligned}
H &  = \frac{1}{2} \sum_p  E(p) f^\dagger_p f_p  \\
& = \frac{1}{2} \sum_{p > 0}\left(  E(p) f^\dagger_p f_p + E(-p) f^\dagger_{-p} f_{-p} \right) \\
& = \frac{1}{2} \sum_{p >0} \left( E(p) f^\dagger_p f_p - E(p) ( 1 - f^\dagger_p f_p ) \right)  \\
& = \sum_{p > 0} E(p) f^\dagger_p f_p  - \frac{1}{2} \sum_{p > 0} E(p),
\end{aligned}
\end{equation}
therefore the model is diagonalised, where the second term is a constant which we can safely ignore.

The ground state of this Hamiltonian depends on the spectrum $E_p$. In the fermionic case, the ground state consisted of the state for which all negative energy single-particle states were occupied, however here we do not have any. If the spectrum contains no zero energy modes, then the ground state is the vacuum state $|0\rangle$, where $f_p |0\rangle = 0$ for all $p$. However, due to the possibility of zero energy modes in the spectrum, the ground state of the model is not unique in general. For this reason, we resort to describing the ground state using a \textit{thermal} state
\begin{equation}
\rho = \frac{1}{Z}e^{-\beta H}, \quad Z = \mathrm{Tr}\left( e^{- \beta H} \right),
\end{equation}
where $\beta = 1/T$ is the inverse temperature, under the assumption that $T \rightarrow 0$. This singles out a ground state on physical grounds.

\subsection{Correlation matrix}
We define the correlation matrix as
\begin{equation}
C_{mn} \equiv i \langle c_m c_n \rangle = i \mathrm{Tr}\left(\rho c_m c_n \right), 
\end{equation}
We can invert the fermionic mode expansion of Eq.~(\ref{eq:appendix_majorana_mode_expansion}) to give us the mode expansion of the Majorana modes as
\begin{equation}
c_n = \sqrt{2} \sum_{p > 0} \left(  U_{np} f_p + U_{np}^* f_p^\dagger \right).
\end{equation} 
Expanding this out we have
\begin{equation}
\begin{aligned}
ic_n c_m & = 2i \sum_{p,q > 0} ( U_{np} f_p + U^*_{np} f^\dagger_p ) ( U_{mq} f_q + U^*_{mq} f^\dagger_q ) \\
& = 2i \sum_{p,q > 0} \left( U_{np} U_{mq} f_p f_q + U_{np} U^*_{mq} f_p f^\dagger_q + U^*_{np} U_{mq} f^\dagger_p f_q + U^*_{np} U^*_{mq} f^\dagger_p f^\dagger_q \right).
\end{aligned}
\end{equation}
Therefore, the correlation matrix is given by
\begin{equation}
C_{nm} = 2i \sum_{p,q > 0 } \left( U_{np} U^*_{mq} \langle  f_p f^\dagger_q\rangle + U^*_{np} U_{mq} \langle  f^\dagger_p f_q  \rangle \right),
\end{equation}
where we note that $\langle f^\dagger_p f^\dagger_q \rangle = \langle f_p f_q \rangle = 0$. Using the thermal state, we have
\begin{align}
\langle f^\dagger_p f_q \rangle & = \frac{1}{Z} \mathrm{Tr}\left( e^{-\beta H} f^\dagger_p f_q \right) = \frac{1}{e^{\beta E(p)} + 1} \delta_{pq} \\
\langle f_p f^\dagger_q \rangle & = \frac{1}{Z} \mathrm{Tr}\left( e^{-\beta H} f_p f^\dagger_q \right) = \frac{e^{\beta E(p)}}{e^{\beta E_p} + 1} \delta_{pq}.
\end{align}
Therefore, we have
\begin{equation}
\begin{aligned}
C_{nm} & = 2i \sum_{p > 0}\left[ \frac{1}{e^{\beta E(p)} + 1} \left( U_{np} U_{mp}^* e^{\beta E(p)} + U^*_{np}U_{mp} \right) \right] \\
& = 2i \sum_{p > 0} \left(  \frac{e^{\beta E(p)} - 1}{e^{\beta E(p)} + 1} \right)  U_{np} U_{mp}^*  \\
& = 2i \sum_{p > 0} \tanh\left( \frac{\beta E(p)}{2} \right)U_{np} U_{mp}^*,
\end{aligned}
\end{equation}
where we used the fact that $U_{np}^* U_{mp} = -U_{np}U^*_{mp}$. For the ground state, we take the limit that $T \rightarrow 0$ which gives us the final result
\begin{equation}
C_{nm} = 2i \sum_{p:E(p)>0} U_{np} U_{mp}^*.
\end{equation}
\section{Tight-binding model of zig-zag nanotubes}

\begin{figure}
\begin{center}
\includegraphics[valign=t,scale=1]{nanotube_boundaries.pdf}
\includegraphics[valign=t,scale=1]{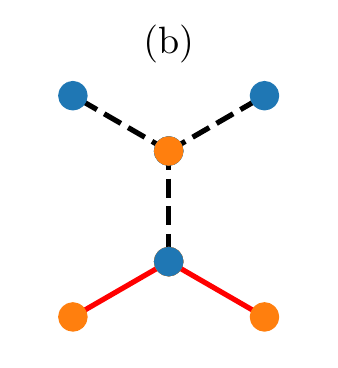}
\includegraphics[valign=t,scale=1]{boundary_unit_cell_bottom.pdf}
\end{center}
\caption{(a) The representation of the zig-zag nanotube. We have a finite length in the $x^1$ direction and roll up the lattice in the $x^2$ direction to form a cylinder with a zig-zag boundaries, represented by the red links (b)-(c) Due to the choice of unit cell as the $z$-links, the unit cells overlap with the outside of the system, which sets the zig-zag boundary conditions $\psi_A = 0$ at the top and $\psi_B = 0$ at the bottom, where the dashed lines represent the links emanating from the unit cells that are removed. These dashed links must are removed in the tight-binding Hamiltonian by the factors $x_\mathbf{r}$, $y_\mathbf{r}$ and $z_\mathbf{r}$.}
\label{fig:appendix_nanotube_boundaries}
\end{figure}

In order to numerically simulate the zig-zag carbon nanotube, we must modify the graphene Hamiltonian of Eq.~(\ref{eq:C1_graphene_ham}) slightly to take into account the open boundaries of the system. We take the Hamiltonian
\begin{equation}
H  =  -t \sum_{\mathbf{r} \in \Lambda } a^\dagger_\mathbf{r} \left( x_\mathbf{r} b_{\mathbf{r} - \mathbf{n}_1 + \mathbf{n}_2} + y_\mathbf{r} b_{\mathbf{r} - \mathbf{n}_1} +  z_\mathbf{r} b_\mathbf{r}  \right) + \text{H.c.} , \label{eq:C1_ham_tb_nanotube}
\end{equation}
where $x_{\mathbf{r}}, y_{\mathbf{r}}, z_{\mathbf{r}} \in \{0,1\}$ are numerical factors that take into account the top and bottom boundaries of the system. These terms ``switch off" the external $x$-, $y$- and $z$-links  of the Hamiltonian respectively, as seen in Fig.~\ref{fig:appendix_nanotube_boundaries}, to ensure the nanotube has zig-zag boundaries represented by the red links.

In order to fix which band we are in numerically, we derive the corresponding tight-binding model. We Fourier transform with respect to the $\mathbf{n}_2$ direction only with
\begin{equation}
a_\mathbf{r} =  \frac{1}{\sqrt{N}}  \sum_{p_2} e^{i \frac{a}{2 \pi} p_2 \mathbf{G}^2 \cdot \mathbf{r}} a(x_1,p_2) \equiv \frac{1}{\sqrt{N}} \sum_{p_2} e^{i a p_2 x_2} a(x_1,p_2),
\end{equation}
which defines a set of modes $a(x_1,p_2)$ which have a mixed position-momentum dependence. Substituting this into the tight-binding Hamiltonian Eq.~(\ref{eq:C1_ham_tb_nanotube}) gives us
\begin{equation}
H = \sum_{p_2} H(p_2), \quad H(p_2) = - t \sum_{i= 0}^L a^\dagger_{i} \left[  z_i b_{i} + \left( y_i + x_i e^{ip_2} \right) b_{i-1} \right] + \text{H.c.},
\end{equation}
where $H(p_2)$ are a set of one-dimensional tight-binding Hamiltonians, where $a_i \equiv a(x_i,p_2)$ and similarly for $b_i$, where the index $i \in \mathbb{N}$ labels the lattice sites of our one-dimensional chain.

In order to describe a nanotube, we now impose the constraint $p_2 = 2n \pi/N$ to give us $N$ Hamiltonians $H_n \equiv H(2 n \pi/N)$ which describe each band $n$ of the nanotube. In order to encode this numerically, we need the single-particle Hamiltonian. We can write this Hamiltonian as
\begin{equation}
H_n = \sum_{i,j} a^\dagger_i (h_n)_{ij} b_j + \mathrm{h.c.}, \quad (h_n)_{ij} = - t \left[ z_i \delta_{ij} + \left(1 + e^{i \frac{2n \pi}{N}} \right) \delta_{i-1,j} \right].
\end{equation}
Note that $x_i$ and $y_i$ are no longer needed here, as the factor of $\delta_{i-1,j}$ removes the $x$- and $y$-links on the top of the cylinder as required.

If we define the $2(L+1)$-dimensional spinor $\psi = (a_0,a_1,\ldots a_L ;b_0,b_1,\ldots b_L)^\mathrm{T} \equiv (\mathbf{a},\mathbf{b})^\mathrm{T}$, then the many-body Hamiltonian can be written as
\begin{equation}
H_n = \begin{pmatrix} \mathbf{a}^\dagger & \mathbf{b}^\dagger \end{pmatrix} \begin{pmatrix} 0 & h_n \\ h^\dagger_n & 0 \end{pmatrix} \begin{pmatrix} \mathbf{a} \\ \mathbf{b} \end{pmatrix} \equiv \psi^\dagger \mathcal{H}_n \psi.
\end{equation}
Numerically, we diagonalise the matrix $2(L+1) \times 2(L+1)$-dimensional matrix $\mathcal{H}_n$ and find its eigenvectors and corresponding eigenvalues. In our basis, the first (last) $L+1$ components will be the wavefunctions $\psi_A$ ($\psi_B$).
\end{appendix}

\singlespacing

\addcontentsline{toc}{chapter}{Bibliography}
\printbibliography

\end{document}